\documentclass[11pt,a4paper]{article}
\usepackage{lmodern}
\usepackage{amsmath,amssymb,bbm}
\usepackage[top=3.11cm,bottom=4.11cm,right=3.11cm,left=3.11cm]{geometry}
\usepackage{physics}
\usepackage{slashed}
\usepackage{xcolor, graphicx}
\usepackage{subfigure}
\usepackage{etoolbox}
\usepackage{bbold}
\usepackage{xspace}

\usepackage[utf8]{inputenc}  \usepackage[T1]{fontenc}
\usepackage{contractions}
\numberwithin{equation}{section}
\usepackage{hyperref} \usepackage[capitalise]{cleveref}

\usepackage{tabularx}
\newcolumntype{C}{>{\centering\arraybackslash}X}
\newcolumntype{R}{>{\raggedleft\arraybackslash}X}
\newcolumntype{L}{>{\raggedright\arraybackslash}X}

\usepackage{multirow, bigdelim}

\allowdisplaybreaks
\newcommand{\Ks}{K}

\def\ghat{{\hat{g}}}

\def\gbar{{\bar{g}}}
\def\HV{{\scshape hv}\xspace}
\def\FDH{{\scshape fdh}\xspace}
\def\DRED{{\scshape dred}\xspace}
\def\CDR{{\scshape cdr}\xspace}

\newcommand{\psibar}{{\overline\psi}}

\newcommand{\cbar}{\bar{c}}
\newcommand{\alphadot}{{\dot\alpha}}

\newcommand{\La}{{\cal L}}
\newcommand{\intx}{\int d^4x}

\def\dg#1{\frac{\delta\Gamma}{\delta#1}}

\title{{\bf Introduction to Renormalization Theory and\\ Chiral Gauge Theories in Dimensional Regularization with Non-Anticommuting $\gamma_5$}}
\newcommand{\email}{}
\author{Hermès Bélusca-Maïto,$^a$\thanks{\email{hbelusca@phy.hr}}\quad
Amon Ilakovac,$^a$\thanks{\email{ailakov@phy.hr}}\\
  Paul Kühler,$^b$\thanks{\email{paul.kuehler@tu-dresden.de}}\quad
Marija Mađor-Božinović,$^a$\thanks{\email{mmadjor@phy.hr}}\\
Dominik  Stöckinger,$^b$\thanks{\email{Dominik.Stoeckinger@tu-dresden.de}}\quad
Matthias
Weißwange,$^b$\thanks{\email{matthias.weisswange@tu-dresden.de}}\\[2em]
$^a$
Department of Physics, University of Zagreb,\\ Bijeni\v{c}ka cesta 32, HR-10000 Zagreb, Croatia\\
$^b$Institut für Kern- und Teilchenphysik, TU Dresden,\\ Zellescher Weg 19, DE-01069 Dresden, Germany}

\begin{document}
\thispagestyle{empty}
\maketitle

\setcounter{footnote}{0}
\vspace{2ex}
\begin{abstract}
This review provides a detailed introduction to chiral gauge
theories, renormalization theory, and the application of dimensional
regularization with the non-anticommuting BMHV scheme for
$\gamma_5$.

One goal is to show how chiral gauge theories can be
renormalized despite the spurious breaking of gauge invariance and how
to obtain the required symmetry-restoring counterterms.  A second goal is
to familiarize the reader with the theoretical basis of
the renormalization of chiral gauge theories, the theorems that
guarantee the existence of renormalized chiral gauge
theories at all orders as consistent quantum theories. Relevant topics include
BPHZ    renormalization, Slavnov-Taylor identities, the BRST formalism
and algebraic renormalization, as well as the theorems
  guaranteeing that dimensional regularization is a consistent
  regularization/renormalization scheme. All of these, including their
  proofs and interconnections, are explained and discussed in
  detail.

  Further, these theoretical concepts are illustrated in practical applications
with the example of an Abelian and a non-Abelian chiral gauge theory.
Not only the renormalization procedure for such chiral gauge theories
is explained step by step, but also the results of all counterterms,
including the symmetry-restoring ones, necessary for the consistent
renormalization are explicitly provided.
\end{abstract}
\vspace{0.5cm}
\centerline

\newpage
\setcounter{page}{1}

\tableofcontents\newpage

\section{Introduction}
\label{sec:introduction}

Except for gravity, all known fundamental particles and interactions in nature are described by
quantum gauge theories. The Standard Model (SM) of particle physics
combines the theories for  electromagnetic, weak and strong
interactions. It is based on the gauge group
SU(3)$\times$SU(2)$\times$U(1) and includes fermionic fields
describing spin 1/2 quarks and leptons and bosonic fields describing
the Higgs boson and electroweak symmetry breaking.

Exact solutions for quantum gauge theories rarely exist. Often, SM
predictions can be successfully evaluated in a perturbative treatment. Based on known exact solutions of the free
non-interacting quantum field theory, higher-order corrections can be
evaluated step by step. The higher-order corrections lead to Feynman
diagrams with closed loops and momentum integrations which lead to
ultraviolet divergences. Therefore, the higher-order amplitudes have
to be regularized and renormalized. Equivalently, a mathematically
rigorous treatment has to inductively construct higher orders from
lower orders, where the construction has to respect fundamental
requirements such as causality, Lorentz invariance and unitarity of
the time evolution. Practical regularization/renormalization
prescriptions that agree with such a rigorous approach are called
consistent schemes.

For so-called vector gauge theories, in which left-handed and
right-handed fermions have the same gauge interactions, an essentially
perfect regularization/renormalization framework is provided by
dimensional regularization \cite{tHooft:1972tcz,Bollini:1972ui,Ashmore:1972uj}. It is not only consistent in the sense
above, but it also manifestly preserves the fundamental gauge
invariance at all steps of calculations. Further, a useful practical
tool is provided by the validity of the quantum action
principle \cite{Breitenlohner:1977hr}, which enables the
straightforward study of symmetries and 
equations of motion on the level of Green functions. Alternative
consistent schemes such as analytic 
renormalization or Pauli-Villars regularization break gauge
invariance. For the status of further modern developments of
alternative schemes we refer to Ref.\ \cite{Gnendiger:2017pys}.

However, a fundamental discovery of elementary particle physics is that electroweak interactions act on chiral fermions, i.e.,\ they treat left-handed and
right-handed fermions differently. Accordingly, the SM and all its
extensions for potential new physics are 
chiral gauge theories, in which left-handed and right-handed fermions
interact differently with gauge bosons.
The presence of such chiral fermions and chiral interactions is
manifested through phenomena such as non-conservation of  parity and
charge conjugation invariance of the weak interactions.
    Connected with chiral fermions is the possibility of chiral anomalies
\cite{Adler:1969gk,Bell:1969ts,Adler:1969er},
i.e.\ the possibility that classically conserved currents are not
conserved in the full quantum theory. 
Chiral anomalies lead to observed phenomena such as neutral pion decay
into two photons. Chiral gauge theories, however, can only be
consistently renormalized if chiral anomalies in currents coupling to gauge
fields cancel. Although the cancellation is valid in the SM
\cite{Bouchiat:1972iq,Gross:1972pv,Geng:1989tcu}, the potential
  presence of chiral anomalies makes it impossible to
  define a consistent regularization/renormalization procedure that manifestly preserves
  all symmetries involving chiral fermions. A particularly transparent
  analysis can be given in terms of the non-invariance of the fermion
  path integral measure \cite{Fujikawa:1979ay,Fujikawa:1980eg}.

Within chiral models, dimensional regularization schemes meet the
so-called ``$\gamma_5$-problem'', which is a consequence of the fact
that $\gamma_5$  (similarly the Levi-Civita tensor
$\epsilon_{\mu\nu\rho\sigma}$) is an 
intrinsically $4$-dimensional quantity. The three basic properties,
anticommutativity of $\gamma_5$ with other $\gamma^\mu$
matrices, cyclicity of traces, and the non-zero
trace of products of $\gamma_5$ with four different $\gamma^\mu$ matrices, cannot be simultaneously
retained without spoiling the consistency of the scheme. The usage of
the naive scheme \cite{Chanowitz:1979zu}, including the
$\gamma_5$ anticommutativity,  
is the most common in practical calculations, but it is restricted to
subclasses of diagrams
\cite{Chanowitz:1979zu,Jegerlehner:2000dz}, and within it the
$\gamma_5$-matrix is ambiguously defined.
Giving up the cyclicity of the trace one has to introduce a
consistent reading prescription defining combinations of reading points
for evaluations of noncyclic traces
\cite{Kreimer:1989ke,Korner:1991sx,Kreimer:1993bh}, which makes the
mathematical consistency of higher orders less transparent
and questionable. Abandoning the anticommutativity of the
$\gamma_5$-matrix
\cite{tHooft:1972tcz,Cicuta:1972jf,Bollini:1972ui,Akyeampong:1973vk,Akyeampong:1973xi}
leads to the mathematically most rigorously established dimensional regularization
scheme, the so-called Breitenlohner-Maison/'t Hooft-Veltman (BMHV) scheme,   
for which all basic quantum field theory properties were proven to be
valid
\cite{Speer:1974cz,Breitenlohner:1977hr,Breitenlohner:1975hg,Breitenlohner:1976te}.

Unfortunately, in the BMHV scheme with non-anticommuting $\gamma_5$,
some of the advantages of dimensional regularization are lost. In
particular gauge invariance is not manifestly valid in chiral gauge
theories, reflecting the possibility of anomalies. Even if the actual
anomalies cancel, as in the SM, gauge invariance is broken in
intermediate steps, and the breaking has to be compensated by a more
complicated renormalization procedure. Instead of the typical textbook
approach of generating a bare Lagrangian and counterterms by a
renormalization transformation of fields and parameters, specific
symmetry-restoring counterterms of  a more general structure need to
be found and included. Several recent works have begun to
systematically investigate the
practical application of the BMHV scheme to chiral gauge theories and
determine such counterterms
\cite{Belusca-Maito:2020ala,Belusca-Maito:2021lnk,Cornella:2022hkc,Belusca-Maito:2022wem},
see also Ref.\ \cite{Belusca-Maito:2022usw} for a compact summary. 

The present review provides a detailed introduction into chiral gauge
theories, dimensional regularization, renormalization theory and the
application of the BMHV scheme to chiral gauge theories. Its
intentions and motivations can be summarized as follows:
\begin{itemize}
\item
  We aim for a pedagogical review, starting at the level of typical
  quantum field theory textbooks and containing detailed step-by-step
  explanations and illustrative examples.
\item
  On a practical level, we show how chiral gauge theories can be
  renormalized employing the BMHV scheme for $\gamma_5$ and how the
  required symmetry-restoring counterterms can be obtained and
  used. Thus we also provide an introduction to the recent
  literature mentioned above. The general motivation is an increasing need for high-precision 
  (multi-)loop calculations in the SM and beyond, and an increasing
  interest in mathematically rigorous treatments which avoid pitfalls
  such as inconsistencies, ambiguities or incorrect results.
\item
  On a conceptual level, we discuss the theoretical basis of
  the renormalization of chiral gauge theories. The
  existence of renormalized quantum gauge theories at all orders,
  together with their physics interpretation, is a major result in
  theoretical physics. It is based on a large set of complicated
  theorems and formalisms, ranging from BPHZ  theorems on causal and unitary
  renormalization to Slavnov-Taylor identities and the BRST formalism,
  the theorems of algebraic renormalization, and to the theorems
  guaranteeing that dimensional regularization is a consistent
  regularization/renormalization scheme. All these relevant theorems,
  their role and their interconnections are discussed and explained in
  detail. The proofs are either given or illustrated and explained.
\item
  In line with the pedagogical goals, we use extensive cross-referencing between sections. Wherever
  possible, introductory sections develop intuition and expectations
  of later steps, and later sections refer back to simpler, more qualitative
  explanations and illustrations.
  In our citations we cite not only original
works, but
  wherever possible we also cite textbooks or other reviews, where
  further details can be found. References to the remarks made at the
  beginning of this Introduction can be found in the appropriate sections.
\end{itemize}
In the following we present an extensive outline of the individual
sections.

\medskip
\noindent
In {\bf Section \ref{sec:setup}} the basic knowledge necessary for a discussion of chiral
gauge theories in dimensional regularization is presented:
\begin{itemize}
\item
 Beginning
with key ingredients, first, non-Abelian Yang--Mills gauge
theories and
spinors, chirality and chiral fermions are introduced, including
required notions from Lie group theory and Poincar\'e group
representations.
BRST
invariance and a corresponding Slavnov-Taylor identity are discussed
in detail already at the
classical level. Turning to the quantum level, the notions needed for discussions of  
Green functions and their generating functionals are
introduced. Then  Slavnov-Taylor 
identities for Green functions and generating functionals  are
introduced, derived from the path integral and interpreted in
detail. The concluding subsection considers the case of an Abelian
gauge theory, and simplifications and additionally
valid equations compared to the non-Abelian case are shown.
\end{itemize}
{\bf Section \ref{sec:DReg}} gives a detailed introduction to dimensional
regularization as a mathematically well-defined regularization procedure which 
allows efficient computations and preserves  basic properties of quantum
field theory:
\begin{itemize}
\item
  As a preview and to set the stage,  the
general structure of dimensional regularization,
renormalization and the counterterms as well as corresponding notations are presented.
Then $D$-dimensional extensions of  $4$-dimensional quantities
are discussed, starting with the notion of the quasi $D$-dimensional
space. The core of the method are $D$-dimensional integrals. After
listing their properties relevant for practical calculations, they
are mathematically constructed in two ways, using parallel and
orthogonal spaces as well as via  
Schwinger parametrization. The definition and properties of the metric tensor and its
inverse are given. Of
particular importance for chiral gauge theories are the definitions
and properties of $D$-dimensional $\gamma$
matrices. Here an explicit construction of quasi-$D$-dimensional
$\gamma$ matrices is provided which is optimized for the study of
chiral gauge theories. The extension to $D$ dimensions leads
to the well-known  $\gamma_5$
problem; this problem is explained and the BMHV scheme is presented
together with its definitions and properties of
the $\gamma_5$ matrix and the $\epsilon_{\mu\nu\rho\sigma}$ symbol.
\item
In addition to defining the regularization and constructing its basic
elements, the relationship of regularized Feynman rules to
Lagrangians in  $D$ dimensions via a $D$-dimensional Gell-Mann-Low formula
is discussed.
Special emphasis is put on the relation between kinetic terms and
corresponding propagators and chiral fermion-gauge boson
interactions. As an outlook and somewhat orthogonal topic, the
variants HV, CDR, DRED and FDH of 
dimensional regularization 
schemes are briefly discussed. Their distinctions are of particular
importance in the context of 
infrared divergences and in the context of supersymmetric gauge theories.
\end{itemize}
In {\bf Section \ref{sec:QAPDReg}} the quantum action principle and
regularized quantum action principle in dimensional regularization are
introduced. This is a set of relations
between variations of the classical action and variations of the
Green functions of the resulting quantum theory, which allow to express
symmetries and symmetry violations of the
regularized or renormalized theory:
\begin{itemize}
\item
  First an instructive but formal
derivation from the path integral is given, sidestepping the need for
regularization and renormalization.
Then an exact proof of the
regularized quantum action principle  within dimensional
regularization is presented. This validity constitutes an important
advantage of dimensional regularization. Its role is illustrated by
proving rigorously the all-order validity of the
Slavnov-Taylor identity for QCD, and explaining the extent of the validity
of supersymmetry in the DRED scheme.
\end{itemize}
{\bf Section \ref{sec:renormalization}} is devoted to general
renormalization theory, focusing on aspects not yet specific to gauge
theories. One goal is to explain the rigorous theorems
guaranteeing that regularization, renormalization and cancellation of
divergences is possible and physically sensible quantum field theories
can be constructed at all orders. A second goal is to analyze
conditions for consistent regularization/renormalization procedures
and to
show how we know that dimensional regularization is one such
consistent procedure:
\begin{itemize}
  \item
Renormalization is introduced as a mathematical construction of
time-ordered products of free field operators in agreement with
unitarity and causality of the perturbative S-matrix. The ``main
theorem'' of renormalization relates the construction and its
ambiguities to
reparametrizations. Importantly, the ambiguities and the
reparametrizations are local in a well-defined sense. The
relationships between the BPH approach and the $R$-operation, the BPHZ
approach and the forest formula and the usual counterterm approach are
explained. Further,  analytic regularization is discussed as a conceptually
interesting non-dimensional regularization scheme which can facilitate
all-order proofs.
\item
In the second subsection, the main theorem on dimensional regularization is
reviewed. First an extensive discussion of the main statements is
given --- most important is the applicability of dimensional
regularization as a consistent regularization/renormalization
framework. Then the proof is sketched in detail.
The first steps set up Feynman graph theoretical
notions, an organization of the loop integrations and an optimized
forest formula. Then the resulting integrals are investigated in
detail, and an inductive proof can be given. All steps are explained
and illustrated with examples.
\end{itemize}
With the fundamentals of regularization and renormalization thus established,
{\bf section  \ref{sec:algren}} goes on to consider the case of quantized
gauge theories and their renormalization. It focuses on the 
compatibility of BRST invariance and 
Slavnov-Taylor identities, which are vital for the correct physical
interpretation of gauge theories, and the
regularization/renormalization procedure, which may in general spoil 
symmetries:
\begin{itemize}
  \item
Revisiting first the familiar textbook case of a
symmetry-preserving 
regularization such as in QED or QCD reminds the reader of practically
important concepts such as renormalization transformations and puts
into context the symmetry-breaking case which is
the central topic of this review.
\item
Focusing on this case of interest, the
theory of algebraic renormalization is reviewed as the framework in
which rigorous and elegant proofs of the renormalizability of gauge theories can
be carried out, even if regularization procedures break symmetries. The quantum action
principle of BPHZ renormalization emerges as the main 
theoretical tool of this framework; hence a brief
exposition of this tool is given, and  its connection to the quantum
action principle in dimensional regularization is explained. The section
then illustrates the inductive all-order proof of the restoration of
the spuriously broken symmetry by symmetry-preserving finite
counterterms. It also includes a brief discussion of anomalies, their
cancellation conditions and an outlook on further applications of
algebraic renormalization.
\item
  Finally,  coming to the practical goal of
this review,
the formalism is specialized to dimensional regularization. Here explicit
equations for the computation of symmetry-preserving counterterms are
derived and the resulting structure of the counterterm Lagrangian is
discussed.
\end{itemize}
{\bf Section \ref{sec:ApplicationsChiralQED}} gives a detailed
illustration of the treatment of chiral gauge theories in the BMHV
scheme, using concrete examples:
\begin{itemize}
\item
It focuses mainly on an Abelian example, a chiral QED model, discusses its structure,
symmetry breaking as the result of the scheme, and the required
counterterm structure. It explains and compares several
ways to determine the required symmetry-restoring
counterterms in practical calculations.
\item The symmetry 
restoration is illustrated in detail for the photon self-energy
case, where it becomes apparent how the quantum 
action principle and Ward identities have a crucial practical role in
the calculations.  
\item
For the chiral QED model, the calculations are generalized to the full one-loop and the full
  two-loop level, and the new features arising at the two-loop level
  are discussed.
  \item
  Finally,  a detailed comparison of the Abelian chiral QED and
a chiral non-Abelian Yang-Mills theory is given at the one-loop level. 
\end{itemize}

\newpage
\section{Setup}
\label{sec:setup}
  In this section we collect background information on the main
  theoretical concepts needed to discuss the renormalization of chiral
  gauge theories in dimensional regularization. We begin with the
  general notions of Yang-Mills gauge theories and of spinors,
  $\gamma^\mu$-matrices and chirality. On the level of classical field
  theory, gauge invariance is then extended to BRST invariance,
  including gauge fixing and Faddeev-Popov ghosts, and BRST invariance
  is formulated as a Slavnov-Taylor identity
  (Secs.\ \ref{sec:YMGaugeTheoriesSetup}, \ref{sec:chiralfermions}, \ref{sec:BRSTSTI}). In 
  Sec.\ \ref{sec:GreenFunctions} the basic objects of quantized field
  theories, Green functions and generating functionals, are
  defined. Sec.\ \ref{sec:STIformal} discusses the role and
  interpretation of the Slavnov-Taylor
  identity on the level of the quantum field theory. Finally,
  Sec.\ \ref{sec:AbelianPeculiarities} discusses the case of abelian
  gauge theories, which involves additional identities.
Much of the material of this section can also be found in standard 
textbooks such as
Refs.\ \cite{Cheng:1984vwu,Weinberg:1995mt,Weinberg:1996kr,Peskin:1995ev,Bohm:2001yx,Srednicki:2007qs,Schwartz:2018}.

\subsection{Yang--Mills Gauge~Theories}

\label{sec:YMGaugeTheoriesSetup}
We begin by summarizing the construction of general Yang-Mills gauge
theories with simple gauge group 
such as SU($N$) or SO($N$) and with generic matter fields. 

The first ingredient is the gauge group. It is a Lie group in which
all group elements can be written as continuous functions of a
certain number $N_{\text{gen}}$ of parameters. The Lie group can be associated with a Lie
algebra with $N_{\text{gen}}$ generators, called $t^a$, $a=1\ldots
N_{\text{gen}}$. The generators satisfy the commutation relations
\begin{align}
[t^a,t^b] & =  if^{ab}_ct^c
\label{LieAlgebra}
\end{align}
with antisymmetric structure constants
$f^{ab}_c$. There exists a set of generators for which the structure
constants are totally antisymmetric, such that we write
$f^{ab}_{c}\equiv f_{abc}\equiv f^{abc}$. This is the case for sums of simple
compact and $U(1)$ subalgebras, see e.g.\ \cite{Weinberg:1996kr}.
Any set of matrices $T^a$ which satisfy the relation
(\ref{LieAlgebra}) is called a representation of the Lie algebra.

One special representation, the so-called adjoint representation,
always exists. It is defined by
\begin{align}
\label{eq:AdjointReprOfGenerators}
  (T^a_{\text{adj}})_{ij} & = -i f^{aij}
\end{align}
and thus a representation in terms of $N_{\text{gen}}\times
N_{\text{gen}}$ matrices. The commutation relation (\ref{LieAlgebra})
is fulfilled because of the Jacobi identity of
commutators.

For any representation of the Lie algebra, we can form a representation
of the Lie group (at least locally in a region around the identity) by
exponentiation, 
\begin{align}
  U(\theta^a) & = e^{-ig\theta^a T^a}\,,
\end{align}
where the $\theta^a$ are real parameters and where $g$ is the gauge coupling.

Once the Lie group and Lie algebra are defined, we assume the existence
of $N_F$ so-called matter fields $\varphi_i(x)$, $i=1\ldots N_F$. We
collectively denote them as a tuple $\varphi=(\varphi_i)$. We
further assume that there exists a representation of the Lie algebra
in terms of $N_F\times N_F$ matrices $T^a$, and we define {\em (local)
  gauge
  transformations} of the matter fields as
\begin{align}
\varphi_i(x) & \to U(\theta^a(x))_{ij}\varphi_j(x)
\,.
\end{align}
The representation may be reducible or irreducible.
To simplify the notation we will often suppress the indices and arguments
and write the previous equation as
\begin{align}
\varphi & \to U\varphi
\,.
\end{align}

Next, we introduce the central elements of Yang-Mills gauge theories:
the covariant derivative $D_\mu$ and the gauge fields
$A^{a}_\mu$. They are related as
\begin{align}
D_\mu & =  \partial_\mu+ig T^aA^a_\mu
\label{Dmu}
\end{align}
where $g$ is the gauge coupling. As the notation indicates, there is
one vector field $A^a_\mu$ for each generator $a=1\ldots
N_{\text{gen}}$. The relation (\ref{Dmu}) is valid for any
representation, and the vector fields $A^a_\mu$ are independent of the
chosen representation. It is often useful to define the matrix-valued and
representation-dependent gauge field $A_\mu\equiv T^a A^a_\mu$.

The fundamental requirement is that under a gauge transformation the
covariant derivative behaves as
\begin{align}
D_\mu\varphi & \to  UD_\mu\varphi\,.
\end{align}
This is valid if and only if the matrix-valued gauge field transforms as
\begin{align}
A_\mu & \to  UA_\mu U^{-1}-\frac{1}{ig}[\partial_\mu
  U]U^{-1}
\,.
\end{align}

Finally, the field strength tensor can be defined as
\begin{align}
F_{\mu\nu} & = \frac{1}{ig}[D_\mu,D_\nu]
\,.
\label{FmunuDef}
\end{align}
With this definition the field strength tensor is matrix-valued and
dependent on the chosen representation. We can decompose it as
$F_{\mu\nu}=T^aF^a_{\mu\nu}$ and evaluate the previous definition 
with the result
\begin{align}
F^a_{\mu\nu} & =  \partial_\mu A^a_\nu-\partial_\nu A^a_\mu
- g f^{abc} A^b_\mu A^c_\nu
\,.
\end{align}
Here we see that the field strength tensors $F^a_{\mu\nu}$ are
independent of the chosen representation and are generalizations of
the field strength tensor of electrodynamics.

At this point we collect all gauge transformations in
compact and matrix-valued form as
\begin{subequations}
  \begin{align}
\varphi & \to U\varphi\,,
\\  
D_\mu\varphi & \to  UD_\mu\varphi\,,\\
A_\mu & \to  UA_\mu U^{-1}-\frac{1}{ig}[\partial_\mu
  U]U^{-1}\,,
\\
F_{\mu\nu} & \to
U F_{\mu\nu} U^{-1}\,,
\end{align}
\end{subequations}
where the last equation directly follows from the definition
(\ref{FmunuDef}). We also record the gauge transformations for the
fundamental fields in more explicit form, by taking the
parameters $\theta^a$ to be infinitesimal, as
\begin{subequations}
\label{GaugeTransformationInfinitesimal}
  \begin{align}
\varphi & \to  \varphi - ig\theta^a T^a\varphi\,,\\
A_\mu&\to
A_\mu+\partial_\mu \theta                -ig[\theta,A_\mu] \,,\\
A^a_\mu & \to
A^a_\mu+\partial_\mu\theta^a+gf^{abc}\theta^b A^c_\mu\,,
  \end{align}
\end{subequations}
where we also set
$\theta=T^a\theta^a$. The last of the previous equations is
particularly important. It holds universally for any representation.
It also contains the gauge coupling $g$. This is at the heart of the
universality of the gauge coupling, i.e.\ the physical statement that
one single gauge coupling governs all interactions of the gauge bosons
with other gauge bosons and with any matter fields. Note that this
statement relies on the assumption of a simple non-Abelian gauge
group.

The renormalizable gauge invariant Lagrangian for this Yang-Mills
theory can be written as
\begin{subequations}
  \label{YMLagrangian}
  \begin{align}
  \La_{\text{inv}} & = \La_{\text{YM}} + \La_{\text{mat}}\,,\\
  \La_{\text{YM}} & = - \frac14 F^a{}^{\mu\nu} F^a_{\mu\nu}\,, \\
  \La_{\text{mat}} & =   \La_{\text{mat}}(\varphi,D_\mu\varphi)
  \,.
  \end{align}
\end{subequations}
The concrete form of the matter field Lagrangian depends on details
such as the spin of the matter field and interactions between
different matter fields.

\subsection{Chiral Fermions}
\label{sec:chiralfermions}

In this subsection we introduce the next ingredient --- chiral
fermions.
A fundamental discovery of elementary particle physics is that
electroweak interactions fundamentally act on chiral fermions,
i.e.\ they treat left-handed and right-handed fermions
differently. Chiral fermions are also fundamental building blocks in
many extensions of the Standard Model, such as grand unified theories
or supersymmetry.

Here we will first summarize general properties of 4-component, or
Dirac or Majorana
spinors in 4 dimensions and then define the notion of chirality in
this context. Thereafter we also introduce the  2-component Weyl/van
der Waerden 
spinor notation, which allows an efficient understanding of many
important relationships. We will then collect such relationships.

\subsubsection{General Representation-Independent Relations for
 $\gamma$ Matrices and~4-Spinors}

Spinors are defined via their properties under Lorentz
transformations. Therefore, we begin with the reminder that a
Lorentz transformation of ordinary 4-vectors is defined by a matrix
$\Lambda^\mu{}_\nu$ that leaves scalar products of 4-vectors
invariant. Infinitesimal Lorentz transformations are given 
by matrices of the form
$\Lambda^\mu{}_\nu=\delta^\mu{}_\nu+\omega^\mu{}_\nu$ with an
infinitesimal, antisymmetric matrix $\omega_{\mu\nu}$. A representation
of the Lorentz group $U(\Lambda)$ is (at least locally) defined by specifying
\begin{equation}\label{DefJ}
U(\delta+\omega) = 1 -\frac{i}{2}\omega_{\mu\nu}J^{\mu\nu}
\end{equation}
with generators $J^{\mu\nu}$ which must satisfy the commutation relations of
the corresponding Lie algebra,
\begin{align}
 \label{Lorentzalgebra}
\left[J^{\mu\nu}, J^{\rho\sigma} \right] & = 
i(g^{\nu\rho}J^{\mu\sigma} - g^{\mu\rho}J^{\nu\sigma} +
g^{\mu\sigma}J^{\nu\rho} - g^{\nu\sigma}J^{\mu\rho})\,.
\end{align}

Now we can turn to spinors.
The basic building blocks of 4-component  spinor theory are the
$\gamma^\mu$-matrices. They are $4\times4$ matrices satisfying the
defining Clifford algebra relation
\begin{align}
  \label{Clifford}
\{\gamma^\mu, \gamma^\nu\} & = 2g^{\mu\nu}\mathbbm{1} \,.
\end{align}
Here and everywhere else we use the mostly-minus metric.
The fundamental importance of these matrices is that they generate a
representation of the Lorentz group. Indeed, setting
\begin{align}
\label{DefSmunu}
S^{\mu\nu} & = 
\frac{i}{4} \left[\gamma^\mu, \gamma^\nu\right] \,,
\end{align}
one can show that these $S^{\mu\nu}$ satisfy the required commutation
relations (\ref{Lorentzalgebra}). Hence we can now define the notion
of a 4-component (Dirac or Majorana) spinor: a 4-component spinor $\psi$ is an object
whose Lorentz transformation properties are given by
\begin{align}
  \psi \stackrel{\Lambda=\delta+\omega}{\longrightarrow}
\left(1 -\frac{i}{2}\omega_{\mu\nu}S^{\mu\nu}\right)\psi \,. 
\label{SpinorTransform}
\end{align}

In addition to the $\gamma^\mu$-matrices, the $\gamma_5$ matrix and
projection operators $P_{L,R}$ 
are defined as
\begin{align}\label{gamma5epsilonDef}
  \gamma_5 & = i\gamma^0\gamma^1\gamma^2\gamma^3 =
  -\frac{i}{4!}{\epsilon}_{\mu\nu\rho\sigma}
  {\gamma}^\mu
  {\gamma}^\nu
  {\gamma}^\rho
  {\gamma}^\sigma \ ,&
  P_{L,R} & = \frac12(\mathbbm{1}\mp\gamma_5) \,,
\end{align}
with the totally antisymmetric Levi-Civita (pseudo-)tensor
$\epsilon_{\mu\nu\rho\sigma}$ with $\epsilon_{0123}=-1$.
These matrices satisfy the additional equations
\begin{align}
  \{\gamma^\mu,\gamma_5\} & = 0 \,,
  &
  (\gamma_5)^2 & = \mathbbm{1} \,,
  &
  (P_{L,R})^2 & = P_{L,R} \,,
  &
  P_LP_R & = 0 \,.
\end{align}
Though not required in general, in many representations (including the
chiral representation introduced below) the relations
\begin{align}
  (\gamma^\mu)^\dagger & = \gamma^0\gamma^\mu\gamma^0 \,,
  &
  (\gamma^\mu)^* & = \gamma^2\gamma^\mu \gamma^2 \,,
  &
  (\gamma^\mu)^T & = -C^{-1}\gamma^\mu C \,,
  &
  C & = i\gamma^0\gamma^2
\label{gammaproperties}
\end{align}
hold. In particular $\gamma^2$ is the only imaginary matrix. We will assume these relations in the following.

For any 4-spinor $\psi$ we can define an adjoint spinor $\bar\psi$ and
a charge-conjugated spinor $\psi^C$ by
\begin{align}
  \bar\psi & = \psi^\dagger\gamma^0 \,, &
  \psi^C & = C\bar\psi^T \,.
  \label{psibarCproperties}
\end{align}
In this way, $\psi^C$ is also a 4-spinor satisfying the transformation
rule (\ref{SpinorTransform}), and $\bar\psi$ transforms with the
inverse matrix. One can show that  bilinear expressions
such as $\bar\psi_1\psi_2$, 
$\bar\psi_1\gamma^\mu\psi_2$ transform as Lorentz scalars and
4-vectors, respectively.

\subsubsection{Chirality and Chiral~Fermions}

At the level of 4-component spinors, the concept of chirality is
related to the $\gamma_5$ matrix and the projectors $P_{L,R}$. Let us
define for any 4-spinor $\psi$  so-called left-handed and
right-handed spinors by
\begin{align}
  \psi_L & = P_L\psi \,, &
    \psi_R & = P_R\psi \,.
\end{align}
Then we can make three observations:
\begin{itemize}
\item
  The matrix $\gamma_5^2=\mathbbm{1}$. Hence the eigenvalues of
  $\gamma_5$ are $\pm1$.
\item
  The spinors $\psi_L$ and $\psi_R$ are eigenspinors of $\gamma_5$
  with eigenvalues $-1$, $+1$, respectively.
\item
  The matrix $\gamma_5$ and the projectors $P_{L,R}$ commute with the Lorentz generators
  $S^{\mu\nu}$. 
\end{itemize}
Hence the left-handed and right-handed spinors are
  proper spinors in the sense of Eq.\ (\ref{SpinorTransform}), and
  they form two distinct invariant subspaces of the Lorentz
  representation: the representation defined by
  Eqs.\ (\ref{DefSmunu},\ref{SpinorTransform}) is reducible.

We refer to the eigenvalue of $\gamma_5$ as chirality; the left-handed
and right-handed spinors are chiral, or chirality
eigenstates. In view of the above,  chirality is a Lorentz invariant
property and its existence is linked to the structure of the
Lorentz group representation theory. For the general analysis we refer
to Ref.\ \cite{Wigner:1939cj} and, in particular, the textbooks by
Weinberg, Srednicki and Ryder
\cite{Weinberg:1995mt,Srednicki:2007qs,Ryder:1985wq}. The spaces of
the 
left-handed and right-handed spinors each define an irreducible
representation of the Lorentz group --- these are the simplest
nontrivial representations, which are commonly known as the
$(\frac12,0)$ and $(0,\frac12)$ representations.

Slightly reformulating the previous statements,
we may say that the left-handed and right-handed spinors have
different, independent Lorentz transformation properties. Hence in a
Lorentz invariant field theory, left-handed or right-handed spinor
fields may appear independently. Specifically, gauge theories may be
constructed in which left-handed or right-handed spinor fields appear
with different gauge group representations. This is precisely what
happens in case of the electroweak interactions.
Chiral fermions are the fermions described by such field theories
based on chiral spinor fields.

\subsubsection{Chiral Representation and 2-Component Spinor~Formalism}
\label{sec:2spinors}
Although many important relationships hold  independently of any specific
representation of the $\gamma^\mu$-matrices,  it is useful to
introduce here
the so-called chiral 
representation, which is given as follows by $2\times2$ block
matrices,
\begin{align}\label{gammachiralrep}
\gamma^\mu & =  \left(\begin{array}{cc} 0 & \sigma^\mu \\ \overline{\sigma}^\mu &
0 \end{array}\right),&
\gamma^5   & =  \left(\begin{array}{cc} -1 & 0 \\ 0 &
1\end{array}\right),&
P_L  =  &  \left(\begin{array}{cc}1&0\\0&0\end{array}\right),&
P_R  =  &  \left(\begin{array}{cc}0&0\\0&1\end{array}\right).
\end{align}
This representation uses the Pauli matrices
\begin{equation}\label{Paulimatrizen}
\sigma^1  = \left(\begin{array}{cc}0&1\\1&0\end{array}\right),\quad
\sigma^2  = \left(\begin{array}{cc}0&-i\\i&0\end{array}\right),\quad
\sigma^3  = \left(\begin{array}{cc}1&0\\0&-1\end{array}\right) 
\end{equation}
and the following 4-vectors of $2\times2$ matrices
\begin{align}
  \sigma^\mu  & = (1, \sigma^k)\,, &
\overline\sigma^{\mu}& = (1, -\sigma^k)\,.
\end{align}

In this representation of $\gamma$-matrices, the Lorentz generators
(\ref{DefSmunu}) 
take the form
\begin{align}
  S^{\mu\nu} & = 
 \left(\begin{array}{cc}\frac{i}{4}(\sigma^\mu\overline\sigma^\nu-\sigma^\nu\overline\sigma^\mu)
 & 0 \\
                   0 & \frac{i}{4}(\overline\sigma^\mu\sigma^\nu-\overline\sigma^\nu\sigma^\mu)\end{array}\right)
 \,.
 \label{eq:ExplicitFormOfSmunu}
 \end{align}
The block structure of all these matrices makes manifest that the
Lorentz representation is reducible and that the left-handed and
right-handed spinor spaces are invariant under Lorentz
transformations.

This block structure of the chiral representation suggests introducing individual
2-component spinors for the left-handed and right-handed parts. In the
following we will briefly introduce the corresponding 2-component spinor
formalism, which allows a very transparent formulation for many
important and useful equations. We mention that a systematic theory of
the Lorentz group representations automatically leads first to such
2-component spinors as the natural spinors for the $(\frac12,0)$ and
$(0,\frac12)$ representations and that in such a context the
4-component spinors appear as secondary objects. We also refer
to the review \cite{Dreiner:2008tw} for an excellent account of
2-spinors and relationships between formalisms and relationships
between different conventions.\footnote{  In our presentation we have chosen to start from the
  4-component spinors despite the fundamental nature of 2-component
  spinors.
  Our most important reason is that we aim
  to consider DReg, where there is the $\gamma_5$-problem which
  precisely means that the treatment of chirality and specifically
  2-component spinors is problematic, while the treatment of ordinary
  $\gamma^\mu$-matrices remains possible.}

To avoid confusion, in the remainder of the present subsection we will always denote
4-component spinors with capital Greek letters such as $\Psi$  and
2-component spinors with lower-case Greek letters such as $\chi$, $\eta$.
The relationship between a 4-component spinor $\Psi$ and 2-component
spinors is given by the decomposition
\begin{align}
  \Psi = {\chi_\alpha \choose \bar\eta^{\dot\alpha}}\,.
\end{align}
Here the indices $\alpha=1,2$ and $\dot\alpha=1,2$, and $\chi_\alpha$
and $\bar\eta^{\dot\alpha}$ are two distinct 2-component spinors.
For the 2-component spinors we define Hermitian conjugation as
\begin{align}
  \bar\chi^{\dot\alpha} & = (\chi^\alpha)^\dagger \,,
  &
\overline\chi_{\dot\alpha} & =
(\chi_\alpha)^\dagger \,,
\end{align}
and raising and lowering of indices as
\begin{align}
\chi_\alpha  & =
\epsilon_{\alpha\beta}\chi^\beta \,,
&
\chi^\alpha  & =
\epsilon^{\alpha\beta}\chi_\beta \,,
&
\overline\chi_{\dot\alpha} & =
\epsilon_{\dot\alpha\dot\beta}\overline\chi^{\dot\beta}
\,,
&
\overline\chi^{\dot\alpha} & =
\epsilon^{\dot\alpha\dot\beta}\overline\chi_{\dot\beta}
\,,
\end{align}
with the antisymmetric symbol
\begin{align}
  \label{DefEpsilon}
\epsilon_{\alpha\beta}  =  - \epsilon_{\beta\alpha},\quad
\epsilon_{\dot\alpha\dot\beta}  = 
- \epsilon_{\dot\beta\dot\alpha}
,\quad
 \epsilon^{\alpha\beta} = \epsilon_{\beta\alpha}
,\quad\epsilon^{\dot\alpha\dot\beta} = \epsilon_{\dot\beta\dot\alpha}
,\quad
\epsilon^{12} =1
,\quad 
\epsilon^{\dot1\dot2} = 1.
\end{align}
 
The Lorentz transformations of the original 4-spinors induce how the
2-spinors transform. For an infinitesimal Lorentz transformation
matrix $\Lambda=\delta+\omega$ we can define the $2\times2$ matrix
\begin{align}
  M(\delta+\omega)_\alpha\ ^\beta & \equiv 
1-\frac{i}{2}\omega_{\mu\nu}\left(\frac{i}{4}(\sigma^\mu\overline\sigma^\nu-\sigma^\nu\overline\sigma^\mu)
\right)_\alpha\ ^\beta
\end{align}
in accordance with the general Eq.\ (\ref{DefJ}). The explicit form of
$S^{\mu\nu}$ in Eq.\ (\ref{eq:ExplicitFormOfSmunu}) shows that this is the transformation
matrix for 2-spinors $\chi_\alpha$. The matrix $M$ is a general
complex invertible matrix with $\text{det}(M)=1$, i.e.\ $M$ is an
element of the group SL(2,$\mathbb{C}$). Elementary computations
involving raising and lowering of indices and
inspection of $S^{\mu\nu}$ show that in total, the four kinds of
2-spinors transform as follows:
\begin{subequations}
  \begin{align}
\chi_\alpha & \rightarrow  (M)_\alpha\ ^\beta\ \chi_\beta, \\
\bar\eta^{\dot\alpha} & \rightarrow 
(M^{-1\dagger})^{\dot\alpha}\ _{\dot\beta}\ \bar\eta^{\dot\beta}, 
\\
\chi^\alpha & \rightarrow  \chi^\beta\  (M^{-1})_\beta\ ^\alpha, \\
\bar\eta_{\dot\alpha} & \rightarrow 
\bar\eta_{\dot\beta}\ (M^{\dagger})^{\dot\beta}\ _{\dot\alpha}.
\end{align}
\end{subequations}
These relations highlight explicitly that the four types of spinors
have four different Lorentz transformation rules. The efficiency of the 2-component
spinor formalism is strongly related to this use of the index notation
to denote the different Lorentz representations. The four
representations are different but not all inequivalent:
The fact that
$\epsilon^{\alpha\beta}M_\beta\, ^\gamma\,\epsilon_{\gamma\delta} =
(M^{-1})_\delta\ ^\alpha$ shows that the spinors $\chi_\alpha$ and
$\chi^\alpha$ transform in equivalent (i.e.\ unitarily related)
transformations --- the $(\frac12,0)$ representation. Analogously, the
representations for $\bar\eta^{\alphadot}$ and $\bar\eta_\alphadot$
are both equivalent to the general $(0,\frac12)$ representation.

The Lorentz transformation properties also suggest the following
definitions for an index-free notation for spinor products:
\begin{subequations}
  \begin{align}
  \chi\eta & = \chi^\alpha\eta_\alpha \,,
  &
  \bar\chi\bar\eta & = \bar\chi_\alphadot\bar\eta^\alphadot \,,
  \\
  \chi\sigma^\mu\bar\eta & =
  \chi^\alpha\sigma^\mu_{\alpha\alphadot}\bar\eta^\alphadot \,,
  &
  \bar\chi\bar\sigma^\mu\eta & =
  \bar\chi_\alphadot\bar\sigma^\mu{}^{\alphadot\alpha}\eta_\alpha \,.
\end{align}
\end{subequations}
The expressions in the first line are clearly Lorentz-invariant scalar
quantities, and a calculation shows that the expressions in the second
line transform as Lorentz 4-vectors. The index-free notation and the
conventions to denote the matrix indices of the $\sigma^\mu$ and
$\bar\sigma^\mu$-matrices in this way reflect the Lorentz
transformation properties of all these objects.

As announced, we will now use the 2-component formalism to write
useful spinor relations in  a transparent way. We begin with the
spinors and their conjugates,

\begin{align}
\label{PsiPsiCDef}
  \Psi & = {\chi_\alpha \choose \bar\eta^{\dot\alpha}} \,,
  &
\overline\Psi & = \left(\eta^\alpha \ \overline\chi_{\dot\alpha}\right)
\,, &
\Psi^C & = {\eta_\alpha \choose \overline\chi^{\dot\alpha}} \,,
&
\overline{\Psi^C} & = \left(\chi^\alpha \ \bar\eta_{\dot\alpha}\right)
\,,
\end{align}
i.e.\ these conjugations simply exchange the 2-component spinors and
the index positions.
Chiral spinors take the forms
\begin{align}
  \Psi_L & = {\chi_\alpha \choose 0} \,, &
  \overline{\Psi_L} & = ({0 \ \bar\chi_\alphadot}) \,, &
  \Psi_R & =  {0 \choose \bar\eta^{\dot\alpha}} \,, &
    \overline{\Psi_R} & = ({\eta^\alpha \ 0 }) \,.
\end{align}
Examples of useful bilinear expressions for anticommuting spinors (which allow
rearrangements such as $\chi\eta = \eta\chi$ in view of $\eta^\alpha\chi_\alpha =
-\chi_\alpha\eta^\alpha$)  are
\begin{eqnarray}
\overline\Psi_1         P_L  \Psi_2 
& = & \overline{\Psi^C_2} P_L  \Psi^C_1
= \eta_1 \chi_2
,\\
\overline\Psi_1         P_R  \Psi_2 
& = & \overline{\Psi^C_2} P_R  \Psi^C_1
= \overline\chi_1\bar\eta_2
,\\
\overline\Psi_1          \gamma^\mu P_L \Psi_2 
& = & \overline{\Psi^C_2} (-P_L\gamma^\mu)  \Psi^C_1
= \overline\chi_1\overline\sigma^\mu\chi_2
= - \chi_2\sigma^\mu\overline\chi_1
,\\
\overline\Psi_1         \{1,\gamma_5, \gamma^\mu,\gamma^\mu\gamma_5 \}   \Psi_2 
& = & \overline\Psi^C_2 \{1,\gamma_5,-\gamma^\mu,-\gamma_5\gamma^\mu \}  \Psi^C_1.
\label{FlipRegeln}
\end{eqnarray}
Using the Hermiticity relations for 2-spinors
\begin{equation}
\overline\psi^{\dot\alpha} = (\psi^\alpha)^\dagger,\quad
\overline\psi_{\dot\alpha} = (\psi_\alpha)^\dagger,\quad
(\psi_1\psi_2)^\dagger = \overline\psi_2\overline\psi_1, \quad
(\psi_1\sigma^\mu\overline\psi_2)^\dagger
= \psi_2\sigma^\mu\overline\psi_1,
\end{equation}
directly leads to the following equations for Hermitian conjugation of
4-component bilinears:
\begin{eqnarray}
(\overline\Psi_1      P_L\Psi_2)^\dagger
& = & \overline\Psi_2 P_R\Psi_1
,\\
(\overline\Psi_1      \gamma^\mu P_L\Psi_2)^\dagger
& = & \overline\Psi_2 P_R\gamma^\mu \Psi_1
,\\
(\overline\Psi_1      \gamma^\mu P_R\Psi_2)^\dagger
& = & \overline\Psi_2 P_L\gamma^\mu \Psi_1
,\\
  (\overline\Psi_1      \{1, \gamma_5,\gamma^\mu,\gamma^\mu\gamma_5\}\Psi_2)^\dagger
& = & \overline\Psi_2 \{1,-\gamma_5,\gamma^\mu,-\gamma_5\gamma^\mu\}\Psi_1.
\end{eqnarray}
At this point we stress again that all equations of this section are
valid in strictly 4-dimensional Minkowski spacetime. Later we will use
dimensional regularization in which 2-component spinors are not
directly defined. However, all equations for 4-component spinors
written in this section have been written in such a way that they
remain valid on the $D$-dimensional regularized level.

\subsection{BRST Invariance and Slavnov-Taylor Identity}
\label{sec:BRSTSTI}
Though the construction of the Yang-Mills Lagrangian
(\ref{YMLagrangian}) is elegant and predictive, the Lagrangian cannot
directly be quantized. On the level of canonical quantization, the
canonical conjugate momentum field corresponding to $A^a_0$
identically vanishes; on the level of path integral quantization, the
naively defined path integral is ill-defined due to the integration over
infinitely many gauge equivalent field configurations.

The well-known proposal by Faddeev and Popov modifies the path integral
definition of the quantum theory by separating off this divergent
factor \cite{Faddeev:1967fc}. Via a clever manipulation the path integral can then
be written in terms of a modified Lagrangian which contains a gauge
fixing term as well as terms with Faddeev-Popov ghost fields. The
interactions of the Faddeev-Popov ghosts are determined by the choice
of the gauge fixing. This path integral formulation also allowed to
  derive Slavnov-Taylor identities which could then be used in the
  first proofs of renormalizability of Yang-Mills theories, as
  discussed later in Sec.\ \ref{sec:algren}. 

  Historically, it was observed afterwards that the resulting
Faddeev-Popov Lagrangian is invariant under a new symmetry, the
so-called BRST invariance \cite{Becchi:1974xu,Becchi:1974md,Becchi:1975nq,Tyutin:1975qk}. Here we will directly start with
this BRST invariance, which can be intrinsically motivated and which
provides an efficient formalism for setting up the quantization of
Yang-Mills theories. Our presentation has similarities to the
presentation of the Kugo/Ojima formalism in Ref.\ \cite{Kugo:1979gm} and
the presentations of the BRST and Batalin/Vilkovisky formalisms in
Refs.\ \cite{Henneaux:1992ig,Weinberg:1996kr}.

The main idea is that the concept of local gauge invariance means that
physics is described by equivalence classes. Precisely speaking on the
classical level, field configurations which are related by local gauge
transformations by definition describe the same physical state. The
BRST formalism implements this idea in an elegant way. It first
introduces the notion of ghost
number $N_{\text{gh}}$. All fields introduced so far have
vanishing ghost number, but we shall introduce objects with positive or
negative ghost number later. The BRST formalism further postulates
the existence of an operator $s$, the BRST operator,  which acts on
classical fields and has the following properties and interpretations:
\begin{itemize}
\item
  It generalizes gauge invariance in the sense that: a field
  configuration $X$ with ghost number zero is ``physical'' if
  \begin{align}
\label{sDef1}    sX & = 0\,.
  \end{align}
\item
  It generalizes gauge transformations and gauge equivalence in the
  sense that: two ``physical'' field configurations $X_1$, $X_2$  with
  ghost number zero are  physically equivalent if some $Y$ exists with 
  \begin{align}
\label{sDef2}    X_1 & = X_2 + sY\,.
  \end{align}
  As a side note, objects $X$ which are total BRST transformations,
  \begin{align}
\label{sDef3}    X & = sY\,,
  \end{align}
  are therefore ``unphysical'' in the sense that they are equivalent
  to the   trivial field configuration where all fields vanish (even if
  they also satisfy $sX=0$).
\item
  It is nilpotent,
  \begin{align}
  \label{eq:NilpotencyOfTheBRSTOperator}
    s^2 & = 0 \,,
  \end{align}
  and this nilpotency is important for the consistency of the previous
  two relations. 
\item
  In general, $s$ acts as a fermionic differential operator which
  increases ghost number by one. Specifically, on products of
  fermionic and bosonic expressions $F_i$, $B_i$ it satisfies the
  product rules corresponding to a so-called graded algebra,
\begin{subequations}
  \begin{align}
    s(B_1B_2) &= (sB_1)B_2+B_1(sB_2), \\
    s(F_1B_2) &= (sF_1)B_2-F_1(sB_2), \\
    s(F_1F_2) &= (sF_1)F_2-F_1(sF_2).
  \end{align}
\end{subequations}
\end{itemize}

In order to define an operator with these properties, one first
introduces ghost fields $c^a(x)$, which are scalar fields with
fermionic statistics and ghost number $+1$. As for the gauge fields,
there is one such ghost
field for each gauge group generator $a=1\ldots N_{\text{gen}}$, and
we can also write $c=T^a c^a$ with representation matrices $T^a$. On
the ordinary fields, the BRST operator is then defined as an
infinitesimal gauge transformation, see
Eq.\ (\ref{GaugeTransformationInfinitesimal}), but with the
replacement 
$\theta^a\to c^a$,
\begin{subequations}
\begin{align}
  s A_\mu(x) & =  \partial_\mu c(x) - ig[c(x),A_\mu(x)]\,,\\
  s A^a_\mu(x) & =  \partial_\mu c^a(x)
  +gf^{abc}c^b(x)A^c_\mu(x)=(D_\mu c(x))^a\,,\\
s \varphi(x)  & =  -ig c(x) \varphi(x)\,.
\label{matterBRST}
\end{align}
\end{subequations}
Here we also used the covariant derivative acting on ghost
fields, which is defined by using the adjoint representation for the
generators. 
The BRST transformation of the ghost fields themselves is defined via
the structure constants of the Lie algebra,
\begin{subequations}
\begin{align}
s c^a(x) & =  \frac12 gf^{abc} c^b(x) c^c(x)\,,\\
s c(x) & =  -ig c(x)^2\,.
\end{align}
\end{subequations}
In this way, the BRST operator is indeed nilpotent if it acts on any
combination of these fields, and it clearly generalizes the original
gauge transformations.

In this formalism introducing gauge fixing and associated ghost
interaction terms becomes very natural and transparent. The existence
of two further kinds of fields  is
postulated, the antighosts $\cbar^a$ and the
Nakanishi-Lautrup auxiliary fields $B^a$ (with ghost number $-1$ and
$0$, respectively). From the present point of view these fields 
essentially have the sole purpose of allowing 
the formulation of a gauge fixing. They form a so-called BRST doublet, which
means the following very simple BRST transformations
\begin{subequations}
  \label{cbarBBRST}
  \begin{align}
s\cbar^a(x) & =  B^a(x)\,,\\
sB^a(x) & = 0\,,
\end{align}
\end{subequations}
which are again consistent with nilpotency.  It is known that introducing such a
  BRST doublet does not change the cohomology classes of the BRST
  operator \cite{Piguet:1995er}. In terms of the interpretation specified above this means
  that introducing the BRST doublet does not not change the physical
  content of the theory.

With these ingredients we can discuss Lagrangians of the type
\begin{align}
  \La_{\text{fix,gh}} & = s\left[\cbar^a X^a\right]
\end{align}
with some ghost-number zero object $X^a$. Evaluating the BRST
transformation on the right-hand side produces terms of ghost number
zero, which are allowed terms in a Lagrangian. Given the interpretations
listed above, such Lagrangians are ``unphysical'' since they are total
BRST transformations. Similarly, adding such a Lagrangian to the
original gauge invariant Yang-Mills Lagrangian
$\La_{\text{inv}}+\La_{\text{fix,gh}}$ does not change the physical
content.

Hence we may use this possibility to design a Lagrangian of this type
that can be used for gauge fixing, allowing straightforward
quantization of the theory. The common choice is
\begin{align}
\La_{\text{fix,gh}} & =  s \left[\cbar^a 
          \left((\partial^\mu A^a_\mu) + \frac{\xi}{2}B^a\right)\right]\nonumber\\
& =  
B^a (\partial^\mu A^a_\mu) + \frac{\xi}{2} (B^a)^2 
- \cbar^a\partial^\mu (D_\mu c)^a
\,.
\label{GaugeFixingGeneral}
\end{align}
The $B$-fields are auxiliary fields in the sense that they have no
kinetic term and have purely algebraic equations of motion. They can hence
be eliminated by their equations of motion
\begin{align}
\label{Beqofmotion}
  B^a & = -\frac{1}{\xi}\partial^\mu A^a_\mu, \\
\La_{\text{fix,gh}} & =  
-\frac{1}{2\xi} (\partial^\mu A^a_\mu)^2
- \cbar^a\partial^\mu (D_\mu c)^a
\,.
\label{GaugeFixingWithoutB}
\end{align}
In this way the Lagrangian contains the usual $\xi$-dependent gauge
fixing term, and the way it was constructed led to corresponding ghost
kinetic terms and ghost--antighost--gauge boson interactions. The
result of this construction is the same as the result of the
Faddeev-Popov approach.

Before turning to quantization, there is one final useful extension of
the classical Lagrangian. We note that most of the BRST
transformations are local products of fields, i.e.\ constitute
non-linear field transformations. In a non-Abelian gauge
theory, the only exceptions are the BRST transformations $s\cbar^a$
and $sB^a$, which are linear or zero. In an Abelian theory (where
$f^{abc}$ would vanish), also the BRST transformations of $c^a$ and
$A^a_\mu$ would be linear. In the quantized theory such field products
will define composite operators that require dedicated
renormalization. It is useful to introduce ``sources'' for these
composite operators, i.e.\ classical fields $\rho^a{}^\mu(x)$,
$\zeta^a(x)$, $Y_i(x)$,\footnote{   These sources are not quantized
  and not integrated over in the path integral. These sources are also
  called ``external sources'' or 
  ``external fields'' or ``antifields''.
 One may also regard
  them as local, $x$-dependent parameters of the Lagrangian.}
which couple to the
composite operators in the Lagrangian. We therefore define
\begin{align}
  \label{LextBRST}
  \La_{\text{ext}} & = \rho^a{}^\mu sA^a_\mu + \zeta^a sc^a + {Y}_i s{\varphi_i} \,.
\end{align}
Each source has negative ghost number such that the Lagrangian has
total zero ghost number, and each source has the opposite statistics of
the original field, such that the Lagrangian is bosonic. The
dimensions of the sources are such that the Lagrangian has dimension
4. Specifically, the sources $\rho^a{}^\mu$ are fermionic with ghost
number $-1$ and dimension $3$, the sources $\zeta^a$  are bosonic with
ghost number $-2$ and dimension $4$. By convention, the BRST
transformation of all sources vanishes.

In total, we can then define the full classical Lagrangian, which will
be the basis of quantization, as follows:
\begin{align}
  \La_{\text{cl}} & =
  \La_{\text{inv}} + \La_{\text{fix,gh}} + \La_{\text{ext}}\,.
\label{FullClassicalLagrangian}
\end{align}
Each of the three parts is individually BRST invariant. The first part
is the gauge invariant physical Lagrangian. It depends only on ordinary
fields, on which BRST transformations act like gauge
transformations. The second part contains the gauge fixing and ghost
terms which allow quantization of the theory. Together they are a
total BRST transformation and hence BRST 
invariant and unphysical. The third part is BRST invariant in view of
the nilpotency $s^2=0$. In total,
\begin{align}
\label{BRSTLcl}
  s\La_{\text{cl}} & = 0\,.
\end{align}

The same statement can be rewritten in functional form. Defining the
classical action
\begin{align}
  \Gamma_{\text{cl}} & = \intx \La_{\text{cl}}
\label{ClassicalAction}
\end{align}
allows to rewrite Eq.\ (\ref{BRSTLcl}) as the Slavnov-Taylor identity
\begin{align}
\label{SofGammacl}
  \mathcal{S}(\Gamma_{\text{cl}}) & = 0
\end{align}
with the Slavnov-Taylor operator
\begingroup
\makeatletter\def\f@size{10}\check@mathfonts
\def\maketag@@@#1{\hbox{\m@th\normalsize\normalfont#1}}\begin{align}
\label{eq:SofFDefinition}
    \mathcal{S}({\cal F}) =
    \intx \left(
        \frac{\delta {\cal F}}{\delta \rho^a{}^\mu(x)} \frac{\delta {\cal F}}{\delta A^a_\mu(x)} +
        \frac{\delta {\cal F}}{\delta \zeta^a(x)} \frac{\delta {\cal F}}{\delta c^a(x)} +
        \frac{\delta {\cal F}}{\delta {Y}_i(x)} \frac{\delta {\cal F}}{\delta \varphi_i(x)} +
        B^a(x) \frac{\delta {\cal F}}{\delta \cbar^a(x)} \right)\,.
\end{align}
\endgroup
The Slavnov-Taylor identity (\ref{SofGammacl}) is the ultimate
reformulation of gauge invariance of the classical action after introducing gauge fixing,
ghost terms and external sources for composite operators. This
identity will be a crucial ingredient in the renormalization
procedure.\footnote{  \label{BFMfootnote}
  We remark that the choice of gauge fixing used in the present review
  is not the only option. Other options include physical gauges such
  as axial gauge where no ghosts are required, or the background field
  gauge, see e.g.\ Refs.\ \cite{Weinberg:1996kr,Bohm:2001yx} for
  textbook discussions. Of particular interest for the present
  discussion is the application of the background field gauge to the
  electroweak SM which includes 
  chiral fermions (and electroweak symmetry breaking)
  \cite{Denner:1994xt}. Later, in Sec.\ \ref{sec:OutlookAlgRen} we
  will further comment on proofs of renormalizability and physical
  properties such as charge universality in these different gauges.
  The central point of the present review is the application of the
  BMHV scheme for non-anticommuting 
  $\gamma_5$ to chiral gauge theories. Here it is
  noteworthy that this application is essentially unchanged regardless
  whether the gauge fixing of the main text or the background field
  gauge is used.  The
  corresponding discussion
  and the required computation of symmetry-restoring counterterms were
   carried out in Ref.\ \cite{Cornella:2022hkc}. The main
  technical difference to the formalism presented here is that the
  dominant role of the Slavnov-Taylor identity is replaced by a Ward
  identity reflecting gauge invariance with respect to background
  fields; the overall logic and detailed calculational steps are
    essentially the same.
  }

\subsection{Green Functions in Quantum Field Theory}
\label{sec:GreenFunctions}

In this subsection we introduce basic notation for quantum field
theory required for our discussion of higher orders and regularization
and renormalization.
We consider a generic quantum field theory with dynamical fields
$\phi_i(x)$ (these may be the gauge fields, matter fields,
ghost or antighost fields introduced in earlier subsections) and a
Lagrangian $\La$. 

Fundamental objects of the full, interacting quantized theory are 
Green functions, i.e.\ time-ordered expectation values of
Heisenberg-picture field operators $\phi_i^H$ in the full vacuum $|\Omega\rangle$ of
the interacting theory:
\begin{align}
G_{i_1\ldots i_n}(x_1,\ldots,x_n) & = 
\langle\Omega|T\phi^H_{i_1}(x_1)\ldots\phi^H_{i_n}(x_n)|\Omega\rangle
\,.
\end{align}
We also consider Green functions involving  composite local operators
$\mathcal{O}$, 
\begin{align}
G_{i_1\ldots i_n}^{k_1\ldots k_m}(y_1,\ldots,y_m, x_1,\ldots,x_n) & =
\langle\Omega|T\mathcal{O}^H_{k_1}(y_1)\ldots\mathcal{O}^H_{k_m}(y_m)\phi^H_{i_1}(x_1)\ldots\phi^H_{i_n}(x_n)|\Omega\rangle\nonumber\\
 & \equiv \langle T\mathcal{O}_{k_1}(y_1)\ldots\mathcal{O}_{k_m}(y_m)\phi_{i_1}(x_1)\ldots\phi_{i_n}(x_n)\rangle
\,.
\end{align}
Here $\phi$ denotes a generic quantum field, and the above expressions may contain different kinds of such fields.
Where unambiguous we shall write $\phi_{i_1}(x_1)\equiv\phi_{i_1}$.
The second line here introduces an alternative short-hand notation for
such Green functions, where the explicit symbols for the vacuum state
and for the Heisenberg picture are suppressed. We will often use this
short-hand notation in the following.
 
Generally,  Green functions are important since they encapsulate the
essential information of a given quantum field theory. We briefly
remark how they particularly allow constructing important observable
quantities. The
physical rest masses of one-particle states are reflected in the poles
of momentum-space 
two-point functions, as a result of the K\"allen-Lehmann
representation.
S-matrix elements for scattering processes between asymptotically free
states are obtained via the Lehmann-Symanzik-Zimmermann reduction
formalism, which can be derived from Haag-Ruelle scattering theory
(see e.g.\ the textbooks by Srednicki and Peskin/Schroeder
\cite{Srednicki:2007qs,Peskin:1995ev} and 
the monograph by Duncan \cite{Duncan:2012aja} for a particularly
detailed account).\footnote{An important subtlety is that Green functions are
  particularly defined in momentum space for off-shell
  momenta. Physical observables are related to the on-shell limits,
  where Green functions may develop infrared divergences. In the
  present review we will not discuss the specifics of the on-shell
  limits of Green functions.}

A very useful tool for general discussions is the
generating functional $Z(J,\Ks)$ for the most general Green functions with
elementary fields and composite operators. It can be written by
introducing sources (or ``external fields'', i.e.\ fields which always
remain classical and never are quantized)
$J_i(x)$ for the elementary fields and $\Ks_i(x)$ for the composite
operators such that
\begingroup
\makeatletter\def\f@size{11}\check@mathfonts
\def\maketag@@@#1{\hbox{\m@th\normalsize\normalfont#1}}\begin{align}
\label{GreenFunctionFromZ}
&G_{i_1\ldots i_n}^{k_1\ldots k_m}(y_1,\ldots,y_m, x_1,\ldots,x_n)\\
&\hspace{1cm} = \frac{1}{Z(0,0)}\ 
\frac{\delta^{m+n} Z(J,\Ks)}
 {\delta i\Ks_{k_1}(y_1)\ldots \delta i\Ks_{k_m}(y_m)\ldots\delta iJ_{i_1}(x_1)\ldots \delta
   iJ_{i_n}(x_n)\ldots}
 \Bigg|_{J=\Ks=0}
\,.\nonumber
\end{align}
\endgroup

In perturbation theory the Green functions are given by Feynman
diagrams obtained from the well-known Gell-Mann-Low
formula. Specifically, in perturbation theory the Lagrangian is split as
$\La=\La_{\text{free}}+\La_{\text{int}}$, where the free part
$\La_{\text{free}}$ is
bilinear in the quantum fields, allowing quantization as a free field
theory. This quantization then leads to free field operators which we
denote as $\phi_i$ without the superscript, and to a free vacuum
$|0\rangle$. The
Gell-Mann-Low formula for the perturbative evaluation of Green functions
then yields an explicit construction of the generating functional:
\begin{align}
Z(J,\Ks) & =
  \frac{
\langle0|T
 \exp\left(i\intx (\La_{\text{int}}+J_i\phi_i+\Ks_i\mathcal{O}_i)\right)|0\rangle}
{\langle0|T\exp\left(i\intx \La_{\text{int}}\ \right)|0\rangle}
\,.
\label{GellMannLow}
\end{align}
The evaluation of this formula via Wick contractions leads to Feynman
rules and Feynman diagrams. In Eq.\ (\ref{GellMannLow}) we also
introduce a short-hand notation which we will often use: all appearing
fields
and sources $J_i,\phi_i,\Ks_i,{\cal O}_i$ and the Lagrangian
$\La_{\text{int}}$ have the spacetime argument $x$, which is
suppressed. Further there is a summation over the index $i$, and the
summation range extends over all quantum fields in the term
$J_i\phi_i$ and over all composite operators with sources in the term
$\Ks_i{\cal O}_i$.

Another representation of the generating functional is given by the
path integral
\begin{align}
Z(J,\Ks) & = \int{\cal D}\phi\ e^{i\intx (\La+J_i\phi_i+\Ks_i\mathcal{O}_i)}
\,,
\label{pathintegral}
\end{align}
where ${\cal D}\phi$ is the measure of the integration over all field
configurations and the quantities in the exponent are number-valued
fields (either sources or path integral integration variables). The
same short-hand notation suppressing the arguments is used.
We stress that both equations (\ref{GellMannLow}) and (\ref{pathintegral}) are formal
and not yet fully defined: the literal application of the
Gell-Mann-Low formula leads to divergences unless the theory is
regularized, and the path integral formula requires a precise
definition of the path integral measure. Both formulas will become well-defined via the process of
regularization and renormalization (this process can also be regarded
as a 
constructive definition of the path integral measure).

The full Green functions discussed so far are described by the most
general Feynman diagrams which are allowed to contain several
disconnected components. It is possible to define a second generating
functional $Z_c$ which directly generates only connected Green
functions, i.e.\ the sums of connected Feynman diagrams.
The relation is given by
\begin{align}
Z(J,\Ks) & = e^{iZ_c(J,\Ks)} \,.
\end{align}
For a proof that this generates precisely the connected Green
functions see e.g.\ Refs.\ \cite{Zinn-Justin:1989rgp,Itzykson:1980rh}.\footnote{  The conventions for the generating functionals differ slightly
  between most references. Our conventions are essentially the same as
  in Ref.\ \cite{Peskin:1995ev} except that our connected functional
  $Z_c=-E$ there.
}

For renormalization, one-particle irreducible (1PI) Feynman diagrams are
most useful since they are the smallest building blocks that suffice
to discuss ultraviolet divergences and counterterms. The corresponding 1PI Green
functions can also be generated by a generating functional. This 1PI
generating functional is 
called $\Gamma$, or effective action. It is defined by a Legendre
transform of $Z_c$ which replaces the sources by classical fields.

In order to prepare for the introduction of this 1PI generating
functional $\Gamma$ we make two remarks:
First we note that there is a mapping between the sources $J_i$ and
expectation values of field operators $\phi_i$. Specifically
the first derivatives of the generating
functional $Z_c$ have the special interpretation as the expectation values
of the field operators,
\begin{align}
  \label{phiclDef}
  \phi_{i}^{\text{class}}(x) & \equiv \frac{\delta Z_c}{\delta J_i(x)}
= \langle\phi_i(x) 
\rangle^{J,\Ks}
\,.
\end{align}
In contrast to Eq.\ (\ref{GreenFunctionFromZ}), we have not set the sources to zero.
Each choice of the sources $J_i(x)$ (for fixed $\Ks_i(x)$) thus defines  expectation 
values of the quantum field operators.
These expectation values are number-valued, ``classical'' fields
$\phi_i^{\text{class}}(x)$. We may regard these classical fields as
functionals of the sources $J_i(x)$ (for fixed $\Ks_i(x)$), or we may invert the relationship and
regard the sources as functionals of the classical fields.
In the following we will always assume that the vacuum expectation
values of the operators $\phi_i$ vanish. Here this means that $J=0$ is
mapped to $\phi^{\text{class}}=0$ and vice versa (for $\Ks=0$):
\begin{align}
\frac{\delta Z_c}{\delta J_i(x)}\Bigg|_{J=\Ks=0} & = 0
\,.
\end{align}
The second remark is the following:  In the classical limit, the
path integral is dominated by the classical field configuration
minimizing the classical action. Hence, in the classical limit (``cl.lim.'') and up
to an irrelevant constant, we have
\begin{align}
  Z(J,\Ks) = e^{iZ_c(J,\Ks)} & \stackrel{\text{cl.lim.}}{\longrightarrow} e^{i(\Gamma_{\text{cl}}(\phi^{\text{class}},\Ks)+\intx
    J_i\phi_i^{\text{class}})}\Big|_{0=\frac{\delta \Gamma_{\text{cl}}}{\delta \phi^{\text{class}}}
  \pm J}\,,
  \label{Zclassicalapprox}
\end{align}
where $  \Gamma_{\text{cl}}=\intx(\La+\Ks_i\mathcal{O}_i)$ is the
classical action (including source terms for composite operators), and
where the $\pm$ signs apply for bosonic/fermionic fields $\phi$,
respectively.

This motivates the definition of a new functional $\Gamma$ via the
analogous, exact relation
\begin{align}
  \label{LegendreZc}
  Z_c(J,\Ks) & = \Gamma(\phi^{\text{class}},\Ks) + 
 \intx J_i\phi^{\text{class}}_i 
 \Big|_{J=\mp\frac{\delta \Gamma}{\delta \phi^{\text{class}}}}\,.
\end{align}
This relation is a Legendre transformation, which can be inverted to
\begin{align}
  \label{LegendreGamma}
  \Gamma(\phi^{\text{class}},\Ks) & = Z_c(J,\Ks) - 
 \intx J_i\phi^{\text{class}}_i 
 \Big|_{\phi^{\text{class}}=\frac{\delta Z_c}{\delta J}}\,.
\end{align}
In the Legendre transformation the sources $\Ks_i$ for composite
operators act as spectators, such that the relation
\begin{align}
  \frac{\delta\Gamma(\phi^{\text{class}},\Ks)}{\delta \Ks_i(x)} & = \frac{\delta Z_c(J,\Ks)}{\delta \Ks_i(x)}
\label{Kspectators}
\end{align}
holds.

The functional $\Gamma$ defined in this way has two very important
properties.
First, it
is equal to the classical action plus quantum
corrections, i.e.\
\begin{align}
  \label{GammaEffectiveAction}
  \Gamma(\phi^{\text{class}},\Ks) & =
  \Gamma_{\text{cl}}(\phi^{\text{class}},\Ks)+{\cal O}(\hbar) \,,
\end{align}
where we reinstate explicit powers of $\hbar$ to count the number of loops.
This justifies the name ``effective action''. Second, $\Gamma$
generates one-particle irreducible (1PI) Green functions.
For the full proofs of these statements see e.g.\ the textbooks by
Zinn-Justin or Itzykson/Zuber \cite{Zinn-Justin:1989rgp,Itzykson:1980rh},
and for detailed discussions including
subtleties in cases with spontaneous symmetry
breaking see e.g.\ the textbooks by Weinberg or Brown \cite{Weinberg:1996kr,Brown_1992}.

Let us introduce further useful notation related to Green functions
and $\Gamma$. First, in the following and in general we simplify the notation for $\Gamma$
and write only $\phi_i$ instead of $\phi_i^{\text{class}}$ for its
arguments if no misunderstanding is possible.

Next we introduce notation for specific 1PI Green
functions. Such concrete 1PI Green functions in position space are obtained from derivatives of $\Gamma$ with respect to the classical fields as
\begin{align}
  \label{ConcreteGamma}
  \Gamma_{\phi_i\phi_j\ldots}(x_1,x_2,\ldots) & = 
\dg{\phi_i(x_1)\delta\phi_j(x_2)\ldots}\Big|_{\phi=0}
=
-i \langle \phi_i(x_1) \phi_j(x_2)\ldots \rangle^\text{\,1PI}
\,.
\end{align}
In terms of Feynman diagrams, $i  \Gamma_{\phi_i\phi_j\ldots}$
corresponds to the set of 1PI diagrams with the indicated external
fields.
When passing to momentum space via a Fourier transform,
we split off a $\delta$-function
corresponding to momentum conservation; symbolically
\begin{align}
  \Gamma_{\phi_i\phi_j\ldots}\Big|^{\text{F.T.}}(p_1,p_2,\ldots) & =
  \Gamma_{\phi_i\phi_j\ldots}(p_1,p_2, \ldots)
  (2\pi)^4 \delta^{(4)}({\textstyle \sum_{j=1}^{n} p_j})\,.
\end{align}

Equations (\ref{Zclassicalapprox},\ref{GammaEffectiveAction}) show
that naturally the source terms for composite
operators combine with the Lagrangian; hence it is motivated to absorb
these source terms into the Lagrangian.  This is precisely what was done in Sec.\ \ref{sec:BRSTSTI}
for certain important operators corresponding to nonlinear BRST
transformations, see Eq.\ (\ref{LextBRST}). In this way, the renormalization of
such composite operators is fully integrated into the standard
renormalization procedure.

Sometimes, special operators need to be considered only in the simpler
context of single operator insertions. Let $\mathcal{O}$ be such an
operator and $\Ks_{\mathcal{O}}$ the corresponding source, treated as in
Eqs.\ (\ref{GellMannLow}) or (\ref{pathintegral}) or absorbed
into the Lagrangian. The sources for all remaining operators
are collectively called $\Ks$. Then, for single insertions of $\mathcal{O}$ a 
special notation is defined:
\begin{subequations}
\begin{align}
  {\cal O}(x)\cdot Z(J,\Ks) & = \frac{\delta
  Z(J,\Ks,\Ks_{\mathcal{O}})}{\delta(i\Ks_{\mathcal{O}}(x))}\Big|_{\Ks_{\mathcal{O}}=0},\\
{\cal O}(x)\cdot \Gamma(\phi,\Ks) & = \frac{\delta
  \Gamma(\phi,\Ks,\Ks_{\mathcal{O}})}{\delta \Ks_{\mathcal{O}}(x)}\Big|_{\Ks_{\mathcal{O}}=0}.
\end{align}
\end{subequations}
For particular 1PI Green functions with a single operator insertion we
can write 
\begin{align}
\left(    \mathcal{O}(x) \cdot
\Gamma\right)_{\phi_i\phi_j\ldots}(x_1,x_2,\ldots) & =
-i \langle \mathcal{O}(x)\phi_i(x_1) \phi_j(x_2)\ldots \rangle^\text{\,1PI}
 \,.\label{OdotGammaDef}
\end{align}
In terms of Feynman diagrams, $i\left(    \mathcal{O}(x) \cdot
\Gamma\right)_{\phi_i\phi_j\ldots}$ corresponds to 1PI diagrams with
the indicated external fields and one insertion of a vertex
corresponding to $i\mathcal{O}(x)$, where the factor $i$ results as
usual from
the exponential function in the Gell-Mann-Low formula (\ref{GellMannLow}).

  An important consequence is the lowest-order behaviour of the
operator insertion into $\Gamma$,
\begin{align}
{\cal O}\cdot \Gamma(\phi) & = {\cal O}{}^{\text{class}} +
{\cal O}(\hbar)\,,
\label{insertionlowestorder}
\end{align}
where ${\cal O}^{\text{class}}$ is the classical field product
corresponding to the operator ${\cal O}$. This is
in line with the interpretation of $\Gamma$
as the effective action.

\subsection{Slavnov-Taylor identities for Green functions and
  their interpretation}
\label{sec:STIformal}

In Sec.\ \ref{sec:BRSTSTI} we introduced BRST invariance as a substitute for
gauge invariance in presence of a gauge fixing, and we found the
BRST invariant classical action. The question is now: How is this BRST
invariance reflected in the full quantum theory? The most general
answer is that the off-shell Green functions introduced in
Sec.\ \ref{sec:GreenFunctions} satisfy so-called Slavnov-Taylor
identities. Here, we provide a formal derivation of these
Slavnov-Taylor identities. This derivation is simple and elegant and
allows an efficient understanding and interpretation of the structure of
the Slavnov-Taylor identities. It is however formal in the sense that
it ignores the procedure of regularization and renormalization; hence we
will later, in Sec.\ \ref{sec:algren}, need to discuss how this procedure might change the
identities. There, we will also discuss the important role of the
Slavnov-Taylor identities in establishing the renormalizability of
Yang-Mills theories, including the decoupling of unphysical degrees of
freedom and the unitarity of the physical S-matrix.

We start from the BRST invariance of the classical action, which was
already expressed by Eq.\ (\ref{BRSTLcl}) and rewritten as the
Slavnov-Taylor identity (\ref{SofGammacl}). Here we rewrite it as an
invariance relation
\begin{align}
  \label{GammaclInvariance}
  \Gamma_{\text{cl}}(\phi,\Ks) & = 
  \Gamma_{\text{cl}}(\phi+\delta\phi,\Ks) 
\end{align}
where
$\phi$ denote all dynamical fields ($A_\mu$, $\varphi_i$, $c$,
$\bar{c}$, $B$) and $\Ks$ denote all sources ($\rho^\mu$, $Y_i$,
$\zeta$) and where the field transformations are given as
\begin{align}
\label{clsymtransform}
  \delta\phi & = \theta s\phi
\end{align}
with an infinitesimal fermionic parameter
$\theta$ such that $\delta\phi$ always has the same bosonic/fermionic
statistics as $\phi$ itself. Eq.\ (\ref{GammaclInvariance}) is meant at first order in
$\theta$ and at this order it is clearly equivalent to both
Eqs.\ (\ref{BRSTLcl},\ref{SofGammacl}).

Now we use this invariance as a starting point and derive
the Slavnov-Taylor identities for the generating functional
(\ref{pathintegral})  in the path
integral formulation. We assume that the path integral measure is
invariant under the same symmetry transformation
$\phi\to\phi+\delta\phi\equiv\phi'$
and therefore write
\begin{align}
  Z(J,\Ks) & = \int{\cal D}\phi'\ e^{i(\Gamma_{\text{cl}}(\phi',\Ks)+\intx
    J_i\phi_i')} \nonumber\\
  & = 
  \int{\cal D}\phi\ e^{i(\Gamma_{\text{cl}}(\phi,\Ks)+\intx
    J_i\phi_i + J_i\delta\phi_i)}
\,.
\end{align}
The variation  $\delta\phi$ only appears in the exponent. We can
expand the right-hand side at first order in $\delta\phi$ and subtract
it from the left-hand side to obtain
\begin{align}
  \label{STIBasic}
  0 & = 
  \int{\cal D}\phi\ \left(\intx J_i\delta\phi_i\right)e^{i(\Gamma_{\text{cl}}(\phi,\Ks)+\intx
    J_i\phi_i )}
\,.
\end{align}
This is already one basic version of the Slavnov-Taylor identity. We can
rewrite it in several ways to familiarize us with its
interpretation.
\begin{itemize}
  \item
A first way is to replace the path integral with its interpretation as an
operator expectation value, in line with Eq.\ (\ref{phiclDef}). Then
we obtain
\begin{align}
  0 & = 
\intx J_i\langle\delta\phi_i\rangle^{J,\Ks}
\,.
\end{align}
This can be further rewritten by replacing the sources $J_i$  in
terms of derivatives of $\Gamma$, the effective action or generating
functional of 1PI Green functions, via the Legendre transform
(\ref{LegendreZc}) such that
\begin{align}
  0 & = 
\intx \langle\delta\phi_i\rangle^{J,\Ks}\frac{\delta\Gamma}{\delta \phi_i}
\,,
\label{eq:STIwithGammaUsingLegendreTrafo}
\end{align}
where again the sum over all fields $i$ is implied and where the order
of the factors was exchanged to compensate the $\pm$ signs in the
relation for $J_i$ in Eq.\ (\ref{LegendreZc}).
Both of these equations have the forms of typical infinitesimal
invariance relations. We may also rewrite the previous equation as
\begin{align}
  \label{GammafullInvariance}
  \Gamma(\phi,\Ks) & = 
  \Gamma\left(\phi+\langle\delta\phi\rangle^{J,\Ks},\Ks\right)
  \,,
\end{align}
valid to first order in the variation. This equation is directly
analogous to the starting point (\ref{GammaclInvariance}). It
clarifies the interpretation of the Slavnov-Taylor identity as an
invariance relation for the full effective action $\Gamma$ under
symmetry transformations given by
$\langle\delta\phi_i\rangle^{J,\Ks}$.
An important distinction can now be made about these symmetry
transformations.
In general the $\delta\phi_i$ are nonlinear
products of fields (i.e.\ composite operators), and generally the
expectation value of a product is different from the product of
expectation values. In other words the
symmetry transformations may receive nontrivial quantum
corrections.  Hence the symmetry transformation in
Eq.\ (\ref{GammafullInvariance}) is in general different from the
classical expression  $\delta\phi_i$ which one might
have expected to appear.\footnote{In the previous section we would have used the more
explicit notation $\delta\phi_i^{\text{class}}$ for the expression
where all fields are replaced by their classical versions, i.e.\ their
expectation values. }
Only in the case where all $\delta\phi_i$ are
linear in the dynamical fields, the symmetry relation
(\ref{GammafullInvariance}) corresponds to the same invariance as
Eq.\  (\ref{GammaclInvariance}).
\item
  A second way to rewrite the Slavnov-Taylor identity (\ref{STIBasic})
  is by taking derivatives with respect to the sources as in
  Eq.\ (\ref{GreenFunctionFromZ}) to obtain 
  identities for specific Green functions. In this way,
  Eq.\ (\ref{STIBasic}) leads to infinitely many identities of the kind 
\begin{align}
  0 & = \delta
  \langle T \phi_{i_1}(x_1)\ldots\phi_{i_n}(x_n)\rangle^{J,\Ks}
\label{deltaAbbrev}
  \\
  &\equiv
  \langle T (\delta\phi_{i_1}(x_1))\ldots\phi_{i_n}(x_n)\rangle^{J,\Ks}
  +\ldots+
  \langle T (\phi_{i_1}(x_1))\ldots\delta\phi_{i_n}(x_n)\rangle^{J,\Ks}
\,,\nonumber
\end{align}
  where the first line is defined as  an abbreviation for the
  second line and the uniform $+$ signs of all terms are correct
  because the transformation $\delta$ as defined by Eq.\ (\ref{clsymtransform}) is
  of bosonic nature.
  In these identities Green functions
  involving ordinary fields $\phi_i$ and the symmetry transformation
  composite operators $\delta\phi_i$ appear. In this form,
  Slavnov-Taylor identities may be checked explicitly by computing
  Feynman diagrams for such Green functions. We can illustrate this
  with a simple but important example. Taking the Yang-Mills theory of
  the previous subsections with fermionic matter fields $\psi$,
  we can consider $\delta\langle
  \cbar\psi_i\bar\psi_k\rangle$ and use the BRST transformations in
  Eqs.\ (\ref{matterBRST},\ref{cbarBBRST})  to obtain 
  \begin{align}
    0 & =
    \langle T B \psi_i\bar\psi_k  \rangle +ig
    \langle T \cbar (c\psi)_i\bar\psi_k  \rangle - ig
    \langle T \cbar \psi_i (\bar\psi c)_k  \rangle 
  \end{align}
  where the brackets indicate local composite operators. The auxiliary
  field $B$ will  effectively be replaced by $\partial^\mu A_\mu$
  via Eq.\ (\ref{Beqofmotion}).  In abelian
  QED, the ghosts are free and can be factored out of the matrix
  elements. Hence in QED this identity simply   leads to the familiar
  Ward identity between the electron self energy and the
  electron--electron--photon vertex function. In non-abelian
  Yang-Mills theories, the identity also relates the fermion self
  energy and the fermion--fermion--gauge boson three-point function,
  but the relationship is more complicated and involves nontrivial
  composite operators which need to be renormalized.
\item
  A final way to rewrite the Slavnov-Taylor identity is to write it as
  functional equations for the generating functionals $Z$, $Z_c$ or
  $\Gamma$.
  Since we have coupled the nonlinear classical symmetry transformation
  (\ref{clsymtransform}) to the sources $\Ks$ in the classical action (\ref{LextBRST}),
  the expectation values of nonlinear composite operators appearing in the
  previous equations may be rewritten in terms of functional
  derivatives with respect to $\Ks$. A slight technical complication is
  that there are also linear symmetry transformations which we have
  not coupled to sources, such as the BRST transformations of the $\cbar$
  and $B$ fields. Precisely we can therefore replace the nonlinear
  $\delta\phi_i$ by $\delta/\delta (i\Ks_i)$ in the Slavnov-Taylor identity (\ref{STIBasic}),
  but the linear $\delta\phi_i$ remain. If we express the path
  integral in terms of the connected functional, Eq.\ (\ref{STIBasic})
  takes the schematic form
  \begin{align}
    \label{STIpathintegralZc}
    0 & = \intx \sum_{\delta\phi_i=\text{nonlinear}}J_i \frac{\delta
      Z_c(J,\Ks)}{\delta \Ks_i}
    +
    \intx \sum_{\delta\phi_i=\text{linear}}J_i \langle\delta \phi_i\rangle^{J,\Ks} \,,
  \end{align}
  where the expectation value in the last term really is a linear
  combination of expectation values of fundamental fields, i.e.\ a
  linear combination of $\phi_j^\text{class}$ as used in
  Eq.\ (\ref{LegendreGamma}) and thus equal to what we mean
  by $\delta\phi_i^{\text{class}}$, where the index $^{\text{class}}$
  will be dropped again. The previous equation can be
  rewritten as an equation for the 1PI functional $\Gamma$ by
  replacing the sources $J_i$ via the Legendre transformation to
  $\Gamma$ and by using that the sources $\Ks$ are unaffected by the
  Legendre transformation as expressed by Eq.\ (\ref{Kspectators}). In
  this way we obtain
  \begin{align}
    \label{STIpathintegralGamma}
    0 & = \intx \sum_{\delta\phi_i=\text{nonlinear}} \frac{\delta
      \Gamma(\phi,\Ks)}{\delta \Ks_i}\frac{\delta \Gamma(\phi,\Ks)}{\delta\phi_i}
    +
    \intx \sum_{\delta\phi_i=\text{linear}}\delta \phi_i \frac{\delta
      \Gamma(\phi,\Ks)}{\delta\phi_i}\,. 
  \end{align}
  This is literally the same equation as the Slavnov-Taylor identity
  for the classical action with the Slavnov-Taylor operator (\ref{eq:SofFDefinition}), but rewritten for the full
  effective action,
  \begin{align}
    \label{SofGammapreview}
     \mathcal{S}(\Gamma) & = 0\,.
  \end{align}
  This explains the reason why we rewrote the BRST invariance of the
  classical action in section \ref{sec:BRSTSTI} as the Slavnov-Taylor identity
  using Eq.\ (\ref{eq:SofFDefinition}): This equation has the
  potential of remaining valid without modification in the full
  quantum theory, provided the above formal manipulations survive the
  regularization and renormalization procedure.
\end{itemize}
Finally we comment on the validity of our derivation. The derivation
assumed the classical action to be symmetric, the path integral to be
well defined and the path integral measure to be invariant under the
symmetry. A full treatment must define the quantum theory via the
procedure of regularization and renormalization, which may be viewed
as a constructive definition of the path integral and its measure, and
which might change the action e.g.\ by counterterms. An essential
result of algebraic renormalization theory (see below in
Sec.\ \ref{sec:algren}) is that the above derivations are essentially
correct up to local terms in the following sense: If the above
Slavnov-Taylor identity (\ref{SofGammapreview}) is valid at some given loop
order, then at the next loop order it can at most be violated by a
local functional of the fields. Hence there is a chance that any such
local violation can be cancelled by adding local, symmetry-restoring
counterterms. If this is possible the Slavnov-Taylor identity indeed
can be established at all orders in the renormalized theory.

In the
present review  we mainly work in
dimensional regularization. In this context the above derivation
acquires a more literal meaning. In Sec.\ \ref{sec:QAPDReg} we will
discuss the so-called regularized quantum action principle, which
essentially states that all derivations remain literally valid in
dimensional regularization if all quantities are defined via
regularized Feynman diagrams in $D\ne4$ dimensions. In that case,
however, it becomes questionable whether the $D$-dimensional version
of the classical action satisfies the same symmetry
(\ref{GammaclInvariance}) as the original 4-dimensional version. If
this is not the case, there is again a violation of the Slavnov-Taylor
identity at the regularized level, which needs to be studied and which
may be cancelled by introducing symmetry-restoring counterterms.

\subsection{Peculiarities of Abelian Gauge~Theories}
\label{sec:AbelianPeculiarities}

So far the discussions above focused on the non-Abelian case. However,
there are some peculiarities in the Abelian case that will be
highlighted in this subsection.\footnote{In the absence of spontaneous
  symmetry breaking.} Obviously, in an Abelian gauge theory there are
less interactions than in the non-Abelian case, with corresponding
implications for higher order corrections. However, there are also
less restrictions by the gauge group, which leads to the need for an
additional symmetry condition to ensure a consistent renormalization
of the Abelian coupling constant, as discussed below. For further
information of Abelian theories in this context we refer the reader to
\cite{Becchi:1974md,Kraus:1995jk,Haussling:1996rq,Haussling:1998pp,Grassi:1997mc},
where they focused, in contrast to the present section, on the Abelian
case with spontaneous symmetry breaking, whereas the more general case
of the Standard Model and extensions was discussed in
Refs.\ \cite{Kraus:1997bi,Grassi:1999nb,Hollik:2002mv}. For a general
overview we refer to the textbook by Piguet/Sorella \cite{Piguet:1995er}.

Starting with the classical Lagrangian of the Abelian gauge theory of quantum electrodynamics; using the notation of section \ref{sec:BRSTSTI} we may write it in the same form as in Eq.\  (\ref{FullClassicalLagrangian}), this time, however, with
\begin{align}
\La_{\text{inv}} & =  i \, \overline{\psi}_{i} \slashed{D}_{ij} \psi_{j} - \frac{1}{4} F^{\mu\nu}F_{\mu\nu}
\,,
\end{align}
with the covariant derivative $D^{\mu}_{ij} = \partial^{\mu} \delta_{ij} + i e Q_{i} \delta_{ij} A^{\mu}$ and the field strength tensor $F_{\mu\nu} = \partial_{\mu} A_{\nu} - \partial_{\nu} A_{\mu}$, with the gauge-fixing and ghost Lagrangian
\begin{align}
\La_{\text{fix,gh}} & =  s \left[\cbar 
          \left((\partial^\mu A_\mu) + \frac{\xi}{2}B\right)\right] =  
B (\partial^\mu A_\mu) + \frac{\xi}{2} B^2 
- \cbar\partial^\mu \partial_\mu c
\,,
\label{AbelianGhostGaugeFixingLagrangian}
\end{align}
with $B = - (\partial_{\mu} A^{\mu}) /\xi$ and with the Lagrangian of the external sources
\begin{align}
  \La_{\text{ext}} & = \rho^{\mu} sA_{\mu} + \bar{R}^{i} s\psi_{i} + R^{i} s\overline{\psi}_{i}\,,
\end{align}
where we used the concrete name $R^i$ for the matter field
sources instead of the generic name $Y_{i}$ of Sec.\ \ref{sec:BRSTSTI}.
The classical action is then again given by (\ref{ClassicalAction}).

The BRST transformations in the Abelian case, already used in (\ref{AbelianGhostGaugeFixingLagrangian}), are provided by
\begin{subequations}
\label{AbelianBRSTTrafos}
\begin{align}
sA_{\mu}(x) & = \partial_{\mu} c(x)\,,\\
s\psi_{i}(x) & = - i e Q_{i} c(x) \psi_{i}(x)\,,\\
s\overline{\psi}_{i}(x) & = i e Q_{i} c(x) \overline{\psi}_{i}(x)\,,\\
sc(x) & = 0\,, \\
s\cbar(x) & =  B(x)\,,\\
sB(x) & = 0\,.
\end{align}
\end{subequations}
It can be seen that in the Abelian case, except from the BRST transformations for the fermions $\psi_{i}$ and $\overline{\psi}_{i}$, all other BRST transformations are linear in dynamical fields.
Recall that for a linear classical symmetry of the form
\begin{align}
  \delta \phi_i(x) = v_i(x) + \int d^{4}y \, \, t_{ij}(x,y) \phi_j(y)
\label{eq:linearSymmetryTrafos}
\end{align}
with number-valued kernel $t_{ij}$, its expectation value is identical
to the classical symmetry transformation (see also the discussions
around
Eqs.\ (\ref{GammafullInvariance},\ref{STIpathintegralZc},\ref{STIpathintegralGamma})), i.e.
\begin{align}
  \langle\delta \phi_i(x)\rangle^{J,\Ks} = v_i(x) + \int d^{4}y \, \, t_{ij}(x,y) \langle\phi_j(y)\rangle^{J,\Ks} = \delta \phi_i^\text{class}.
\label{eq:ExpectationValueLinearSymmetryTrafos}
\end{align}
Hence, on the basis of equations (\ref{eq:STIwithGammaUsingLegendreTrafo}) and (\ref{GammafullInvariance}) from section \ref{sec:STIformal}, the full effective quantum action $\Gamma$ is invariant under such linear classical symmetries as they do not receive nontrivial quantum corrections. In other words, linear symmetry transformations of the classical action $\Gamma_{\text{cl}}$ are automatically symmetry transformations of the full effective quantum action $\Gamma$.

In particular, the BRST-transformation of the photon $A_{\mu}$ is linear, and hence $sA_{\mu}$ does not receive quantum-corrections and the expectation value $\langle sA_{\mu} \rangle^{J,\Ks}$ is identical with the classical expression $(sA_{\mu})^\text{class}$. 

Further, $R^{i}$ and $\bar{R}^{i}$ are external sources and the Abelian Fadeev-Popov ghost and antighost completely decouple from the rest of the theory (cf. (\ref{AbelianGhostGaugeFixingLagrangian})). Hence, neither $R^{i}$ and $\bar{R}^{i}$ nor the ghost $c$ and antighost $\cbar$ can occur in loops, they can only appear as external legs, as there are no corresponding interactions and the external sources are not dynamical fields, and thus cannot propagate. Consequently, none of the Abelian BRST transformations obtain quantum corrections, or in other words, in the Abelian case the BRST transformations do not renormalize.

In a theory with a non-Abelian simple gauge group $G$ with gauge
coupling $g$, the generators $T^a$ are uniquely determined by choosing
a representation. For this reason, the couplings of all matter fields
to the gauge fields $\propto g T^a$ and of all gauge boson
self-interactions $\propto g f^{abc}$ are uniquely determined up to
one common, universal gauge coupling $g$.

In contrast to this, in an Abelian gauge theory every diagonal matrix
would be a representation of the corresponding Lie algebra. Thus, the
corresponding charges $Q_{i}$ of the respective fermions could in
principle be arbitrary real numbers. Group theory alone would
allow these charges to obtain quantum-corrections, i.e. they could renormalize, and could thus even take different values at every order in the perturbation theory. Hence, due to the fact that the group structure of an Abelian gauge group is not as powerful as the one of a non-Abelian gauge group the Abelian couplings need to be determined, in all orders, by an additional symmetry condition to the full effective quantum action, either by the local Ward identity or by the so called antighost equation. 

The special simplicity of Abelian gauge theories and the existence of
additional all-order identities is technically reflected in several
field derivatives of the classical action.
We begin with the antighost equation
\begin{align}
  \frac{\delta \Gamma_{\text{cl}}}{\delta c(x)} = \Box \cbar(x) + \partial_{\mu} \rho^{\mu}(x) - i e Q_{i} \bar{R}^{i}(x) \psi_{i}(x) + i e Q_{i} \overline{\psi}_{i}(x) R^{i}(x)\,.
\label{eq:AntiGhostEquationAbelianCase}
\end{align}
Additionally, varying $\Gamma_{\text{cl}}$ w.r.t.\ the antighost and
the external source of the photon yields
\begin{align}
  \frac{\delta \Gamma_{\text{cl}}}{\delta \cbar(x)} & = - \Box c(x)\,,
  &
  \frac{\delta \Gamma_{\text{cl}}}{\delta \rho_{\mu}(x)} &= sA^{\mu}(x) = \partial^{\mu}c(x)\,,
\end{align}
which can be combined to obtain the so-called ghost equation
\begin{align}
  \left( \frac{\delta}{\delta \cbar} + \partial_{\mu} \frac{\delta}{\delta \rho_{\mu}}\right) \Gamma_{\text{cl}} = 0\,.
\label{eq:GhostEquationAbelianCase}
\end{align}
The gauge fixing condition is obtained by varying $\Gamma_{\text{cl}}$ w.r.t.\ the Nakanishi-Lautrup field $B$
\begin{align}
  \frac{\delta \Gamma_{\text{cl}}}{\delta B(x)} = \xi B(x) + \partial_{\mu} A^{\mu}(x)\,.
\label{eq:GaugeFixingAbelianCase}
\end{align}
Importantly, it can be seen that all of the above equations
(\ref{eq:AntiGhostEquationAbelianCase}) to
(\ref{eq:GaugeFixingAbelianCase}) are linear in dynamical fields,
e.g. $\delta \Gamma_{\text{cl}}/\delta c(x) = (\text{linear
  expression})$.
In contrast, all other functional derivatives of the classical action
\begin{subequations}
\begin{align}
  \frac{\delta \Gamma_{\text{cl}}}{\delta R^{i}(x)} & = s\overline{\psi}_{i}(x) = i e Q_{i} c(x) \overline{\psi}_{i}(x)\,,\\
  \frac{\delta \Gamma_{\text{cl}}}{\delta \psi_{i}(x)} & = i
  \partial_\mu\overline{\psi}_{i}(x) \gamma^\mu + e Q_{i} \overline{\psi}_{i}(x) \slashed{A}(x) + i e Q_{i} \bar{R}^{i}(x) c(x)\,,\\
  \frac{\delta \Gamma_{\text{cl}}}{\delta \bar{R}^{i}(x)} & = s\psi_{i}(x) = - i e Q_{i} c(x) \psi_{i}(x)\,,\\
  \frac{\delta \Gamma_{\text{cl}}}{\delta \overline{\psi}_{i}(x)} & = i \slashed{\partial} \psi_{i}(x) - e Q_{i} \slashed{A}(x) \psi_{i}(x) - i e Q_{i} R^{i}(x) c(x)\,,
\label{eq:nonlinearfunctionalderivatives}
\end{align}
\end{subequations}
are non-linear in dynamical fields.
The special feature of linear equations (\ref{eq:AntiGhostEquationAbelianCase}) to (\ref{eq:GaugeFixingAbelianCase}) is that there are no quantum corrections
expected which could spoil these linear relations.\footnote{For loop
  corrections we need interactions, and thus at least three dynamical
  fields which is not the case here.}
Hence, we may require that these identities hold at all orders as part of the definition of the theory, meaning that they also hold for the full effective quantum action $\Gamma$, i.e.\footnote{In case of an Abelian gauge theory with spontaneous symmetry breaking, not all of these identities are valid, but one may introduce background fields which allow obtaining a valid local Ward identity and/or an Abelian antighost equation, see Refs.\ \cite{Haussling:1996rq,Grassi:1997mc}.}
\begin{align}
    \frac{\delta \Gamma}{\delta c(x)} & \stackrel{!}{=} \frac{\delta \Gamma_{\text{cl}}}{\delta c(x)},
    &
    \frac{\delta \Gamma}{\delta \cbar(x)} &\stackrel{!}{=} \frac{\delta \Gamma_{\text{cl}}}{\delta \cbar(x)},
    &
    \frac{\delta \Gamma}{\delta \rho_{\mu}(x)} & \stackrel{!}{=} \frac{\delta \Gamma_{\text{cl}}}{\delta \rho_{\mu}(x)},
    &
    \frac{\delta \Gamma}{\delta B(x)}  &\stackrel{!}{=} \frac{\delta \Gamma_{\text{cl}}}{\delta B(x)}.
\label{RquireAllOrdersRelations}
\end{align}
The charges $Q_{i}$ of all fields explicitly occur in the antighost equation (\ref{eq:AntiGhostEquationAbelianCase}), and (\ref{RquireAllOrdersRelations}) thus fixes the charges of the fields to all orders.

Additionally, we can derive the aforementioned Ward
identity.\footnote{In fact, in the present case without spontaneous
  symmetry breaking and in presence of the identities
  (\ref{RquireAllOrdersRelations})    the following Ward identity is equivalent to the
  Slavnov-Taylor identity.}
 Starting with the Slavnov-Taylor identity for the abelian case 
\begin{align}
  0 = \mathcal{S}({\Gamma}) = \intx \left( \frac{\delta \Gamma}{\delta \bar{R}^{i}} \, \frac{\delta \Gamma}{\delta \psi_{i}} + \frac{\delta \Gamma}{\delta R^{i}} \, \frac{\delta \Gamma}{\delta \overline{\psi}_{i}} + \frac{\delta \Gamma}{\delta \rho^{\mu}} \, \frac{\delta \Gamma}{\delta A_{\mu}} + B \, \frac{\delta \Gamma}{\delta \cbar} \right),
\label{eq:STIAbelianCase}
\end{align}
cf. (\ref{eq:SofFDefinition}) for the non-Abelian case, varying it w.r.t.\ the Fadeev-Popov ghost
$c(x)$, i.e. 
\begingroup
\makeatletter\def\f@size{11}\check@mathfonts
\def\maketag@@@#1{\hbox{\m@th\normalsize\normalfont#1}}\begin{align}
    0 & = \frac{\delta \mathcal{S}({\Gamma})}{\delta c(x)}
    \nonumber\\
    & = \int d^{4}y \Bigg[ \left(\frac{\delta}{\delta c(x)} \frac{\delta \Gamma}{\delta \bar{R}^{i}(y)}\right) \frac{\delta \Gamma}{\delta \psi_{i}(y)} + \frac{\delta \Gamma}{\delta \bar{R}^{i}(y)} \left(\frac{\delta}{\delta c(x)}\frac{\delta \Gamma}{\delta \psi_{i}(y)}\right)
    \nonumber\\
    &\hspace{1.4cm} + \left(\frac{\delta}{\delta c(x)} \frac{\delta \Gamma}{\delta R^{i}(y)}\right) \frac{\delta \Gamma}{\delta \overline{\psi}_{i}(y)} + \frac{\delta \Gamma}{\delta R^{i}(y)} \left(\frac{\delta}{\delta c(x)} \frac{\delta \Gamma}{\delta \overline{\psi}_{i}(y)}\right)
    \nonumber\\
    &\hspace{1.4cm} + \left(\frac{\delta}{\delta c(x)}\frac{\delta \Gamma}{\delta \rho^{\mu}(y)}\right)\frac{\delta \Gamma}{\delta A_{\mu}(y)} - \frac{\delta \Gamma}{\delta \rho^{\mu}(y)} \left(\frac{\delta}{\delta c(x)} \frac{\delta \Gamma}{\delta A_{\mu}(y)}\right)
    \nonumber\\
    &\hspace{1.4cm} + B \left(\frac{\delta}{\delta c(x)} \frac{\delta \Gamma}{\delta \cbar(y)}\right) \Bigg]
\label{STIabelianWIderivation}    \\
    & = \int d^{4}y \Bigg[ \left(\frac{\delta}{\delta \bar{R}^{i}(y)} \frac{\delta \Gamma}{\delta c(x)}\right) \frac{\delta \Gamma}{\delta \psi_{i}(y)} - \frac{\delta \Gamma}{\delta \bar{R}^{i}(y)} \left(\frac{\delta}{\delta \psi_{i}(y)} \frac{\delta \Gamma}{\delta c(x)} \right)
    \nonumber\\
    &\hspace{1.4cm} + \left(\frac{\delta}{\delta R^{i}(y)} \frac{\delta \Gamma}{\delta c(x)} \right) \frac{\delta \Gamma}{\delta \overline{\psi}_{i}(y)} - \frac{\delta \Gamma}{\delta R^{i}(y)} \left(\frac{\delta}{\delta \overline{\psi}_{i}(y)} \frac{\delta \Gamma}{\delta c(x)} \right)
    \nonumber\\
    &\hspace{1.4cm} - \left(\frac{\delta}{\delta \rho^{\mu}(y)} \frac{\delta \Gamma}{\delta c(x)} \right)\frac{\delta \Gamma}{\delta A_{\mu}(y)} - B \left(\frac{\delta}{\delta \cbar(y)} \frac{\delta \Gamma}{\delta c(x)} \right) \Bigg]
    \nonumber\\
    & = - i e Q_{i} \psi_{i}(x) \frac{\delta \Gamma}{\delta \psi_{i}(x)} + i e Q_{i} \bar{R}^{i}(x) \frac{\delta \Gamma}{\delta \bar{R}^{i}(x)} + i e Q_{i} \overline{\psi}_{i}(x) \frac{\delta \Gamma}{\delta \overline{\psi}_{i}(x)}
    \nonumber\\
    &\hspace{0.44cm} - i e Q_{i} R^{i}(x) \frac{\delta \Gamma}{\delta R^{i}(x)} - \partial_{\mu} \frac{\delta \Gamma}{\delta A_{\mu}(x)} - \Box B(x)\,,
    \nonumber
\end{align}
\endgroup
where we used the fact that fermionic objects anti-commute and that
$\delta/\delta c(x)$ is a fermionic functional derivative. After the
third equality we have moved $\delta/\delta c(x)$ past the other
respective functional derivative and utilized the antighost
equation\footnote{This is possible because the antighost equation is
  valid to all orders, see Eq.\ (\ref{RquireAllOrdersRelations}).} (\ref{eq:AntiGhostEquationAbelianCase}). We dropped the penultimate term of the second equality, as the RHS of the antighost equation (\ref{eq:AntiGhostEquationAbelianCase}) does not contain a term depending on $A_{\mu}$.
Rearranging the last line we obtain
the functional form of the local Abelian Ward identity
\begin{align}
  \left( \partial_{\mu} \frac{\delta}{\delta A_{\mu}(x)} + i e Q_{i} \sum_{\Psi} (-1)^{n_{\Psi}} \, \Psi(x) \frac{\delta}{\delta \Psi(x)} \right) \Gamma = - \Box B(x) \,,
\label{eq:FunctionalAbelianWardId}
\end{align}
with $\Psi\in\{\psi_{i},\overline{\psi}_{i},R^{i},\bar{R}^{i}\}$ and $n_{\Psi}\in\{0,1,0,1\}$. The well known Ward identity for the relation of the electron self energy and the electron-electron-photon interaction vertex may then be deduced from this equation.\footnote{Further discussions will be made later in Sec.\ \ref{sec:ApplicationsChiralQED} for the example of chiral QED.} Again, the charges $Q_{i}$ of all fields are fixed as (\ref{eq:FunctionalAbelianWardId}) is established to all orders. Consequently, the above statements imply a non-renormalization of the field charges $Q_{i}$, which means that a single counterterm is sufficient to renormalize the Abelian coupling to all orders of the perturbation theory, thus guaranteeing a consistent renormalization of the coupling constant.

The above identities, viewing them as part of the definition of the theory, constrain the regularization and renormalization procedure. On the one hand, symmetry-preserving (field and parameter) renormalization constants are constrained by the equations (meaning in particular that certain combinations such as the gauge fixing term, or terms such as $\bar{R}^{i}s\psi_{i}$ do not renormalize). On the other hand, particularly the local Ward identity (\ref{eq:FunctionalAbelianWardId}) will be of interest in determining symmetry-restoring counterterms. It can be used to interpret the breaking and restoration of the Slavnov-Taylor identity.
\newpage
\section{Dimensional Regularization}
\label{sec:DReg}

In a perturbative quantum field theory, Feynman diagrams  with closed
loops correspond to higher orders in $\hbar$. They hence represent
genuine quantum corrections and are of fundamental interest. Such
loop diagrams, however, are known to give rise to ultraviolet (UV)
divergences which need to be handled. The reason for this can easily
be understood by imagining a loop made of a propagator with coinciding
end points. Since the propagator is a distribution, one may expect
this object to be ill-defined, as is the product of distributions at
the same space-time point in general. In fact, such loops correspond
to the exchange of virtual particles whose momenta are integrated over
and which may run up to infinity; hence the possibility of divergent
integrals in momentum space.
In essence, the purpose of renormalization is to  remove all
divergences and assign a meaning to such ill-defined expressions and
ultimately   to define physically meaningful results. 

In practice, this means that we first need to isolate the aforementioned
divergences before they can be subtracted.
In the typical setting, isolating divergences is
achieved via regularization, while their subtraction is performed via
counterterms which are added to the Lagrangian. The entire procedure
constitutes the renormalization. Hence, in
order to obtain meaningful results at the quantum level,
i.e.\ including higher order corrections, one needs regularization and
renormalization, as already mentioned at the end of Sec.\ \ref{sec:STIformal}.

There are several regularization schemes; here we focus on dimensional
regularization (DReg).  
In this present section \ref{sec:DReg} and the subsequent section \ref{sec:QAPDReg} we provide an overview of the main properties
of DReg and of how to perform calculations using this regularization
procedure.

Dimensional regularization and its variants are the most common
regularization schemes in relativistic quantum field theories. These
schemes have several key advantages that make them particularly
useful in practical, concrete computations. The structure of integrals
in formally $D$ dimensions is essentially unchanged, allowing
efficient integration techniques. The divergent terms appear as
$1/(D-4)$ poles and can be isolated in a transparent way. 
Lorentz invariance and gauge invariance of non-chiral gauge theories
is essentially kept manifest. Furthermore, fundamental properties such
as equivalence to BPHZ renormalization, consistency with the unitarity and
causality of quantum field theory, and consistent applicability at all
orders are rigorously established. The key disadvantage is the
problematic treatment of the $\gamma_5$ matrix and the
$\epsilon_{\mu\nu\rho\sigma}$ symbol. As a result, gauge invariance is
manifestly broken in chiral gauge theories. The treatment of such theories
is the main topic of the present review.

The previous statements are discussed in detail later in
Sec.\ \ref{sec:renormalization}. That section will explain that based
on DReg, local counterterms exist which can subtract the UV
divergences. It will also explain how the
regularization/counterterm/renormalization procedure in DReg amounts
to a rigorous and physically sensible construction of higher orders.
Then in section \ref{sec:algren}, we will consider DReg applied to
gauge theories and see that
(under certain conditions where chiral gauge anomalies are absent) the
Slavnov-Taylor identity can be established at all orders in the
renormalized, finite theory. In case DReg breaks the symmetry in
intermediate steps, the existence of symmetry-restoring counterterms
is then guaranteed.

The basic idea of DReg is to replace the 4-dimensional spacetime and
the 4-dimensional momentum space by formally $D$-dimensional ones.\footnote{With parametrisation $D=4-2\epsilon$.} In
this way all integrals become formally $D$-dimensional. 
DReg was put forward in several works by 't Hooft and Veltman \cite{tHooft:1972tcz},
by Bollini and Giambiagi \cite{{Bollini:1972ui}} and by Ashmore \cite{Ashmore:1972uj}. Specifically
Ref.\ \cite{tHooft:1972tcz} already highlighted all key advantages and
disadvantages mentioned above and showed how to compute 1-loop and
2-loop Feynman diagrams using DReg.

In the following we begin the section by introducing our notation for the dimensionally regularized and renormalized effective quantum action and schematically sketch its construction. This provides a short overview of the general structure of dimensional regularization and renormalization (subsection\ \ref{sec:DRegGamma}).

Then we  explain what the properties of $D$-dimensional integrals
are and how these integrals can be consistently defined (subsection
\ref{sec:integrals}). Together with the integrals, many other quantities have to be
formally continued to  $D$ dimensions, in particular momenta, vector
fields, metric tensors, and $\gamma$ matrices. Subsection \ref{sec:gammamatrices} 
focuses on such quantities and delineate to what extent a purely
$D$-dimensional treatment is correct and at which points a distinction
of 4-dimensional and $D$-dimensional quantities needs to be made in
calculations. In particular, it introduces the BMHV scheme for
non-anticommuting $\gamma_5$.

Subsection \ref{sec:relationL} describes an important feature of DReg which is not
shared by all regularization methods: the precise expressions of regularized
Feynman diagrams in $D$ dimensions may be encoded in a formally
$D$-dimensional Lagrangian, from which Feynman rules are obtained in
the usual way. This relation is obviously useful in the study of symmetries of
regularized Feynman diagrams since properties of diagrams can be
obtained from properties of the regularized Lagrangian. In subsection
\ref{sec:variants} we discuss several variants of DReg such as regularization by
dimensional reduction and further sub-variants. We  discuss
relationships between the variants on the level of the regularized
Lagrangians and on the level of Green functions and S-matrix elements.
\subsection{General structure of Dimensional Regularization and Renormalization}
\label{sec:DRegGamma}
Before we discuss properties of $D$-dimensional integrals and how to formally continue certain quantities to $D$ dimensions, and thus 
perform calculations in DReg, we briefly introduce our notation
w.r.t.\ the dimensionally regularized and renormalized effective
quantum action, the key quantity of the theory, and sketch its construction.

As mentioned above, UV divergences  in loop integrals  are isolated as
$1/(D-4)$ poles in DReg. These divergences must be subtracted using
counterterms in order to renormalize the theory. In general, such
counterterms may not only contain these UV divergent but also finite
contributions.\footnote{The general counterterm structure of a
  dimensionally regularized theory using the BMHV scheme makes use of
  further subdivisions of counterterms. This will be
  presented in Sec.\ \ref{sec:algren} and illustrated in a practical
  example in Sec.\ \ref{sec:ApplicationsChiralQED}.}
Here we sketch the renormalization procedure and introduce useful
notation.

The perturbative expansion is  organized in terms of orders in
$\hbar$, equivalent to orders in loops. The classical action  of order $\hbar^0$ defining
the theory is denoted $S_{0}\equiv \Gamma_{\text{cl}}$, the counterterm action is denoted as
$S_{\text{ct}}$; the sum of the two is called the bare action
$S_{\text{bare}}$.
In the following, symbols without an upper index denote all-order quantities, while for perturbative expressions, an upper index $i$ labels quantities of precisely order $i$, whereas quantities up to and including order $i$ are labelled with an upper index $(i)$. Using this notation, the bare and the counterterm actions may be written as
\begin{align}
\label{eq:IntroductionOfSbareAndSct}
  S_{\text{bare}} &= S_{0} + S_{\text{ct}}, 
  &
  S_{\text{ct}} &= \sum_{i=1}^{\infty} S_{\text{ct}}^{i},
  &
  S_{\text{ct}}^{(i)} &= \sum_{j=1}^{i} S_{\text{ct}}^{j}.
\end{align}
In dimensional regularization and renormalization the perturbative construction of the effective action is performed iteratively at each order of $\hbar$, i.e.\ at each loop order, starting from the tree-level action $S_{0}$. Then, a counterterm action $S_{\text{ct}}^{i}$ needs to be constructed at each higher order $i\geq1$ which has to satisfy the two conditions that the renormalized theory is UV finite and in agreement with all required symmetries.

The subrenormalized quantum action of order $i$ is denoted by
\begin{align}
  \Gamma^{i}_{\text{subren}}
\end{align}
and obtained at order $i$ by using
Feynman rules from the tree-level action and counterterms up to order
$i-1$. The counterterms $S_{\text{ct}}^i$ to be constructed at the order $i$ are
subdivided into singular counterterms (which by
definition contain only pole terms in $(D-4)$ and are denoted by
subscript $_\text{sct}$) and finite counterterms
(finite in the limit $D\to4$  and  denoted by
subscript $_\text{fct}$). 
By constructing and including singular counterterms  of the order $i$ we obtain 
\begin{align}
\label{subrenplussct}  \lim_{D \to 4} \, \Big( \Gamma^i_{\text{subren}} + S_{\text{sct}}^i \Big) = \text{finite},
\end{align}
which determines the singular counterterms unambiguously. If necessary
we may then also include additional finite counterterms.
Once the finite counterterms are determined, we obtain
\begin{align}
  \Gamma_{\text{DRen}}^i \equiv \Gamma^i_{\text{subren}} + S_{\text{sct}}^i + S_{\text{fct}}^i.
\end{align}
This quantity $\Gamma_{\text{DRen}}^i$ is finite and essentially
renormalized, but it may still contain the variable $\epsilon = (4-D)/2$
and so-called evanescent quantities, which vanish in strictly
$D=4$ dimensions. Thus, the completely renormalized quantum action is
obtained by taking the limit $D \rightarrow 4$ and setting all
evanescent quantities to zero. This procedure is denoted
by\footnote{We will sometimes synonymously refer to the completely renormalized and $4$-dimensional quantum action as $\Gamma^i_{\text{ren}}$, i.e.\ $\Gamma^i \equiv \Gamma^i_{\text{ren}}$, in order to emphasize that it is completely renormalized.}
\begin{align}
  \label{DefLIM}
  \Gamma^i \equiv \mathop{\text{LIM}}_{D \, \to \, 4} \, \Gamma_{\text{DRen}}^i. \end{align}

Some comments on the finite counterterms are in order.
They can have two purposes.
On the one hand, it may happen that regularized quantum corrections spoil a
symmetry of the theory, such that e.g.\ the Slavnov-Taylor
identity is invalid on the level of
Eq.\ (\ref{subrenplussct}). If the symmetry is part of the definition
of the theory, finite counterterms must be found and added such that
the symmetry is valid on the renormalized level (\ref{DefLIM}).
The purpose
of counterterms is then not solely to remove UV divergences but also to restore symmetries if necessary (and if possible).
If no finite counterterms can be found that restore the symmetry, the
symmetry is lost. This situation is called an anomaly, or anomalous
symmetry breaking. It signals an irreconcilable clash of the symmetry and the quantum
theory.\footnote{  If the symmetry is part of the definition of the theory or required
  for the consistency of the theory, the theory must be abandoned.}
The later Sec.\ \ref{sec:algebraicrenormalizationDetails} will provide a
detailed discussion of the symmetry restoration using finite counterterms.

On the other hand, the finite counterterms can also be used in order to fulfil
certain renormalization conditions. In general, the choice of the
finite counterterms (beyond symmetry restoration) is called a
renormalization scheme. Popular examples of renormalization schemes
are on-shell or (modified) minimal subtraction schemes. In the present
review we will not further discuss renormalization schemes. For 
textbook-level discussions of this important topic we refer to
the books by B\"ohm/Denner/Joos and Srednicki
\cite{Bohm:2001yx,Srednicki:2007qs}.\footnote{Although the main focus of the review is on the renormalization of Green
functions, we provide here a remark on the extraction of physical S-matrix
elements via LSZ reduction as mentioned  in
Sec.\ \ref{sec:GreenFunctions}. LSZ reduction involves the need for
so-called wave function renormalization, which ties in with the
discussion of finite counterterms and renormalization schemes. In
order to obtain properly normalized 
S-matrix elements, Green functions need to be divided by $\sqrt{z_i}$
for each external line, where $z_i$ is the residue of the
corresponding two-point  function   at the pole corresponding to the
rest mass of the considered external particle $i$. This may be
automatically achieved by choosing an on-shell renormalization scheme
for renormalized fields, where all such residues are equal to unity, see e.g.\ the discussion
in Ref.\ \cite{Bohm:2001yx}. If a different renormalization scheme is
chosen, the wave function factor $\sqrt{z_i}$ may be different from
unity and needs to be explicitly taken into account, such as in the
scheme proposed in Ref.\ \cite{Bohm:1986rj} for the electroweak
Standard Model. In practical computations in DReg, it is actually
often possible to carry out the renormalization programme only
partially, such that quantum fields remain unrenormalized and the
residue factors
$\sqrt{z_i}$ remain divergent. After LSZ reduction and proper wave
function renormalization, nevertheless finite and correct S-matrix
elements can be obtained.
  }

Finally, we reiterate that we have only sketched the general procedure and
introduced notation, but we have not yet proven that this procedure
actually works. This will be done in the later sections
\ref{sec:renormalization} and \ref{sec:algren}, and it is exemplarily
illustrated in section \ref{sec:ApplicationsChiralQED} for the case
where finite symmetry-restoring counterterms are
required.\footnote{In textbooks and in practical computations,
  counterterms are often
  obtained by applying a so-called renormalization transformation onto
  the tree-level action. Section \ref{sec:QAPexamples} and, in more
  generality, Sec.\ \ref{sec:ctsymmetrypreserving} will also
  explain under which conditions this procedure  is possible.}

\subsection{Integrals in $D$ Dimensions}
\label{sec:integrals}

In this subsection we will discuss momentum integrations in
DReg. 
As explained above, in DReg we replace $4$-dimensional spaces by formally $D$-dimensional ones. In this way all integrals become formally $D$-dimensional
and we can
schematically write for the loop integration measure
\begin{align}
  \int \frac{d^4k}{(2\pi)^4}
  & \to
  \mu^{4-D}\int \frac{d^Dk}{(2\pi)^D}
  \,,
\label{DRegbasicidea}
\end{align}
where $\mu$ denotes a new, artificial mass scale, the dimensional
regularization scale.
Though the basic idea \cite{tHooft:1972tcz,Bollini:1972ui,Ashmore:1972uj} is simple, care is needed to avoid
incorrect or inconsistent results. After first detailed discussions in
Ref.\ \cite{Wilson:1972cf,Speer:1970ss,Speer:1974cz}, very systematic definitions and analyses of
$D$-dimensional integrals were given by Breitenlohner and Maison
\cite{Breitenlohner:1977hr} and by Collins \cite{Collins_1984}.

\subsubsection{Quasi-$D$-Dimensional~Space}
\label{sec:QDS}
Before discussing integrals we discuss the simpler concept of a
$D$-dimensional space. Let us denote the original 4-dimensional
Minkowski space as 4S and the formal, or quasi-$D$-dimensional space
as Q$D$S. The question is which properties Q$D$S can have and what its
relationship to the original space 4S can be.

Clearly, even on the regularized level we need the usual properties of
linear combinations. If two momenta $p^\mu$ and $q^\mu$ are elements
of Q$D$S, then also $a p^\mu + b q^\mu$ is an element of Q$D$S for any
real or complex $a$ and $b$, with the usual properties of linear
combinations. Hence Q$D$S must constitute a proper mathematical vector
space.
However, there do not exist mathematical vector spaces with
dimensionality $D$ if $D$ is a non-integer real or complex number.

The crucial observation \cite{Wilson:1972cf} is that on the regularized level
we need to accept that arbitrary sets of momentum vectors may have to
be treated as linearly independent. Hence we need to accept that Q$D$S
must actually be an infinite dimensional vector
space. Correspondingly, what we call $D$-dimensional momentum vectors
are actually elements of Q$D$S with infinitely many components (of
course, in the case of physical momenta, only four of them will be
nonzero). It turns out to be possible to define objects and operations
on Q$D$S with the desired properties which resemble $D$-dimensional
behavior, justifying the name quasi-$D$-dimensional space.

An important consequence for practical applications is that the
original space 4S is always a subspace of Q$D$S,
\begin{align}
  \text{4S} \subset \text{Q$D$S}\,,
\label{4Ssubspace}
\end{align}
regardless whether $D>4$ or $D<4$ or $D$ is complex. Assuming the
opposite relation leads to mathematical inconsistencies, which will be
discussed in the context of dimensional reduction below in
Sec.\ \ref{sec:variants}.

\subsubsection{Properties of $D$-Dimensional~Integrals}
\label{sec:DdimIntegrals1}
Now we turn to integrals over functions of vectors defined on Q$D$S.
Clearly, the plethora of successful calculations and
available multi-loop techniques (see e.g.\ the book \cite{Smirnov:2004ym}) provides
ample evidence of the existence  of $D$-dimensional integrals and of
the consistency of their evaluations.
Still, as stressed in Ref.\ \cite{Collins_1984}, it is important to
establish the existence of $D$-dimensional integrals in general, and
to prove the uniqueness of the results.
In the literature, different constructive
definitions have been proposed. Here we will describe the construction
by Collins \cite{Collins_1984}, which extends earlier work by
Wilson \cite{Wilson:1972cf}. 

We begin by listing important properties of $D$-dimensional
integration given in Ref.\ \cite{Collins_1984}. 
It is generally sufficient to discuss the case of
Euclidean metric. $D$-dimensional
Minkowski spacetime can then be treated as one fixed time dimension
combined with $(D-1)$-dimensional Euclidean space, and in quantum
field theory applications Minkowski space integrals can be converted
to Euclidean space integrals via Wick rotation. Depending on the
context either Minkowski space or Euclidean space notation can be more
convenient. For the following integrals we assume Euclidean space, with
Euclidean metric for scalar products of vectors.
\begin{description}
\item[Property a)] Linearity: for all functions $f_{1,2}$ and
  coefficients $a,b$,
  \begin{align}
    \label{linearity}
    \int d^Dk \left(a f_1(\vec{k}) + b f_2(\vec{k})\right) & = 
    a\int d^Dk  f_1(\vec{k}) + 
    b\int d^Dk  f_2(\vec{k}) \, .
  \end{align}
\item[Property b)] Translation invariance: for all vectors $\vec{p}\in$Q$D$S,
  \begin{align}
    \label{translationinvariance}
    \int d^Dk f_1(\vec{k}+\vec{p}) & =     \int d^Dk f_1(\vec{k}) \,.
  \end{align}
\item[Property c)] Scaling: for all numbers $s$,
  \begin{align}
    \label{scaling}
    \int d^Dk f_1(s\vec{k}) & =    s^{-D} \int d^Dk f_1(\vec{k}) \,.
  \end{align}
\item[Property d)] The $D$-dimensional Gaussian integral in $D$-dimensional
  Euclidean metric has the value
  \begin{align}
    \int d^Dk e^{-\vec{k}^2  } & = \pi^{D/2} \,.
    \label{gaussintegral}
  \end{align}
  Using $D$-dimensional spherical coordinates to evaluate this
  rotationally symmetric integral, $\int d^Dk \to \int d^{D-1}\Omega
  \int_0^\infty dk k^{D-1} e^{-k^2}$, implies the result for the
  surface of $D$-dimensional sphere 
  \begin{align}
    \Omega_D & \equiv \int d^{D-1}\Omega = \frac{2\pi^{D/2}}{\Gamma(D/2)}
    \label{OmegaD}
  \end{align}
  which depends on the well-known $\Gamma$-function defined as
  $\Gamma(z)=\int_0^\infty t^{z-1}e^{-t}dt$ for $\text{Re}(z)>0$ and by
  analytic continuation otherwise.
\item[Remark: ]
  Properties a,b,c,d may also be viewed as axioms on the
  integration. Taken together, they uniquely fix the integration
  \cite{Wilson:1972cf}. 
\item[Property e)] Commutation with differentiation
  \begin{align}
    \label{commutedifferentiation}
    \frac{\partial}{\partial \vec{p}}\int d^Dk f_1(\vec{k},\vec{p}) &
    =
    \int d^Dk     \frac{\partial}{\partial \vec{p}}f_1(\vec{k},\vec{p})\,.
  \end{align}
\item[Property f)] Partial integration:
  The previous equation, together with translation invariance
  (\ref{translationinvariance}), implies the possibility for  partial
  integration
  \begin{align}
    \label{partialintegration}
    \int d^Dk     \frac{\partial}{\partial \vec{k}}f_1(\vec{k}) & = 0
    \,.
  \end{align}
\item[Property g)] Two different integrations can be interchanged
  \begin{align}
    \label{Ddimintegralinterchange}
    \int d^Dp\int d^Dk f(\vec{p},\vec{k}) & =
    \int d^Dk\int d^Dp f(\vec{p},\vec{k}) \,.
  \end{align}
\item[Property h)] 
  If an integral is finite in 4 dimensions, the $D$-dimensional version is analytic in
  a region for  $D$ around $D=4$ and in the external momenta, and it
  reproduces the original value for $D=4$.
\item[Remark: ]
  The explicit construction of Refs.\ \cite{Collins_1984,Wilson:1972cf} guarantees the
  existence of the $D$-dimensional integration and allow to establish
  general properties. Uniqueness together
  with existence implies ``consistency'' in the sense that  one
  initial expression in DReg will always 
  lead to one unique final expression, no matter how and in which
  order calculational steps are organized.
\end{description}

\subsubsection{Uniqueness and Construction of $D$-Dimensional Integrals Using
 Parallel and Orthogonal~Spaces}

For the full proofs of the properties listed above and for further properties we refer to
Ref.\ \cite{Collins_1984}.
In the following we summarize the uniqueness proof  
and then sketch the integral constructions of Refs.\ \cite{Wilson:1972cf,Collins_1984}.

We begin with the uniqueness of the $D$-dimensional
integral. It is sufficient to assume Euclidean metric,  such that
scalar products are given by
$\vec{p}\cdot\vec{k} = p_1k_1+p_2k_2+\ldots$ for $D$-dimensional
vectors  $\vec{p},\vec{k}$. Ref.\ \cite{Wilson:1972cf}
starts from the observation that any 
function of the form $f(\vec{p}_1\cdot \vec{k}, \ldots ,
\vec{p}_n\cdot 
\vec{k}, \vec{k}^2)$ can be 
obtained from suitable combinations of derivatives\footnote{  Derivatives with respect to $\vec{p}$ and $s$ generate arbitrary
  polynomials in all components of $\vec{k}$ and $\vec{k}^2$, multiplied
  by $g(s,\vec{p},\vec{k})$. Ignoring convergence questions, any
  function can be sufficiently approximated in this way.}
of the generating
function
\begin{align}\label{generatingfunctionWilson}
  g(s,\vec{p},\vec{k}) & \equiv e^{-s \vec{k}^2 + \vec{p}\cdot \vec{k}} \,.
\end{align}
Using linearity (\ref{linearity}) it is sufficient to prove uniqueness
of the integral over the 
generating function $g(s,\vec{p},\vec{k})$. Using 
translation invariance to complete the square, scaling, and the $D$-dimensional Gaussian
integral we obtain
\begin{subequations}
\begin{align}
  \int d^Dk g(s,\vec{p},\vec{k}) & \stackrel{(\ref{translationinvariance})}{=}
  \int d^Dk e^{-s \vec{k}^2  + \vec{p}^2/4s} \\
  & \stackrel{(\ref{scaling})}{=}
  s^{-D/2}  e^{\vec{p}^2/4s}\int d^Dk e^{-\vec{k}^2  } \\
  & \stackrel{(\ref{gaussintegral})}{=}
  s^{-D/2}  e^{\vec{p}^2/4s} \pi^{D/2} \,.
\end{align}
\end{subequations}
The integral over the generating function is uniquely fixed given the
four properties
(\ref{linearity}, \ref{translationinvariance}, \ref{scaling}, \ref{gaussintegral}),
establishing general uniqueness of the integral.

Now we sketch the $D$-dimensional integral construction proposed by
Refs.\ \cite{Wilson:1972cf,Collins_1984}. Suppose the function
$f(\vec{p}_1\cdot \vec{k}, \ldots \vec{p}_n\cdot
\vec{k}, \vec{k}^2)$ is to be integrated over  $\vec{k}$, and we take
seriously that all these vectors are elements of Q$D$S, which is
actually infinite dimensional. The result
will depend on the $n$ ``external momenta''
$\vec{p}_1\ldots\vec{p}_n$, and these span a subspace which is at most
$n$-dimensional. The basic idea is then that the space of all
$\vec{k}$ can be split into a ``parallel''
space and an ``orthogonal'' space. The parallel space is defined such
that it contains all $n$ external vectors
$\vec{p}_1\ldots\vec{p}_n$. It has a finite,
integer dimensionality $n_p$. Once the parallel space is fixed we can
uniquely decompose any loop 
momentum and its scalar products as 
\begin{align}
  \vec{k} & = \vec{k}_\parallel + \vec{k}_\perp
  &
  \vec{p}_i\cdot\vec{k} & = \vec{p}_i\cdot\vec{k}_\parallel
  &
  \vec{k}^2 & = \vec{k}_\parallel^2 + \vec{k}_\perp^2 \,.
\end{align}
For this reason the $\vec{k}$ dependence of the integrand may be
abbreviated as 
\begin{align}
 f(\vec{p}_1\cdot \vec{k}, \ldots \vec{p}_n\cdot
 \vec{k}, \vec{k}^2) & \equiv
 f(\vec{k}_\parallel, \vec{k}_\perp^2)
 \,,
\end{align}
i.e.\ the $\vec{k}$ dependence is separated: the vector
$\vec{k}_\parallel$ appears explicitly but it is an element of a
finite-dimensional vector 
space where ordinary integrals are defined. The orthogonal components
appear only as the square $\vec{k}_\perp^2$.
This is the crucial simplification, which
allows the two-step definition where first the integral is split as
\begin{align}
  \int d^Dk f(\vec{p}_1\cdot \vec{k}, \ldots \vec{p}_n\cdot
  \vec{k}, \vec{k}^2) & \equiv
  \int d^{n_p}k_\parallel \int
  d^{D-n_p}k_\perp  f(\vec{k}_\parallel, \vec{k}_\perp^2)
\label{DefStep1}
\end{align}
and second the $D-n_p$-dimensional integral on the right-hand side is
defined via spherical coordinates, using Eq.\ (\ref{OmegaD}),
\begin{align}
  \int
  d^{D-n_p}k_\perp  f(\vec{k}_\parallel, \vec{k}_\perp^2) & \equiv
  \Omega_{D-n_p}\int_0^\infty dk k^{D-n_p-1} f(\vec{k}_\parallel,k^2) \,.
\label{DefStep2}
\end{align}
In these two steps the original $D$-dimensional integral has been
defined in terms of a series of ordinary integrals in one dimension
and in $n_p$ dimensions.
The effect of the regularization becomes manifest as the
$D$-dependence in the exponent, which governs the behaviour of the
integrand at large $k$ and at small $k$. If the function $f$ has at
most a power-like divergence at large/small $k$, there is a range of
$D$ for which the $k$-integral is well defined. Its value for
arbitrary $D$ is then defined by analytical continuation.

Ref.\ \cite{Collins_1984} provides detailed discussions of the independence
of the choice of the parallel space and its dimensionality $n_p$, of
the analytical continuation in the variable $D$, and of more
general integrals.

We will now discuss the computation of such integrals with two
examples, which will illustrate several important general points.
The examples are (we again work in Euclidean space and use a
dimensionless integration variable $\vec{k}$)
\begin{subequations}
  \label{DefIallI1}
\begin{align}
  I_\text{all}(D) & = \int d^Dk \vec{k}^2\delta(\vec{k}^2-1)\,,
  \\
  I_\text{1}(D) &  = \int d^Dk k_1^2\delta(\vec{k}^2-1)\,.
\end{align}
\end{subequations}
Both integrals only depend on the dimensionality $D$. In both cases we
essentially integrate over the surface of the unit sphere, in the
first case multiplied by $\vec{k}^2$ and in the second case multiplied
by $k_1^2$. Since no direction is special, the second integral would
not change if we replaced $k_1^2$ by any other $k_i^2$ with a fixed
index $i$. We will discover a useful relationship between the two
integrals.

The first integral may immediately be computed by treating the entire
$\vec{k}$ as $\vec{k}_\perp$. We can apply the definition
(\ref{DefStep2}) and evaluate the integral as
\begin{align}
  I_\text{all}(D) & = \frac{\Omega_D}{2} \,.
\end{align}
For the second integral we treat the first component as special and
align the parallel space along this first component (the explicit
component $k_1$ might also be regarded as the scalar product
$\vec{p}\cdot\vec{k}$ with a vector that happens to be
$\vec{p}=(1,0,0,\ldots)$). Then the integral
becomes by definition
\begin{align}
  I_\text{1}(D) & = \int_{-\infty}^{\infty}dk_1 k_1^2 \int d^{D-1}k_\perp
  \delta(\vec{k}_\perp^2 - (1-k_1^2)) \,.
\end{align}
The $D-1$-dimensional integral is now of the same type as
$I_\text{all}$ except in reduced dimensionality, and it is only
nonzero if $|k_1|\le1$. Applying standard
substitutions we obtain
\begin{align}
  I_\text{1}(D) & = \int_{-1}^{1}dk_1 k_1^2
  \frac{\Omega_{D-1}}{2}(1-k_1^2)^{(D-3)/2} \,.
\end{align}
The remaining integral can be related to the definition of the
Beta-function  $B(3/2,(D-1)/2)$ by the substitution $x=k_1^2$, and the result is
\begin{align}
  I_\text{1}(D) & = \frac{\Omega_{D-1}}{2}\frac{\Gamma(3/2)
    \Gamma(D/2-1/2)}{\Gamma(D/2+1)} \,.
\label{Result1I1}
\end{align}
As announced these results illustrate important general points:
\begin{itemize}
\item
  The result (\ref{Result1I1}) can be simplified by using the explicit
  result $\Gamma(3/2) = \sqrt{\pi}/2$, the recursion relation
  $z\Gamma(z)=\Gamma(z+1)$ and the explicit result for $\Omega_D$ in
  Eq.\ (\ref{OmegaD}). After simplification we obtain
\begin{align}
  I_\text{1}(D) & = \frac{\Omega_{D}}{2D}\,,
\label{Result1I2}
\end{align}
where the $(D-1)$-dimensional surface volume is replaced by the
$D$-dimensional one.
\item
  As a result we simply obtain the relation
  \begin{align}
    \label{ResultIallI1}
    I_\text{all}(D) & = D I_\text{1}(D)\,,
  \end{align}
  which agrees with the naive expectation from a $D$-dimensional space
  with $D$ vector components despite the construction of Q$D$S as an
  infinite dimensional vector space.
\item
  These two integrals $I_\text{all}$ and $I_{1}$ and their relationships will allow defining
  metric tensors on the quasi-$D$-dimensional space Q$D$S with
  appropriate properties resembling $D$-dimensional behavior.
\item
  Similar relationships are also the essence of the proof of the
  independence of the choice of the parallel space in defining the
  integrals \cite{Collins_1984}.
\end{itemize}

\subsubsection{Construction of $D$-Dimensional Loop Integrals via
 Schwinger~Parametrization}

\label{sec:Schwingerparametrization}

In addition to the integral construction via parallel and orthogonal
spaces, we also sketch a second way to construct $D$-dimensional
integrals. This second way was carried out and used in particular
in Refs.\ \cite{Speer:1974cz,Breitenlohner:1977hr}. It also realizes the four basic
properties of linearity, translation invariance, scaling and the
generalization of the Gaussian integral
(\ref{linearity}, \ref{translationinvariance}, \ref{scaling}, \ref{gaussintegral}),
but otherwise it
is formulated specifically for loop
integrals in Minkowski space
quantum field theory. It is based on the well-understood
Schwinger parametrization, which has been developed for arbitrary loop
integrals and used e.g.\ in BPHZ renormalizability proofs in
Refs.\ \cite{Hepp:1966eg,Anikin:1973ra,Bergere:1974zh}
and in the context of analytical regularization \cite{Speer:1971fub}. For
general accounts, see also the books
Refs.\ \cite{Smirnov:2004ym,nakanishi1971graph}. We present here first a simple example
and then indicate the general case.

The example is a standard one-loop two-point function with loop
integrand
\begin{align}
  \frac{i^2 e^{i(u_1^\mu (k+p)_\mu+u_2^\mu k_\mu)}
  }{[(k+p)^2-m^2+i\varepsilon][k^2-m^2+i\varepsilon]} \equiv \frac{i^2
    e^{i(u_1^\mu (k+p)_\mu+u_2^\mu k_\mu)}}{D_1D_2}
  \label{exampleintegrand}
\end{align}
with loop integration momentum $k$ and external momentum $p$, two
equal masses and the
customary 
$+i\varepsilon$ prescription. We also allowed for a generating
function in the numerator similar to Eq.\ (\ref{generatingfunctionWilson}) with two vector-like parameters $u_1^\mu$,
$u_2^\mu$ such that derivatives at $u_{1,2}=0$ can generate arbitrary polynomials of
propagator momenta in the numerator.
The Schwinger parametrization, or  $\alpha$-parametrization, uses the
following replacement for generic propagators,
\begin{align}
  \frac{1}{[p^2-m^2+i\varepsilon]^\nu}
  & =
  \frac{1}{i^\nu\Gamma(\nu)}
  \int_0^\infty d\alpha \alpha^{\nu-1}
  e^{i(p^2-m^2+i\varepsilon)\alpha} \,,
\end{align}
which is derived by substitution and by using the definition of the 
$\Gamma$ function. In this way the integrand (\ref{exampleintegrand})
becomes
\begin{align}
  \int_0^\infty d\alpha_1 d\alpha_2
  e^{i(D_1\alpha_1+D_2\alpha_2)}
  e^{i(u_1\cdot (k+p)+u_2\cdot k)}
\end{align}
and the appearing exponent is a quadratic polynomial in the loop
momentum which, up to the factor $i$, can be written as\footnote{  Note that in this particular case, the quantity $M$ is a number,
  while in the general case of multiloop integrals $M$ will be a matrix.}
\begin{align}
  k^2 M + 2 k^\mu J_\mu + K + K'\,,
\end{align}
or, by completing the square, as
\begin{align}
  k'{}^2 M - J^2 M^{-1} + K + K'\,,
\end{align}
with
\begin{subequations}
\begin{align}
  k'_\mu & = k_\mu+M^{-1}J_\mu \,,\\
  M & = \alpha_1+\alpha_2 \,,\\
  J_\mu & = p_\mu \alpha_1 + \frac{1}{2}(u_1+u_2)_\mu\,,\\
  K & = p^2 \alpha_1 + u_1\cdot p\,,\\
  K' & = (i\varepsilon-m^2 )(\alpha_1+\alpha_2)\,.
\end{align}
\end{subequations}
Using this rearrangement in the exponent, the loop integral over $k$
becomes essentially a Gaussian integral over $e^{ik'{}^2M}$. Using
translation
invariance and the scaling property
(\ref{translationinvariance},\ref{scaling}) and employing Minkowski
metric we obtain
\begin{align}
  \int \frac{d^Dk}{(2\pi)^D}  e^{i(k'{}^2+i\varepsilon)M} & =
  (4\pi)^{-D/2} i^{1-D/2} M^{-D/2}\,.
\end{align}
The previous steps have transformed the integrand
(\ref{exampleintegrand}) into a product of a purely Gaussian integrand
and a remainder which does not depend on the integration
momentum. This leads to the following definition
\begin{align}
  \int \frac{d^Dk}{(2\pi)^D} \frac{i^2  e^{i(u_1\cdot (k+p)+u_2\cdot k)}
}{D_1D_2} =&
  (4\pi)^{-D/2} i^{1-D/2} \nonumber\\&\times
  \int_0^\infty d\alpha_1 d\alpha_2 
  M^{-D/2} e^{i(- J^2 M^{-1} + K + K')}
  \,.
  \label{oneloopresultalpha}
\end{align}
In this way the $D$-dimensional integral is defined in terms of two
standard integrals over $\alpha_{1,2}$. The integrand depends on
$\alpha_{1,2}$ via the exponential function and via the term
$M^{-D/2}$, where the $D$-dependence enters.

This example can be generalized to arbitrary loop integrals, and it
may be generalized to numerator polynomials in the integration
momentum. We provide here the result for the general case of a 1PI
graph $G$ with $L$
loops, loop momenta $k_i$ and $I$ internal lines with momenta
$\ell_k$, a generating function with parameters $u_k$ and a derivative
operator $Z(-i\partial/\partial_u)$ with respect to all the $u_k$ in the numerator (see
e.g.\ \cite{Breitenlohner:1977hr,Smirnov:2004ym}) 
\begin{align}
  {\cal T}_{G} & =
  \int d^Dk_1\ldots d^Dk_L  Z(-i\partial/\partial_u)\frac{i^I e^{iu_k\cdot\ell_k}}{D_1\ldots D_I}
\Bigg|_{u=0}  \,.
\label{loopintegralgeneral}
\end{align}
Selecting specific choices of the operator $Z$ and setting $u=0$ after
taking the derivative produces specific numerators.
Going through similar steps as before the integrand can be rearranged into the
form of pure Gaussian integrals,
leading to the result and $D$-dimensional definition
\begin{subequations}
\begin{align}
  {\cal T}_{G} & =
   c_D^L
  \int_0^\infty d\alpha_1\ldots d\alpha_I
  Z(-i\partial/\partial_u){\cal U}^{-D/2} e^{iW}
  \Bigg|_{u=0}
  \label{alpharesultgeneral}
  \,, \\
  c_D & =   i^{1-D/2} (4\pi)^{-D/2} \,.
\end{align}
\end{subequations}
By definition the variables $u$ have to be set to zero before performing
the $\alpha$ integration.
The formula clearly corresponds to the one-loop example where $L=1$, $I=2$ and
$Z=1$ and
\begin{subequations}
  \label{UWexample}
  \begin{align}
  {\cal U} & = M = \alpha_1+\alpha_2 \,, \\
  W & = \frac{p^2\alpha_1\alpha_2-\alpha_1 u_2\cdot p+\alpha_2
    u_1\cdot p-\frac14(u_1+u_2)^2}{{\cal U}} + K'\,.
\end{align}
\end{subequations}
In the general case, the quantities in the result
(\ref{alpharesultgeneral}) have the following properties: 
\begin{itemize}
\item
  ${\cal U}$ is a so-called Symanzik polynomial in the $\alpha$'s of
  degree $L$. All its 
  terms have unity coefficient, hence inside the $\alpha$-integration 
  range ${\cal U}$ is positive.
\item
 The ultraviolet divergences (including subdivergences) of the original
  loop integral are mapped to 
  singularities of the $\alpha$ integrals at small $\alpha$. As some
  of the $\alpha$'s approach zero, ${\cal U}$ vanishes with a  certain
  power-like behavior, depending on the original power counting of
  the Feynman diagram. The $D$-dependence of ${\cal U}^{-D/2}$ then
  effectively regularizes the divergences.
\item
  The exponent $W$  is a rational function in the $\alpha$'s and
  depends on the external momenta, the masses, and the $u_k$
  variables.
\end{itemize}
The definition of the general loop integral
(\ref{loopintegralgeneral}) via Eq.\ (\ref{alpharesultgeneral}) 
provides not only a second constructive definition of $D$-dimensional
integration (which is of course equivalent to the one in
Sec.\ \ref{sec:DdimIntegrals1} 
thanks to the uniqueness theorem), but it also provides a starting
point for practical computations, and it allows rigorous proofs of
renormalizability and further renormalization properties \cite{Speer:1974cz,Breitenlohner:1977hr}.

For completeness we present here briefly the full computation of the
one-loop example (\ref{oneloopresultalpha}) for the scalar numerator case where $u_{1,2}=0$. With the substitutions
$\alpha=\alpha_1+\alpha_2$ and $\beta=\alpha_1/\alpha$ we obtain
\begin{align}
  \text{(\ref{oneloopresultalpha})} & =
  (4\pi)^{-D/2} i^{1-D/2}  \int_0^\infty d\alpha \int_0^1 d\beta 
  \alpha^{1-D/2} e^{-i\alpha Q(\beta)}
\end{align}
with
\begin{align}
  Q(\beta) & =- p^2 \beta(1-\beta) + m^2-i\varepsilon \,.
\end{align}
The $\alpha$-integration is given by the $\Gamma$ function up to a
substitution, so we obtain the final expression
\begin{align}
  \text{(\ref{oneloopresultalpha})} & =
  -i(4\pi)^{-D/2} \Gamma(2-D/2) \int_0^1 d\beta
  Q(\beta)^{D/2-2} \,,
\end{align}
which is the well-known one-dimensional integral representation of the
result.

\subsection{Metric Tensors, $\gamma$ Matrices, and Other Covariants in
 $D$ Dimensions}
\label{sec:gammamatrices}

In this subsection we will discuss covariant objects used in DReg
calculations, such as momentum vectors $k^\mu$, vector fields 
$A^\mu(x)$, $\gamma^\mu$ matrices and the metric tensor
$g^{\mu\nu}$. We will first provide a summary of the basic properties
which are often sufficient in practical calculations. Afterwards we
will give details on the explicit construction of the required objects
on the quasi-$D$-dimensional space Q$D$S. As in the case of integrals,
the explicit construction is important to guarantee the consistency of
the calculational rules.

In the context of Eq.\ (\ref{4Ssubspace}) we have seen that the original
4-dimensional Minkowski space is necessarily a subspace of
Q$D$S. Hence strictly 4-dimensional objects always exist in addition
to the quasi $D$-dimensional ones, and we will discuss the
relevant relationships. At the end of the subsection we will discuss
the objects $\gamma_5$ and $\epsilon_{\mu\nu\rho\sigma}$, which are
tied to strictly four dimensions.

\subsubsection{Properties of $D$-Dimensional Covariants and $\gamma$ Matrices}
\label{sec:propertiesgammaD}

  We begin with the main properties that can be used in calculations:
\begin{itemize}
\item
Vectors or more general objects $X^\mu$ on Q$D$S with upper indices such as
$k^\mu$, $A^\mu(x)$, $\gamma^\mu$ and $g^{\mu\nu}$ can be defined by the
explicit values of their components. The index $\mu$ takes infinitely
many values and runs from $0,1,2,\ldots
$ to infinity.
\item
  Indices can be lowered and raised with the $D$-dimensional metric
  tensor $g_{\mu\nu}$ and $g^{\mu\nu}$ as
  \begin{align}
    X_\mu & = g_{\mu\nu} X^\nu
    &
    X^\mu & = g^{\mu\nu} X_\nu \,.
  \end{align}
  We reiterate that we use a mostly-minus metric.
\item
  The $D$-dimensional metric tensor with a mostly-minus signature
  satisfies the expected relations 
  \begin{subequations}
    \label{gmunurelations}
    \begin{align}
      \label{gmunuindexvalues}
    g^{\mu\nu} & = g_{\mu\nu} = \left\{\begin{array}{ll}
    +1 & \text{ for }\mu=\nu=0\\
    -1 & \text{ for }\mu=\nu=1,2,\ldots\\
    0 & \text{ for } \mu\ne\nu
    \end{array}\right.
    \\
    g_{\mu\nu}g^{\mu\nu} & = D \,.
  \end{align}
  \end{subequations}
  These two relations extend the most important and obvious properties
  of the metric tensor to $D$ dimensions. They however seem
  contradictory since the indices take 
  infinitely many values and naively one might expect the contraction
  in the second equation to diverge. The solution is to regard a
  contraction with the lower-index $g_{\mu\nu}$ as a linear mapping,
  acting on upper-index quantities, instead of defining it via
  summation over explicit index values. Below we will show in detail
  how this idea reconciles the two equations (\ref{gmunurelations})
  and gives meaning to general lower-index quantities.
\item
  Contraction with $g_{\mu\nu}$ commutes with $D$-dimensional
  integration, as e.g.\ in
  \begin{align}
    \label{gmunuintegral}
    g_{\mu\nu} \int d^Dk k^\mu k^\nu f(k) & =
    \int d^Dk g_{\mu\nu} k^\mu k^\nu f(k) = \int d^Dk k^2 f(k) \,, 
  \end{align}
  and if a tensor $T^{\mu\nu}$ has only a finite number of
  nonvanishing entries, the expected result with an explicit summation
  is obtained,
  \begin{align}
    \label{Tmunufinitecontraction}
    g_{\mu\nu} T^{\mu\nu} & = \sum_{\mu,\nu=0}^\infty     g_{\mu\nu}
    T^{\mu\nu} = T^{00} - \sum_{i=1}^\infty T^{ii} \,.
  \end{align}
\item
  The $\gamma^\mu$ matrices may also be defined on Q$D$S, i.e.\ for
  $\mu=0,1,2,\ldots$ up to infinity such that they satisfy the basic
  relations
  \begin{align}
    \{\gamma^\mu,\gamma^\nu\} &= 2g^{\mu\nu}\mathbbm{1} \,,&
    \gamma_\mu\gamma^\mu & = D \mathbbm{1}\,.
    \label{GammaDproperties}
  \end{align}
  A representation exists which satisfies the same relations for
  complex conjugation, hermitian conjugation and charge conjugation as the ones of
  Eqs.\ (\ref{gammaproperties}) also for all $\mu$.
Hence it is also possible to define spinors on Q$D$S and to use the
definitions (\ref{psibarCproperties}) for adjoint and charge
conjugated spinors in $D$ dimensions.

As a result, the following relations hold for bilinear expressions of
anticommuting spinors on Q$D$S:
\begin{subequations}
  \begin{align}
\bar\psi_1\Gamma\psi_2 &= \overline{\psi_2^C}\Gamma^C\psi_1^C&
\mbox{ with }&&
\Gamma^C&=-C\Gamma^T C
\\
\left(\bar\psi_1\Gamma\psi_2\right)^\dagger &= \bar{\psi_2}\overline{\Gamma}\psi_1&
\mbox{ with }&&
\overline{\Gamma}&=\gamma^0\Gamma^\dagger\gamma^0
\end{align}
\end{subequations}
and
\begin{subequations}
  \begin{align}
\{1,\gamma_5,\gamma^\mu,\gamma^\mu\gamma_5
\}^C =
\{1,\gamma_5,-\gamma^\mu,-\gamma_5\gamma^\mu
\},
\\
\overline{\{1,\gamma_5,\gamma^\mu,\gamma^\mu\gamma_5
\}} =
\{1,-\gamma_5,\gamma^\mu,-\gamma_5\gamma^\mu
\}.
  \end{align}
\end{subequations}
For more details on the $\gamma_5$ matrix see Sec.\ \ref{sec:gamma5}.
\item
    The quasi-$D$-dimensional space actually is infinite dimensional
    and hence contains the original 4-dimensional Minkowski space, as
    expressed in Eq.\ (\ref{4Ssubspace}). On the level of covariants
    we therefore can define the purely 4-dimensional metric tensor
    $\gbar^{\mu\nu}$ by the 4-dimensional entries
    $\gbar^{00}=-\gbar^{ii}=+1$ for $i=1,2,3$ and $\gbar^{\mu\nu}=0$
    in all other cases. This tensor acts as a projector on the
    original Minkowski space. It also allows defining a complementary
    projector, the metric tensor of the $(D-4)$-dimensional complement
    as $\ghat^{\mu\nu}=g^{\mu\nu}-\gbar^{\mu\nu}$. In summary, all
    these tensors satisfy the following equations:
\begin{align}
    \text{$D$-dim.}:\ g^{\mu\nu} & = \gbar^{\mu\nu}+\ghat^{\mu\nu} &
    \text{$4$-dim.}:\ \gbar^{\mu\nu} &&
    \text{$(D-4)$-dim.}:\ \ghat^{\mu\nu} 
\end{align}
   with the dimensionalities expressed by
\begin{align}
    g_{\mu\nu} g^{\mu\nu} &= D \; , \;\;&
    \bar{g}_{\mu\nu} \bar{g}^{\mu\nu} &= 4 \; , \;\;&
    \hat{g}_{\mu\nu} \hat{g}^{\mu\nu} &= D-4 
\end{align}
   and the following contraction rules, expressing the projection and subspace
   relationships,
\begin{subequations}
\begin{align}
    \bar{g}_{\mu\nu} \bar{g}^{\nu\rho} & = \bar{g}_{\mu\nu}
    g^{\nu\rho} = g_{\mu\nu} \bar{g}^{\nu\rho}  =  \bar{g}_{\mu\;}^{\;\rho}\; , \;\;\\
    \hat{g}_{\mu\nu} \hat{g}^{\nu\rho} & = \hat{g}_{\mu\nu}
    g^{\nu\rho} = g_{\mu\nu} \hat{g}^{\nu\rho} = \hat{g}_{\mu\;}^{\;\rho} \; , \\
    \bar{g}_{\mu\nu} \hat{g}^{\nu\rho} & = \hat{g}_{\mu\nu}
    \bar{g}^{\nu\rho} = 0 \,.
\end{align}
\end{subequations}
\item
  Since the metric tensors $\gbar^{\mu\nu}$ and $\ghat^{\mu\nu}$ act as
projectors on the 4-dimensional and $(D-4)$-dimensional subspaces we
can generally decompose any vector $X^\mu$ as
\begin{align}
  X^\mu & = \bar{X}^\mu + \hat{X}^\mu &
  \bar{X}^\mu & = \bar{g}^\mu_{\;\nu}X^\nu &
  \hat{X}^\mu & = \hat{g}^\mu_{\;\nu}X^\nu \,,
  \label{XbarXhatDef}
\end{align}
such that e.g.\ squares and scalar products behave as
\begin{align}
  X^2 & = \bar{X}^2 + \hat{X}^2 &
  X_\mu Y^\mu & = \bar{X}_\mu \bar{Y}^\mu + \hat{X}_\mu \hat{Y}^\mu &
  \bar{X}_\mu \hat{Y}^\mu & = 0 \,.
\end{align}
Similar relationships can be defined for tensors in obvious ways.
\item
  As in Eq.\ (\ref{XbarXhatDef}) we can define  4-dimensional and
  $(D-4)$-dimensional versions $\bar{\gamma}^\mu$ and
  $\hat{\gamma}^\mu$ respectively,  which satisfy
\begin{subequations}
\label{eq:GammaDreg}
\begin{align}
   \{\gamma^\mu,\bar{\gamma}^\nu\}  =
   \{\bar{\gamma}^\mu,\bar{\gamma}^\nu\} & =
   2\bar{g}^{\mu\nu}\mathbbm{1}
   &
   \gamma_\mu
   \bar{\gamma}^\mu = \bar{\gamma}_\mu \bar{\gamma}^\mu & =
   4\,
   \mathbbm{1} \, ,
   \\
   \{\gamma^\mu,\hat{\gamma}^\nu\} =
   \{\hat{\gamma}^\mu,\hat{\gamma}^\nu\} & =
   2\hat{g}^{\mu\nu}\mathbbm{1}
   \, ,
   &
   \gamma_\mu \hat{\gamma}^\mu = \hat{\gamma}_\mu
   \hat{\gamma}^\mu & = (D-4) \mathbbm{1} \, ,
   \\
   \{\bar{\gamma}^\mu,\hat{\gamma}^\nu\} & = 0 \, ,
   &
   \bar{\gamma}_\mu \hat{\gamma}^\mu & = 0 \, .
\end{align}
\end{subequations}
Traces of $\gamma$-matrices are defined such that
\begin{align}
  \text{Tr}(\mathbbm{1}) & = 4 &
  \text{Tr}(\gamma^\mu) & = 0 \,.
\end{align}
With these relations all other traces of products of $\gamma$-matrices
can be calculated.
\item
  The properties of $\gamma_5$ and $\epsilon_{\mu\nu\rho\sigma}$ are
  discussed below in Sec.\ \ref{sec:gamma5}.
\item
  Generally, objects (covariants or operators) which vanish in purely
  4 dimensions are called evanescent. Examples of evanescent objects
  are all contractions with $\ghat^{\mu\nu}$ such as $\ghat^{\mu\nu}$
  itself,  $\hat{\gamma}^\mu$, or products such as $\hat{\gamma}^\mu\hat{\gamma}^\nu$,
  $\hat{\gamma}^\mu\bar{\gamma}^\nu$. Later we will see that many
  objects related to
  $\gamma_5$ or related to Fierz identities are also evanescent. 
\end{itemize}

\subsubsection{Construction of $D$-Dimensional Covariants and $\gamma$ Matrices}

Now we describe how objects may be defined which satisfy these
relations. The main difficulties are to define the lower-index metric
tensor and its contraction rules, and the $\gamma^\mu$-matrices. We
essentially follow Collins \cite{Collins_1984} in the construction of all
these quantities.

As mentioned above, at first sight it appears difficult to reconcile the different
properties (\ref{gmunurelations}) of the $D$-dimensional metric tensor
$g_{\mu\nu}$. The basic idea is that fundamentally tensors with lower
indices can be viewed as 
multilinear forms, i.e.\ mappings of objects with upper indices to
numbers. In the case of infinite dimensional vector spaces it is not
always sufficient to specify their component values. For Euclidean
metric and for a general tensor $T$ with components $T^{ij}$ Collins proposed
the definition of $\delta_{ij}T^{ij}$ as an abbreviation of a
mapping $\delta(T)$. This mapping can be defined via a $D$-dimensional integral \cite{Collins_1984}
\begin{align}
\delta_{ij} T^{ij}   & = \delta (T) = A\int d^Dk T^{ij}k_i k_j \delta(\vec{k}^2 - 1)
\label{deltaDefinition}
\end{align}
with normalization constant $A=D\Gamma(D/2)/\pi^{D/2}$. For the
integration momentum we simply take $k_i=k^i$ such that $\delta^{ij}k_ik_j=\vec{k}^2$.
The crucial point is that by definition the index contraction is performed
before evaluating the integral. As a special case, the definition
also contains a
definition of the individual components
\begin{align}
  \delta_{ij} & = A\int d^Dk k_i k_j \delta(\vec{k}^2 - 1) \,.
\end{align}
The calculations of the integrals in Eqs.\ (\ref{DefIallI1}) leading to
Eq.\ (\ref{ResultIallI1}) then show that
\begin{subequations}
\begin{align}
  \delta^{ij}\delta_{ij} & = D \,,\label{Ddimdelta}\\
  \delta_{ij} & = \delta^{ij} \,.\label{deltaijindexvalues}
\end{align}
\end{subequations}
The first of these relations demonstrates the effective
$D$-dimensional behavior of the metric tensor, and the second holds
component-wise and  shows that
the individual components have the usual values. However, the equations
also show again that contraction with $\delta_{ij}$ is not defined by
summation over explicit component values but via the integral
(\ref{deltaDefinition}), where contraction and integration cannot be
interchanged. Clearly,
\begin{align}
  \sum_{i,j=1}^\infty \delta^{ij}\delta_{ij} & = \infty 
\end{align}
in contrast to the correct equation (\ref{Ddimdelta}).

By treating the space-like components of $g_{\mu\nu}$ analogously to
the definition of $\delta_{ij}$ discussed above it is clear that we
can define a metric tensor which indeed fulfills the announced
equations (\ref{gmunurelations}).
General tensor contractions of the form $T^{\mu\nu}g_{\mu\nu}$ are
defined via integrals such as  Eq.\ (\ref{deltaDefinition}) and not via
explicit summation over component values --- in general summation over indices
does not commute with integration (which here defines contraction). The exception are cases of tensors
with only a finite number of nonvanishing components, in which case
Eqs.\ (\ref{gmunuindexvalues},\ref{deltaijindexvalues}) immediately
establish the relation (\ref{Tmunufinitecontraction}). 
In addition, the definition via an integral benefits from the fact
that different $D$-dimensional integrations
can be interchanged, see Eq.\ (\ref{Ddimintegralinterchange}). Therefore, $g_{\mu\nu}$ may be pulled inside or
outside integrals as exemplified in Eq.\ (\ref{gmunuintegral}).
In this way we have established all desired properties of the
$D$-dimensional metric tensor by explicit construction.

Next, we discuss the construction of $\gamma^\mu$-matrices which
satisfy the formally $D$-dimensional relations
(\ref{GammaDproperties}).
We define them similarly to Ref.\ \cite{Collins_1984}. We start from any standard
representation for the usual 4-dimensional $\gamma^\mu$-matrices such
as the representation  (\ref{gammachiralrep}) and denote these
$4\times4$-matrices now as
$\gamma_{[4]}^\mu$, $\mu=0,1,2,3$. The usual 4-dimensional
$\gamma_5$-matrix is now denoted as
$\gamma_{[4]}{}_5=i\gamma_{[4]}^0\gamma_{[4]}^1\gamma_{[4]}^2\gamma_{[4]}^3$.
We assume a representation such as (\ref{gammachiralrep}) in which the
properties (\ref{gammaproperties}) hold, such that only $\gamma^2$ is
imaginary and all others are 
  real.

Then  the formally $D$-dimensional $\gamma^\mu$-matrices can be
defined as
infinite-dimensional block matrices. Adapting the construction of
Ref.\ \cite{Collins_1984}, 
we first set for $\mu=0,1,2,3$ 
\begin{align}
  \gamma^\mu & = \left(\begin{array}{cccc}
    \gamma_{[4]}^\mu & 0 & 0 & \cdots\\
    0 & \gamma_{[4]}^\mu & 0 & \cdots\\
    0 & 0 & \gamma_{[4]}^\mu & \cdots\\
    \multicolumn{4}{c}{\ldots}
  \end{array}\right)&(\mu=0,1,2,3)
  \,,
\label{gammamuDdim}
\end{align}
where each entry corresponds to a $4\times4$ submatrix.
To construct $\gamma^\mu$ with $\mu>3$, we define the intermediate matrices
$\hat{\gamma}_{(4^k)}$ by 
\begin{align}
  \hat{\gamma}_{(4)}&=\gamma_{[4]}{}_5 &  \hat{\gamma}_{(4^{k+1})}&=
    \begin{pmatrix}
        \hat{\gamma}_{(4^k)} & 0 & 0 & 0\\ 0 & -\hat{\gamma}_{(4^k)} &
        0 & 0\\
        0 & 0 & - \hat{\gamma}_{(4^k)} & 0\\
        0 & 0 & 0 & \hat{\gamma}_{(4^k)}
    \end{pmatrix}&(k\ge1)
\,.\end{align}
In this way, $\hat{\gamma}_{(4^k)}$ is a real, hermitian,
$4^k$-dimensional matrix which consists of
$\pm\gamma_{[4]}{}_5$-blocks on the diagonal and which satisfies
$(\hat{\gamma}_{(4^k)})^2=1$. Using these matrices, we define, for any $\mu\ge4$,
the $2^{2\mu+1}$-dimensional real, anti-hermitian block matrix
\begin{align}
    \gamma^{\mu}_{(2^{(2\mu+1)})} &=
    \begin{pmatrix}
        0 & \hat{\gamma}_{(4^\mu)} \\ -\hat{\gamma}_{(4^\mu)} & 0
    \end{pmatrix}&(\mu\ge4)
\end{align}
and finally the infinite-dimensional block matrix
\begin{align}
  \gamma^{\mu} &=
    \begin{pmatrix}
        \gamma^{\mu}_{(2^{2\mu+1})} & 0 & \ldots
        \\
        0 & \gamma^{\mu}_{(2^{2\mu+1})} & {}
        \\
        \vdots & {} & \ddots
    \end{pmatrix}&(\mu\ge4)\,.
\label{gammarest}
\end{align}
The $\gamma^\mu$ matrices defined in
Eqs.\ (\ref{gammamuDdim},\ref{gammarest}) satisfy all properties
announced in Sec.\ \ref{sec:propertiesgammaD}; with the exception of
the commutation relations of $\gamma_5$ (see below) these are
identical to the purely 4-dimensional properties listed in
Eqs.\ (\ref{Clifford},\ref{gamma5epsilonDef},\ref{gammaproperties},\ref{psibarCproperties}).\footnote{  The construction of Ref.\ \cite{Collins_1984} is different in that
  the hermiticity/reality/charge conjugation properties of the
  $\gamma^\mu$ matrices are   different from
  Eqs.\ (\ref{gammaproperties}). Our construction corresponds
  essentially to a subset of the $\gamma^\mu$ matrices of
  Ref.\ \cite{Collins_1984}.}
\subsubsection{Definition of $\gamma_5$ and $\epsilon_{\mu\nu\rho\sigma}$
 in DReg}
\label{sec:gamma5}

A particularly problematic issue is the definition $\gamma_5$ and the
$\epsilon_{\mu\nu\rho\sigma}$ symbol in DReg --- the issue is often
referred to as the ``$\gamma_5$-problem of DReg''.
In 4 dimensions, three properties hold for the $\gamma_5$-matrix and traces:
\begin{subequations}
  \label{gamma54Dproperties}
  \begin{align}
  \{\gamma_5,\gamma^\mu\}&=0,\label{anticommutativity}\\
  \label{traceformula}
  \text{Tr}(\gamma_5\gamma^\mu\gamma^\nu\gamma^\rho\gamma^\sigma)&=-4i\epsilon^{\mu\nu\rho\sigma},\\
  \text{Tr}(\Gamma_1\Gamma_2)&=\text{Tr}(\Gamma_2\Gamma_1)\,.
\end{align}
\end{subequations}
The last equality means that traces are cyclic. In $D\ne4$
dimensions, it is inconsistent to require these properties
simultaneously, and one has to give up one of them.
To exhibit the problem we consider the trace
$t_{\mu_1\ldots\mu_4}={\rm
  Tr}(\gamma_{\mu_1}\ldots\gamma_{\mu_4}\gamma_5) $ and employ the
following series of steps, making use of equations
(\ref{gamma54Dproperties}). 
\begin{align}
D t_{\mu_1\ldots\mu_4} & = \text{Tr}\big(\gamma^\alpha\gamma_\alpha
\gamma_{\mu_1}\ldots\gamma_{\mu_4}\gamma_5\big)\nonumber\\
& = 
\text{Tr}\big((2\gamma^\alpha g_{\alpha\mu_1}
-\gamma^\alpha\gamma_{\mu_1}\gamma_{\alpha})
\ldots\gamma_{\mu_4}\gamma_5\big)
\nonumber\\
& = \ldots\nonumber\\
& = 8t_{\mu_1\ldots\mu_4}+
\text{Tr}\big(\gamma^\alpha
\gamma_{\mu_1}\ldots\gamma_{\mu_4}\gamma_\alpha\gamma_5\big)\nonumber\\
& = (8-D)t_{\mu_1\ldots\mu_4}\,.
\end{align}
In the first step, the $D$-dimensional contraction rule is used,
leading to the factor $D$, in
the intermediate steps the $\gamma^\mu$ anticommutation rule is used four times,
leading to the factor $8$. In the last step cyclicity and the
anticommutation relation (\ref{anticommutativity}) are used to
relate all terms to the initial trace. The outcome is that
\begin{align}
(4-D)t_{\mu_1\ldots\mu_4} & = 0 \,,
\end{align}
hence either $D=4$ or the trace must vanish. In other words, for
$D\ne4$ two of the equations (\ref{gamma54Dproperties}) imply that the
third equation is wrong.
In order to set up a consistent regularization which allows a
continuous limit to 4 dimensions we need both $D\ne4$ and a
non-vanishing trace at the same time, and therefore we need to give up
the validity of some of the equations (\ref{gamma54Dproperties}).

As a result there
is a plethora of proposals how to treat $\gamma_5$. The standard one,
which is known to be mathematically well-defined and consistent, is
the so-called BMHV scheme
\cite{tHooft:1972tcz,Breitenlohner:1977hr}. This scheme gives up the 
anticommutation property of $\gamma_5$; it is consistent in the sense
that it is compatible with unitarity and causality of quantum field
theory, but it does not manifestly lead to the correct
conservation/non-conservation properties of currents and does not
manifestly preserve gauge invariance of chiral gauge theories. 

In the BMHV scheme, $\gamma_5$ is defined in the identical way as in
four dimensions,
\begin{align}
  \gamma_5 & = i\gamma^0\gamma^1\gamma^2\gamma^3
  \,.
\end{align}
This clearly treats the first, original four dimensions differently
from the remaining $(D-4)$ dimensions. Accordingly, we obtain the
modified anticommutation relations
\begin{subequations}
  \label{modifiedGamma5rules}
  \begin{align}
\{\gamma^\mu,\gamma_5\} =  \{\hat{\gamma}_\mu,\gamma_5\}
& = 2\hat{\gamma}_\mu\gamma_5 \,,\\
\{\bar{\gamma}_\mu,\gamma_5\} & = 0 \,, \\
[\hat{\gamma}_\mu,\gamma_5] & = 0 \,,
\end{align}
\end{subequations}
where as in Eq.\ (\ref{eq:GammaDreg}) the split
$\gamma^\mu=\bar{\gamma}^\mu+\hat{\gamma}^\mu$ into the 4-dimensional
and $(D-4)$-dimensional parts was used. Only the original
matrices $\bar{\gamma}^\mu$ fully anticommute with $\gamma_5$.
In this way,  $D$-dimensional Lorentz invariance is effectively broken
by the regularization. Similarly, this modification leads to a
breaking of gauge invariance in chiral gauge theories on the
regularized level in DReg. This is clearly a drawback and a central
topic of the present review.

Similarly, the Levi-Civita $\epsilon_{\mu\nu\rho\sigma}$ symbol,
defined as a fully antisymmetric object with four indices is only well
defined in purely 4 dimensions. Hence, using the split notation we may
write, as stressed in Ref.\ \cite{Breitenlohner:1977hr},
\begin{align}
  \epsilon_{\mu\nu\rho\sigma} & = \bar{\epsilon}_{\mu\nu\rho\sigma}
  \,, &
  \hat{\epsilon}_{\mu\nu\rho\sigma} & = 0 \,,
\end{align}
and rewrite the definition of $\gamma_5$ as
\begin{align}
  \gamma_5 & = -\frac{i}{4!}\bar{\epsilon}_{\mu\nu\rho\sigma}
  \bar{\gamma}^\mu
  \bar{\gamma}^\nu
  \bar{\gamma}^\rho
  \bar{\gamma}^\sigma \,,
\end{align}
with the sign convention
\begin{align}
  \epsilon^{0123}= - \epsilon_{0123} = +1 \,,
\end{align}
which was already used in Eq.\ (\ref{gamma5epsilonDef}).
In practical computations often combinations of two $\epsilon$-symbols
appear. The following 4-dimensional identity remains valid,
\begin{align}
  \bar{\epsilon}^{\mu\nu\rho\sigma}\bar{\epsilon}_{\alpha\beta\gamma\delta}
& =  - \bar{g}^\mu{}_\alpha\bar{g}^\nu{}_\beta
\bar{g}^\rho{}_\gamma\bar{g}^\sigma{}_\delta \pm
\ldots
\label{epsepsrule}
\end{align}
where the dots denote 23 further similar terms leading to total
antisymmetrization in the indices. Some calculations, e.g.\ the
prescription by Larin \cite{Larin:1993tq} propose to elevate this identity to
the level of $D$ dimensions i.e.\ to assume the validity of the
corresponding identity with formally $D$-dimensional metric tensors,
i.e.\ effectively without the bars.
Let us remark that such a $D$-dimensional identity can ultimately lead
to inconsistencies in the sense that one initial expression could
lead to different answers. To make this inconsistency explicit we
denote the right-hand side of Eq.\ (\ref{epsepsrule}) in $D$
dimensions as $p^{\mu\nu\rho\sigma}_{\alpha\beta\gamma\delta}$. Then
consider the product of four $\epsilon$-symbols
\begin{align}
  {\epsilon}^{\mu\nu\rho\sigma}{\epsilon}_{\alpha\beta\gamma\delta}
  {\epsilon}_{\mu\nu\rho\sigma}{\epsilon}^{\alpha\beta\gamma\delta}
\,.
\label{epsiloninconsistency}
\end{align}
This can be evaluated in two ways with the two results
\begin{align}
  \text{either\ }& p^{\mu\nu\rho\sigma}_{\alpha\beta\gamma\delta}
  p_{\mu\nu\rho\sigma}^{\alpha\beta\gamma\delta} &
  \text{or\ }& p^{\mu\nu\rho\sigma}_{\mu\nu\rho\sigma}
  p_{\alpha\beta\gamma\delta}^{\alpha\beta\gamma\delta} \,.
\end{align}
In strictly 4 dimensions, both expressions give $24^2=576$ so there is
no inconsistency. However, assuming validity of these equations in $D$
dimensions and using $D$-dimensional metric tensors in the
contractions, the two results are different:
\begin{align}
  \text{either\ } & 24 D(D-1)(D-2)(D-3) &
  \text{or\ }& [D(D-1)(D-2)(D-3)]^2 \,.
  \label{epsiloninconsistency2}
\end{align}
Hence in an amplitude involving such contractions of
$\epsilon$-symbols, the result is ambiguous, except for the leading
poles in $1/(D-4)$. For this reason in a fully consistent treatment only the 4-dimensional
version of the identity (\ref{epsepsrule}) is valid \cite{Breitenlohner:1977hr}.

In view of the drawbacks of the BMHV scheme, many  alternative versions of DReg have been
proposed in the literature. For instance, Ref.\ \cite{Chanowitz:1979zu} has
proposed that a fully anticommuting $\gamma_5$ may be used in certain
Feynman graphs, in spite of the inconsistency between
Eqs.\ (\ref{gamma54Dproperties}) mentioned above.  Similarly,
Refs.\ \cite{Trueman:1995ca,Chetyrkin:1997gb} derived that in specific
applications correct results can be 
also be obtained using a simpler schemes with anticommuting
$\gamma_5$. A well-known review of the situation was given by
Jegerlehner \cite{Jegerlehner:2000dz}, where further arguments were presented
that the ``naive'' anticommuting $\gamma_5$ may be used in many
cases. Kreimer et al \cite{Korner:1991sx} have proposed a different kind of
alternative to BMHV: out of the three equations (\ref{gamma54Dproperties}),
the cyclicity of the trace is given up, but the anticommutativity is
kept. In this case, special attention must be paid to ``subdiagram
consistency'' as described in Ref.\ \cite{Chetyrkin:1997gb}: ``It should give unique results
independently of whether some diagram is considered as a subdiagram,
and independently of the order in which subdiagrams are
calculated. Otherwise subdivergences could not be properly subtracted
in multiloop diagrams.''  Ref.\ \cite{Korner:1991sx} introduces
 so-called ``reading-point'' prescriptions to deal with this
difficulty. 

All these alternative proposals have in common that their general
applicability to all cases has not been established; hence the
all-order proofs of renormalizability properties of e.g.\
Refs.\ \cite{Speer:1974cz,Breitenlohner:1977hr,Bonneau:1979jx,Bonneau:1980zp} do
not apply to them.

We also briefly comment on two recent investigations of the $\gamma_5$-problem in
alternatives to DReg.
Ref.\ \cite{Gnendiger:2017rfh} considered dimensional schemes
in various slightly different implementations
(e.g.\ the so-called four-dimensional helicity
(\FDH) scheme discussed in more detail below in Sec.\ \ref{sec:variants})
from the point of view of practical one- and two-loop
calculations.  At the two-loop level, there is no single scheme that
stands out as computationally most efficient.
Ref.\ \cite{Bruque:2018bmy} considered strictly
4-dimensional schemes as alternatives to dimensional regularization,
in the hope that these schemes might offer practical advantages with respect to the
treatment of $\gamma_5$. The considered class of schemes is wide and
general but contains only schemes which  do not break gauge
invariance as immediately as e.g.\ the Pauli-Villars scheme. This
reference showed clearly that all these schemes have very similar
problems for $\gamma_5$ as dimensional schemes. The reason is that in
those schemes the regularization is essentially performed by replacement
rules, and those replacement rules do not necessarily commute with
applying, e.g.,\ the cyclicity of traces.

\subsection{Relation to the Lagrangian in $D$ Dimensions}
\label{sec:relationL}

This subsection is devoted to a seemingly simple statement, which
however constitutes another important advantage of DReg. DReg can
already be formulated at the level of the Lagrangian, and regularized
Feynman diagrams can literally be obtained from a $D$-dimensional
version of the Gell-Mann-Low formula with a $D$-dimensional
Lagrangian.
This fact allows a very efficient investigation of properties of
regularized Green functions. Examples are the all-order proof of the
regularized quantum action principle (see Sec.\ \ref{sec:QAPproof}) and the
textbook derivation of renormalization group $\beta$ functions and
anomalous dimensions from divergences in the counterterm Lagrangian
(see e.g.\ the textbook by Srednicki \cite{Srednicki:2007qs}).

The explicit construction of formally $D$-dimensional objects in DReg
provides all objects needed to formulate a $D$-dimensional
Lagrangian. Fields $\phi(x)$ are defined as functions of
$D$-dimensional vectors $x^\mu$, i.e.\ of elements of the
quasi-$D$-dimensional space Q$D$S. Metric tensors, derivatives, vector 
fields, and $\gamma$-matrices have all been
extended to $D$ dimensions as well. The construction of
$\gamma$-matrices implies also a definition of $D$-dimensional
extensions of  4-spinor fields (which have infinitely
many components in view of Eq.\ (\ref{gammamuDdim})).
For this reason any Lagrangian of a 4-dimensional quantum field theory
involving such fields can be naturally extended to $D$ dimensions.
\footnote{  Unfortunately, the 2-component spinor notation described in
  Sec.\ \ref{sec:2spinors} is not known to be extendable to $D$ dimensions since
  it is explicitely tied to the representation theory of the
  4-dimensional Lorentz group. 2-component spinor Lagrangians need to
  be rewritten in terms of 4-component spinors before an extension to
  $D$ dimensions and an application of DReg becomes possible.}

If a Lagrangian involves the $\gamma_5$ matrix or the
$\epsilon_{\mu\nu\rho\sigma}$ symbol, e.g.\ in case of chiral fermion
interactions, an extension to $D$ dimensions remains possible, but the
$D$-dimensional version involves e.g.\ $\gamma_5$ with its modified
anticommutation relations (\ref{modifiedGamma5rules}). Hence in such cases
the resulting $D$-dimensional Lagrangian will not be invariant under
formally $D$-dimensional Lorentz transformations. This, however, does
not preclude the application of DReg.\footnote{  In particular, even in such cases it remains true that 4-dimensional Lorentz invariance
  is manifestly preserved.}

This issue illustrates a more general point. Though there is often a
preferred choice, the extension of any
Lagrangian to $D$ dimensions is in principle never unique. It is
always possible to change so-called evanescent terms in the
Lagrangian, i.e.\ terms that vanish in 4 dimensions. If $\gamma_5$ is
present, this possibility is obvious. E.g.\ a 4-dimensional expression
$\bar\psi \gamma^\mu P_L \psi$ may be extended
to the following 
three inequivalent $D$-dimensional choices
\begin{align}
  \overline{\psi} \gamma^\mu P_L \psi \, , &&\text{ or }
  \overline{\psi} P_R \gamma^\mu \psi \, , &&\text{ or }
  \overline{\psi} P_R \gamma^\mu P_L \psi \, .
\label{eq:inequivalentchoices}
\end{align}
In 4 dimensions these terms are all equal but in $D$ dimensions they
are different due to the modified anticommutation relations. But even
independently of $\gamma_5$, one may extend e.g.\ an interaction
term between a vector and a scalar field as
\begin{align}
  \phi^\dagger A^\mu \partial_\mu \phi \, ,&&\text{ or }
  \phi^\dagger \bar{A}^\mu \bar{\partial}_\mu \phi \, ,
\end{align}
where the second possibility involves only the purely 4-dimensional
part of the derivative.

Despite the non-uniqueness, clearly any field theory Lagrangian can be
extended to a $D$-dimensional version. This Lagrangian $\La^{(D)}$ can
then be split into a free part and a remainder (the ``interaction''
part)
\begin{align}
  \La^{(D)} & =   \La^{(D)}_{\text{free}} +  \La^{(D)}_{\text{int}} \,,
\end{align}
where the free part must be bilinear in the fields and contain the
appropriate kinetic terms. The non-uniqueness affects mainly the
``interaction'' part; a constraint we will always impose is that
the kinetic terms involve strictly $D$-dimensional derivatives.
A reason for this constraint will be illustrated below. It essentially
fixes the ``free'' part of the Lagrangian, such
that we may schematically write the free Lagrangian as
\begin{align}
  \La^{(D)}_{\text{free}} & = \frac12 \phi_i {\cal D}^{(D)}_{ij}\phi_j
  \label{LaDfree}
\end{align}
with some differential operator ${\cal D}^{(D)}_{ij}$ involving
$D$-dimensional derivatives. The notation is meant in a general sense,
including the familiar expressions for complex scalar fields, spinor fields or
vector fields.
Standard free field theory quantization then leads to the
$D$-dimensional propagators 
\begin{align}
  {\cal P}^{(D)}_{jk} & = \langle0|T\phi_j\phi_k|0\rangle
\label{DPropagators}
\end{align}
which are the Green functions of the differential operators,
i.e.\ which satisfy the inverse relation
\begin{align}
  \tilde{\cal D}^{(D)}_{ij}\tilde{\cal P}^{(D)}_{jk}&=i\delta_{ik}
\label{PropagatorsInverse}
\end{align}
in momentum space in $D$ dimensions.

Let us exemplify these relations and highlight the related
subtleties. E.g.\ for spinor fields we take the straightforward
$D$-dimensional free Lagrangian
$\bar\psi(i\gamma^\mu\partial_\mu-m)\psi\equiv\bar\psi{\cal D}^{(D)}\psi$, leading to the
momentum-space propagator
\begin{align}\label{propagatorDexample}
\tilde{\cal P}^{(D)}=  \langle0|T\psi\bar\psi|0\rangle^{\text{F.T.}} & =
  \frac{i}{\slashed{p}-m} = \frac{i(\slashed{p}+m)}{p^2-m^2}
\end{align}
where F.T.\ denotes Fourier transformation of the respective
expression  ($x$-arguments are suppressed); the argument of the
Fourier transformation is the momentum $p$; all appearing momenta are
$D$-dimensional and the $+i\varepsilon$ prescription in the propagator
denominator is suppressed. Such propagator Feynman rules lead to loop
integrals such as the ones of Sec.\ \ref{sec:Schwingerparametrization}
and denominator structures as in the example
(\ref{exampleintegrand}). The propagator (\ref{propagatorDexample}) is
indeed the inverse of the momentum-space differential operator of the
Lagrangian,
\begin{align}
  \tilde{\cal D}^{(D)} &=(\slashed{p}-m).
\end{align}
Taking instead the
purely 4-dimensional derivative $\bar\partial_\mu$ in the free
Lagrangian would lead to
\begin{align}
  \langle0|T\psi\bar\psi|0\rangle^{\text{F.T}} & =
  \frac{i}{\bar{\slashed{p}}-m} =
  \frac{i(\bar{\slashed{p}}+m)}{\bar{p}^2-m^2}
  \,, 
\end{align}
which involves only the purely 4-dimensional momentum in the
denominator. The problem of this choice is that loop integrals would
not be regularized, hence such a
choice is not permitted. Similarly, one may propose a recipe where
Dirac propagators are regularized as
\begin{align}
  \langle0|T\psi\bar\psi|0\rangle^{\text{F.T}} & \to
  \frac{i(\bar{\slashed{p}}+m)}{{p}^2-m^2}
  \,, 
\end{align}
which involves the purely 4-dimensional momentum in the numerator and
the $D$-dimensional momentum in the denominator. Such a recipe cannot
arise from a $D$-dimensional Lagrangian; it will not be used and
statements such as  the regularized quantum action principle would not
necessarily be
valid.

As illustrated by this example, the general $D$-dimensional
relationships for the free Lagrangian and the propagators
(\ref{LaDfree},\ref{DPropagators},\ref{PropagatorsInverse}) can always
be realized, they will always be assumed, and they   are nontrivial.

Once the free Lagrangian  is chosen in agreement with the mentioned
constraint, and the interaction Lagrangian is fixed, $D$-dimensional
regularized Feynman diagrams can be defined via the standard
Gell-Mann-Low formula, suitably written in $D$ dimensions. One way to
write it is to take the original formula (\ref{GellMannLow}) and
replace the integrations by $D$-dimensional ones. In this case the
parameters and fields must have appropriately modified
dimensionalities, see e.g.\ \cite{Neubert:2019mrz} for a presentation
that makes extensive use of this possibility. A
second way is to write
\begin{align}
Z(J,\Ks) & =
  \frac{
\langle0|T
 \exp\left(i\mu^{D-4}\int d^Dx (\La^{(D)}_{\text{int}}+J_i\phi_i+\Ks_i\mathcal{O}_i)\right)|0\rangle}
{\langle0|T\exp\left(i\mu^{D-4}\int d^Dx \La^{(D)}_{\text{int}}\ \right)|0\rangle}
\,,
\label{GellMannLowDReg}
\end{align}
where the regularization scale $\mu$ is introduced such that the
regularized Lagrangian has mass-dimension 4. Either way, if the
Gell-Mann-Low formula is evaluated via Wick contractions and Fourier
transformed, the correct DReg expressions for regularized Feynman
diagram amplitudes are obtained. The variant (\ref{GellMannLowDReg})
also generates a factor $\mu^{4-D}$ accompanying each loop
integration, as indicated by Eq.\ (\ref{DRegbasicidea}).

As mentioned in the beginning, this relation between the Lagrangian
and regularized Feynman diagrams has important consequences, some of
which we will discuss in subsequent sections. Here we remark that the
present discussion allows the possibility that the Lagrangian contains
$1/(D-4)$ poles in coefficients; in particular the discussion is
unaffected if the interaction Lagrangian $\La^{(D)}_{\text{int}}$ is
defined to include counterterms that are defined order by order to
cancel divergences or to restore symmetries.

\subsection{Variants: Dimensional Reduction and \CDR, \HV, and \FDH~Schemes}
\label{sec:variants}

DReg as defined so far still leaves room for different options, and
there are other variants of dimensional schemes which share the idea
of $D$-dimensional integrals. Here we give a brief overview of several
schemes used in the literature. The overview essentially follows the
review \cite{Gnendiger:2017pys}, and we refer to this review for
more details and original references.

We remark that the following distinction between the schemes does not
have much influence on the discussion of chiral fermions and the
treatment of $\gamma_5$ in DReg. The remarks of Sec.\ \ref{sec:gamma5}
apply to all the following schemes, and different alternative
treatments of $\gamma_5$ have been employed in the literature. In the
following discussion we focus on aspects independent of $\gamma_5$.

All the following schemes treat integrals always in $D$ dimensions. They
differ in their treatment of vector fields. 
In order to consistently define the different schemes it has
turned out useful \cite{Stockinger:2005gx,Signer:2008va} to introduce the following spaces
extending the original 4-dimensional space 4S. In
Sec.\ \ref{sec:QDS} we already introduced the quasi-$D$-dimensional space
Q$D$S, on which objects such as formally $D$-dimensional momenta
$p^\mu$ and momentum integrations
are defined. The explicit construction 
showed that this space necessarily is infinite dimensional and contains
the original space 4S. Now we introduce an even bigger space
Q$D_s$S (later, $D_s=4$ will be taken, so this is often called a
``quasi-4-dimensional'' space). It contains Q$D$S and is formally
$D_s$-dimensional. The relationships are thus
\begin{align}
\label{DREDsubspaces}
  \text{4S} \subset \text{Q$D$S} \subset \text{Q$D_s$S}
\end{align}
regardless of the values of $D$ and $D_s$. 

Before describing the scheme definitions we note that vector fields
can appear in different roles in Feynman diagrams:
\begin{itemize}
\item
  There are vector fields appearing in propagators in loop diagrams or
  as propagators or external fields in phase space regions which lead
  to infrared, soft or collinear singularities. We call such vector
  fields {\em singular} vector fields. They may be treated in either
  4S, Q$D$S, or Q$D_s$S.
\item
  All other vector fields appear outside of 1PI diagrams and outside
  singular phase space regions. We call them {\em regular}, and they
  may be treated differently from singular vector fields.
\end{itemize}
To motivate the concrete scheme choices we further list two simple
observations.
\begin{itemize}
\item
  Gauge invariance relies on the gauge covariant derivative $D_\mu$,
  which combines the ordinary derivative (which is always
  $D$-dimensional) and vector fields. In order not to directly break
  gauge invariance on the regularized level, there should be at least
  a fully $D$-dimensional covariant derivative. Hence the singular
  vector fields should be treated at least as $D$-dimensional.
\item
  Supersymmetry relies on an equal number of fermionic and bosonic
  degrees of freedom. The number of spinor degrees of freedom is
  essentially fixed via $\text{Tr}\mathbbm{1}=4$. Hence in order not to
  directly break supersymmetry, singular vector fields should be
  treated as 4-dimensional.
\end{itemize}
It appears difficult to reconcile the requirements of gauge invariance
and supersymmetry,  and the
different schemes are motivated by focusing on different aspects.

Now we list the
four schemes and refer to Tab.\ \ref{tab:schemes} for a summary.
\begin{itemize}
\item Dimensional regularization has two subvariants, called \HV and \CDR ('t
  Hooft/Veltman and Conventional Dimensional Regularization). Both
  variants treat singular vector fields as $D$-dimensional, i.e.\ in
  Q$D$S. This is in line with $D$-dimensional gauge
  invariance\footnote{    We stress again that here our definitions of the four schemes only
    refer to the treatment of vector fields. In principle, in either
    scheme one would also have different options of treating
    $\gamma_5$, of which the non-anticommuting one is the most
    rigorous. The agreement with gauge invariance is meant on a
    superficial level. The existence of a $D$-dimensional covariant
    derivative by itself does not prove the all-order
    preservation of gauge invariance, and clearly gauge invariance of
    chiral gauge theories can be broken in dimensional schemes. For an example
    rigorous statement on the preservation of gauge invariance see
    later Sec.\ \ref{sec:QAPexamples}.}
  but leads to a direct breaking of supersymmetry. The \HV scheme
  treats regular vector fields without regularization, i.e.\ in 4S,
  and the \CDR scheme treats all vector fields in Q$D$S. The space
  Q$D_s$S is not used.
\item
  The other class of choices is dimensional reduction, originally
  introduced in the context of supersymmetry \cite{Siegel:1979wq}.  It also has two
  subvariants, called \FDH and \DRED (Four-dimensional helicity scheme
  and Dimensional Reduction). Singular vector fields are treated as
  $D_s$-dimensional, and in practical calculations $D_s$ is eventually
  set to $D_s=4$. Hence singular vector fields are essentially treated as
  quasi-4-dimensional, but the quasi-4-dimensional space contains the
  $D$-dimensional subspace, such that both gauge invariance and
  supersymmetry are not immediately broken. \FDH is analogous to \HV and
  treats regular vector fields as strictly 4-dimensional, and \DRED
  treats all vector fields in Q$D_s$S.
\end{itemize}

\begin{table}
\begin{center}
\begin{tabular}{l|cccc}
&\CDR&\HV&\FDH&\DRED\\
\hline
singular vector field&$g_{[D]}{}^{\mu\nu}$&$g_{[D]}{}^{\mu\nu}$&
$g_{[D_s]}{}^{\mu\nu}$&$g_{[D_s]}{}^{\mu\nu}$\\
 regular vector field&$g_{[D]}{}^{\mu\nu}$&$g_{[4]}{}^{\mu\nu}$&
$g_{[4]}{}^{\mu\nu}$&$g_{[D_s]}{}^{\mu\nu}$
\end{tabular}
\end{center}
\caption{\label{tab:schemes}
Treatment of singular and regular vector fields in the four different
schemes. The table indicates  which metric tensor is to be used in propagator
numerators and polarization sums. This table is adapted from
Refs.\ \cite{Signer:2008va,Gnendiger:2017pys}.} 
\end{table}

Technically the schemes are expressed and summarized by
Tab.\ \ref{tab:schemes} by specifying which metric tensor is to be
used in propagator numerators or in polarization sums for squared
matrix elements. In the table and in the remainder of this subsection
we use a more explicit notation for metric tensors on the different
spaces and use the symbols $g_{[\text{\em dim}]}{}^{\mu\nu}$ where 
$\text{\em dim}$ denotes the respective space, i.e.\ $\text{\em
  dim}=4, D, D_s$ or $\text{\em dim}=D-4, D_s-D$. Our previous
notation is rewritten as
\begin{align}
  \gbar^{\mu\nu} &\equiv g^{\mu\nu}_{[4]}\, ,&   \ghat^{\mu\nu}
  &\equiv g^{\mu\nu}_{[D-4]}  \,,&
  g^{\mu\nu}  & \equiv g^{\mu\nu}_{[D]}
\,.
\end{align}

The scheme differences for singular vector fields (which are
sufficient for 1PI Green functions) can be well explained by comparing
the gauge covariant derivatives. In the \CDR and \HV schemes, a generic
covariant derivative is purely $D$-dimensional,
\begin{align}
D_{[D]}{}^\mu & = \partial_{[D]}{}^\mu + ig A_{[D]}{}^\mu \,,
\end{align}
and the regularized vector field $A_{[D]}{}^\mu$ plays the role of a
$D$-dimensional gauge field. In contrast, a covariant derivative in
the \DRED and \FDH schemes can be split as
\begin{align}
D_{[D_s]}{}^\mu & = \partial_{[D]}{}^\mu + ig A_{[D]}{}^\mu + i g_e A_{[D_s-D]}{}^\mu \,.
\label{DmuDRED}
\end{align}
From a $D$-dimensional spacetime point of view, only the part $A_{[D]}{}^\mu$
acts as a $D$-dimensional gauge and vector field. In contrast, the
field components $A_{[D_s-D]}{}^\mu$ are extra fields which behave
like scalar fields in $D$ dimensions; they are often referred to as
``$\epsilon$-scalars''. The behavior under renormalization reflects
this difference, and in general the two coupling constants $g_e, g$
renormalize differently.

In practical calculations it is often not required to write the
covariant derivative as explicitly as in Eq.\ (\ref{DmuDRED}). Often
it is sufficient to set $D_s=4$ and $g_e=g$ such that the vector field
in the covariant
derivative in \DRED and \FDH behaves essentially
4-dimensionally. If
this is possible it constitutes an advantage of these schemes. Specifically in supersymmetric theories the symmetry
leads to $g=g_e$.  In
general, however, the 
split (\ref{DmuDRED}) is in principle always possible and sometimes required.
In the literature, the split was often useful to understand scheme
behaviours, to resolve
inconsistencies and to derive scheme translation rules (for references
to examples see Ref.\ \cite{Gnendiger:2017pys}).

We now give a brief overview of the theoretical status of the
\DRED and \FDH schemes. For a more practical description with example
calculations in all schemes we refer to Ref.\ \cite{Gnendiger:2017pys}.
\DRED was introduced with the goal to preserve supersymmetry on the
regularized level \cite{Siegel:1979wq,Capper:1979ns}. Over time, however, several
inconsistencies were reported in the literature. Ref.\ \cite{Siegel:1980qs}
found a mathematical inconsistency in the simultaneous application of
4-dimensional and $D$-dimensional algebra. The inconsistency is very
similar to
Eqs.\ (\ref{epsiloninconsistency},\ref{epsiloninconsistency2}). It
turned out that the inconsistency is due to the assumption that the
$D$-dimensional space is a proper subspace of the original
4-dimensional space. If one distinguishes between the original
4-dimensional space and the quasi-4-dimensional space Q$D_s$S and uses
the relationships (\ref{DREDsubspaces}), the inconsistency is resolved
\cite{Stockinger:2005gx}. 

An important result is the all-order equivalence between all the
schemes \cite{Jack:1993ws,Jack:1994bn} (the proof was given for Green functions without
infrared divergences and hence
does not distinguish \CDR/\HV or \DRED/\FDH). For this proof, the split
(\ref{DmuDRED}) and the independent renormalization of couplings such
as $g_e$ and $g$ is essential. In this way, another 
inconsistency reported in Ref.\ \cite{vanDamme:1984ig} was resolved. In
that reference, couplings 
such as $g_e$ and $g$ were always assumed to be identical and it was
shown that unitarity  of the S-matrix can be violated at higher
orders.
This necessity of the split (\ref{DmuDRED}) and its role for
renormalization, finiteness and unitarity has also been stressed and
exemplified by explicit calculations in
Refs.\ \cite{Harlander:2006xq,Kilgore:2011ta}. In summary, \DRED is established
as  a fully consistent and applicable regularization for UV
divergences.

The scheme properties for infrared divergences have also been
investigated, in particular focusing on the computation of real and
virtual higher-order corrections to physical processes. In the context
of such calculations the different treatments of regular vector fields
becomes important. The schemes \HV and, in particular, \FDH are motivated
by the potential to carry out much of the algebra in strictly 4
dimensions, allowing e.g.\ powerful spinor and helicity methods.  It
was shown that
the \CDR, \HV, and \FDH schemes are equivalent at the next-to-leading
(NLO) level, and elegant scheme transition rules were derived \cite{Kunszt:1993sd,Catani:1996pk,Catani:2000ef}.
In a parallel development, several references observed an apparent
inconsistency in \DRED with infrared factorization
\cite{Beenakker:1988bq,Beenakker:1996dw,Smith:2004ck}.\footnote{  As discussed in Ref.\ \cite{Signer:2008va}, these important results were
  somewhat obscured by the fact that 
  different authors used different names for equivalent schemes, and
  sometimes the same names for different schemes: The schemes called
  DR (dimensional reduction) in Refs.\ \cite{Kunszt:1993sd,Catani:1996pk,Catani:2000ef} are
  actually equivalent to the \FDH scheme
\cite{Bern:1991aq}; but 
Refs.\ \cite{Beenakker:1988bq,Beenakker:1996dw,Smith:2004ck} used the term dimensional reduction in the
same sense as we define \DRED here. }
The
resolution of this inconsistency \cite{Signer:2005iu,Signer:2008va} is again based on the
observation that the split (\ref{DmuDRED}) and a separate treatment of
$D$-dimensional gauge fields and $\epsilon$-scalars is in general necessary. 
Further higher-order extensions of these analyses were presented
in Refs.\ \cite{Kilgore:2012tb,Broggio:2015ata,Broggio:2015dga}.

Finally we comment on the question of supersymmetry preservation. In
dimensional regularization (regardless whether \CDR or \HV) the number
of bosonic and fermionic degrees of freedom on the regularized level
is different. This immediately leads to a violation of supersymmetry
relations already at the one-loop level. Dimensional regularization
may still be used, but specific finite supersymmetry-restoring
counterterms have to be added to the Lagrangian.
Such  counterterms  were evaluated and documented in
Refs.~\cite{Martin:1993yx,Mihaila:2009bn,Stockinger:2011gp}. 

For dimensional reduction (\DRED or \FDH), many studies have confirmed the
compatibility with SUSY and the absence of non-SUSY
counterterms. Overviews of results can be found e.g.\ in
Refs.\ \cite{Jack:1993ws,Jack:1994bn,Stockinger:2005gx,Hollik:2005nn}. Refs.\ \cite{Stockinger:2005gx,Avdeev:1981vf}
made clear that in the consistent versions of \DRED/\FDH,
supersymmetry will 
eventually be broken. The reason is that the regularized Lagrangian is
formulated not in the actual 4-dimensional space but in Q$D_s$S, where
Fierz identities do not hold. The quantum action principle in \DRED
\cite{Stockinger:2005gx} then implies a supersymmetry breaking on the level of
Green functions; the reasoning applied in Ref.\ \cite{Stockinger:2005gx} is
essentially the same as the strategy described in the present review
for restoring gauge invariance in chiral gauge theories. Because of
this general statement, supersymmetry of \DRED must be investigated on
a case-by-case basis, and it has turned out that for a large set of
relevant multi-loop calculations, supersymmetry is preserved \cite{Hollik:2005nn,Harlander:2006xq,Harlander:2009mn,Stockinger:2018oxe}.

\newpage

\section{Quantum Action Principle in~DReg}

\label{sec:QAPDReg}

If Green functions of a quantum field theory are defined via the path
integral (\ref{pathintegral}) or the Gell-Mann-Low formula (\ref{GellMannLow}), the
properties of Green functions clearly reflect the properties of the
underlying Lagrangian. Example properties are the Ward or
Slavnov-Taylor identities already discussed in
Secs.\ \ref{sec:STIformal}, \ref{sec:AbelianPeculiarities} which
reflect symmetry properties of the Lagrangian.

This section is devoted to a related but more general relationship ---
the so-called quantum action principle, specifically the regularized
quantum action principle in DReg. This is a very useful
relationship, allowing e.g.\ rigorous derivations of Slavnov-Taylor
identities or their breakings. The quantum action principle might appear
obvious or straightforward, and sometimes its validity is taken for
granted. However, actually its validity and also its precise meaning depends on the
chosen regularization and renormalization procedure.
For DReg, it was proven in \cite{Breitenlohner:1977hr} both on the regularized and
on the renormalized level; the proof was extended to the consistent
version of dimensional reduction in Ref.\ \cite{Stockinger:2005gx}.
We remark that there is also a regularization-independent quantum action principle,
established in the context of BPHZ-renormalization in
Refs.\ \cite{Lowenstein:1971vf,Lowenstein:1971jk,Lam:1972mb,Lam:1972fzg,Lam:1973qa,Clark:1976ym}. We
will discuss it and its relation to the regularized quantum action
principle of DReg later in Sec.\ \ref{sec:AlgRenormalization}.

Here in this section we will begin with a formal derivation to
motivate the statement, to highlight its simplicity and to fix its
interpretation (Sec.\ \ref{sec:formalQAP}). Then we will present a full
proof of the regularized quantum action principle in DReg
(Sec.\ \ref{sec:QAPproof}). Finally Sec.\ \ref{sec:QAPexamples} will
illustrate how to use this regularized quantum action principle to establish symmetry
properties.

\subsection{Formal Derivation of the Quantum Action~Principle}
\label{sec:formalQAP}

The quantum action principle is a simple relation between
the properties of the Lagrangian and the full
Green functions.
Here we will present  a formal derivation using the path integral
(allowing general dimension $D$)
\begin{align}
Z(J,\Ks) & = \int{\cal D}\phi\ e^{i\int d^Dx
  (\La+J_i\phi_i)} \,,
\end{align}
where possible composite operator terms coupled to sources $\Ks$ have
been absorbed into the Lagrangian $\La$.
Similarly to Sec.\ \ref{sec:STIformal} we consider a variable
transformation
\begin{align}
\label{fieldtransformationQAP}
  \phi&\to\phi+\delta\phi \,,
\end{align}
however here we do not assume that the action is invariant, but
instead we allow a change of the Lagrangian
\begin{align}
  \La & \to \La + \delta\La \,.
\end{align}
By assuming the path integral measure to be invariant under the
transformation, steps analogous to the ones of
Sec.\ \ref{sec:STIformal} lead to
\begin{align}
0 &=  \int{\cal D}\phi \left(
 \int d^Dx \, i(\delta{\cal L} + J_i\delta\phi_i)\right)
e^{i\int d^Dx
  (\La+J_i\phi_i)}.
\label{QAPbasic}
\end{align}
This is the most important basic version of the quantum action principle. 

In an even simpler way one may derive the following relations for
derivatives with respect to an external field $\Ks(x)$ or to a parameter
$\lambda$ appearing in the Lagrangian:
\begin{subequations}
\begin{align}
  \frac{\delta Z(J,\Ks)}{\delta \Ks(x)} & =
   \int{\cal D}\phi \left(
 \frac{\delta}{\delta \Ks(x)}\int d^Dx \, i\La\right)
e^{i\int d^Dx
  (\La+J_i\phi_i)} \,,\\
  \frac{\partial Z(J,\Ks)}{\partial \lambda} & =
   \int{\cal D}\phi \left(
 \frac{\partial}{\partial \lambda}\int d^Dx \, i\La\right)
e^{i\int d^Dx
  (\La+J_i\phi_i)} \,.
\end{align}
\end{subequations}
These are further variants of the quantum action principle.

Similar to Sec.\ \ref{sec:STIformal} it is instructive to rewrite
the quantum action principle in various ways. First, identities for
explicit  Green
functions can be obtained by taking suitable derivatives of the above
identities with respect to sources $J$. In summary,  the three
variants of the quantum action principle then read as follows:
\begin{itemize}
\item{Variation of quantum fields:} $\delta=\int d^Dx
  \delta\phi_i(x)\frac{\delta}{\delta\phi_i(x)}$. 
\begin{equation}
i\,\delta\langle T\phi_1\ldots\phi_n\rangle
= \langle T\phi_1\ldots\phi_n\Delta\rangle\,,
\label{QAPbasicGreenfunctions}
\end{equation}
where $\Delta=\int d^Dx \,\delta{\cal L}$ and the left-hand   side is
an abbreviation of Green functions involving $\delta\phi_i$ as in
Eq.\ (\ref{deltaAbbrev}).\footnote{ Generally, in the present section we use a compact notation and
 suppress field arguments in a self-explanatory way such that
 e.g.\  $\phi_k\equiv\phi_k(x_k)$ and $\int d^Dx
  J_i\phi_i\equiv\int d^Dx
  J_i(x)\phi_i(x)$, etc.
}
\item{Variation of an external (non-propagating) field $\Ks(x)$:}
\begin{equation}
-i\,\frac{\delta}{\delta \Ks(x)}\langle T\phi_1\ldots\phi_n\rangle
= \langle T\phi_1\ldots\phi_n\Delta\rangle\,,
\end{equation}
with $\Delta=\frac{\delta}{\delta \Ks(x)}\int d^Dx {\cal L}$.
\item{Variation of a parameter $\lambda$:}
\begin{equation}
-i\frac{\partial}{\partial\lambda}
\langle T\phi_1\ldots\phi_n\rangle
= \langle T\phi_1\ldots\phi_n\Delta\rangle\,,
\end{equation}
with $\Delta=\frac{\partial}{\partial\lambda}\int d^Dx {\cal L}$.
\end{itemize}
An important further way to rewrite the quantum action principle is in
terms of the generating functional $\Gamma$. By suitable Legendre
transformation and expressing $\delta\phi$ in Eq.\ (\ref{QAPbasic}) by
derivatives with respect to sources $\Ks$ we obtain, in particular, the form
\begin{align}
\label{QAPSofGamma}
  \mathcal{S}(\Gamma) & = \Delta\cdot\Gamma \,,
\end{align}
where $\mathcal{S}(\Gamma)$ is a Slavnov-Taylor operator as in
Eq.\ (\ref{SofGammapreview}) or (\ref{SofGammacl})
and where $\Delta =\mathcal{S}\left(\int
d^Dx\La\right)$. Interestingly, this identity
relates the Slavnov-Taylor identity for full Green functions on the LHS with the
Slavnov-Taylor identity for the action appearing in the path
integral on the RHS. 

\subsection{Proof of the quantum action principle in DReg}
\label{sec:QAPproof}

The derivation presented above is only heuristic
because the path integral measure ${\cal D}\phi$ was assumed to be
invariant under 
the variable transformation. This is precisely the point where the
regularization and renormalization enters. Hence, the quantum action
principle has to be 
established separately for each regularization. Here we consider what
is called the regularized
quantum action principle  in DReg and present its proof. The proof
was first given in Ref.\ \cite{Breitenlohner:1977hr}; here we follow
the presentation of Ref.\ \cite{Stockinger:2005gx}, where the proof
was extended to dimensional reduction.

Put simply, on the regularized level in DReg, all identities presented above are
literally valid, provided all equations are
interpreted as identities between Feynman diagrams regularized in DReg
in $D$ dimensions. A possible interpretation of this validity is that
DReg provides a concrete perturbative definition of the path integral
in which the measure is invariant under all field transformations of
the form (\ref{fieldtransformationQAP}).

For the proof, we focus on the most basic and most complicated case,
Eq.\ (\ref{QAPbasic}) equivalently rewritten for explicit Green
functions in 
Eq.\ (\ref{QAPbasicGreenfunctions}). All
other identities can be treated similarly.
To precisely formulate the statement we rewrite
Eq.\ (\ref{QAPbasicGreenfunctions})  as an identity of 
Feynman diagrams regularized in DReg. As stressed in Sec.\ \ref{sec:relationL} the Feynman
diagrams regularized in DReg can be obtained from the Gell-Mann-Low
formula in $D$ dimensions. We call the regularized Lagrangian simply
$\La$, omitting the superscript $^{(D)}$, and split it
again as
\begin{align}
  \La & =   \La_{\text{free}} +  \La_{\text{int}} \,,
\end{align}
where $\La_{\text{free}}$ determines the propagators in Feynman
diagrams and $\La_{\text{int}}$ may contain terms coupling composite
operators to sources $\Ks$, and it may contain counterterms involving
coefficients with $1/(D-4)$ poles. Then
Eq.\ (\ref{QAPbasicGreenfunctions}) is rewritten as
  \begin{align}
\sum_{k=1}^n      i\,\langle T\phi_1\ldots(\delta\phi_k)\ldots\phi_n\exp(i\textstyle\int d^Dx {\cal L}_{\rm int})\rangle 
& = \langle T\phi_1\ldots\phi_n\Delta\exp(i\textstyle\int d^Dx {\cal
  L}_{\rm int})\rangle\
\label{tobeshown}
\end{align}
with
\begin{align}
\Delta & = \int d^Dx (\delta\La_{\text{free}}+\delta\La_{\text{int}})
\,,
\end{align}
 where both sides of equation (\ref{tobeshown}) are to be evaluated via Wick
contractions in dimensional regularization. This is the statement which
needs to be proven.

Let us write down the three parts of Eq.\ (\ref{tobeshown}) at some
specific order with $N$ powers of $\La_{\text{int}}$. Each term on the
left-hand 
side becomes
\begin{align}
  \frac{ i}{N!}\,\bigg\langle T\phi_1\ldots(\delta\phi_k)\ldots\phi_n
  \underbrace{(i\textstyle\int
 d^Dx_1 {\cal L}_{\rm int})\ldots
 (i\textstyle\int d^Dx_N {\cal L}_{\rm int})}_{\text{$N$ factors}}\bigg\rangle   
\label{tobeshownLHS}
\end{align}
and the term involving $\delta \La_{\text{int}}$ on the right-hand
side becomes
\begin{align}
  \frac{1}{(N-1)!}  \bigg\langle T\phi_1\ldots\phi_n
  (\textstyle\int d^Dx \delta{\cal
  L}_{\rm int})
  \underbrace{(i\textstyle\int d^Dx_1 {\cal
  L}_{\rm int})\ldots(i\textstyle\int d^Dx_{N-1} {\cal
  L}_{\rm int})}_{\text{$N-1$ factors}}\bigg\rangle
\label{tobeshownRHS1}
\end{align}
For the term involving $\delta\La_{\text{free}}$ the discussion of
Sec.\ \ref{sec:relationL} is crucial. The free Lagrangian in DReg
contains fully $D$-dimensional derivative operators and can be
schematically written as $  \La_{\text{free}}  = \frac12 \phi_i
{\cal D}^{(D)}_{ij}\phi_j$, such that $\delta\La_{\text{free}} =
\delta\phi_i{\cal D}^{(D)}_{ij}\phi_j$. Hence the corresponding term
in Eq.\ (\ref{tobeshown}) becomes
\begin{align}
 \frac{ 1}{N!}\,\bigg\langle T\phi_1\ldots\begC2{\phi_k}\conC{\ldots\phi_n
  (\textstyle\int d^Dx }  \begC1{\delta\phi_i}\conC{{\cal D}^{(D)}_{ij}}\conC{\hspace{.5ex}}\endC1{\vphantom{\phi}}\hspace{-.5ex}\endC2{\phi_j}\hspace{-.5ex}\begC3{\vphantom{\phi}}\conC{\, )}
  \endC3{\underbrace{(i\textstyle\int
 d^Dx_1 {\cal L}_{\rm int})\ldots
 (i\textstyle\int d^Dx_N {\cal L}_{\rm int})}_{\text{$N$ factors}}}\bigg\rangle   
  \,.
  \label{tobeshownRHS2}
\end{align}
Each term must be evaluated using Wick contractions. It will be
sufficient to consider 
all possible kinds of Wick contractions for the special field operator
$\phi_j$ in $\delta\La_{\text{free}}$ as indicated in
Eq.\ (\ref{tobeshownRHS2}). This field operator can be Wick contracted
either with $\delta\phi_i$ at the same spacetime point (contraction
(a)), or with an external field operator $\phi_k$ (contraction (b)), or with a field
operator inside one of the $\La_{\text{int}}$ factors (contraction (c)).

For each contraction we can use the crucial property
(\ref{PropagatorsInverse}), which means that the Feynman diagram
propagators are the inverse of the kinetic operators appearing in the
regularized Lagrangian,
\begin{align}
  {\cal D}^{(D)}_{ij}{\cal P}^{(D)}_{jk}&=i\delta_{ik}
  \,.
\end{align}
This relation establishes the relationship between the Lagrangian and
Feynman rules and is the core reason why the quantum action principle
holds. Using this relation, contraction (a) produces a single loop integral over
${\cal D}^{(D)}_{ij}$ times the propagator ${\cal P}^{(D)}_{jl}$ from $\phi_j$ to some field
$\phi_l$ within the composite operator $\delta\phi_i$. The loop
integrand is therefore simply a constant $\delta_{il}$, hence
scaleless and therefore zero.

Contraction (b) with the external field $\phi_k$ produces the
combination ${\cal D}^{(D)}_{ij}{\cal P}^{(D)}_{jk}=i\delta_{ik}$. In
this way the $\int d^Dx$ integral is effectively cancelled and the
external field operator $\phi_k$ is replaced by $i\delta\phi_k$. Hence
the contractions of type (b) in Eq.\ (\ref{tobeshownRHS2}) yield exactly the same
as Eq.\ (\ref{tobeshownLHS}), and we have proven the first required
cancellation.

Finally,
a contraction of type (c) between $\phi_j$ and some field $\phi_l$ within one
of the ${\cal L}_{\rm int}$ factors results in the
product
${\cal D}^{(D)}_{ij}{\cal P}^{(D)}_{jl} \frac{\delta{\cal L}_{\rm
    int}}{\delta \phi_l}$.
Using
the inverse relation for the propagators again, we find that
all contractions of type (c) in Eq.\ (\ref{tobeshownRHS2}) lead to
\begin{align}
 \frac{i^2 N}{N!}\,\bigg\langle T\phi_1\ldots\phi_n
  (\textstyle\int d^Dx \delta\phi_l\frac{\delta{\cal L}_{\rm
    int}}{\delta \phi_l}   )
  \underbrace{(i\textstyle\int
 d^Dx {\cal L}_{\rm int})\ldots
 (i\textstyle\int d^Dx {\cal L}_{\rm int})}_{\text{$N-1$ factors}}\bigg\rangle   
  \,.
\end{align}
This is precisely the negative of Eq.\ (\ref{tobeshownRHS1}). In
total, we have therefore shown the equality
(\ref{tobeshownLHS})$=$(\ref{tobeshownRHS1})$+$(\ref{tobeshownRHS2}),
and we have established the quantum action principle (\ref{tobeshown}).

In the same way it is possible to prove all other identities presented
in Sec.\ \ref{sec:formalQAP}. The essential point in the proof is the
possibility to express Feynman diagrams in DReg via the Gell-Mann-Low
formula together with the relationship between regularized propagators
and the regularized free Lagrangian. Ref.\ \cite{Breitenlohner:1977hr}
gave the proof using the
$\alpha$-representation of all diagrams explained in Sec.\ \ref{sec:Schwingerparametrization},
where the relationship for the propagators is less
obvious. Ref.\ \cite{Stockinger:2005gx} extended the proof to the
consistent version of dimensional reduction.

\subsection{Examples of Applications of the Quantum Action~Principle}
\label{sec:QAPexamples}

The quantum action principle is a very powerful tool to study symmetry 
properties of Green functions. 
Here we provide two example applications which illustrate this. The
examples are very important in their own right, but they also provide
a blueprint for the analysis of chiral gauge theories discussed later.

The first example is gauge invariance in non-chiral gauge theories
such as QED or QCD. The gauge invariant Lagrangian of QCD with one
quark flavour is given by
\begin{subequations}
\begin{align}
  \La_{\text{inv}} & = - \frac14 F^a{}^{\mu\nu} F^a_{\mu\nu}
  + \bar\psi i\slashed{D}\psi
  \,, \\
  D_\mu & =  \partial_\mu+ig T^aA^a_\mu
  \,,
\end{align}
\end{subequations}
where the generators $T^a$ correspond to the triplet representation of
SU(3) and the field strength tensor is defined as in
Eq.\ (\ref{FmunuDef}). The full Lagrangian including gauge fixing and
ghost terms and source terms for BRST transformations is given by
\begin{align}
  \La_{\text{cl}}  =& \La_{\text{inv}} +
  B^a (\partial^\mu A^a_\mu) + \frac{\xi}{2} (B^a)^2 
- \cbar^a\partial^\mu (D_\mu c)^a
\nonumber\\
&+ \rho^a{}^\mu sA^a_\mu + \zeta^a sc^a + {Y}_\psi s\psi+ Y_{\psibar}
s\psibar
\,,
\end{align}
where the BRST transformations are given as in Sec.\ \ref{sec:BRSTSTI}.

All ingredients of the QCD Lagrangian can be interpreted as
$D$-dimensional quantities without any changes in the algebraic
relations. The $D$-dimensional version of $\La_{\text{inv}}$ is still
fully gauge invariant, and the full BRST invariant classical
Lagrangian $\La_{\text{cl}}$ is BRST invariant in $D$
dimensions. Likewise, the Slavnov-Taylor identity (\ref{SofGammacl})
is satisfied in $D$ dimensions.

We therefore have
\begin{align}
  \mathcal{S}\left(\int d^Dx \La_{\text{cl}}\right) & = 0 \,,
\end{align}
for the $D$-dimensional regularized theory. Now we can use the quantum
action principle in the form of Eq.\ (\ref{QAPSofGamma}), where now
the breaking term $\Delta=0$. Accordingly,
the symmetry of the $D$-dimensional classical action implies that the
regularized Green functions represented by the generating functional $\Gamma_{\text{DReg}}$ satisfy the Slavnov-Taylor
identity $\mathcal{S}(\Gamma_{\text{DReg}})=0$ at all orders.

This is the precise form of the statement that DReg preserves gauge
invariance of QCD manifestly. The analogous statement is also true for
QED or other non-chiral gauge theories.
One can go one step further and discuss the renormalized level. If
counterterms are generated from the classical Lagrangian by the
standard procedure of field and
parameter renormalization, the bare Lagrangian
$\La_{\text{bare}}=\La_{\text{cl}}+\La_{\text{ct}}$ still satisfies
the Slavnov-Taylor identity, $\mathcal{S}(\int d^Dx \La_{\text{bare}})=0$. 
For this reason even the renormalized, finite functional
$\Gamma_{\text{DRen}}$ in the notation of Sec.\ \ref{sec:DRegGamma}
which is obtained from $\La_{\text{bare}}$ satisfies the
Slavnov-Taylor identity without the need for special
symmetry-restoring counterterms. The manifest preservation of gauge/BRST
invariance at all steps of the construction of QCD dramatically
simplifies practical calculations as well as all-order proofs.

As our second example we briefly sketch the situation of supersymmetry
in regularization by dimensional reduction. As explained in
Sec.\ \ref{sec:variants}, the dimensional reduction scheme treats
vector fields in quasi 4 dimensions and should therefore be better
compatible with supersymmetry. Without going into the details,
supersymmetry can also be expressed in terms of a Slavnov-Taylor
identity. If a supersymmetric Lagrangian is defined in dimensional
reduction as  $\La_{\text{SUSY}}^{\text{DRed}}$ and this scheme is
defined mathematically consistently, it 
does not remain supersymmetric. Instead, applying the corresponding
Slavnov-Taylor operator yields $\mathcal{S}(\int
d^Dx\La_{\text{SUSY}}^{\text{DRed}})=\Delta\ne0$. The value of
$\Delta$ for a general supersymmetric gauge theory was provided
in Ref.\ \cite{Stockinger:2005gx}. The reason for the non-vanishing value of
$\Delta$ is that the quasi-4-dimensional space does not permit using
Fierz identities. The non-vanishing value of $\Delta$ implies that
ultimately supersymmetry is not preserved by dimensional reduction at
all orders.

Nevertheless, dimensional reduction preserves supersymmetry to a very
large extent, and the quantum action principle provides a succinct
method to check the validity of supersymmetry in concrete cases: The
evaluation of concrete Green functions with an insertion of the
breaking, $\Delta\cdot\Gamma$, directly determines the potential
breaking of the supersymmetric Slavnov-Taylor identity in a concrete
sector. This method was used e.g.\ to verify that supersymmetry indeed
is conserved in a variety of important cases, including
phenomenologically important 2-loop and 3-loop contributions to the
Higgs boson mass prediction in the Minimal Supersymmetric Standard
Model \cite{Hollik:2005nn,Stockinger:2018oxe}.

\newpage
\section{Renormalization in the Context of DReg}
\label{sec:renormalization}
In this section we review basic
renormalization theory in the context of perturbative relativistic
quantum field theories, from the point of view of applications of
dimensional regularization.
Renormalization has both technical and physical aspects.
On the most technical level, renormalization is a procedure to remove
ultraviolet divergences and generate finite Green functions, S-matrix
elements and other quantities of interest.
It effectively provides a definition of each term in the Gell-Mann-Low
formula (\ref{GellMannLow}) and may be viewed as a definition of the path
integral measure (\ref{pathintegral}). 
The removal of ultraviolet divergences is not arbitrary but subject to
important physical constraints such as unitarity and
causality. In more physical terms, renormalization can be viewed as a
reparametrization. This is reflected by the ``main theorem of renormalization''\footnote{  This name was coined in Ref.\ \cite{Popineau:2016lhf}, where also a very general
  proof is given which essentially relies on the physical
  causality constraint.} which states that all
allowed renormalization procedures differ by nothing but
reparametrizations. It is also reflected by the customary
practical procedure of first regularizing 
the theory, then introducing counterterms which depend on the
regularization and cancel the divergences. These counterterms can be
viewed as arising from 
reparametrizations, or renormalizations, of Lagrangian parameters and
fields.

The need for renormalization and the possibility of renormalization to
generate a finite theory also reflect further deep physical
properties of quantum field theories.
The existence of ultraviolet divergences and the resulting need for
subtractions and a renormalization procedure result in the
possibility of so-called anomalies. These are breakings of symmetries
which are valid in the classical theory but broken on the quantum
level via the regularization and renormalization
procedure. Fundamentally anomalies arise if the unitarity and
causality constraints on renormalization are incompatible with the
symmetry in question.

The possibility to successfully carry out the renormalization
programme and its relation to reparametrizations reflects the physical
phenomenon of decoupling. Physics at a certain distance and energy
scale is insensitive to physics at much smaller distance and higher
energy scales, leading to the important concepts of effective field
theories and the renormalization group. Ultra-short
distance details influence  long-distance physics only  via
their effect on long-distance parameters. Since any regularization
effectively changes the short-distance behavior of the theory in a
cutoff-dependent (but unphysical) way, it is not too surprising that
the cutoff-dependencies, including divergences, can be compensated by
reparametrizations such that a finite and regularization-independent
limit exists.

In the present section, we provide a brief review of the general
theorems governing the previous statements; this discussion has a
strong focus on the so-called BPHZ approach to renormalization, and an
outcome is that the customary 
regularization/renormalization procedure  is correct. Then we review
the main theorem stating that dimensional regularization may be
employed  as one such consistent regularization/renormalization framework.

\subsection{General Renormalization Theory and Constraints from
 Unitarity and~Causality}
\label{sec:generalrenormalization}
Here we review basic properties of renormalization as a means to
eliminate ultraviolet (UV) divergences
and to generate finite relativistic quantum field theories. The
discussion is organized along four questions: What are required
properties of any renormalization procedure? Which procedures satisfy
these properties? What are possible differences between different
allowed renormalization procedures? And how does the usual procedure
of regularization and counterterms fit into the fundamental analysis
of renormalization?

As we will discuss, all these questions have rigorous and positive
answers, first obtained by Bogoliubov/Parasiuk \cite{Bogoliubov:1957gp},
Hepp \cite{Hepp:1966eg}, with important additional
developments by Speer \cite{Speer:1971fub,Speer:1974cz,Speer:1974hd} and Zimmermann \cite{Zimmermann:1969jj} and
Epstein/Glaser \cite{Epstein:1973gw}.
We refer to lectures by Hepp \cite{Hepp:1971bda} (contained in Ref\ \cite{DeWitt:1971kda}) for very detailed and pedagogical
explanations and to Ref.\ \cite{Piguet:1980nr} for an overview.

We begin by explaining the fundamental requirements on any
renormalization procedure. A minimal requirement would be that
perturbative S-matrix elements become UV finite; a very strong
requirement would be the nonperturbative construction of well-defined
products of interacting field operators. Following the analysis of the
mentioned references we choose an intermediate approach. In this
approach a {\em renormalization} is a procedure which constructs all
possible time-ordered products of
free field operators,
or equivalently a renormalization is a mapping that maps any Feynman
diagram to a well-defined and UV finite expression. In detail, the
requirement can be efficiently formulated by writing an interaction
Lagrangian 
\begin{align}
  \label{LWigi}
  \La_{\text{int}}(x) & = \sum_i W_i(x)g_{W_i}(x) \,,
\end{align}
where $W_i(x)$ are all local field monomials of interest (including
all monomials appearing in the actual Lagrangian of interest, but also
possible further composite operators of interest, similarly to the
discussion of composite operators in sections \ref{sec:BRSTSTI} and \ref{sec:GreenFunctions}),
and where $g_{W_i}(x)$ are number-valued test functions (acting like
the sources $K_i$ in Sec.\ \ref{sec:BRSTSTI} and \ref{sec:GreenFunctions} or in Eq.\ (\ref{LextBRST}) or like
localized coupling constants). This Lagrangian generates a
perturbative scattering operator $S(g)$, where the argument $g$ denotes the
functional dependence on all the
$g_{W_i}$,
\begin{align}
  S(g) & = 1+ \sum_{n=1}^\infty\frac{i^n}{n!}
  \int \sum_{i_1\ldots i_n} T_{i_1\ldots i_n}(x_1,\ldots,x_n)
  g_{W_{i_1}}(x_1)\ldots  g_{W_{i_n}}(x_n)d^4x_1\ldots d^4 x_n \,,
\label{SEpsteinGlaser}
\end{align}
where formally the appearing $T$-products would be given by
\begin{align}
  T_{i_1\ldots i_n}(x_1,\ldots,x_n) & =
  T(W_{i_1}(x_1)\ldots W_{i_n}(x_n)) \,.
  \label{Tproducts}
\end{align}
However, the expressions in Eq.\ (\ref{Tproducts}) are in general
ill-defined if $n>1$ and several
of the $x_i$ coincide. Hence a {\em renormalization} is a construction
of the $T$-products and thus of Eq.\ (\ref{SEpsteinGlaser}) which
satisfies the following properties, adapted from
Refs.\ \cite{Epstein:1973gw,Popineau:2016lhf}: 
\begin{description}
\item[Initial conditions: ]
\begin{subequations}
  \begin{align}
    S(0) & = 1 \,,
    \\
    T_{i}(x) & = W_i(x) \,.
  \end{align}
\end{subequations}
\item[Unitarity: ]
  \begin{align}
    S(g)^\dagger S(g) & = S(g) S(g)^\dagger = 1 \ 
  \end{align}
  for all hermitian $W_ig_{W_i}$. Here $S(g)^\dagger$ must be written in
  terms of anti-$T$-products $\bar{T}_{i_i\ldots i_n}$ which also must
  be constructed.
\item[Translational invariance: ]
  \begin{align}
    U(1,a) S(g) U(1,a)^\dagger & = S(g_a) \,,
  \end{align}
  where $U(1,a)$ is the representation of translations on the
  respective free Fock space and $g_a(x)=g(x-a)$.
\item[Causality: ]
  \begin{align}
    S(g+h) & = S(g)S(h) &\text{if supp}(g)\gtrsim\text{supp}(h) \,,
  \end{align}
  where $\text{supp}(g)\gtrsim\text{supp}(h)$ means that all points in
  the support of $h$ are outside the support of $g$ and its
  future lightcone, such that the points in supp$(h)$ cannot be
  causally influenced by the points in supp$(g)$.
\end{description}
Via the expansion (\ref{SEpsteinGlaser}) these requirements translate
into constraints on the $T$-products and 
$\bar{T}$-products. For instance, the causality requirement is
particularly powerful \cite{Popineau:2016lhf} and translates into the relation
\begin{align}
  T_{i_1\ldots i_n}(x_1,\ldots,x_n) =&
  T_{i_{j_1}\ldots i_{j_m}}(x_{j_1},\ldots,x_{j_m})
  T_{i_{j_{m+1}}\ldots i_{j_n}}(x_{j_{m+1}},\ldots,x_{j_n})
   \nonumber\\&\text{if
  }\{x_{j_1},\ldots,x_{j_m}\}\gtrsim\{x_{j_{m+1}},\ldots,x_{j_n}\}
\nonumber\\  &\text{and }
\{{j_1},\ldots,{j_n}\}=\{1,\ldots,n\}
\label{Tcausality}
\end{align}
for $T$-products.

A construction fulfilling all these constraints
thus amounts to a construction of all $T$-products of possible field
monomials $W_i$ and thus of all terms appearing in the Gell-Mann-Low
formula (\ref{GellMannLow}) and ultimately of Feynman diagrams and
Green functions, including Green functions of composite
operators. Similar sets of requirements 
can also be found in the Bogoliubov/Shirkov textbook 
\cite{Bogoliubov1980} and, for  Feynman diagrams, in Hepp's lectures
\cite{Hepp:1971bda}.

Let us briefly comment on the central role of unitarity and
causality. Both requirements allow expressing $T$-products with a
certain number of operator factors in terms of $T$-products (or
$\bar{T}$-products) with fewer factors, such as in
Eq.\ (\ref{Tcausality}). Hence higher-order 
$T$-products and thus the entire renormalization procedure are not
arbitrary but largely fixed. The only ambiguity arises when all
arguments are equal, $x_{i_1}=\ldots=x_{i_n}$, in which case causality
and unitarity do not imply a relation to lower-order
$T$-products.

This clarifies that renormalization is not unique and there can be
different {\em renormalization schemes} with different choices to fix these
ambiguities. However it also gives an indication that the ambiguities
affect only local terms, such that different schemes differ only by
reparametrizations of local terms in the Lagrangian (\ref{LWigi}).
Further it is
in line with the fact that UV divergences are local in position space
and can be cancelled (in the presence of a regularization) by adding local
counterterms to the Lagrangian.

The local nature of the ambiguities and possible scheme differences
can be formulated as a rigorous theorem: The statement is that any two
constructions satisfying all requirements listed above differ only in
a finite reparametrization (often called {\em finite renormalization} in the
original literature);
conversely, if an allowed renormalization is 
changed by a finite reparametrization, another allowed renormalization
is obtained. In our formulation, two different
renormalizations may be expressed as
$S_{T}(g)$ and $S_{T'}(G)$,
where $T$ and $T'$ denote the two different constructions of
$T$-products, and $g$ and $G$ represent two different sets of the
prefactors $g_{W_i}$ in the Lagrangian (\ref{LWigi}). A finite
reparametrization may be written as \cite{Popineau:2016lhf}
\begin{align}
  G_{W_i}(x) & = g_{W_i}(x) + \sum_{n=1}^\infty G_{W_i,n}(g,Dg)(x)\,,
\label{finiteRen}
\end{align}
which is a reparametrization
of couplings  expressed in terms of $G_{W_i,n}$, which are local
functions of all $g_{W_j}(x)$ and their
derivatives. The index $n$ denotes the order in perturbation theory.
On this level, the statement is that if both $S_{T}(g)$ and $S_{T'}(G)$,
are allowed renormalizations, then they can be related as
\begin{align}
      S_{T}(g)&=S_{T'}(G)\label{Sequality}
\end{align}
with a
suitable finite reparametrization of the form (\ref{finiteRen}), and
conversely if
$S_{T'}(G)$ is allowed, then any finite reparametrization of the form
(\ref{finiteRen}) effectively defines another allowed renormalization via
requiring (\ref{Sequality}). Ref.\ \cite{Popineau:2016lhf} gave a very general
proof based directly on the causality requirement of renormalizations,
and Ref.\ \cite{Hepp:1971bda} gave a proof on the level of
Feynman diagrams.

Since reparametrizations do not change the physical content of a
theory, this also shows that any two allowed renormalizations are
equivalent, i.e.\ describe the same physics.

Now we turn to the question about which renormalization procedures
exist and how they are related to the counterterm approach
often used in practical computations, giving a brief survey of
approaches and results. Historically, the BPH theorem
constitutes the first rigorous proof 
that all the above properties can be established \cite{Hepp:1966eg,Bogoliubov:1957gp}. These
references used a recursive, so-called $R$-operation and an intermediate
regularization. Though successful, Hepp \cite{Hepp:1971bda} 
assessed the approach as ``hideously'' complicated and noted that a
cleaner approach is provided by analytic regularization
\cite{Speer:1970ss,Speer:1971fub}. Working on the level of Feynman diagrams, the
idea of analytic regularization is to replace the propagator denominator
of any internal line with index $k$ as
\begin{align}
  \frac{1}{\ell_k^2-m_k^2+i\varepsilon}
  & \to
  \frac{1}{(\ell_k^2-m_k^2+i\varepsilon)^{\lambda_k}}
\end{align}
with complex parameters $\lambda_k$. Similar to DReg, there is a
domain for $\lambda_k$ where all integrals are well defined, and
analytic continuation leads to poles at the physical value
$\lambda_k=1$. It is then possible to define the renormalized
expressions via Laurent expansion in $(\lambda_k-1)$ and keeping only
the zeroth-order term.

In this approach the finiteness of the construction as well as the
validity of all required properties including causality and unitarity
are comparatively easy to prove \cite{Hepp:1971bda}.
The equivalence to the counterterm method was at first only established
indirectly by using the equivalence to BPH, but later also directly
\cite{Speer:1971fub}. A drawback of analytic regularization is that the
relation to the Lagrangian is obscured. In contrast to e.g.\ DReg (see
Sec.\ \ref{sec:relationL}) the regularization cannot be expressed in terms of a
regularized Lagrangian.

Though technically more complicated, the BPH approach and the BPH
theorem are very instructive, most importantly since they establish
the connection with the customary procedure of regularization and
counterterms. In this approach, first every Feynman diagram is
regularized, e.g.\ using the Pauli-Villars prescription.
Then, the renormalization procedure is carried out via the so-called
recursive $R$-operation. For any 1PI graph $G$, a subrenormalized amplitude is defined by
\begin{align}
  \overline{\cal R}_G & = G + \sum_{H_1\ldots H_s}
  {G/_{H_1\cup\ldots\cup H_s}}\cdot C(H_1)\ldots
  C(H_s) \,,
\label{Roperation1}
\end{align}
where the sum runs over all possible sets of disjoint 1PI subgraphs $H_i$ of
$G$ (excluding $G$ itself). The object in the sum denotes the
amplitude for the graph
where all the disjoint subgraphs $H_1\ldots H_s$ are shrunk
to points and replaced by the counterterms $C(H_1)\ldots
C(H_s)$. 
The fully renormalized
result and the counterterms are defined as
\begin{subequations}
  \label{Roperation2}
  \begin{align}
  {\cal R}_G & = \overline{\cal R}_G + C(G) \,,\\
  C(G) & = - T \overline{\cal R}_G \,,
\end{align}
\end{subequations}
where $T$ denotes the operation to extract the divergent
part. In the BPH approach, $T$ is defined via a Taylor expansion in
external momenta of a graph and therefore by construction a polynomial
in momentum space.

The BPH theorem \cite{Hepp:1966eg,Bogoliubov:1957gp} states that the renormalized graphs ${\cal
  R}_G$ are finite (in the sense of distributions in momentum space)
and that all required properties are valid. The difficult part of the
proof is the proof of finiteness.
The big advantage of the $R$-operation is its relationship to the
usual counterterm approach.
Indeed it is easy to see that the formula (\ref{Roperation1})
combinatorically corresponds to the
prescription to add to $G$ all possible counterterm Feynman diagrams with all
possible insertions of counterterm vertices; furthermore the
counterterms are local in position
space  and therefore can be obtained from a local counterterm
Lagrangian. For a detailed 
discussion of the $R$-operation and a full proof of its relationship 
to counterterm diagrams and counterterm Lagrangians we also refer to
the monography \cite{Collins_1984}, chap.\ 5.7.

Since both the BPH procedure and
analytic regularization constitute allowed renormalizations they must
be physically equivalent in the sense defined above, i.e.\ they differ
only by reparametrizations/finite renormalization. This equivalence
has also been directly established in Refs.\ \cite{Hepp:1971bda,Speer:1971fub},
where it was also shown that the required finite renormalization only
involves counterterms whose power-counting degree is bounded by the
superficial degree of divergence of the original Feynman
diagrams.\footnote{  Such a renormalization is called ``minimal'' in
Ref.\ \cite{Hepp:1971bda}, but we stress that this is a different
notion 
of minimality than, e.g.\ in the so-called minimal subtraction
prescription within DReg.}

A further instructive and important renormalization procedure was
developed by Zimmermann \cite{Zimmermann:1969jj}, leading to the
notion of BPHZ renormalization. Its main virtue is that it
completely eliminates the need for any regularization but directly
constructs finite momentum-space loop integrals. Its technical tool is
the famous forest formula, which is a direct solution of the
recursive $R$-operation. It allows constructing loop integrals via
repeated applications of Taylor subtractions on the integrand level. A
technical obstacle is that care must be taken to avoid ambiguities
from different loop momentum routings in case the same subdiagram is
inserted into different higher-order diagrams. While the proof of the
finiteness of the construction is highly nontrivial, the proof of
equivalence to the BPH approach is rather straightforward if an
intermediate regularization is employed.

Already Ref.\ \cite{Hepp:1966eg} on the BPH theorem and
Refs.\ \cite{Speer:1970ss,Speer:1971fub,Hepp:1971bda} on analytic
regularization made essential use of the
$\alpha$-parametrization (see Sec.\ \ref{sec:Schwingerparametrization}) in their proofs.
The idea of using the
$\alpha$-parametrization was combined with the forest formula
in Refs.\ \cite{Anikin:1973ra,Speer:1974hd,Bergere:1974zh} to strongly
simplify the 
finiteness proof. These
references applied subtractions via Taylor expansions with respect to
the $\alpha$'s such that directly finite $\alpha$ integrals were
obtained.

\subsection{Theorem on Divergences and Renormalization in~DReg}

\subsubsection{Statement of the~Theorem}
\label{sec:DRegTheoremStatement}
Here we discuss the central theorem of dimensional regularization,
most rigorously established as Theorem 1 in the paper by
Breitenlohner/Maison, Ref.\ \cite{Breitenlohner:1977hr}. In
essence it implies the following: renormalization of relativistic quantum field
theories can be performed using DReg as an intermediate regularization,
the renormalized answer is correct and equivalent to the results from
other consistent schemes discussed in the previous subsection,
and the required
subtractions can be implemented as counterterm Lagrangians.

In more detail it can be formulated as follows. Let $G$ be a 1PI
Feynman graph (in Ref.\ \cite{Breitenlohner:1977hr} a theory without massless particles
is required, Refs.\ \cite{Breitenlohner:1975hg,Breitenlohner:1976te} consider the case with massless
particles). The corresponding regularized Feynman integral ${\cal T}_G$ is
defined as discussed in Sec.\ \ref{sec:DReg}, making use of the
consistently constructed formally
$D$-dimensional covariants and $D$-dimensional
integrals. Ref.\ \cite{Breitenlohner:1977hr} specifically employs the
$\alpha$-parametrization introduced in
Sec.\ \ref{sec:Schwingerparametrization}.

Then it is possible to apply a subtraction algorithm to the graph
that defines first a subrenormalized Feynman integral
$\overline{\cal T}_G$ and finally a fully renormalized Feynman integral
${\cal R}_G$. Assuming 4-dimensional quantum field theory and writing
$D=4-2\epsilon$, these objects have the following properties:
\begin{itemize}
\item
  The regularized but not yet renormalized amplitude ${\cal T}_G$ is a meromorphic
  function of $D$ or equivalently  of $\epsilon$.
\item
  The subrenormalized amplitude $\overline{\cal T}_G$ may have
  singularities in $\epsilon$ which are poles of the form
  \begin{align}
    \label{poletermsDReg}
    \frac{1}{\epsilon}P_G^{(1)}+\ldots+\frac{1}{\epsilon^{L_G}}P_G^{(L_G)} \,, 
  \end{align}
  where $L_G$ is the number of closed loops in the graph $G$. The
  coefficients $P_G^{(k)}$ are polynomials in the external momenta and
  the masses appearing in $G$ (corresponding to local terms in
  position space). The degree of all these polynomials is
  bounded by the superficial power-counting degree of the graph
  $\omega_G=4L_G-2I_G+r_G$ with $I_G$ the number of internal lines in
  $G$ and $r_G$ the power-counting degree of the numerator.
\item
  ${\cal R}_G$ is finite, i.e.\ it is an analytic function of
  $\epsilon$ in a region around $\epsilon=0$.  
\end{itemize}
The theorem provides several crucial additional details:
\begin{itemize}
\item
  The subtraction is organized according to a forest formula which is
  equivalent to Bogoliubov's recursive $R$-operation (we also refer to
  the monograph \cite{Collins_1984} for a detailed explanation).
  For this reason
  the subtraction algorithm is equivalent to adding counterterm
  Feynman diagrams.
\item
  The subtractions corresponding to subgraphs $H$ of $G$, called $C_H$, are
  given by $\overline{\cal T}_H$ with analogous properties to $\overline{\cal
    T}_G$ as explained above.
\item
  The subtractions corresponding to a subgraph $H$ are independent of
  the surrounding graph $G$ --- they really only depend on $H$ itself
  (and of course its subgraphs).
\item
  The renormalized results for all graphs ${\cal R}_G$ are equivalent
  to the results obtained in the BPHZ framework (before Ref.\ \cite{Breitenlohner:1977hr},
  this point had been established also in Ref.\ \cite{Speer:1974cz}).
  This means they
  differ from BPHZ results at most by finite, local counterterms at
  each order, in line with the general theorem discussed around
  Eqs.\ (\ref{finiteRen},\ref{Sequality}). 
\end{itemize}
The previous rather technical details have very important consequences
for practical calculations and physical interpretations:
\begin{itemize}
\item
  The
  combinations of all subtractions of all graphs can be written as a counterterm
  Lagrangian which is local and contains only terms of
  dimensionalities limited by the power-counting of the original
  graphs.
\item
  The renormalized amplitudes constructed in DReg provide a finite
  quantum field theory which is 
  consistent with unitarity and causality in the sense analyzed by
  Refs.\ \cite{Hepp:1966eg,Bogoliubov:1957gp,Zimmermann:1969jj,Epstein:1973gw,Bogoliubov1980}.
\end{itemize}
We provide even further details:
\begin{itemize}
\item
  Initially, all propagators in the integrals are defined via the
  $+i\varepsilon$ prescription in momentum space (which corresponds to
  time-ordering in position space) with $\varepsilon>0$. As long as
  $\varepsilon>0$, the dependence of ${\cal R}_G$ on external momenta
  and masses is infinitely differentiable, i.e.\ of the $C^\infty$
  type. After the limit $\varepsilon\to0$ has been taken, the
  dependencies take the character of tempered distributions. In this
  regard, DReg behaves identically to e.g.\ BPHZ \cite{Hepp:1966eg}.
\item
  The setup of the subtractions requires that all $1/\epsilon$ poles
  are subtracted, even if the coefficients happen to be evanescent in
  the sense defined in Sec.\ \ref{sec:relationL}. In the coefficients
  $P_G^{(k)}$ in Eq.\ (\ref{poletermsDReg}) a 4-dimensional limit is
  not permitted during the subtraction procedure. For the
  counterterm Lagrangian this implies that evanescent operators
  (operators that have no 4-dimensional counterpart since they would
  vanish either in view of Fierz identities or $\gamma_5$ identities
  or because of contractions with $\ghat^{\mu\nu}$)   must
  be included in case they are needed to cancel $1/\epsilon$ poles.
\end{itemize}

\subsubsection{Overview of the~Proof}

The full proof of the theorem explained above requires many
ingredients which need to be analyzed in detail. Most of them are
largely independent of the regularization scheme but related to
Feynman graph theory, relationships between graphs and subgraphs and
structural properties of the 
$\alpha$-parametrization. Several key ideas for the proof are common
to proofs for BPHZ renormalization. The specific aspects of DReg enter
in very localized form.

Here we first list the most important ingredients of
the proof:
\begin{itemize}
\item
  The $\alpha$-parametrized integral can be decomposed into sectors.
\item
  A particularly elegant forest formula holds for each sector of the
  $\alpha$-parametrization. 
\item
  In each sector, clever variable substitutions can be made which lead
  to an explicit
  general formula for the integral.
\item
  There is  a general relationship between the integrand for a certain
  graph and the integrands for corresponding subgraphs and reduced graphs.
\item
  There are a few simple observations for typical integrals and
  functions encapsulating the $1/\epsilon$ poles.
\end{itemize}
  The following subsections will illustrate each of these
ingredients with the help of suitable examples, and will motivate the
general statements, which can all be found in Ref.\ \cite{Breitenlohner:1977hr}. A
further subsection will sketch the essential steps 
of the proof by induction.

\subsubsection{Ingredient 1: Sectors of the $\alpha$ integration}
\label{sec:sectorsalpha}

In Eq.\ (\ref{oneloopresultalpha}) we already considered a simple
one-loop integral transformed into \mbox{Schwinger-,} or
$\alpha$-parametrization. For each internal line of the diagram there
is one $\alpha_l$ parameter, and all $\alpha_l$ are integrated in the
range from $0$ to $\infty$.
It was easy to compute one integral explicitly, and the second
integral could be computed as well. For the general proof of
renormalization we neither want nor need an explicit computation of
all loop integrals. We rather need to transform all integrals into a
uniform structure from which we are able to read off the required
properties. It turns out that decomposing the $\alpha$ integrations
into sectors is extremely helpful in this regard.

\begin{figure}[h]
  \caption{\label{6LoopDiagSubGraphs}
    The same diagram as in Eq.\ (\ref{6LoopDiag}), with additional
    indications of subdiagrams $H_i$ $(i=1\ldots6)$ which define a
    example maximal forest.}
\includegraphics[width=0.99\textwidth]{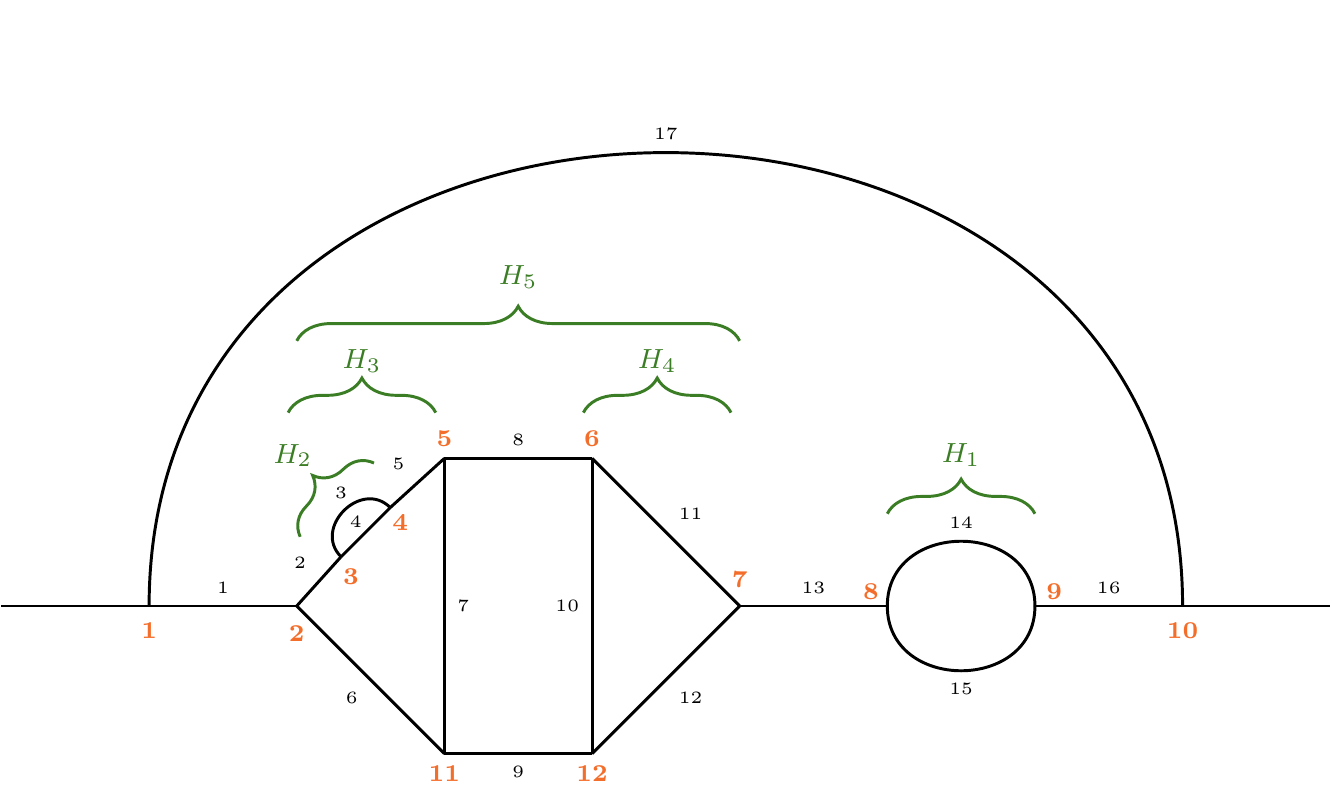}
\end{figure}

The strategy of similar sector decompositions of the $\alpha$ integrations has
been employed also in the important proof of the BPH theorem
in Ref.\ \cite{Hepp:1966eg} and in simplified proofs of the BPHZ theorem in
Refs.\ \cite{Anikin:1973ra,Bergere:1974zh} and is the basis of modern numerical
evaluations of multiloop integrals
\cite{Binoth:2000ps,Binoth:2003ak}. For the integral
(\ref{oneloopresultalpha}) the sector decomposition is very simple:
\begin{subequations}
  \begin{align}
    \int_0^\infty d\alpha_1 d\alpha_2 & = \int_{\text{sector 1}}
    d\alpha_1 d\alpha_2 +  \int_{\text{sector 2}}
    d\alpha_1 d\alpha_2\,,\\
\intertext{where the two sectors are defined as}
    \text{sector 1} & = \{ \alpha_1 \le \alpha_2\} \,,\\
    \text{sector 2} & = \{ \alpha_2 \le \alpha_1\} \,.
\end{align}
\end{subequations}
Let us describe the sector decomposition used for the proof in
Ref.\ \cite{Breitenlohner:1977hr} with the following 6-loop example diagram,
\begin{align}
  \label{6LoopDiag}
\includegraphics[width=0.99\textwidth]{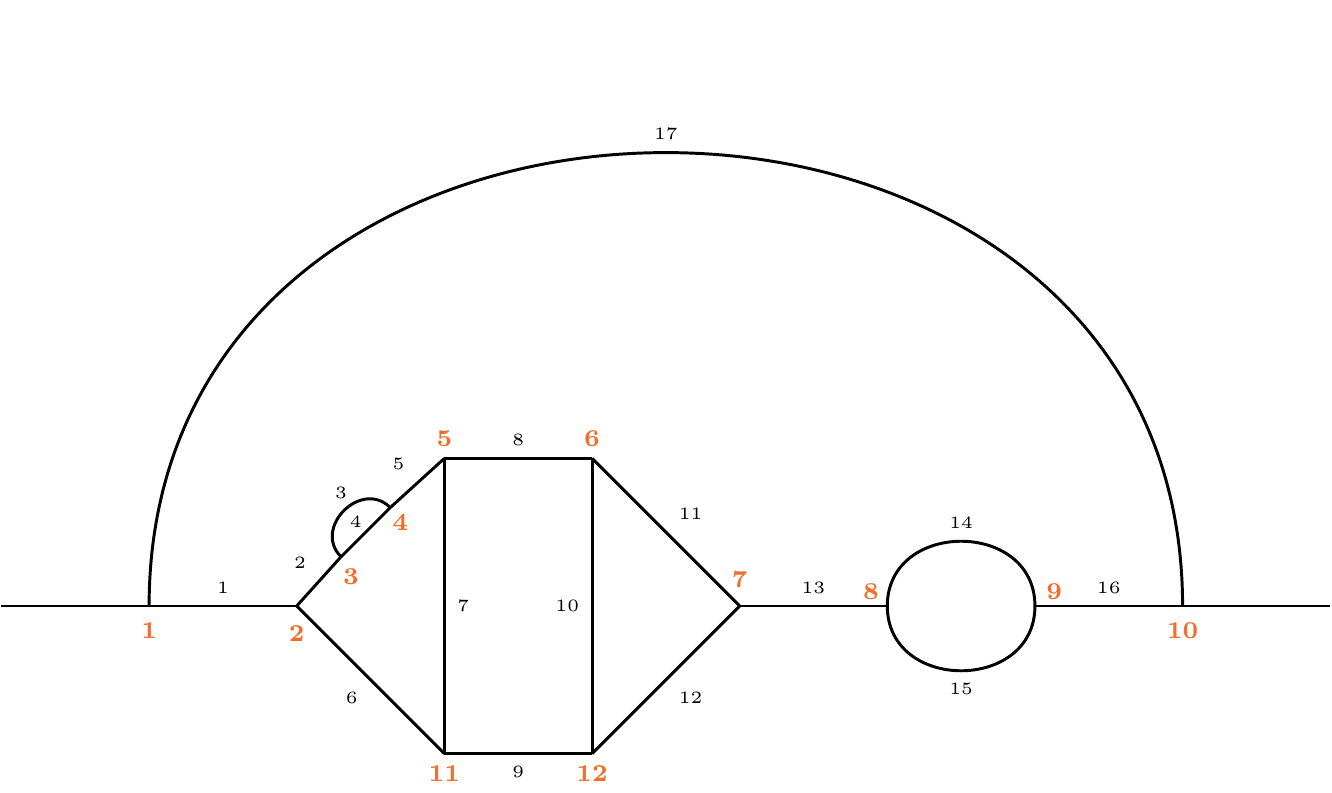}
\end{align}
with line labels and vertex labels as indicated.
One particular sector is constructed by the following algorithm. First
we choose one particular 1-loop subdiagram. As an example we choose
the diagram consisting of the lines $14,15$ and call it $H_1$. Next we
choose either a second, disjoint 1-loop subdiagram, or a 2-loop
subdiagram which contains $H_1$. Let us choose the diagram consisting
of lines $3,4$ and call it $H_2$. Next we choose a subdiagram $H_3$
such that $H_3$ either contains $H_1$ and/or $H_2$ or is disjoint and
such that overall the union of $H_{1,2,3}$ contain three loops. We
might choose $H_3$ as the 2-loop diagram with lines $2,3,4,5,6,7$. We
continue this way until we reach the 6-loop diagram $H_6\equiv G$ itself. An
example choice of subgraphs ${\cal C}=\{H_1,H_2,\ldots,H_5,H_6\}$ is
illustrated in the diagram of Fig.\ \ref{6LoopDiagSubGraphs}.
In this way we can generally construct what is called a {\em maximal
  forest}. In general, the definition of a forest is a set of 1PI
subgraphs of $G$ which are non-overlapping, i.e.\ either disjoint or
nested. A maximal forest is thus a maximal set of 1PI subgraphs which are
non-overlapping. The above construction illustrates how one can
construct all such maximal forests, and it illustrates that each
maximal forest contains as many elements as there are loops in $G$.

\begin{figure}[h]
  \caption{\label{6LoopDiagHighlighted}
    The same diagram as in Fig.\ \ref{6LoopDiagSubGraphs}, with additional
    indications of the labelled lines  $\sigma(H_i)$ for each
    subgraph, according to Eq.\ (\ref{labellingExample}).}    
\includegraphics[width=0.99\textwidth]{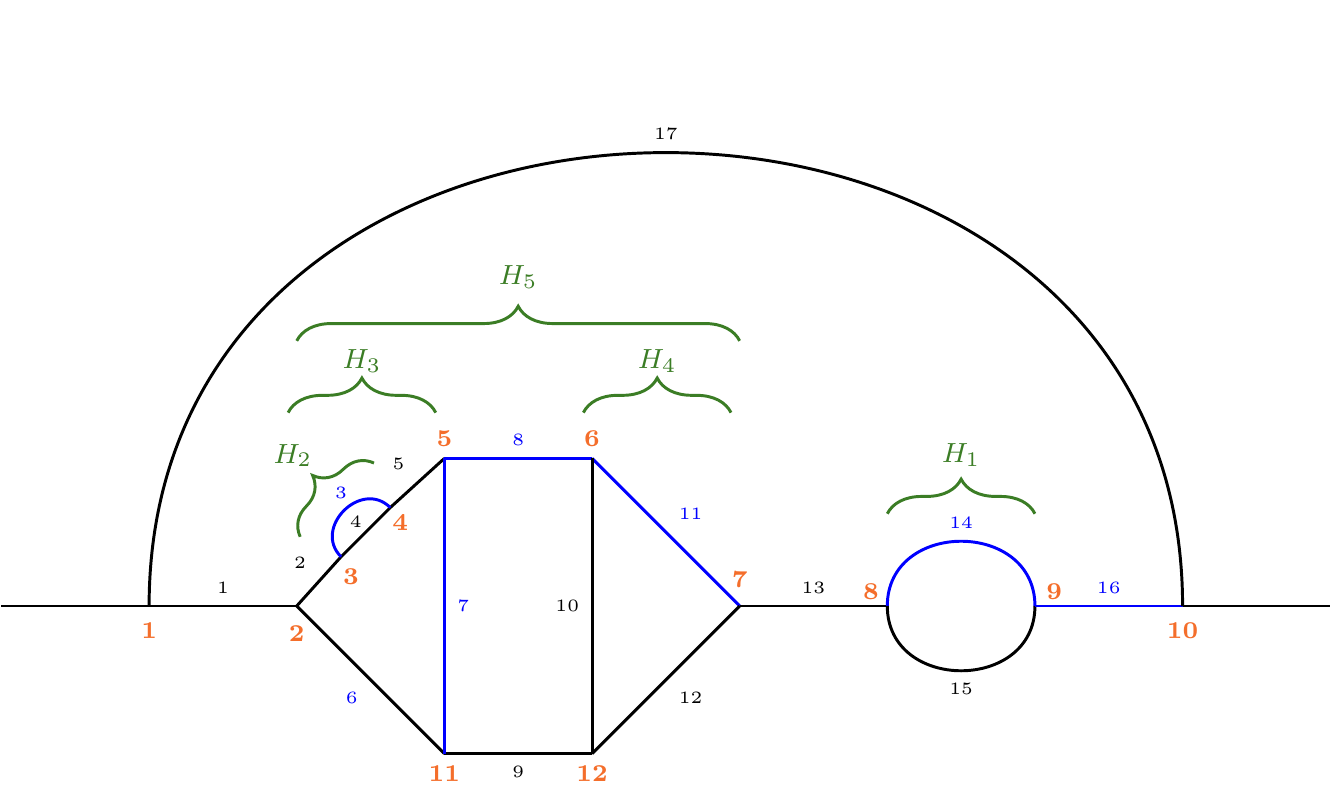}
\end{figure}

The example also illustrates that each subgraph $H_i$ in a maximal forest
contains at least one line that is specific to it, i.e.\ that is not
contained in any smaller subgraphs of the maximal forest. We may
define a mapping, called ``labelling'' in Ref.\ \cite{Breitenlohner:1977hr}, of the form
\begin{align}
  H_i & \mapsto \sigma(H_i) = \text{one of the lines specific to subgraph $H_i$.}
\end{align}
In the example, we can choose
\begin{subequations}
  \label{labellingExample}\begin{align}
  \sigma(H_1) & = 14 & 
  \sigma(H_2) & = 3 & 
  \sigma(H_3) & = 7 \\
  \sigma(H_4) & = 11 & 
  \sigma(H_5) & = 8 & 
  \sigma(H_6) & = 16.
\end{align}
\end{subequations}
The labelled lines are illustrated in blue colour in the diagram of
Fig.\ \ref{6LoopDiagHighlighted}.
For any such choice of a maximal forest together with a labelling for
specific lines, $(\cal C,\sigma)$, we define an integration sector for
the $\alpha_l$ variables in the following way: in each subgraph $H_i$,
the $\alpha$ for the specific labelled line is the largest,
i.e.
\begin{align}\alpha_l&\le\alpha_{\sigma(H_i)}\ \forall\ l\in H_i \,.
\end{align}
For the example the integration sector defined by $(\cal C,\sigma)$ is
\begin{subequations}
  \label{6loopexamplesector}
  \begin{align}
  \alpha_{15} &\le\alpha_{14} &
  \alpha_{4} &\le\alpha_{3} &
  \alpha_{2,3,5,6} &\le\alpha_{7} \\
  \alpha_{10,12} &\le\alpha_{11} &
  \alpha_{7,11,9} &\le\alpha_{8} &
  \alpha_{1,14,8,13,17} &\le\alpha_{16} \,.
\end{align}
\end{subequations}
Note that this does not imply a fixed ordering of all the $\alpha_l$.

It is elementary to prove a variety of useful properties of maximal forests and
labellings. In particular, this way of defining sectors leads to a
partitioning of the  entire $\alpha$ integration region of any Feynman
graph loop integral,
\begin{align}
  \int_0^\infty d\alpha_1\ldots d\alpha_I & =
  \sum_{(\cal C,\sigma)} \int_{(\cal C,\sigma)} d\alpha_1\ldots
  d\alpha_I \,.
\end{align}
Using the notation ${\cal T}_G$ for the regularized amplitude of the
graph G, we can therefore write
\begin{align}
  {\cal T}_G & =   \sum_{(\cal C,\sigma)} {\cal T}_{G,(\cal C,\sigma)}
  \,,
  \label{alphasectors}
\end{align}
with an obvious meaning and where the sum extends over all maximal
forests of $G$ and all possible labellings $(\cal C,\sigma)$.
This construction of sectors is the essential content of Lemma 3 in
Ref.\ \cite{Breitenlohner:1977hr}.

\subsubsection{Ingredient 2: Forest formula after decomposition into
  sectors}
\label{sec:forestformula}
In the all-order investigation of renormalization, the graphical
language of Feynman
diagrams with counterterms has to be formalized in terms of subtractions of
divergent integrals. 
In the historical development of the rigorous BPHZ renormalization,
this formalization was first performed via Bogoliubov's recursive definition of
the so-called $R$-operation. This recursive definition was later
rewritten into Zimmermann's forest formula
\cite{Zimmermann:1969jj}. The sector decomposition described above
permits a very elegant and powerful alternative version of the forest
formula, which simplifies the proof. Such simplified forest formulas
were also discussed and applied in the context of BPHZ e.g.\ in
Refs.\ \cite{Anikin:1973ra,Speer:1974hd,Bergere:1975ei}.

To explain these relations we begin with the recursive $R$-operation,
defined in Eqs.\ (\ref{Roperation1},\ref{Roperation2}). We recall the
main equation, the definition of  a subrenormalized amplitude,
\begin{align}
  \overline{\cal R}_G & = G + \sum_{H_1\ldots H_s}
  {G/_{H_1\cup\ldots\cup H_s}}\cdot C(H_1)\ldots
  C(H_s) \,,
\end{align}
where the counterterms are defined as $C(H)=- T \overline{\cal R}_H$.

This $R$-operation is recursive because the definition of the
subrenormalized amplitude depends on lower-order counterterms which in
turn are defined via lower-order subrenormalized amplitudes. One may
work out the recursion and obtain a direct, non-recursive formula. To
illustrate this, consider the case where the full graph $G$ has one
2-loop subgraph $\gamma_2$, which in turn has a 1-loop subgraph
$\gamma_1$. Then one term in $\overline{\cal R}_G$ is given by
\begin{align}
  \overline{\cal R}_G & = \ldots +  {G/_{\gamma_2}} \cdot C(\gamma_2)
  \nonumber
  \\
  & = \ldots + {G/_{\gamma_2}} \cdot [-T\overline{\cal R}(\gamma_2)]
  \nonumber
  \\
  & = \ldots + {G/_{\gamma_2}} \cdot
    [-T({\gamma_2}+{\gamma_2/_{\gamma_1}}\cdot C(\gamma_1)+\ldots)]
  \nonumber
  \\
  & = \ldots + {G/_{\gamma_2}} \cdot
    [-T{\gamma_2}+T({\gamma_2/_{\gamma_1}}\cdot T{\gamma_1})+\ldots]
    \,.
\label{motivateforestformula}
\end{align}
Hence, working out the recursion leads to subtraction operators $T$
acting on unrenormalized (potentially multiloop) graphs like
$\gamma_2$ and to iterated subtractions. 
If we introduce a new notation $T_\gamma \cdot G\equiv
G/_{\gamma}\cdot T\gamma$ for the operation ``replace $\gamma$ within
$G$ by $T\gamma$'', where products are defined as e.g.\ $T_{\gamma_2}
\cdot T_{\gamma_1} \cdot G=G/_{\gamma_2}\cdot T(T_{\gamma_1}\cdot\gamma_2)$, then we can rewrite the above terms as
\begin{align}
  \ldots + G - T_{\gamma_2} \cdot G + T_{\gamma_2} \cdot T_{\gamma_1} \cdot G
  \,.
\label{Tproductexample}
\end{align}
We note that both the subgraph $\{\gamma_2\}$ as well as the chain of subgraphs $\{\gamma_1,\gamma_2\}$ constitute  forests in
the sense defined above.

In general, if $\gamma_1$ and $\gamma_2$ are subgraphs and elements of
a forest of $G$, we
define
\begin{subequations}
  \label{Tproductdefinition}
  \begin{align}
    \text{$\gamma_2\supsetneq\gamma_1$:}&&T_{\gamma_2}
\cdot T_{\gamma_1} \cdot G&=G/_{\gamma_2}\cdot
T(T_{\gamma_1}\cdot\gamma_2)
\\
\text{$\gamma_1$, $\gamma_2$ disjoint:}&& T_{\gamma_2}
\cdot T_{\gamma_1} \cdot G&=G/_{\gamma_1\cup\gamma_2}(T\gamma_1)( T\gamma_2)
  \end{align}
\end{subequations}
while the product $T_{\gamma_2}\cdot T_{\gamma_1}$ is undefined for
the case when $\gamma_2$ is subgraph of $\gamma_1$. Working out the
recursion formula in general leads to the 
following forest formula \cite{Zimmermann:1969jj}
\begin{align}
  {\cal R}(G) & =   \sum_{\begin{array}{c}\scriptstyle{\cal F}=\text{forest}\\\scriptstyle\text{of $G$}
    \end{array}}
  \prod_{\gamma_i\in {\cal F}}(-T_{\gamma_i})\cdot G \,,
\end{align}
where the forests may contain the full graph $G$ and where also the
empty set is an allowed forest ${\cal F}=\emptyset$.
The formula for
$\overline{\cal R}(G)$ is similar but the forests may not contain the full
graph $G$. The $T_{\gamma_i}$-operators are by definition always ordered 
as in Eqs.\ (\ref{Tproductexample},\ref{Tproductdefinition}) according
to nesting. Simply put: operators with
bigger subgraphs act on the left, operators with  subgraphs contained in
the bigger subgraphs on the
right. The forest formula can be easily 
proven by noting that every
forest that does not contain $G$ itself has certain disjoint maximal
elements $M_1\ldots M_s$ and can 
be partitioned into forests of the $M_1\ldots M_s$. Based on this, the
equivalence to the recursive formula can be established by induction
over the number of loops.

Now we turn to the announced elegant simplification of the forest
formula due to the sector decomposition. We need to know one
additional statement about sectors relevant for combinations like
\begin{align}
  T_\gamma\cdot G & =
  G/_{\gamma} \cdot T(\gamma) =
  \sum_{\begin{array}{c}(\scriptstyle{\cal
      C}_1,\sigma_1)\\\scriptstyle\text{for }G/_{\gamma}
    \end{array}}
  \sum_{\begin{array}{c}\scriptstyle({\cal
      C}_2,\sigma_2)\\\scriptstyle\text{for }{\gamma}
    \end{array}}
  (G/_{\gamma})_{({\cal
      C}_1,\sigma_1)} \cdot T(\gamma_{({\cal
      C}_2,\sigma_2)}) \,.
\end{align}
The statement is that there is a one-to-one correspondence between
such combinations for sectors $({\cal      C}_1,\sigma_1)$, $({\cal
      C}_2,\sigma_2)$ for the graphs $G/_{\gamma}$ and $\gamma$ and
sectors $({\cal C},\sigma)$ for the full graph with the constraint
that $\gamma\in{\cal C}$. Then we can split the forest formula into
sectors as
follows,
\begin{align}
  {\cal R}(G) & = 
  \sum_{\begin{array}{c}\scriptstyle{\cal F}=\text{forest}\\\scriptstyle\text{of $G$}
  \end{array}}
  \sum_{\begin{array}{c}\scriptstyle({\cal C},\sigma) \text{ for $G$
        which}\\\scriptstyle\text{contain all $\gamma\in{\cal F}$}
    \end{array}}
  \prod_{\gamma_i\in {\cal F}}(-T_{\gamma_i}|_{\text{subsector
  }})\cdot G|_{{\text{subsector}}} \,,
\end{align}
where it is used that every sector $({\cal C},\sigma)$ with the given
constraint generates appropriate subsectors for all subtraction operators
$T_{\gamma_i}$ and the remaining reduced graph, and that all possible
subsectors are generated in this way.
Abbreviating slightly we can then rearrange as
\begin{align}
  {\cal R}(G) & = 
  \sum_{{\cal F}}\quad 
  \sum_{{\cal C}\supseteq{\cal F}} \quad
  \prod_{\gamma_i\in {\cal F}}(-T_{\gamma_i})\cdot G
  \nonumber\\
  & = 
  \sum_{{\cal C}} \quad
  \sum_{{\cal F}\subseteq{\cal C}}\quad 
  \prod_{\gamma_i\in {\cal F}}(-T_{\gamma_i})\cdot G
  \nonumber\\
  & =
  \sum_{{\cal C}} \quad
  \prod_{\gamma_i\in {\cal C}}(1-T_{\gamma_i})\cdot G \,.
\end{align}
The last step has used that the sum over all possible forests ${\cal
  F}$ which are contained in ${\cal C}$ effectively generates the
power set of ${\cal C}$, i.e.\ the set of all possible subsets of
${\cal C}$. This simply leads to the last line, which contains only a
summation over all maximal forests ${\cal C}$ and the factors
$(1-T_{\gamma_i})$. 
In this way the forest formula becomes
\begin{subequations}
  \label{sectorforestformula}
  \begin{align}
  {\cal R}(G) & = \sum_{({\cal C},\sigma)} {\cal R}(G)_{({\cal
      C},\sigma)}
  \,,\\
  {\cal R}(G)_{({\cal
      C},\sigma)}
  & =
  \prod_{\gamma_i\in {\cal C}}(1-T_{\gamma_i}) \cdot G|_{({\cal
      C},\sigma)} \,.
\end{align}
\end{subequations}
The ordering of the $(1-T_{\gamma_i})$-operators is as in the original
forest formula, according
to the nesting of subgraphs.

This represents an important improvement. The operators
$(1-T_{\gamma_i})$ have the effect of replacing an object by the one
without the subdivergences from the subgraph $\gamma_i$ (in the
appropriate sector). Intuitively, every such operator improves the
finiteness. On a more technical level, consider what any specific
$T_\gamma$ for a multiloop subgraph $\gamma$ acts onto. In the
original forest formula, there are terms 
such as $T_\gamma\cdot G$ which lead to $G/_{\gamma}\cdot
T(\gamma)$. The $T(\gamma)$ is the divergence of the unrenormalized
multiloop graph $\gamma$, which is typically a very complicated
expression, non-polynomial in momentum space, or non-local in position
space. In contrast, in the forest formula modified for sectors, any
such multiloop $T_\gamma$ only acts onto expressions where all
subdivergences corresponding to subgraphs of $\gamma$ have already
been subtracted:
\begin{align}
T_\gamma\prod_{\gamma_i\in {\cal
      C},\gamma_i\subsetneq\gamma}(1-T_{\gamma_i})\cdot G_{({\cal
      C},\sigma)}
  & =
  T\left(\prod_{\gamma_i\in {\cal
      C},\gamma_i\subsetneq\gamma}(1-T_{\gamma_i})\cdot \gamma\right)\cdot G_{({\cal
      C},\sigma)}
\end{align}
Hence here the left-most $T$ actually acts on the fully
subrenormalized expression $\overline{\cal R}(\gamma)$ in the appropriate
subsector, which can be hoped to have simpler, polynomial/local
divergences. These properties of the forest formula help in setting up 
an inductive proof of renormalization.

\subsubsection{Ingredient 3: Sector Variables and Formula for
 the~Integral}

Introducing sectors into the $\alpha$ integrations required for
Feynman graph integrals has further important advantages. Besides~yielding the
simpler forest formula, the~sectors allow rewriting the actual
integrals such that the power counting and the structure of
divergences are isolated in a quite transparent way. Here we 
illustrate this in a very simple case; then we  provide the general
result and give comments.

Let us focus on the integral (\ref{oneloopresultalpha}),
$
(4\pi)^{-D/2} i^{1-D/2}  \int_0^\infty d\alpha_1 d\alpha_2  
{\cal U}^{-D/2} e^{iW}$
and consider
the sector $\alpha_1\le \alpha_2$. In this sector we introduce
sector-specific variables: the largest $\alpha$ in the sector is
replaced by a new variable $t^2$, the other $\alpha$ is rewritten as
$t^2\beta$ in
terms of a scaling variable $\beta$ which runs from 0 to 1. In total
we carry out the following 
substitution of variables and the integration measure in the sector
\begin{subequations}
  \label{alpharescale1loop}
\begin{align}
  \alpha_2 & = t^2 \,,   \\
  \alpha_1 & = t^2\beta \,,  \\
  \int_{0\le\alpha_1\le\alpha_2 }d\alpha_1 d\alpha_2 & = 2 \int_0^\infty dt t^{(2I-1)}
  \int_0^1 d\beta \,,
\end{align}
\end{subequations}
where $I=2$ is the number of internal lines. The integral
(\ref{oneloopresultalpha}) depends on two functions, the Symanzik polynomial ${\cal U}$ and
the exponent $W$ given in Eq.\ (\ref{UWexample}). After the variable substitution
the Symanzik polynomial takes the value
\begin{align}
  {\cal U} & = M = t^2(1+\beta) \,,
\end{align}
and we observe that we can factor
out the variable $t^2$. This is no accident. As
already mentioned in Sec.\ \ref{sec:Schwingerparametrization}, the
behaviour of ${\cal U}$ if some 
$\alpha$'s vanish  reflects the ultraviolet behaviour of the
original Feynman integral. If all $\alpha$'s simultaneously vanish
$\propto t^2$,
${\cal U}$ generally behaves as $t^{2L}$, where $L$ is the number of
loops in the graph.
We can exhibit this behaviour by defining a new function $\tilde{d}$
\begin{align}
  {\cal U} & = t^2 \tilde{d} \,,
  &
  \tilde{d} & = 1+\beta \ge1\,.
\end{align}
The indicated inequality provides a very important lower bound on the
function $\tilde{d}$.

A second observation is that we can essentially eliminate the $t$-variable from
the exponent $W$ by rescaling the physical variables $p$, $u_{1,2}$ and $m$ as
\begin{subequations}
\label{rescalingpmu}
  \begin{align}
  \tilde{p} & = t\,p \,,\\
  \tilde{m} & = t\,m \,,\\
  \tilde{u}_{1,2} & = t^{-1}\,u_{1,2} \,.
\end{align}
\end{subequations}
The rescaled variables are dimensionless.
In terms of these variables we can write the exponent as
\begin{align}
  W & = \frac{\tilde{p}^2\beta-\beta \tilde{u}_2\cdot \tilde{p}+
    \tilde{u}_1\cdot \tilde{p}-\frac14(\tilde{u}_1+\tilde{u}_2)^2}{(1+\beta)} +
  (it^2\varepsilon-\tilde{m}^2 )(1+\beta)\,,
  \label{Wexamplet}
  \end{align}
where indeed $t$ does not appear explicitly, except in the product $t^2\varepsilon$.

Using all these ingredients we can rewrite the $\alpha$ integral
(\ref{oneloopresultalpha}) in the considered sector as
\begin{align}
  (-i)^2(4\pi)^{-D/2} i^{1-D/2}  2\int_0^\infty dt t^{-D+2I-1}
  \int_0^1 d\beta \tilde{d}^{-D/2} e^{iW} \,,
\label{onelooptbetaresult}
\end{align}
where we record the following observations:
\begin{itemize}
\item
  The power-like behaviour of $\int dt t^{-D+2I-1}$ corresponds to the
  superficial ultraviolet power counting of the original loop integral
  (\ref{oneloopresultalpha}) which behaves like $\int d^Dk k^{-2I}$.
\item
  The remaining integrand $\tilde{d}^{-D/2}e^{iW}$ has essentially no explicit dependence
  on $t$ at all; it only depends on $t$ via the rescaled variables
  (\ref{rescalingpmu}) and via $t^2\varepsilon$.
\item
  If
  $\varepsilon>0$ in the $+i\varepsilon$ prescription, $e^{iW}$
  decreases exponentially for large $t$, and the full dependence of
  the integrand on the rescaled variables 
  (\ref{rescalingpmu}) and on $\beta$ is of $C^\infty$-type.
  The result of the  $\beta$ integration is still
    $C^\infty$ in the rescaled variables.
\end{itemize}

\begin{figure}[h]
  \caption{\label{fig:6loopgraphwithmomenta}
  Illustration of sectors and sector variables  $t$ and $\beta$  in
  Eqs.\ (\ref{tVariableExample},\ref{SubAwareMomenta}). The example is the 6-loop diagram and
  its subdiagrams already used in Eq.\ (\ref{6LoopDiag}) and
  Figs.\ \ref{6LoopDiagSubGraphs}, \ref{6LoopDiagHighlighted}. Here we
  choose six different colours for the reduced subdiagrams $\bar{H}_i$
  $(i=1\ldots6)$ into which the diagram can be partitioned. }
\includegraphics[width=0.99\textwidth]{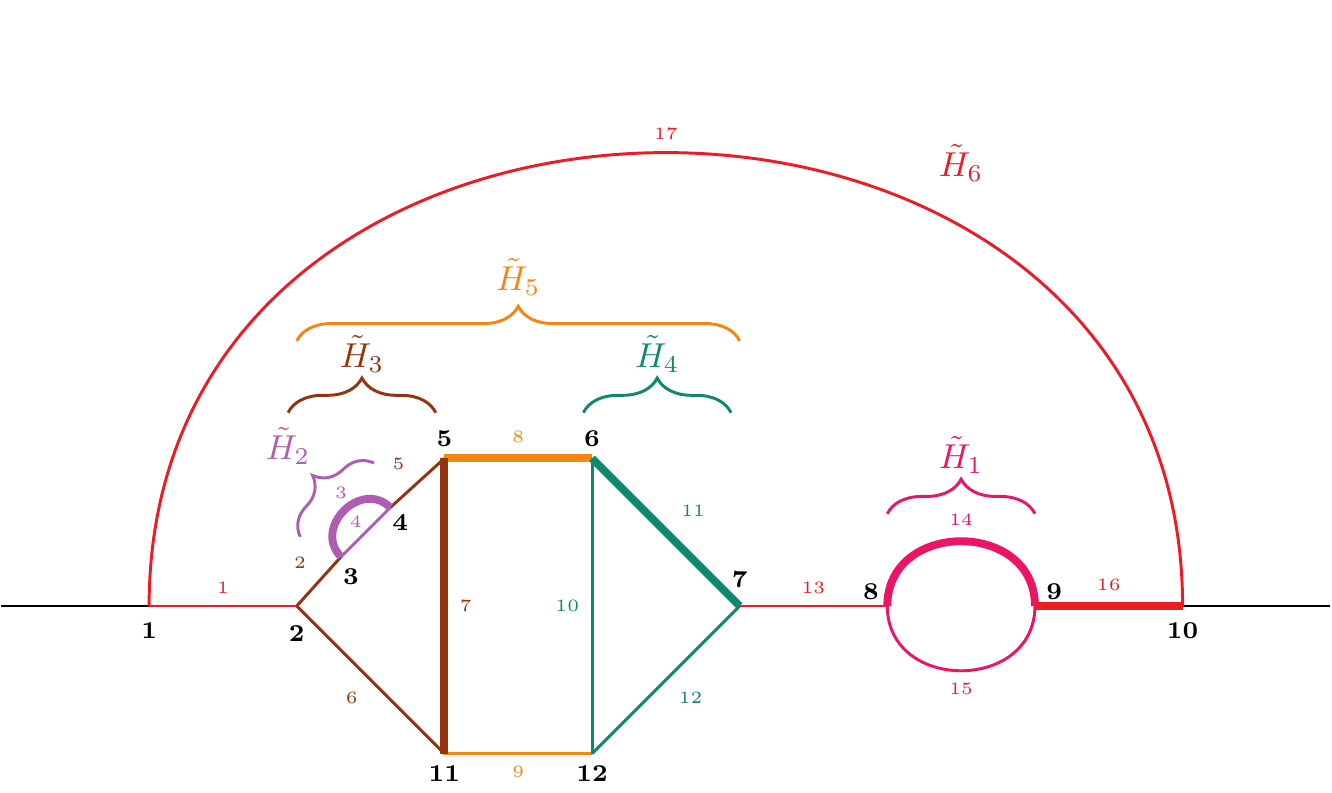}
\end{figure}

We need a second example to shape our understanding of the general
case. Let us consider again the 6-loop diagram of
Sec.\ \ref{sec:sectorsalpha} and fix the same sector $({\cal
  C},\sigma)$ discussed there, see
Eq.\ (\ref{6loopexamplesector}). Which variable substitutions
analogous to Eqs.\ (\ref{alpharescale1loop},\ref{rescalingpmu}) should
we now choose? The sector is defined by a maximal forest with six
subgraphs, each subgraph contains one specific labelled line, and for
each subgraph there is an inequality stating that the labelled
$\alpha$ is the largest. The idea, generalizing the 1-loop case, is to
introduce one $t_i$-variable for each subgraph $H_i$ and to define the
labelled $\alpha$'s in terms of these $t_i$-variables. The $t_6\equiv
t_G$-variable corresponding to the full graph runs from $0$ to
$\infty$, and all the other $t_i$ run from $0$ to $1$. Then all
inequalities for the labelled $\alpha$'s are implemented by the
following scheme:
\begin{subequations}
\label{tVariableExample}  \begin{align}
  \text{subgraph:} &&\text{label}&\text{led $\alpha$ substitution:} &
  \text{re}&\text{write}
  \nonumber
  \\
  H_1 && \alpha_{14} &= t_1^2 t_6^2 && t_1^2\xi_1^2
  \\
  H_2 && \alpha_{3} &= t_2^2t_3^2t_5^2 t_6^2 && t_2^2\xi_2^2
  \\
  H_3 && \alpha_{7} &= \quad t_3^2 t_5^2 t_6^2 && t_3^2\xi_3^2
  \\
  H_4 && \alpha_{11} &= \quad t_4^2 t_5^2 t_6^2 && t_4^2\xi_4^2
  \\
  H_5 && \alpha_{8} &= \quad\quad t_5^2 t_6^2 && t_5^2\xi_5^2
  \\
  H_6 && \alpha_{16} &=  \quad\quad\quad t_6^2 && t_6^2\xi_6^2
\end{align}
\end{subequations}
where also abbreviation variables $\xi_i$ were introduced; they
are products of all the ``other $t_i$'', as appropriate. In the next
step we introduce $\beta_k$-variables for all the remaining,
non-labelled, $\alpha$'s, where the $\beta_k$ all run from $0$ to $1$.
We remark that
$t_6\equiv t_G$ is dimensionful, while all other $t_i$ and $\beta$
variables are dimensionless. In addition we introduce two further useful
notations, illustrated in the graph in
Fig.\ \ref{fig:6loopgraphwithmomenta}. First, for each subgraph in
${\cal C}$ we define a reduced subgraph $\bar{H}_i=H_i/_{{\cal M}(H_i)}$, where ${\cal
  M}(H_i)$ is the set of maximal elements in ${\cal C}$ which are
properly contained in $H_i$. The lines in $\bar{H}_i$ are the lines
specific to $H_i$, i.e.\ the lines contained in $H_i$ but in no
smaller subgraph in ${\cal C}$. Clearly, the full graph is partitioned
into $\bar{H}_i$, i.e.\ every line is in one unique $\bar{H}_i$.
Second, we denote by $\underline{q}_{H_i}$
a set of independent external momenta of $\bar{H}_i$, where we in
principle allow nonzero incoming momenta into all vertices of the
graph (the graph is drawn as if it has only two external momenta, but
the renormalization procedure becomes more systematic if every graph is
generalized to allow arbitrary incoming momenta into all vertices).
This leads to the following scheme:
\begin{subequations}\label{SubAwareMomenta}
  \begin{align}
  \text{red.}&\text{\ subgraph:} &&\text{$\alpha$'s}
  &&\text{indep.\ }\text{ext.\ momenta}\nonumber\\
  \bar{H}_1 &= H_1 &
  \{\alpha_{15},\alpha_{14}\}&=\{\beta_{15},1\}\times t_1^2\xi_1^2 &&
  p_8\\
  \bar{H}_2 &= H_2 &
  \{\alpha_{4},\alpha_{3}\}&=\{\beta_{4},1\}\times t_2^2\xi_2^2 &&
  p_3\\
  \bar{H}_3 &= H_3/_{H_2} &
  \{\alpha_{2,5,6},\alpha_{7}\}&=\{\beta_{2,5,6},1\}\times t_3^2\xi_3^2 &&
  p_2,p_5,p_{11}\\
  \bar{H}_4 &= H_4 &
  \{\alpha_{10,12},\alpha_{11}\}&=\{\beta_{10,12},1\}\times t_4^2\xi_4^2 &&
  p_6,p_7\\
  \bar{H}_5 &= H_5/_{H_3\cup H_4} &
  \{\alpha_{9},\alpha_{8}\}&=\{\beta_{9},1\}\times t_5^2\xi_5^2 &&
  p_6+p_7+p_{12}\\
  \bar{H}_6 &= H_6/_{H_5\cup H_1} &
  \{\alpha_{1,13,17},\alpha_{16}\}&=\{\beta_{1,13,17},1\}\times t_6^2\xi_6^2 &&
  p_1,p_8+p_9,p_{10}
\end{align}
\end{subequations}
The reduced subgraphs $\tilde{H}_i$ are formed solely by the lines proper to them, and lines shared amongst the $H_i$ are shrunk to a point. The subgraphs $H_1, H_2, H_4$ are identical to the reduced ones and hence take the obvious form as depicted in Fig.\ \ref{fig:6loopgraphwithmomenta}. In the case of $H_3, H_5, H_6$, the reduced subgraphs are obtained by shrinking different subgraphs to a point. Let us illustrate this by specifying the form of these reduced subgraphs as follows:
\begin{equation}\centerline{  \includegraphics[width=0.99\textwidth]{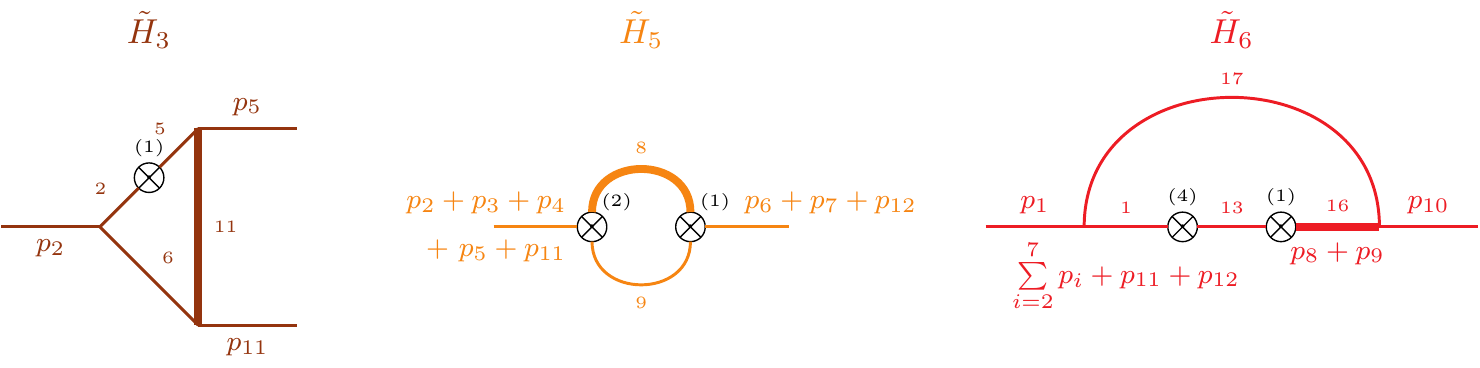}}
\end{equation}
The crossed dots of order $(n)$ denote the counterterm insertion due
to shrinking the respective $n$-loop subgraph to a point. Since we
assume that to each vertex $V_i$ there is associated an entering
momentum $p_i$, shrinking a subgraph comprised of vertices
$V_{i_1},\dots,V_{i_k}$, leads to a combination of incoming momenta
$p_{i_1}+\dots+p_{i_k}$ for that counterterm vertex, as indicated in
the graphs. In choosing independent momenta, we can make use of momentum conservation. For a reduced subgraph with $n$ vertices, it is sufficient to specify $n-1$ incident momenta to the vertices. They uniquely characterize the momenta of a given reduced subgraph. What is more, all momenta of the graph can be reconstructed by linear combinations of these independent external momenta. A specific choice is given in Eq.\ (\ref{SubAwareMomenta}).
Clearly, in this way all inequalities of the sector
(\ref{6loopexamplesector}) are implemented and
the combination of all the selected independent incoming
momenta of the $\bar{H}_i$ span all independent incoming momenta of
the full graph and can be used as independent variables in the result.
The variable substitution leads to the following replacement of the
integration measure, analogous to Eq.\ (\ref{alpharescale1loop}),
\begin{align}
    \int_{({\cal C},\sigma)}d\alpha_1\ldots d\alpha_{17} & = 2^L
    \int_{\substack{t_6=0\ldots\infty\\t_{1\ldots5}=0\ldots 1
      }
      }\prod_{i=1}^6dt_i t_i^{(2I_{H_i}-1)}
  \int_0^1 \prod_k d\beta_k \,,
\end{align}
where $I_{H_i}$ is the number of internal lines in $H_i$.

This example provides us with sufficient information to construct the
general result for the integral representation of a general 1PI graph $G$ in a
specific sector $({\cal C},\sigma)$. As in the example, the sector
defines a chain of subgraphs $H_i$ (as many as there are loops;
one of them is equal to the full graph $G$). The sector also defines
a particular replacement of all $\alpha$'s in terms of $t_i$ and
$\beta_k$; for each subgraph $H_i$ it is also useful to define the
variable $\xi_i$ for the product of all the ``other $t_i$''.
All lines of $G$ are partitioned into lines of the reduced graphs
$\bar{H}_i$, and for each $\bar{H}_i$ one can choose a set of
independent incoming momenta $\underline{q}_{H_i}$, which in total
span all incoming momenta of the full graph. Since each line carries
one mass variable and one $u$-variable, we can also partition these
variables into sets of masses $m_{H_i}$ and sets of $u$'s, $u_{H_i}$,
corresponding to the respective $\bar{H}_i$.

With these variables we can rescale physical quantities, generalizing
Eq.\ (\ref{rescalingpmu}), as
\begin{subequations}
  \label{rescalinggeneral}
  \begin{align}
  \tilde{\underline{q}}_{H_i} & = t_i\xi_i \underline{q}_{H_i} \,, \\
  \tilde{m}_{H_i} & = t_i\xi_i m_{H_i} \,, \\
  \tilde{u}_{H_i} & = (t_i\xi_i)^{-1} u_{H_i} \,.
\end{align}
\end{subequations}
We allow the integral to contain a numerator expressed as a derivative with
respect to $u$-variables as in
Eqs.\ (\ref{loopintegralgeneral},\ref{alpharesultgeneral}), but we
assume that the derivative operator  $Z$ in the numerator is a product
of $Z_{H_i}$, where each $Z_{H_i}$ only depends
on variables specific to $\bar{H}_i$. This is always the case in
actual Feynman diagrams. For simplicity we follow Ref.\ \cite{Breitenlohner:1977hr} and
assume that all $Z_{H_i}$ are homogeneous polynomials in the variables
$\partial/\partial u_{H_i}$ and $m_{H_i}$ of some degree
$r_{\bar{H}_i}$.
Then we can write
\begin{align}
  \tilde{Z}_{H_i}& =(t_i\xi_i)^{r_{\bar{H}_i}}{Z}_{H_i}
\end{align}
where $\tilde{Z}_{H_i}$ is the same homogeneous polynomial expressed
with $\partial/\partial\tilde{u}_{H_i}$ and $\tilde{m}_{H_i}$.
Writing $D=4-2\epsilon$ we can finally define a power-counting degree
of each reduced subgraph 
$\bar{H}_i$ and the complete (sub)graphs $H_i$ as
\begin{subequations}
\begin{align}
  \omega_{\bar{H}_i} & = 4L_{\bar{H}_i} - 2I_{\bar{H}_i} +
  r_{\bar{H}_i} \,,\\
  \omega_{H_i} & = \sum_{\substack{H'\subseteq H_i\\ H'\in {\cal C}}}
  \omega_{\bar{H}'} \,.
\end{align}
\end{subequations}
This clearly corresponds to the superficial power-counting
degree of the original momentum integral.

With these building blocks we can formulate the general result for
the integral specified in 
Eqs.\ (\ref{loopintegralgeneral},\ref{alpharesultgeneral}). Decomposing
the integral in sectors as in Eq.\ (\ref{alphasectors}),
\begin{align}
  {\cal T}_{G} & = \sum_{({\cal C},\sigma)}  {\cal T}_{G,({\cal C},\sigma)}
\,,
\end{align}
and setting again $D=4-2\epsilon$, the result for each sector can be written as
\begin{align}
  {\cal T}_{G,({\cal C},\sigma)} & =
  c_D^L 2^L 
    \int_{\substack{t_L=0\ldots\infty\\t_{1\ldots L-1}=0\ldots 1
      }
      }\prod_{i=1}^L\frac{dt_i}{t_i}
    (t_i\xi_i)^{-\omega_{\bar{H}_i}+2\epsilon} \tilde{Z}_{H_i}
    \nonumber\\
    &\qquad\times
    \int_0^1 \prod_k d\beta_k
    \tilde{d}_G^{-D/2}e^{iW_G} \Bigg|_{u=0}
    \,.
\label{GeneralTGresult}
\end{align}
The properties of the appearing objects are:
\begin{itemize}
\item
  All variables $t_i$, $\xi_i$, $\beta_k$ and the rescaled physical
  variables $\tilde{\underline{q}}_{H_i}$, $\tilde{m}_{H_i}$,
  $\tilde{u}_{H_i}$, and the power-counting degrees
  $\omega_{\bar{H}_i}$ are defined above.
\item
  The explicit powers of $t_i$ correspond to the original superficial
  power-counting degrees of the momentum integrals over the 
  subdiagrams $H_i$. For each $t_i$ integral, a factor
  $(t_i\xi_i)^{2\epsilon}$ was split off which may be viewed as
  the essence of the $D$-dimensional integration measure.
\item
  The remaining integrand $    \tilde{d}_G^{-D/2}e^{iW_G}$ has no
  explicit dependence on $t_L$ at all. It depends on $t_L$ only via
  the rescaled physical variables. The other $t_i$ with $i=1\ldots
  L-1$ typically appear explicitly, however.
\item
  The function $\tilde{d}_G$ is a rescaled Symanzik polynomial which
  satisfies $\tilde{d}_G\ge1$ in the integration region.
\item
  For $\varepsilon>0$ in the $+i\varepsilon$ prescription, the
  function $e^{iW_G}$ is exponentially decreasing for large $t_L$.
\item
  The product $    \tilde{d}_G^{-D/2}e^{iW_G}$, therefore, is analytic
  in $\epsilon$ and $C^\infty$ in $t_i$, $\beta_k$ and the rescaled
  physical variables $\tilde{\underline{q}}_{H_i}$, $\tilde{m}_{H_i}$,
  $\tilde{u}_{H_i}$.
\end{itemize}
This statement is the starting point for the inductive proof of
renormalization in DReg given in Ref.\ \cite{Breitenlohner:1977hr}, and it is a direct
consequence of Lemma 4 of that reference.

\subsubsection{Ingredient 4: Integrand Relation between Graphs and~Subgraphs}
\label{sec:integrandrelations}

An important step in the proof is the application of subtraction operators $T_H$
to a graph $G$. In order to analyze this operation, relationships
between the original graph $G$, the reduced graph $G/_{H}$ and the
subgraph $H$ are needed. These relationships are again essentially
independent of $D$-dimensional treatments. They rely on detailed
analysis of the graphs themselves, and the relationships between
graphs and the $\alpha$-parametrizations.

The required theory involves incidence matrices and graph theoretical
representations of the Symanzik polynomial ${\cal U}$,
or $\tilde{d}_G$, and the exponent $W_G$. Although the theory is
very elegant and not too difficult we do not develop it here. Hence we
only list several important statements without proof. For the proofs
we refer to Ref.\ \cite{Breitenlohner:1977hr} and references therein.
Further discussions were given e.g.\ in Refs.\ \cite{nakanishi1971graph,Itzykson:1980rh,Bogoliubov1982}

Consider the Symanzik polynomial ${\cal U}_G$ for a graph $G$, and let
$H$ be a subgraph of $G$. ${\cal U}$ is a homogeneous polynomial in
all $\alpha$'s of degree $L$. Consider the case where all $\alpha$'s
corresponding to the subgraph $H$ are rescaled by a factor $\rho$,
while all other $\alpha$'s remain fixed. Then, for small $\rho$ we
have
\begin{align}
  {\cal U}_G(\text{$\alpha$'s in $H$ rescaled by $\rho$}) & =
  \underbrace{{\cal U}_{G/_H}}_{\text{$\rho$-independent}}\quad\underbrace{{\cal U}_H}_{\propto\rho^{L_H}}+ {\cal
    O}(\rho^{L_H+1}) \,,
\end{align}
i.e.\ at the lowest non-vanishing order the Symanzik polynomial factorizes
into the two individual Symanzik polynomials for the reduced graph and
the subgraph.
If $G$ and $H$ are part of an integration sector as
defined above, then variables $t_G$, $t_H$ (and possibly further
$t_i$) and $\beta_k$ exist and rescaled Symanzik polynomials
$\tilde{d}$ can be defined for each of these graphs. In this case,
$\tilde{d}_G$ for the full graph cannot depend on $t_G$ but it can
depend on $t_H$, while $\tilde{d}_{G/_H}$ and $\tilde{d}_H$ can
neither depend on $t_G$ nor on $t_H$. Their relationship is the
factorization
\begin{align}
  \tilde{d}_G|_{t_H=0} & = \tilde{d}_{G/_H}\tilde{d}_H \,.
\end{align}

A similar relationship can be established for the exponent $W_G$
appearing in the general result of the integral
(\ref{GeneralTGresult}). Defining $W_H$ and $W_{G/_H}$ 
using the same variable transformations for the sector $({\cal
  C},\sigma)$, suitably adapted to the subgraph and reduced graph, the
relationship is
\begin{align}
  W_G|_{t_H=0} & = W_{G/_H} + W_H|_{t_H=0} \,,
\end{align}
if all these quantities are expressed in terms of rescaled
variables $\tilde{\underline{q}}$, $\tilde{u}$, $\tilde{m}$.
This property can be established in an elementary way once the
exponents are constructed via incidence matrices.

For the same conditions, a further, more intricate property  can also be established and is
important. It is the following property involving derivatives,
\begin{align}
  \frac{d}{dt_H} \tilde{d}_G^{-D/2}e^{iW_G}\Bigg|_{t_H=0}
  & =
  \xi_H U_H\left[\frac{d}{dt_H}\tilde{d}_H^{-D/2}e^{iW_H}\right]_{t_H=0}
  \cdot \tilde{d}_{G/_H}^{-D/2}e^{iW_{G/_H}} \,.
\label{subgraphproperty}
\end{align}
Here $U_H[X]$ denotes an insertion operator which effectively inserts its
argument $X$ as a vertex into a bigger graph. To achieve this insertion,
the external momenta of the argument $X$ must become internal momenta
of the bigger graph, in this case of $G/_H$. Technically, $U_H$ acts by shifting in its
argument the momentum variables $\tilde{\underline{q}}_H$ by terms
involving derivatives with respect to $u$-variables for the bigger
graph $G/_H$.

This is a statement of pivotal importance for the full proof of the
theorem stated in Sec.\ \ref{sec:DRegTheoremStatement}
since it allows relating divergences of a full graph to
divergences of counterterm graphs and thus allows making manifest the
cancellation of subdivergences.
It is essentially the content of Lemma 5 of Ref.\ \cite{Breitenlohner:1977hr}.

\subsubsection{Ingredient 5: Simple Integrals
and Non-Analytic Functions of $D-4$}
\label{sec:nonanalyticfunctions}

Now we discuss several simple integrals and special functions that
arise in DReg due to the $D$-dimensionality of spacetime. They
encapsulate how the regularization acts, how divergences arise as
$1/(D-4)$ poles and how divergences cancel by adding suitable
counterterms. We set again $D=4-2\epsilon$.

First we discuss a simple type of integral, defined as
\begin{align}
  f(z) & = \int_0^\infty dt t^{z-1}g(t) \,, 
\label{simpleintf}
\end{align}
where $z$ is a complex variable and $g(t)$ is a $C^\infty$ function which
either decreases exponentially for $t\to\infty$ or which involves the
step function $\theta(1-t)$ cutting off the integral at $t=1$.
This simple integral appears in the general
result (\ref{GeneralTGresult}) but also in the 1-loop example
(\ref{onelooptbetaresult}). In all these cases the $t$-integration involves
one factor which is of the form $t^{n-1+2\epsilon}$, where $n$ is an
integer. This corresponds to the above form for $z=n+2\epsilon$.
This factor is non-analytic in $t$ around $t=0$. The remaining
$t$-dependences in
Eqs.\  (\ref{GeneralTGresult},\ref{onelooptbetaresult}) are
complicated but are $C^\infty$ functions in $t$ which indeed fulfil
the requirements on $g(t)$ listed above. In case of
the $t_L$ integration, the remaining integrand exponentially
decreases, in case of all other $t_i$ integrations, the integration
stops at $t_i=1$.

The above function $f(z)$ is a generalization of the
$\Gamma$-function, where $g(t)=e^{-t}$. The $\Gamma$-function is known
to have simple poles at $z=0$, $z=-1$, $z=-2$, \ldots. It is easy to
see that the same is true for the more general $f(z)$. Clearly, when for $\text{Re}(z)>0$ the integral defining $f(z)$ converges and defines
an analytic function. To study negative $\text{Re}(t)$ we can add to
and subtract from $g(t)$ a Taylor polynomial $\sum g^{(k)}(0)t^k/k!$, where
$g^{(k)}$ denotes the $k$-th derivative. Integrating
this polynomial from $0$ to $1$ we obtain
\begin{align}
  f(z) & = \int_0^\infty dt t^{z-1}
  \left[g(t)-\theta(1-t)\sum_{k=0}^n\frac{g^{(k)}(0)}{k!}t^k\right]
  +\sum_{k=0}^{n} \frac{g^{(k)}(0)}{k!}\frac{1}{z+k} \,.
\label{fanalyticcont}
\end{align}
For any non-negative integer $n$ and for $\text{Re}(z)>0$ the
value and convergence properties of the integral are not
changed. However, the square bracket behaves like $t^{n+1}$ for small
$t$, hence the integral now converges even for negative $z$, as long
as $\text{Re}(z)>-n-1$. Hence this formula represents an analytic
continuation of $f(z)$ onto the entire complex $z$ plane. It makes
also manifest that this analytically continued $f(z)$ has single poles
at $z=-0$, $z=-1$, $z=-2$, \ldots.

We can rewrite the result in the form of an integration rule for the
typical $t$-integrals appearing in DReg by replacing $z=-n+2\epsilon$
with integer non-negative $n$ and $\epsilon\approx0$. We then have the rule
\begin{align}
  \int_0^\infty dt t^{-n-1+2\epsilon} g(t)& =
  \frac{1}{n!}\left(\frac{d}{dt}\right)^ng(t)\Bigg|_{t=0}\frac{1}{2\epsilon}
  +
  \text{regular expression}\,,
  \label{tintegralrule}
\end{align}
where the form of the regular expression can be read off from
Eq.\ (\ref{fanalyticcont}). The $t$-integrals
in the general formula (\ref{GeneralTGresult}) are to be analytically continued in
this way. Hence this rule immediately shows that any
$t$-integration can only lead to
single $1/\epsilon$-poles and not to more complicated divergences as $\epsilon\to0$.

Next, we consider two special simple classes of non-analytic functions
of $t$. They are defined as the two kinds of sets (for integer $K,L$)
\begin{subequations}
\begin{align}
  K&<L: &
  J_K^L & = \left\{f(t,\epsilon)=\frac{c_1 t^{2\epsilon}+\ldots+ c_L
    t^{2L\epsilon}}{\epsilon^K}
  =\text{finite for }\epsilon\to0\right\}
  \,, \\
  K&\le L: &
  \tilde{J}_K^L & = \left\{f(t,\epsilon)=\frac{c_0+c_1 t^{2\epsilon}+\ldots+ c_L
    t^{2L\epsilon}}{\epsilon^K}
  =\text{finite for }\epsilon\to0\right\}
  \,.
\end{align}
\end{subequations}
In the definitions of the sets, the lower index $K$ refers to the
$\epsilon$-power in the denominator, and the upper index $L$ can be
thought of as the loop number at which the functions become of
interest. The coefficients $c_i$ are arbitrary
except for the constraint that the defined functions are finite for
$\epsilon\to0$.

Let us illustrate how such functions can appear by considering a
2-loop diagram $G$ with a 1-loop subdiagram $H$. We imagine a
calculation not only of the diagrams themselves but  of the entire
renormalization procedure, taking into account suitable counterterm
diagrams cancelling subdivergences. In the imagined calculations we use
the general formula
(\ref{GeneralTGresult}). If the 1-loop diagram 
$H$ is computed in isolation, it involves one $t_1$-integral whose
essential non-analytic part is simply 
\begin{align}
  t_1^{2\epsilon} \in J_0^1 \,,
\end{align}
which is an element of the set $J_0^1$ and which may be attributed to
the $D$-dimensional measure. The result of the
$t_1$-integration via the rule (\ref{tintegralrule}) then leads particularly to a 
$1/(2\epsilon)$ pole, and a counterterm for diagram $H$ can be defined
that cancels this divergence.
In the 2-loop calculation of $G$ the 1-loop diagram $H$ appears as a
subdiagram with corresponding $t_1$ integration. Here the $t_1$
variable is accompanied by $\xi_1$, which is here simply
$\xi_1=t_2$. After the $t_1$ integration, the non-analytic factor
$\xi_1^{2\epsilon}$ remains and combines with the $1/(2\epsilon)$
pole. In the corresponding counterterm diagram, 
where the subdiagram $H$ is replaced by the counterterm cancelling its
$1/(2\epsilon)$ pole there is no $t_1$ integration and no appearance of
the variable $\xi_1$. Therefore, after the $t_1$ integration and after
combining with the counterterm diagram cancelling the subdivergence, a
combined function 
\begin{align}
  \frac{t_2^{2\epsilon}-1}{2\epsilon} \in \tilde{J}_1^1
\end{align}
appears. The finiteness of functions in the set $\tilde{J}_1^1$
reflects the successful cancellation of the subdivergence.
Proceeding with the computation of the 2-loop diagram $G$, this
function is  combined with the measure factor, such that the interesting
non-analytic part of the 
$t_2$-integrand is
\begin{align}
  t_2^{2\epsilon}\frac{t_2^{2\epsilon}-1}{2\epsilon} \in J_1^2\,.
\end{align}

This example illustrates the general idea: 
Functions in $J_K^L$ 
are the functions that actually appear as the non-analytic factors in
the  $t_L$ integrations at the $L$-loop level during the
renormalization procedure. After carrying 
out a $t_L$ integral and after combining with the suitable counterterm
contribution, a function in the set 
$\tilde{J}_{K+1}^L$ appears. At the next loop level the integrand
needs to be prepared by suitable rearrangements and combined with the
measure factor $t_{L+1}^{2\epsilon}$ to produce a function of the set
$J_{K+1}^{L+1}$, and so on.

For this reason it is helpful to study the properties of functions in
these sets on their own, before tackling the actual loop
integrations. Some particularly useful properties are:
\begin{description}
\item[$(i)$] any function $f\in J_K^L$ has the limit
  $f(t,0)=\text{const}\times(\ln t)^K$.
\item[$(ii)$] 
  for a function $f\in J_K^L$, the integral $
    \int_1^t\frac{dt'}{t'}f(t',\epsilon) $  produces an
  element of  the next set $\tilde{J}_{K+1}^L$.
\item[$(iii)$] the converse is also true, i.e.\ every element of
  $\tilde{J}_{K+1}^L$ can be written in terms of such an integral.
\item[$(iv)$]
  a function $f\in J_K^L$ where the first argument is a product can be
  factorized as
$    f(\xi t,\epsilon)  =
    \sum_j f_{1j}(\xi,\epsilon)f_{2j}(t,\epsilon)
    $
  where all functions on the right-hand side are elements of
  $    f_{nj}\in J_{K_{nj}}^L$ where $K_{1j}+K_{2j}=K$. This property is obviously
  important to prepare higher-loop integrands such that $t$ integrals
  act on isolated functions depending only on $t$, not on $\xi$.
\item[$(v)$]
  there is a simple product rule $f_{K_1}^{L_1}f_{K_2}^{L_2}\in
  J_{K_1+K_2}^{L_1+L_2}$ for functions $f_{K_i}^{L_i}\in
  J_{K_i}^{L_i}$. This property is also important on the multiloop
  level in case a multiloop diagram contains two disjoint divergent subdiagrams.
\end{description}
The properties can all be proved using elementary integration tricks
and l'Hopital's rule for limits. 
Such properties of these functions are the content of Lemma 2 of
Ref.\ \cite{Breitenlohner:1977hr}.

\subsubsection{Sketch of Proof by~Induction}

All explained ingredients are important in the full proof of the
central Theorem 1 in Ref.\ \cite{Breitenlohner:1977hr} and stated in
Sec.\ \ref{sec:DRegTheoremStatement}. Here we give a sketch of this
proof.
The proof applies the $\alpha$ parametrization of integrals decomposed
into sectors as in Eq.\ (\ref{alphasectors}). The renormalization
procedure is then expressed in terms of the
forest formula (\ref{sectorforestformula}).
This formula provides the basis for an
inductive proof, where a graph $G$ and a sector are fixed and then all
factors $(1-T_{H_i})$ in the forest formula are successively applied
in the correct ordering.
The base case of the induction is provided by the general formula (\ref{GeneralTGresult}).
The induction step needs to carry out the actual integration over one
$t$ variable
and some $\beta$ variables corresponding to the next $(1-T_{H_i})$
factor. The step uses the properties of the special functions of
$\epsilon$ defined in Sec.\ \ref{sec:nonanalyticfunctions}, and the
relationships between the graph, subgraph and reduced 
graph described in Sec.\ \ref{sec:integrandrelations}.

Obtaining the precise form of the induction hypothesis is highly
nontrivial, but it can be motivated using all the developed
insight. It can be formulated as follows. Consider a 1PI graph $G$ and
a sector $({\cal C},\sigma)$.
All following quantities are specific to this sector but for brevity
we will omit all indices denoting this dependence.
The graph has $L_G$ loops and the sector contains $L_G$ subgraphs
$H_1,\ldots,H_{L_G}$. Without loss of generality we assume the
labelling such that the subgraphs are already ordered according to
their allowed appearance in the forest formula
(\ref{sectorforestformula}), such that if $H_j\supseteq H_i$ then also
$j\ge i$ (the ordering is not unique). Then after evaluating $L\le
L_G$ factors in the forest formula we obtain the expression
(suppressing the dependence on the sector $({\cal C},\sigma)$)
\begin{align}
  {\cal R}_X(G) \equiv
  (1-T_{H_L})\cdot\ldots\cdot(1-T_{H_2})\cdot(1-T_{H_1})\cdot G
  \,.
\end{align}
This represents a partially renormalized
graph where $L$ loops and $L$ subgraphs have already been treated in
previous induction steps.
Sec.\ \ref{sec:forestformula} gave arguments that this
expression should have simple divergence properties when acted upon by
further $T_{H_i}$ operators. Despite this, the partially renormalized
expression on its own clearly can have  very complicated analytical
structure and can still have non-polynomial divergences which the
proof needs to deal with. The label $X$ denotes the set of all
subgraphs which have already been treated, and we also define $X_0$ as the
subset of $X$ which contains only maximal subgraphs, i.e.\
\begin{align}
  X & = \{H_1,\ldots,H_L\} \,, & X_0 & = \{M_1,\ldots,M_S,H_L\} \,,
\label{XDef}
\end{align}
where it is used that $H_L$ itself is necessarily a maximal subgraph
in $X$ and where names have been given to all other elements of $X_0$.

The induction hypothesis states that after evaluating all $t_i$ and $\beta_k$
integrals corresponding to lines in the already treated graphs in $X$, we
obtain
\begin{align}
  {\cal R}_X(G)  = {}&{}
  \text{sum of terms like }
  \nonumber
  \\&
  \int_{\ge L+1}
              \prod_{M\in X_0}\xi_{M}^{-\omega_M}
  \tilde{f}_M(\xi_M,\epsilon) g_{G,X}\Bigg|_{\tilde{u}=0}
  \,,
  \label{inductionhypothesis}
  \intertext{where the integration factors for the remaining integrals
    are abbreviated as}
  \int_{\ge L+1} = {}&
  c_D^{L_G-L}
  \int\prod_{i=L+1}^{L_G} \frac{dt_i}{t_i}
  (t_i\xi_i)^{-\omega_{\bar{H}_i}+2\epsilon}\int \prod_{k\in G/_{X_0}}
  d\beta_k \tilde{Z}_{H_i} \,.
  \label{integralfactor}
\end{align}
Here the integration boundaries of the $t_i$ and $\beta_k$ integrals
are as in Eq.\ (\ref{GeneralTGresult}), and the notation $k\in G/_{X_0}$
corresponds to all indices $k$ corresponding to any line outside the
already treated graphs in the set
$X$. In the product over the maximal subgraphs $M$ (which includes the
case $M=H_L$), each $M$ is equal to one particular $H_{j(M)}$ and for
simplicity we identify the indices $\xi_M\equiv \xi_{{j(M)}}$.

We provide the following comments on the induction hypothesis:
\begin{itemize}
\item
  The ``sum of terms like'' refers to the expression in the integrand
  which really is of the form
  $\sum_{a}\prod_M\tilde{f}_{M,a}g_{G,X,a}$. Since the proof can be carried
  out for each such term we drop the index $a$ and this summation.
\item
  The  integration variables $t_i$ and $\beta_k$ and the $\tilde{u}_k$
  variable for the
  already treated graphs do not exist anymore since they have been
  integrated over/set to zero. Hence the only appearing $t_i$,
  $\beta_k$ and $\tilde{u}_k$ are the ones for $i=L+1,\ldots, L_G$ and
  for $k\in   G/_{X_0}$.
\item
  The sets of physical variables $\tilde{\underline{q}}_{H_i}$,
  $\tilde{m}_{H_i}$ and the remaining $\tilde{u}_{H_i}$ (for
  $H_i\notin X$) are rescaled only by the remaining
  $t_i$'s. I.e.\ Eq.\ (\ref{rescalinggeneral}) applies in a modified form
  where on the right-hand side $t_i=1\,\forall i\le L$ and where the 
  $\tilde{u}_{H_i}$ for $i\le L$ do not exist.
\item
  The particularly nontrivial and interesting part of the statement is
  the integrand in Eq.\ (\ref{inductionhypothesis}). It displays the analytic structure of the partially
  renormalized graph and the result of all the evaluated $t_i$ and
  $\beta_k$ integrals. The result is a product of functions
  $\tilde{f}_M$, which are non-analytic in the remaining $t_i$, and the
  function $g_{G,X}$.
\item
  Each function $\tilde{f}_M$ is an element of a set   $\tilde{J}_K^L$
  with $K\le L$. These functions are thus non-analytic in the remaining $t_i$ but
  have a finite limit for $\epsilon\to0$, reflecting the
  successful subtraction of subdivergences.
  The functional form of each $\tilde{f}_M$ is further specific to the chain of
  subgraphs $X_M=\{H'\subseteq M, H'\in {\cal C}\}$, and does not
  depend on any details of graphs or parts of graphs outside $M$. Only
  the argument $\xi_M$ has a dependence on $t_i$ variables
  corresponding to bigger graphs.
\item
  The function 
  $g_{G,X}$ carries the complicated dependence on all physical
  variables and all other $t_i$ and $\beta_k$ variables. $g_{G,X}$ is
  $C^\infty$ in all these remaining integration variables and all the
  physical variables $\tilde{\underline{q}}_{H_i}$, $\tilde{m}_{H_i}$
  and $\tilde{u}_{H_i}$ rescaled as defined above.
  It is analytic in $\epsilon$, again reflecting
  the cancellation of subdivergences, and it has no explicit
  dependence on $t_{L_G}$ corresponding to the full graph $G$ (except
  for the product $t_{L_G}^2\varepsilon$, similar to
  Eq.\ (\ref{Wexamplet})).
  Its functional form is specific to the full graph $G$ and the treated
  graphs $H_i\in X$.
\end{itemize}

The induction base case is the one where $L=0$ and no subgraph has
been treated yet. In this case the sets $X$ and $X_0$ are empty and
${\cal R}_X(G)$ simply refers to the unrenormalized result ${\cal
  T}_G$. The form of the unrenormalized result is given in
Eq.\ (\ref{GeneralTGresult}) and it directly confirms the induction
hypothesis (\ref{inductionhypothesis}) with
$g_{G,\emptyset}=2^{L_G}\tilde{d}^{-D/2}e^{iW_G}$.

For a sketch of the induction step we assume $L\ge1$ and assume the
partial renormalization was carried out up to loop number $L-1$
and that the induction hypothesis holds at loop number $L-1$. It is then useful to
introduce a notation for the previously treated subgraphs and
previously treated maximal subgraphs. We write
\begin{align}
  X' & =\{H_1,\ldots, H_{L-1}\} &
  X'_0 & = \{m_1,\ldots,m_s\}\cup \{M_1,\ldots,M_S\}\,,
\end{align}
and we keep the definitions of Eq.\ (\ref{XDef}) such that $X=X'\cup
\{H_L\}$ and such that the subgraphs $m_{i}$ are the maximal subgraphs
of $H_L$. The remaining subgraphs are $H_L$ as well as $H_i$ with
$i\ge L+1$, the lines and $\beta_k$ are the ones with $k\in G/_{X'_0}$
or equivalently the ones with $k\in G/_{X_0}$ or with $k\in \bar{H}_L$.
The induction hypothesis for loop number $L-1$ can therefore be cast into
the form
\begin{align}
  {\cal R}_{X'}(G) & = \text{sum of terms like }
\nonumber\\&
\int_{\ge L+1}
\prod_{M\in X_0\setminus \{H_L\}}\xi_{M}^{-\omega_M}
  \tilde{f}_M(\xi_M,\epsilon)
\nonumber\\&
  \times c_D
  \int \frac{dt_L}{t_L}
  (t_L\xi_L)^{-\omega_{\bar{H}_L}+2\epsilon}
  \tilde{Z}_{H_L}    \int \prod_{k\in \bar{H}_L} d\beta_k
  \nonumber\\
  &
  \times\prod_{m_i}\xi_{m_i}^{-\omega_{m_i}}
  \tilde{f}_{m_i}(\xi_{m_i},\epsilon)
  g_{G,X'}\Bigg|_{\tilde{u}=0}
  \,.
  \label{inductionhypothesisstep}
\end{align}
In this way of writing the role of the graph $H_L$ which is to be
treated next is exhibited, while the factors in the first line contain
the same integration factors and almost the same $\tilde{f}_M$ factors
as Eq.\ (\ref{inductionhypothesis}). The physical variables appearing
here inside the $\tilde{Z}_{H_i}$ and $g_{G,X'}$ are rescaled with
all $t_i$ for $i\le L$, and all comments made for the induction
hypothesis apply with suitable modifications.

In the induction step we need to assume the validity of
Eq.\ (\ref{inductionhypothesisstep}) and carry out the next step,
construct ${\cal R}_X(G)$ and prove 
that it takes the form (\ref{inductionhypothesis}) with all listed
properties. The construction involves the evaluation of all integrals
in the last two lines of Eq.\ (\ref{inductionhypothesisstep}).
It also involves the application of the next
subtraction operator $(1-T_{H_L})$, which also only affects the
last two lines of Eq.\ (\ref{inductionhypothesisstep}) in particular because the
integration factors $\int_{L+1}$ stay unchanged if the subgraph $H_L$
is replaced by its counterterm.

We begin with  several 
immediate simplifications of the factors in the last two lines of
Eq.\ (\ref{inductionhypothesisstep}). First we observe that all
the $\xi_{m_i}$ in the last line are all equal to each other, and they
are equal to $\xi_{m_i}=t_L\xi_L$. The reason is that the $\xi_{m_i}$ are
the products of $t_j$ for all subgraphs in ${\cal C}$ which contain
$m_i$ and that the $m_i$ are maximal subgraphs of $H_L$. One
consequence is that the $\omega_{\bar{H}_L}$- and
$\omega_{m_i}$-dependent terms combine simply to
$(t_L\xi_L)^{\omega_{H_L}}$.
A less trivial consequence is that
all non-analytic functions and the measure factor for $t_L$ can
be combined as
\begin{align}
  f_{H_L}(t_L\xi_L,\epsilon )
  & =
  (t_L\xi_L)^{2\epsilon}\prod_{m_i}\tilde{f}_{m_i}(t_L\xi_L,\epsilon) \,, 
\end{align}
which is an element of the set $J_K^L$ for some $K<L$ thanks to the
properties of the functions discussed in
Sec.\ \ref{sec:nonanalyticfunctions}. Second, after the $\beta_k$
integrations and after applying the derivative operator
$\tilde{Z}_{H_L}$ and setting $\tilde{u}_{H_L}=0$, we
obtain
\begin{align}
  \bar{g}_{G,X'} = \tilde{Z}_{H_L}\int\prod_{k\in \bar{H}_L} d\beta_k
  g_{G,X'}\Bigg|_{\tilde{u}_{H_L}=0} \,.
\end{align}
This function is still $C^\infty$ in the remaining variables and
analytic in $\epsilon$. Hence the last two lines of
Eq.\ (\ref{inductionhypothesisstep}) can be written as
\begin{align}
  c_D\int \frac{dt_L}{t_L}(t_L\xi_L)^{-\omega_{H_L}}
  f_{H_L}(t_L\xi_L,\epsilon)
  \bar{g}_{G,X'} \,.
\label{simpletwolines}
\end{align}

The more difficult part of the induction step is the evaluation of the
$t_L$ integral and the application of the $(1-T_{H_L})$ subtraction
operator. Two cases need to be distinguished. The first case is when
the next step is the final step of renormalization, i.e.\ when $L=L_G$
and $H_L=G$. The second case is when $L<L_G$ and $H_L$ is still a
proper subgraph of the full graph $G$.

To sketch the first case with $L=L_G$ and $H_L=G$ we note that in this case the
second line of Eq.\ (\ref{inductionhypothesisstep}) is just the factor
1 since there are no remaining integrations and there are no other
maximal subgraphs $M$. Likewise, the remaining $\xi_L=1$, and from
the induction hypothesis we know that the variable $t_L=t_{L_G}$ does
not appear explicitly in $\bar{g}_{G,X'}$ --- this variable only
enters via rescaled physical variables $\tilde{q}_{H_i}$ and
$\tilde{m}_{H_i}$ i.e.\ via products of $t_L$ and physical momenta and
masses. Plugging in the general
form of the function $f_{H_L}$ yields a sum of terms like
\begin{align}
  \sum_n \frac{c_n t_L^{2n\epsilon}}{\epsilon^K}\bar{g}_{G,X'} \,,
\end{align}
which need to be integrated over $t_L$ with the measure $\int dt_L
t_L^{-\omega_{H_L}-1}$. This integral is performed via the general rule
(\ref{tintegralrule}). This rule leads to a regular expression and a
singular term. The regular expression
can be shown to be analytic in $\epsilon$ and $C^\infty$ in all other
variables. The singular term contains  poles in $\epsilon$ and takes the
form
\begin{align}
  \sum_n \frac{c_n}{2n\epsilon^{K+1}}\frac{1}{\omega_{H_L}!}
  \left[\left(\frac{d}{dt_L}\right)^{\omega_{H_L}}\bar{g}_{G,X'}\right]_{t_L=0}
  \,.
\end{align}
This singular term can be shown to have all
desirable properties. The poles in $\epsilon$ are at most of
degree $1/\epsilon^{L_G}$. The coefficients are polynomials in the
physical variables, masses and momenta, of degree
$\omega_{H_L}$.\footnote{  Here, and in Ref.\ \cite{Breitenlohner:1977hr}, the factor of the dimensional
  regularization scale $\mu^{2\epsilon}$ is omitted from the
  definition of renormalized amplitudes. If this factor is included it
  is also possible to prove that the divergent polynomial is independent of
  $\mu$.}
It is therefore possible to define the subtraction operator $(1-T_G)$
for this sector such that it subtracts this polynomial divergence; the
resulting finite remainder satisfies all properties listed after the
induction hypothesis (\ref{inductionhypothesis}). It is further
possible to define the full divergent part of the full 
diagram, $T\overline{\cal R}_G$, as the sum of all these singular terms
arising in this way in all sectors. This object has all properties
required for a possible contribution to a counterterm Lagrangian: in
position space it
is local, it has the correct power-counting degree, and its value
depends only on the graph $G$ and not on its embedding into bigger graphs.

Finally we also sketch the remaining induction step for the case
$L<L_G$ and $H_L\ne G$.
Here, evaluating the $t_L$ integral and applying the subtraction
operator $(1-T_{H_L})$ to Eq.\ (\ref{simpletwolines}) leads to three
terms: the regular expression from the $t_L$ integration, the singular
expression from the $t_L$ integration and the counterterm contribution
from $T_{H_L}$, where $T_{H_L}$ is defined via the full
renormalization of the graph $H_L$ in isolation. All terms need to be
rearranged by using properties of the $f$ functions discussed in
Sec.\ \ref{sec:nonanalyticfunctions}, in particular of the
factorization property of these functions. Furthermore, the singular
expression of the $t_L$ integration has to be rearranged by using
properties such as (\ref{subgraphproperty}) for the relationships
between graphs, subgraphs and reduced graphs. In these ways it is
possible to show that the combination of all terms acquires the form
of the induction hypothesis (\ref{inductionhypothesis}) and that all
announced properties are fulfilled.

In this way all properties announced in
Sec.\ \ref{sec:DRegTheoremStatement} are established, except for the
equivalence to BPHZ. Illustrating this point requires comparing the
structure of appearing integrals in the DReg and the BPHZ
approaches. For this we refer to the original literature
\cite{Speer:1974cz,Breitenlohner:1977hr}. 

\newpage

\section{Renormalization and Symmetry}
\label{sec:algren}

In the preceding section we have seen how the renormalization programme allows to subtract the divergences from Feynman diagrams. Importantly, the subtraction terms are polynomials in momenta constrained by power counting, and the subtraction is equivalent to adding certain counterterms to the Lagrangian. By choosing a certain renormalization scheme, the remaining ambiguities of finite counterterms can be fixed and the Lagrangian supplemented by those counterterms defines a finite $4$-dimensional theory.

In this section we consider the problem of renormalization in the
presence of symmetries, specifically gauge invariance. On the one hand, symmetries put additional restrictions on certain quantities which allows for simplifications. On the other hand, we also have to ask about the compatibility of symmetries and regularization and whether they can be restored if intermediately broken.
Since regularization may in general spoil the classical symmetry, we shall require its validity as part of the definition of our theory.
The symmetry of interest for us is gauge invariance promoted to BRST invariance as described in Sec.\ \ref{sec:BRSTSTI}. On the level of Green functions, this symmetry is implemented by the Slavnov-Taylor identity as described in Sec.\ \ref{sec:STIformal}. In a more compact notation (cf.\ Eq.\ (\ref{STIpathintegralGamma})), it can be written as
\begin{equation}\label{REQUIRE}
\mathcal{S}(\Gamma_\mathrm{ren})=\int\,\mathrm{d}^4x\,\frac{\delta\Gamma_\mathrm{ren}}{\delta\phi(x)}\frac{\delta\Gamma_\mathrm{ren}}{\delta K_{\phi}(x)}\stackrel{!}{=}0.
\end{equation}
Here we assumed for simplicity that all symmetry transformations, i.e. both linear and non-linear, are coupled to sources $K_\phi$.
The Slavnov-Taylor identity is the pivotal tool in the proof of
renormalizability of quantized Yang-Mills gauge theories, including the proof
that the quantum theory actually is physically sensible.

The first proofs of the renormalizability of non-Abelian gauge theories
were given by 't Hooft, Lee and Zinn-Justin in Refs.\ \cite{tHooft:1971qjg,tHooft:1971akt,Lee:1972fj,Lee:1972ocr,Lee:1972yfa,Lee:1973fn}, all employing various
versions of Slavnov-Taylor identities. These proofs establish not only
the finiteness and validity of the Slavnov-Taylor identity but also
the interpretation of the quantum theory with a unitary and gauge-fixing
independent S-matrix defined on a Hilbert space of quantum states with
positive norm. Later, the proofs were
generalized by Becchi, Rouet, Stora and Tyutin (BRST) to the case where
nothing is known about symmetry 
properties of the employed regularization scheme, establishing the
approach of algebraic renormalization 
\cite{Becchi:1974xu,Becchi:1974md,Becchi:1975nq,Tyutin:1975qk}, see also the 
reviews by Piguet/Rouet and Piguet/Sorella \cite{Piguet:1980nr,Piguet:1995er}. A particularly
satisfactory formulation is achieved with the 
Kugo/Ojima formalism \cite{Kugo:1979gm} where the existence of a nilpotent operator $Q_B$
is derived from the Slavnov-Taylor identity. $Q_B$ generates BRST
transformations on the level of asymptotic states and its role on the
level of quantum states is similar to the role of the BRST operator
$s$ on the classical level, see Eqs.\ (\ref{sDef1}--\ref{eq:NilpotencyOfTheBRSTOperator}). It may be used to
define the physical Hilbert space as the quotient space
\begin{equation}
  \mathcal{H}_{\text{phys}}=(\mathrm{ker}\,Q_B)/(\mathrm{im}\,Q_B)\,.\end{equation}
Hence two states are equivalent if they differ by a total
$Q_B$-variation. A single state is called physical if $Q_B\ket{\psi}=0$,
provided it is not some total variation, i.e. $\ket{\psi}\neq
Q_B\ket{\chi}$ for some $\ket{\chi}$, in which case it would be
equivalent to the zero vector.
The fields act Lorentz covariantly on the whole space including
unphysical states and because of the Slavnov-Taylor identity, $Q_B$ commutes with the
S-matrix. Hence the physical S-matrix defined on the physical Hilbert
space  maps physical states to physical
states, it is Lorentz invariant, unitary and causal.
All these properties can be shown in a very elegant way \cite{Kugo:1979gm}.
We thus see that if we make sure that the Slavnov-Taylor identity is
obeyed after renormalization, we are guaranteed a consistent quantum
field theory.

Hence the logic now is the following. In section \ref{sec:setup} we
defined gauge theories which classically satisfy the BRST
symmetry. Then we established dimensional regularization as a
framework for treating such theories perturbatively in loop
orders. Now we are in a position to define our renormalized theory
with the fundamental Slavnov-Taylor identity intact and study the
possible obstructions posed by regularization. 
To this end we shall first discuss the counterterm structure for
manifestly preserved symmetries during renormalization in
Sec.\ \ref{sec:ctsymmetrypreserving}. Then in
Sec.\ \ref{sec:AlgRenormalization} we give a brief overview of the
field of algebraic renormalization which is the appropriate setting in
which to discuss breaking and restoration of symmetries. Finally we
discuss how the general analysis of algebraic renormalization can be
specialized to the case of dimensional regularization in
Sec.\ \ref{sec:AlgRenInDReg}. 

\subsection{Counterterms in Symmetry-Preserving Regularization}\label{sec:ctsymmetrypreserving}
We first recall the simple case where a symmetry is manifestly
preserved at all steps of the calculation. This is the standard case
often encountered in textbook discussions and practical calculations
using DReg in QED and QCD, for reasons described already in
Sec.\ \ref{sec:QAPexamples}. There one frequently uses so called
renormalization 
transformations of the generic form
\begin{subequations}
\label{RenTrafo}  
\begin{align}
g&\rightarrow g+\delta g\\
\phi_i&\rightarrow\sqrt{Z_{ij}}\,\phi_j,
\end{align}
\end{subequations}
for coupling constants $g$ and  quantum fields $\phi_i$ with
associated parameter and field renormalization constants $\delta g$
and $\delta Z_{ij}=Z_{ij}-\delta_{ij}$. The renormalization constants are to be
understood  as  power series in loop orders or equivalently in the
renormalized parameters.

This procedure is applied onto the classical action $S_0$ and thereby defines a bare action $S_\mathrm{bare}$, cf.\ (\ref{eq:IntroductionOfSbareAndSct}), itself giving rise to the counterterm action
\begin{equation}
S_\mathrm{ct}=S_\mathrm{bare}-S_0.
\end{equation}
The divergent parts of these generated counterterms cancel UV
divergences of loop diagrams, and the finite parts of the counterterms
can be used to fulfil certain renormalization conditions as mentioned
in Sec.\ \ref{sec:DRegGamma}.

In terms of the Slavnov-Taylor identities, the standard case is expressed by the statement
\begin{equation}
\label{STIReg}
\mathcal{S}(\Gamma_{\mathrm{reg}})=0,
\end{equation}
which, as explained in Sec.\ \ref{sec:QAPexamples}, means that the regularized Green
functions already satisfy the Slavnov-Taylor identity. If applicable,
similar equations should hold for other identities such as the ones
discussed in Sec.\ \ref{sec:AbelianPeculiarities} (e.g.\ ghost equation). This is indeed the case
in QED and QCD in DReg at all orders. The basis of this statement was
explained in Sec.\ \ref{sec:QAPDReg}.
The manifest symmetry at the regularized level (\ref{STIReg}) has two implications for the structure of renormalization. First, the possible divergences are restricted by Eq.\ (\ref{STIReg}) which, in turn, also restricts the structure of counterterms needed to cancel divergences. Second, possible finite counterterms are also restricted by Eq.\ (\ref{STIReg}), together with the ultimate requirement (\ref{REQUIRE}) for the renormalized theory. Both implications can be simultaneously evaluated as follows. Assuming that the theory has been renormalized up to order $\mathcal{O}(\hbar^{n-1})$, we are interested in the $\mathcal{O}(\hbar^{n})$-order counterterms $\mathcal{L}_\mathrm{ct}^n$ and the $\mathcal{O}(\hbar^{n})$ divergences of the regularized theory. The renormalized theory at order $\mathcal{O}(\hbar^{n})$ can be written as
\begin{equation}
\label{GammaDecomp}
\Gamma^{(n)}_\text{ren}=\Gamma^{(n)}_\text{reg,fin}+\Gamma^{n}_\text{reg,div}+S^n_\text{ct}.
\end{equation}
For further analysis it is customary to introduce the linearized
Slavnov-Taylor operator $s_\Gamma$, defined by expanding the
Slavnov-Taylor operator $\mathcal{S}(\Gamma)$ for both linearly and non-linearly transforming fields $\phi$ and $\Phi$, respectively, 
\begin{equation}
\mathcal{S}(\Gamma)=\int\,\mathrm{d}^4x\,\frac{\delta\Gamma}{\delta K_{i}(x)}\frac{\delta\Gamma}{\delta\Phi_{i}(x)}+\int\,\mathrm{d}^4x\,s\phi_{i}(x)\frac{\delta\Gamma}{\delta\phi_{i}(x)},
\end{equation}
as follows,
\begin{equation}
\label{Linearize}
\mathcal{S}(\Gamma+\zeta\mathcal{F})=\mathcal{S}(\Gamma)+\zeta s_\Gamma\mathcal{F}+\mathcal{O}(\zeta^2),\end{equation}
for some functional $\mathcal{F}$. Its concrete form is given by
\begin{equation}
s_{\Gamma}=\int\,\mathrm{d}x\,\left(\frac{\delta\Gamma}{\delta K_i(x)}\frac{\delta}{\delta\Phi_i(x)}+\frac{\delta\Gamma}{\delta\Phi_i(x)}\frac{\delta}{\delta K_i(x)}+s\phi_{i}(x)\frac{\delta}{\delta\phi_{i}(x)}\right).
\end{equation}
Of special interest is the case of the classical action $\Gamma_\text{cl}$, for which we define\begin{equation}
\label{bDefinition}b\equiv s_{\Gamma_\text{cl}},\end{equation}
as the linearized Slavnov-Taylor operator based on the classical
action. 
In agreement with the  nilpotency of the BRST operator
(\ref{eq:NilpotencyOfTheBRSTOperator}), the algebraic structure of the
Slavnov-Taylor operator leads to two
nilpotency relations
\begin{align}\label{SNilpotent}
s_{\Gamma}\mathcal{S}(\Gamma)&=0,\\
s_{\Gamma}s_{\Gamma}&=0\quad\text{if}\quad\mathcal{S}({\Gamma})=0.\end{align}
Substituting the decomposition of Eq.\ (\ref{GammaDecomp}) into
Eqs.\ (\ref{STIReg}) and (\ref{REQUIRE}),
      we first get\begin{equation}\label{STIdiv}
b\,\Gamma^n_\text{reg,div}=0.\end{equation}
This establishes the restriction on the possible divergences. Second, we obtain\begin{equation}\label{bOnFin}
b\,S^n_\text{ct}=0,\end{equation}
both for the divergent and the finite parts. The most general solution
of this equation in terms of admissible counterterm actions yields the
counterterm structure which is sufficient to cancel the divergences
and required to establish the symmetry. The corresponding calculations
were carried out in the original references on the
renormalization of Yang-Mills theories cited at the
beginning of this section; textbook discussions can be found e.g.\ in
the textbooks by Zinn-Justin, Weinberg and B\"ohm/Denner/Joos
\cite{Zinn-Justin:1989rgp,Weinberg:1996kr,Bohm:2001yx}.

For most theories of interest
including the SM, the outcome is the familiar statement cited in the beginning (cf. Eq.\ (\ref{RenTrafo})) that all counterterms can be obtained by renormalization transformation of the classical action.
A second related outcome is then that any two consistent
regularization/renormalization prescriptions which both fulfil the
symmetry requirement (\ref{REQUIRE}) can only differ by a
reparametrization of the form (\ref{RenTrafo}).\footnote{  This is a stronger
statement than the one of Eq.\ (\ref{finiteRen}) because a
smaller number of parameters is affected.}
\subsection{Broken Symmetries and Algebraic Renormalization}
\label{sec:AlgRenormalization}
Now we turn to the case of interest for e.g.\ chiral gauge theories in which the symmetry is not manifestly preserved by the regularization. This case is characterized by\begin{equation}\label{BrokenDQAP}
\mathcal{S}(\Gamma_\text{reg})\neq 0,\end{equation}
in contrast to Eq.\ (\ref{STIReg}). Clearly, the required structure of
the counterterms is more complicated. Now, the divergences and required
divergent counterterms may 
be non-symmetric and not fulfil Eq.\ (\ref{STIdiv}). In this case one
has to determine them by explicit calculation of the divergences of
Green functions instead of reading off their structure from a
renormalization transformation such as (\ref{RenTrafo}). In this way
the theory can be rendered finite despite the broken symmetry
(\ref{BrokenDQAP}).

Even on the finite level, the symmetry breaking (\ref{BrokenDQAP})
might still persist. Finite counterterms then have to be determined
such that the fully renormalized theory fulfils the basic requirement
(\ref{REQUIRE}). In some cases it can actually be impossible to
find such counterterms; the symmetry is then said to be broken by an
anomaly. Since we consider the Slavnov-Taylor identity as part of the
definition of the theory, an anomalous breaking of the Slavnov-Taylor
identity means that the theory is inconsistent and not renormalizable.
In  cases without an anomaly it is indeed possible to
recover the symmetry by appropriately chosen finite counterterms. 

Even though the precise form of the symmetry breaking depends on the
regularization, it is possible to study the general case of (\ref{BrokenDQAP}) in a
regularization-independent way. This study is the content of algebraic
renormalization, pioneered by BRST
\cite{Becchi:1974xu,Becchi:1974md,Becchi:1975nq,Tyutin:1975qk}, see also the
reviews \cite{Piguet:1995er,Piguet:1980nr}. The main insight of the
procedure is that the possible breakings are restricted in two
ways. On the one hand, they are restricted by the Slavnov-Taylor
identity itself, similar to the possible divergent structures in
Eq.\ (\ref{STIdiv}). On the other hand, they are restricted by a
regularization-independent version of the quantum action
principle.

Those two restrictions taken together provide a regularization-independent analysis of the renormalization of gauge theories.
In the following we shall first sketch the quantum action principle
in the BPHZ framework of renormalization, where it was originally
established and subsequently used for algebraic analysis, as well as
exhibit a connection to the regularized quantum action principle of
DReg. The central point is then to review how the aforementioned
restrictions can be used to restore the broken symmetry by suitable
counterterms provided there are no anomalies. 
\subsubsection{The Quantum Action Principle in~BPHZ}

As discussed in Sec.\ \ref{sec:generalrenormalization}, the BPHZ
approach to renormalization constituted
one of the first full discussions of all-order  renormalization,
rigorously establishing the possibility to obtain finite Green
functions and S-matrix elements in agreement with basic postulates
such as causality and unitarity.
In this framework Lowenstein and Lam derived various
theorems now summarized as the quantum action principle \cite{Lowenstein:1971jk,Lowenstein:1971vf,Lam:1972fzg,Lam:1972mb,Lam:1973qa}. The
theorems are similar to the regularized quantum action principle in
DReg discussed in Sec.\ \ref{sec:QAPDReg}. The difference is that the
theorems discussed here are valid in strictly $4$ dimensions, for the
fully renormalized theory.

Further, this form of the quantum action principle is generally valid not only
in the BPHZ framework but in all regularization/renormalization
frameworks that are equivalent; hence it also applies to results
obtained using DReg, if the $\text{LIM}_{D\to4}$ defined in
Eq.\ (\ref{DefLIM}) has been taken.
The algebraic method is based on this general formulation and
its results hold for all such equivalent frameworks.
In BPHZ, finite expressions and the Gell-Mann-Low
formula are defined by an iterative operation on momentum space
integrals whereby Taylor series contributions up to some UV-subtraction
degree are subtracted from the integrands giving finite integrals by
power counting. Further, normal products, i.e. products of fields and
their derivatives at the same space-time point, may be defined as
finite parts of certain Wick-ordered insertions into the Green
function. One can derive so called Zimmermann-identities which
linearly relate over-subtracted normal products, i.e. of higher
UV-degree than the canonical operator dimension, to minimally
subtracted ones. These prove a powerful tool in e.g.\ deriving field
equations and studying anomalies. 

A first version of the quantum action principle can be used to
express the relation of 
some infinitesimal variation of Green functions, or equivalently
generating functionals, with the insertion of a normal
product. Ref.\ \cite{Lowenstein:1971jk} considers differential vertex operations (DVO)
which are insertions of integrated normal ordered local field
polynomials into the Gell-Mann-Low formula corresponding to the
respective Green function 
\begin{equation}
\Delta\cdot G_{i_1,\dots,i_n}(x_1,\dots,x_n)=\langle 0|T\int\,\mathrm{d}y\,N[P(y)]\phi^{i_1}(x_1)\dots\phi^{i_n}(x_n)|0\rangle.
\end{equation}Then one can connect the variation of the Green function w.r.t.\ some parameter with those DVO's, i.e.\ taking some infinitesimal variation as $\mathcal{L}_{\text{int}}\rightarrow\mathcal{L}_{\text{int}}+\sum_k\varepsilon_k P_k(x)$, it follows
\begin{equation}
\frac{\partial G^{\varepsilon}}{\partial\varepsilon^k}\bigg\vert_{\varepsilon=0}=i\Delta\cdot G.
\end{equation}
This result is valid for BPHZ renormalized disconnected, connected and
1PI Green functions, and therefore also for the corresponding
generating functionals.

It can be used to derive the renormalized QAP for a generic parameter of the theory $\lambda$,\begin{equation}\frac{\partial\Gamma}{\partial\lambda}=i\Delta_\lambda\cdot G,\end{equation}where $\Delta_\lambda=\int\,\mathrm{d}x\,N[\frac{\partial\mathcal{L}}{\partial\lambda}]$.

There are several further versions of the quantum action principle
with regards to variations of parameters or (external)
fields. In particular Refs.\ \cite{Lam:1972fzg,Lam:1972mb,Lam:1973qa} established a version of the action principle
w.r.t variations of dynamical fields (see e.g.\ Ref.\ \cite{Lam:1972mb}, Eq.\ (5.4)). The
left-hand side being equal to zero due to conservation of some
current, the resulting relation corresponds to Eq.\ (\ref{deltaAbbrev})
for the more general case of a non-invariant Lagrangian
$\delta\mathcal{L}\neq 0$ under some symmetry transformation. It is
rigorously established in terms of the generating functional
for general Green functions renormalized  in the BPHZ framework,  and
it can be connected to the generating 
functional of 1PI Green functions via Legendre transformation.

Thus the finite BPHZ framework is a  setting in which formally derived
identities among generating functionals such as the ones described in
Secs.\ \ref{sec:formalQAP} or \ref{sec:STIformal} can be given a
sensible all-order meaning. 

In addition, in any  regularization/renormalization procedure in
agreement with the basic postulates  there is a way to cancel divergences
and to obtain finite Green functions. These may differ from the ones
obtained in BPHZ (or any other regularization), but in view of the
theorems discussed in Sec.\ \ref{sec:generalrenormalization} the
differences can only amount to local
counterterms at each order.

In the following we summarize important statements
of the quantum action principle valid for any such finite Green
functions defined via any consistent regularization and subtraction of
divergences. The statements can be cast in a variety of forms, similarly
to Sec.\ \ref{sec:formalQAP}. Here we provide the formulation for the
effective action $\Gamma$, as reviewed
in Ref.\ \cite{Piguet:1995er}.
First, equations of motion for the generating functionals can be
written as
 \begin{align}
\frac{\delta\Gamma}{\delta\phi_i(x)}-\Delta_i(x)\cdot\Gamma&=0\,.
\label{QAPBPHZ1}
 \end{align}
For variations with respect to parameters we have
\begin{align}
\frac{\partial\Gamma}{\partial\lambda}&=\int\mathrm{d}x \, \Delta(x)\cdot\Gamma\,.
\end{align}
As
discussed in Secs.\ \ref{sec:STIformal} and \ref{sec:GreenFunctions}, in the case of nonlinear
symmetry transformations it is useful to couple the 
composite operators to some external field, say $\rho^a(x)$. Then one can  arrive at the
following version of the quantum action principle relevant for such nonlinear
symmetry transformations,
\begin{equation}\label{QAPBPHZ3}
\frac{\delta\Gamma}{\delta\rho_a(x)}\frac{\delta\Gamma}{\delta\phi_i(x)}=\Delta^{ai}(x)\cdot\Gamma\,.\end{equation}

In all previous equations (\ref{QAPBPHZ1}--\ref{QAPBPHZ3}), the
quantities $\Delta$ denote insertions of local composite field
operators, whose dimensions are bounded by power counting and whose
tree-level value is fixed in terms of the classical expression
$\Gamma_{\text{cl}}$. For example, in case of Eq.\ (\ref{QAPBPHZ1}),
$\Delta_i$ is a local composite field operator whose 
dimension is bounded by $(D-d_i)$, where $d_i$ denotes the
power-counting dimension of
the corresponding field $\phi_i$, and
\begin{equation}
  \Delta_i=\frac{\delta\Gamma_{cl}}{\delta\phi_i}+\mathcal{O}(\hbar)\,.
\end{equation}

\subsubsection{Comparing Quantum Action Principles in BPHZ and~DReg}

The quantum action principles discussed in the previous subsection for
BPHZ and in Sec.\ \ref{sec:QAPDReg} for DReg are similar but
different. Here we briefly comment on their relationship. The BPHZ
version is valid for any regularization/renormalization procedure,
including DReg. However, it is valid for the finite theory, in DReg for the
theory after taking $\text{LIM}_{D \, \to \, 4}$ as defined in
Eq.\ (\ref{DefLIM}). The definition of this limit includes setting
evanescent quantities (such as the $(D-4)$-dimensional metric
$\hat{g}^{\mu\nu}$) to zero. The insertions $\Delta$ appearing e.g.\ in
Eqs.\ (\ref{QAPBPHZ3}) are always finite, $4$-dimensional normal
product insertions into the finite Green functions.

In contrast, in the DReg case, the counterpart equation
(\ref{QAPbasicGreenfunctions}) is valid for general $D\ne4$, including
evanescent quantities. In addition, if the identity corresponds to a
symmetry such as the Slavnov-Taylor identity which is valid at tree
level and in $4$ dimensions, then the insertion $\Delta$ appearing in
Eq.\ (\ref{QAPbasicGreenfunctions})  is purely evanescent.

It may not be immediately obvious how this can be reconciled with the
purely $4$-dimensional case of BPHZ. This is however important as we
shall be making use of general considerations following from the
algebraic framework while working in DReg. In fact, both versions of
the quantum action principle are valid and useful. The BPHZ version is
useful to establish general existence proofs which we can rely on also
within DReg, but the DReg version is useful for explicit computations
since there the explicit form of the insertion $\Delta$ is known.

The key is
provided by the Bonneau identities established in
Refs.\ \cite{Bonneau:1979jx,Bonneau:1980zp}. These identities precisely state that
the insertion of an evanescent operator in DReg as in Eq.\ (\ref{QAPbasicGreenfunctions})
may in the $\text{LIM}_{D\to4}$ be rewritten as an insertion of a
finite, $4$-dimensional operator as in Eq.\ (\ref{QAPBPHZ3}). In this way, the
BPHZ quantum action principle can also be rederived from the one in
DReg.

On the technical level, the Bonneau relationship
also provides the coefficients in the expansion of  evanescent
operator insertions in terms of $4$-dimensional, finite
insertions. They are given by the
residue of simple $1/(D-4)$ pole of the insertion of the evanescent operator
into Green functions. The proof is essentially achieved by taking
dimensionally renormalized amplitudes $\mathcal{R}_G$ associated 
to a graph $G$ and comparing the vertex insertions
$\hat{g}^{\mu\nu}[\mathcal{O}_{\mu\nu}\cdot\mathcal{R}_G]$ on the one
hand with vertex insertions with
$[\hat{g}^{\mu\nu}\mathcal{O}_{\mu\nu}]\cdot\mathcal{R}_G$ on the other
hand.

At the one-loop level, the Bonneau identities are not surprising since
evanescent quantities can only contribute in the $\text{LIM}_{D\to4}$
if they hit $1/(D-4)$ poles, which at the one-loop level have local
coefficients which may be interpreted as a $4$-dimensional local
operator. However, their validity lies in their all-order nature. We
mention here that Bonneau identities can also be used to obtain
information on renormalization group equations in the presence of
symmetry breakings of the regularization, see
e.g.\ Refs.\ \cite{Martin:1999cc,Belusca-Maito:2020ala,Belusca-Maito:2022wem}. 

\subsubsection{Algebraic Renormalization and Symmetry~Restoration}
\label{sec:algebraicrenormalizationDetails}
With the quantum action principle at our disposal we can now describe
the logic of algebraic renormalization of gauge theories. The starting
point are possible breakings of the Slavnov-Taylor identity (or
similar identities) as given by Eq.\ (\ref{BrokenDQAP}) due to the regularization. The quantum action principle provides a useful tool in restricting the structure of the breaking and in determining whether the symmetry can be restored, i.e. whether there are anomalies. For that we proceed inductively order by order in perturbation theory. The goal is to determine the required finite, symmetry-restoring counterterms $S_\text{ct}^n$ step by step for each $n$.

At lowest order, at the classical level $n=0$, the Slavnov-Taylor
identity is valid by construction. This forms the basis of the
inductive procedure. Let us then suppose the theory is renormalized
completely, hence it is finite and the Slavnov-Taylor identity is
fulfilled at some order $n-1$. In addition, at the next order $n$, the
divergences are already cancelled by appropriate singular
counterterms. Hence we have
\begin{equation}
\label{InductionHypothesis}
\mathcal{S}(\Gamma_{\text{subren}}^{(n),\text{fin}})=\mathcal{O}(\hbar^n),
\end{equation}
where we have introduced the notation
\begin{equation}\label{Gammasubrenfin}
\Gamma_{\text{subren}}^{(n),\text{fin}}=\Gamma^{(n)}_\text{subren}+S^n_\text{sct},
\end{equation}
which denotes the effective action finite at order $n$ after
subrenormalization and adding the necessary divergent $n$-loop
counterterms. This quantity corresponds to the set of finite Green
functions for which the validity of the quantum action principle in
BPHZ has been proven, and it can be defined in any other
regularization scheme equivalent to BPHZ. 

The task is then to study the possible breakings on the RHS of
Eq.\ (\ref{InductionHypothesis}) as well as the possible structure of
counterterms. As mentioned before the breaking is restricted in two
ways. First, we may employ the quantum action principle to
find,
\begin{equation}
\label{InductionStep}
\mathcal{S}(\Gamma_{\text{subren}}^{(n),\text{fin}})=\hbar^n\Delta\cdot\Gamma_{\text{subren}}^{(n),\text{fin}}=\hbar^n\Delta+\mathcal{O}(\hbar^{n+1}).
\end{equation}
The important point is that $\Delta$ is a local polynomial in fields
and derivatives, also restricted by power counting.  This property was
announced in Sec.\  \ref{sec:STIformal}, where the Slavnov-Taylor
identity was formally derived from the path integral.

  Second, applying the linearized BRST operator $s_{\Gamma_\text{cl}}\equiv b$ to Eq.\ (\ref{InductionStep}) using Eq.\ (\ref{SNilpotent}) and extracting the $\mathcal{O}(\hbar^n)$ terms, we arrive at a consistency condition (also called the Wess-Zumino consistency condition),
\begin{equation}\label{WessZuminoCondition}
b\Delta=0.\end{equation}
Hence the possible breaking $\Delta$ is restricted very similarly
(cf.\ Eq.\ (\ref{STIdiv})) to the possible divergences
$\Gamma_\text{div}$ in Sec.\
\ref{sec:ctsymmetrypreserving}. Both $\Gamma_\text{div}$ in Eq.\ (\ref{STIdiv}) and $\Delta$ in Eq.\ (\ref{WessZuminoCondition}) are local polynomials restricted by power counting which are annihilated by $b$, but $\Gamma_\text{div}$ is of ghost number $0$ whereas $\Delta$ has ghost number $1$.
Now one can make a distinction. If $\Delta$ is a $b$-exact term,
i.e. if there exists another local polynomial $\Delta^{'}$
with\begin{equation}\label{TrivialC}\Delta=b\Delta^{'},\end{equation}it
is called a trivial element of the cohomology of the BRST operator. In
this case we can supplement the original action with a new $n$-loop
order counterterm
\begin{equation}
S^n_\text{fct}=S^n_\text{fct,non-inv}+S^n_\text{fct,inv}=-\Delta^{'}+S^n_\text{fct,inv},
\end{equation}
where the last term reflects the freedom to add to the action any finite, symmetric counterterm, obeying $b\,S^n_\text{fct,inv}=0$.
Hence, we end up with
\begin{equation}
\mathcal{S}(\Gamma_{\text{subren}}^{(n),\text{fin}}+\hbar^n
S^n_\text{fct})=\mathcal{S}(\Gamma_{\text{subren}}^{(n),\text{fin}})+b\,\hbar^n
S^n_\text{fct}+\mathcal{O}(\hbar^{n+1})=\mathcal{O}(\hbar^{n+1}),\end{equation}where
the last step follows from the induction hypothesis. Compatibility
with ghost and gauge fixing equation is shown in
Ref.\ \cite{Piguet:1995er}.

Hence, under the condition (\ref{TrivialC}), we can find a counterterm
action $S_\text{fct,non-inv}^n$ which defines finite, non-invariant
counterterms that repair the symmetry. Furthermore, it is possible to add
any number of finite, invariant counterterms to the action as they
satisfy $b\,S_\text{fct,inv}=0$ and hence do not disturb the
STI. These invariant counterterms behave like the finite counterterms
discussed in Sec.\ \ref{sec:ctsymmetrypreserving} and can be used to satisfy certain
renormalization conditions.

One task of the algebraic renormalization programme is therefore to determine the most general solution of the equation $
b\Delta=0$. If all possible solutions are $b$-exact, then this constitutes a proof that the Slavnov-Taylor identity can be established at all orders in the renormalized theory.

However, if we cannot write the breaking $\Delta$ as a $b$-exact term,
the symmetry cannot be repaired. This is an anomaly. In case of the
Slavnov-Taylor identity such an anomaly is disastrous since it
destroys the interpretation of the theory as a sensible quantum
theory, see the discussion at the beginning of the present section.
Anomalies are thus non-trivial elements of the cohomology of the
$b$-operator, i.e.\ expressions which are annihilated by $b$ but are
not $b$-exact.

The previous remarks constitute crucial insights of the BRST formalism
\cite{Becchi:1974xu,Becchi:1974md,Becchi:1975nq,Tyutin:1975qk}. The
analysis of whether a gauge theory is renormalizable, 
i.e.\ whether the Slavnov-Taylor identity can be restored at each
order, can be made on a purely classical level, by finding all
possible solutions of Eq.\ (\ref{WessZuminoCondition}) and checking 
whether they are all $b$-exact.

The actual computation can be found in the original references and in
the reviews \cite{Piguet:1980nr,Piguet:1995er}. It can be sketched as
follows.
From  the Wess-Zumino consistency condition
(\ref{WessZuminoCondition}) and the nilpotency of the BRST operator,
one can derive a set of equations, the so called descent
equations. Solving these gives a general expression of the possible
anomalies of a theory. In the present case of interest for a generic
Yang-Mills theory it can be shown that the consistency condition
simplifies to $s\Delta(G,c)=0$, see e.g.\ \cite{Piguet:1995er}, with
dependence on the gauge and the ghost field only. Writing $\Delta$ as
an integrated local product and solving the descent equations, it
leads to the famous Adler-Bell-Jackiw gauge anomaly first discovered in Refs.\ \cite{Adler:1969gk,Bell:1969ts,Adler:1969er},\footnote{Note the different relative sign of the first term of Eq.\ (\ref{ABJexplicit}) compared to \cite{Piguet:1995er} which comes from a different sign convention in the covariant derivative, see Eq.\ (\ref{Dmu}).}
\begin{equation}
\label{ABJexplicit}
\Delta=L\times\varepsilon_{\mu\nu\rho\sigma}\mathrm{Tr}\int\mathrm{d}^4x\,c_a\partial^{\mu}\left(-gd_{A}^{abc}\partial^{\nu}G_b^{\rho}G_c^{\sigma}+g^2\frac{\mathcal{D}_{A}^{abcd}}{12}G_b^{\nu}G_c^{\rho}G_d^{\sigma}\right),
\end{equation}
where $L$ is a coefficient that can be determined from explicit
calculations and which depends on the theory inputs. The group symbols
are given by,
\begin{equation}
d_{A}^{abc}=\mathrm{Tr}\Big(T^a_{\text{adj}}\big\{T^b_{\text{adj}},T^c_{\text{adj}}\big\}\Big),
\end{equation}
and
\begin{equation}
\label{DabcdRelation}\mathcal{D}_{A}^{abcd}=d_{A}^{nab} f^{ncd}+d_{A}^{nac} f^{ndb}+d_{A}^{nad} f^{nbc},
\end{equation}
where $T^a_{\text{adj}}$ denotes adjoint generators under which
ghosts and gauge fields transform,
cf.\ Eq.\ (\ref{eq:AdjointReprOfGenerators}). Expression
(\ref{ABJexplicit}) must vanish by itself, i.e. cannot be absorbed by
counterterms, for the theory to be consistent. In the case of a single
left-handed fermion it can be shown by a one-loop calculation that the
anomaly is proportional to
\begin{equation}
\label{LFanomaly}
\frac{1}{2} \, d_{A}^{abc} \, \mathrm{Tr}(T^a\{T^b,T^c\}),
\end{equation}
which means that its cancellation depends on an appropriate choice of the matter content of the theory. The famous Adler-Bardeen theorem guarantees that if the gauge anomaly vanishes at one-loop order, it also vanishes at all orders, cf.\ \cite{Piguet:1995er}. The expression in Eq.\ (\ref{LFanomaly}) cannot vanish by itself, but in such a theory with a family of left-handed fermions, their charges may add up to zero as is the case in the SM. For some gauge groups such as SU(2), the above expression vanishes identically due to the vanishing of some group symbols. Hence there can be no anomaly.

In summary we have sketched how algebraic renormalization allows 
identifying the general structure of the breaking of the Slavnov-Taylor
identity. It constitutes a setting in which the restoration of the
symmetry can be proven to all orders for trivial elements of the BRST
cohomology such as spurious breakings introduced by the BMHV
algebra. In the case of non-spurious breakings, e.g.\ the gauge
anomaly, one can derive explicit conditions for its cancellation which
a sensible theory must satisfy. Further, nonrenormalization theorems,
as in the case of the Adler-Bardeen theorem, can be shown and allow
 evaluating the gauge anomaly in a simple way. The main technical tool
which serves to establish these findings is the general quantum
action principle valid in many equivalent subtraction schemes. A key
advantage of the algebraic proof is that there is no need for an
invariant regularization which for e.g.\ chiral gauge theories does
not exist.

\subsubsection{Outlook and Further Remarks on Anomalies and Algebraic~Renormalization}
\label{sec:OutlookAlgRen}

  At this point we interject a brief outlook on anomalies and further applications  of the
  techniques of algebraic renormalization. Next to the perturbative
  chiral gauge anomalies discussed above and discovered in Refs.\
  \cite{Adler:1969gk,Bell:1969ts,Adler:1969er} there exist
  global chiral anomalies \cite{Witten:1982fp} and 
  pertubative  mixed gauge-gravitational anomalies
  \cite{Delbourgo:1972xb,Eguchi:1976db,Alvarez-Gaume:1983ihn}.  A
  chiral gauge model can be renormalized only if all these chiral anomalies cancel, which may be achieved by a proper choice of
fermion representations of for the chiral model, for example see
Ref. \cite{Geng:1989tcu} and references therein. Eq.\ (\ref{ABJexplicit}) is necessary but not
sufficient if gravity and nonperturbative effects are taken into account.

Important theories such as the  Standard Model of particle physics,
are renormalizable. In particular, the electroweak SM was
completely treated in algebraic renormalization 
in Ref.\ \cite{Kraus:1997bi}, establishing the SM as a fully all-order consistent,
renormalizable theory. Ref.\ \cite{Grassi:1997mc} gave a similar proof
using the background field gauge (see footnote \ref{BFMfootnote}), and
Ref.\ \cite{Hollik:2002mv} gave a similar proof for the supersymmetric
SM.
These papers complement earlier extensive discussions of the
renormalization of the
electroweak SM
 by e.g.\ Refs.\ \cite{Aoki:1982ed,Bohm:1986rj}, see
also Ref.\ \cite{Denner:2019vbn}.

The validity of the
Slavnov-Taylor identity  and the techniques of algebraic renormalization
can also be used to establish further 
interesting physics properties of quantum gauge theories such as the
renormalized electroweak SM.
E.g.\ charge universality can be established based on both gauge
choices \cite{Aoki:1982ed,Denner:1994xt}, see also Ref.\ \cite{Denner:2019vbn} for further
discussions. As another example, the renormalization of Higgs vacuum
expectation values in spontaneously broken gauge theories can be
controlled via a suitable Slavnov-Taylor identity
\cite{Sperling:2013eva,Sperling:2013xqa}. 

\subsection{Algebraic Symmetry Restoration in the Context of~DReg}
\label{sec:AlgRenInDReg}

So far in this section we have studied the role of symmetries in the
process of renormalization. If the symmetry is respected by the
regularization, it implies a great simplification for the UV
counterterms. If it is not, algebraic renormalization constitutes a
general setup which allows identifying symmetry violations and restoring
the symmetry.
Here we specialize the general procedure to the case of DReg. We use
the BMHV scheme with non-anticommuting $\gamma_5$ in which gauge invariance
may be broken.

\subsubsection{Formulation of Symmetry and Symmetry Breaking in~DReg}
\label{sec:SymmetryAndSymmetryBreakingInDReg}

The ultimate symmetry requirement is the Slavnov-Taylor identity
expressing BRST invariance of the full renormalized theory,
Eq.\ (\ref{REQUIRE}). In the context of DReg this requirement can be
formulated as
\begin{equation}
\label{SymRegD}
\mathop{\text{LIM}}_{D \, \to \, 4} \, (\mathcal{S}_D(\Gamma_\mathrm{DRen})) = 0.
\end{equation}
As defined in Sec.\ \ref{sec:DRegGamma}, $\Gamma_\mathrm{DRen}$
denotes the renormalized effective action, still in $D$ dimensions but
including all counterterms cancelling $1/\epsilon$ divergences and
restoring symmetries. The limit refers to the operation of letting
$\epsilon\rightarrow0$ as well as putting evanescent quantites such as
$\hat{g}^{\mu\nu}$ to zero.

In order to discuss the inductive procedure, we consider some order $n$
and suppose the theory has been renormalized and all counterterms have
been constructed up to the previous order
$n-1$. This provides us with
\begin{align}
  \Gamma^{(n)}_{\text{subren}}\,,
\end{align}
again using the notation of Sec.\ \ref{sec:DRegGamma}.
At this point we know from Sec.\ \ref{sec:renormalization} that the divergences at
the $n$-th order can be cancelled by adding a local counterterm action
$S_{\text{sct}}^n$. It may or may not be true that the divergences
follow the simple pattern described in
Sec.\ \ref{sec:ctsymmetrypreserving}. In general we can always write
\begin{align}
  S_{\text{sct}}^n &=
  S_{\text{sct,inv}}^n +S_{\text{sct,non-inv}}^n \,,
\end{align}
where the first term corresponds to symmetric counterterms as
described in Sec.\ \ref{sec:ctsymmetrypreserving} and the second term
corresponds to whatever other divergent counterterms are required.

After subtracting these divergences the theory is finite at the order
$n$ and the Slavnov-Taylor identity may be written as
\begin{align}
\label{SdGammasubrenDREG}
  \mathcal{S}_D(\Gamma_\mathrm{subren}^{(n)}+S_{\text{sct}}^n)=\hbar^n\Delta_D+\mathcal{O}(\hbar^{n+1})
  \,,
\end{align}
where $\Delta_D$ is a possible finite breaking term, still evaluated in $D$
dimensions. The subrenormalized and finite effective action introduced
for the algebraic analysis in Eq.\ (\ref{Gammasubrenfin}) is now given
by $\mathop{\text{LIM}}_{D \, \to \, 4}   (\Gamma_\mathrm{subren}^{(n)}+S_{\text{sct}}^n)$, and the counterpart of Eq.\ (\ref{InductionStep}) is given
by the 4-dimensional limit
\begin{align}
\mathop{\text{LIM}}_{D \, \to \, 4} \, \Delta_D &= \Delta_{\text{from Eq.\   (\ref{InductionStep})}}\,.
\end{align}
This finite quantity $\Delta$ is the one constrained by algebraic
renormalization and discussed after
Eq.\  (\ref{InductionStep}). That is, it is a local breaking term
which satisfies the Wess-Zumino consistency conditions and which can
be cancelled by adding suitable counterterms (we assume that there is
no genuine anomaly).

The practical question is then how to obtain first the breaking term
$\Delta$ and then the symmetry-restoring counterterms. There are two
strategies for this. The first, obvious option is to evaluate all
Green functions appearing on the LHS of
Eq.\ (\ref{SdGammasubrenDREG}) including their finite parts, plug them
into the Slavnov-Taylor identity and determine the potentially
non-vanishing breaking. This straightforward procedure is convenient
in that it operates on ordinary Green functions. Its drawback is that
most finite parts of Green functions --- in particular parts
that are non-polynomial in the momenta --- will be in agreement with the symmetry
and hence drop out of Eq.\ (\ref{SdGammasubrenDREG}), such that the
calculation can become unnecessarily complicated. Nevertheless this direct
approach has been used in the literature, e.g.\ in
Refs.\ \cite{Grassi:1999tp,Grassi:2001zz,Hollik:1999xh,Hollik:2001cz,Fischer:2003cb}
on applications on chiral gauge theories and supersymmetric gauge
theories. In the subsequent section \ref{sec:oneloopphotonSE} we will
also illustrate this approach with a concrete example.

A second, alternative approach is provided by using the regularized
quantum action principle in DReg, described in
Sec.\ \ref{sec:QAPDReg}. This theorem guarantees that we can rewrite 
the LHS of Eq.\ (\ref{SymRegD}) as
\begin{equation}
\label{applyQAP}
  \mathcal{S}_D(\Gamma_\mathrm{DRen})=(\widehat{\Delta}+\Delta_{\mathrm{ct}})\cdot\Gamma_\mathrm{DRen}.
\end{equation}
The possible breaking of the Slavnov-Taylor identity is thus rewritten
as an operator insertion of the composite operator
$\widehat{\Delta}+\Delta_{\mathrm{ct}}$, which is defined as
\begin{subequations}
	\label{eq:Deltas}
	\begin{align}
	\widehat{\Delta} & = \mathcal{S}_D( S_0) \, ,
	\label{eq:DelpertE0}\\
	\widehat{\Delta}+	\Delta_\text{ct} & =\mathcal{S}_D( S_0+S_\text{ct}) \, ,
	\label{eq:DelpertEtotal}		\end{align}
\end{subequations}
In this approach, the breaking $\Delta$ may be computed in terms of
the RHS of (\ref{applyQAP}). The advantage lies in significantly
restricting possible non-vanishing contributions. In particular,
$\widehat\Delta$ is evanescent; hence it can contribute in the
$\mathrm{LIM}_{D\rightarrow4}$ only in combination with $1/\epsilon$
singularities of Feynman diagrams.

The RHS of (\ref{applyQAP}) can be expanded in loop orders as
\begin{equation}
  \widehat{\Delta}+\sum_{i=1}^{\infty}\hbar^i\left(\widehat{\Delta}\cdot\Gamma^i_\mathrm{DRen}+\sum_{k=1}^{i-1}\Delta^k_\mathrm{ct}\cdot\Gamma^{i-k}_\mathrm{DRen}+\Delta^i_\mathrm{ct}\right).
\end{equation}
Plugging the previous definitions into Eq.\ (\ref{SymRegD}), we arrive
at an equation expressing the symmetry requirement exactly at the  order $n$,
\begin{equation}
\label{LIMcond}
\mathop{\text{LIM}}_{D \, \to \, 4} \, \left(\widehat{\Delta}\cdot\Gamma_\mathrm{DRen}^n+\sum_{k=1}^{n-1}\Delta^k_\mathrm{ct}\cdot\Gamma^{n-k}_\mathrm{DRen}+\Delta^n_\mathrm{ct}\right)=0,
\end{equation}
for all $n\geq1$. The individual terms in this equation have divergent
and finite parts, but by construction the entire expression is finite;
hence the cancellation of divergences may be used as a consistency
check of practical calculations. For the determination of
symmetry-restoring counterterms,
Eq.\ (\ref{LIMcond}) should be viewed as follows. At the order $n$ and
after subrenormalization and adding divergent $n$-loop counterterms,
everything in Eq.\ (\ref{LIMcond}) is already known except the finite
counterterms of order $n$. They enter via $\Delta^n_\mathrm{ct}$,
which in turn depends on the to-be-determined counterterms. The
following subsubsection will make the dependence explicit.
Hence Eq.\ (\ref{LIMcond}) can be regarded as the optimized defining relation for the
symmetry-restoring counterterms in DReg. 

We close with the remark that Eq.\ (\ref{LIMcond}) does not fully
determine all finite counterterms. It only determines the required
form of counterterms in order to restore the symmetry. However,
Eq.\ (\ref{LIMcond})  is blind to several types of counterterms:
finite and symmetric counterterms (which often correspond to a
renormalization transformation as described in
Sec.\ \ref{sec:ctsymmetrypreserving}) drop out; 
such counterterms can therefore still be adjusted at will  e.g.\ to satisfy the
renormalization conditions corresponding to an on-shell or a different
desirable renormalization scheme. In addition,  evanescent and finite
counterterms also drop out and may be added to optimize the
counterterm action.

\subsubsection{Practical Restoration of the~Symmetry}
\label{sec:practicalrestorationgeneral}
Here we illustrate the blueprint for the practical restoration of the
symmetry, if  Eq.\ (\ref{LIMcond}) is used as a basis.

We begin at the one-loop level and start from the regularized but
unrenormalized effective action $\Gamma^{(1)}$.\footnote{We slightly simplify the notation and use $\Gamma^{(1)}$  in
the following equations of this subsubsection to denote the
unrenormalized  effective action up to one-loop order. According to the general
notational scheme defined in Sec.\ \ref{sec:DRegGamma}, this could
also be called $\Gamma^{(1)}_{\text{subren}}$.\label{footnoteSimpleNotation}
  }
At the one-loop level the regularized action plus counterterms as well
as the
symmetry breaking induced by the counterterms at one-loop order are given by
\begin{subequations}
	\begin{align}
	\Gamma_\text{DRen}^{(1)} &= \Gamma_{}^{(1)} + S_\text{sct}^1 + S_\text{fct}^1 \, ,
	\label{eq:GammaDR1_2}
	\\\label{Delta1rearrange}
	\Delta_\text{ct}^1 &= \mathcal{S}_D (S_0+S_\text{ct})^1
        = b_D S_\text{sct}^1 + b_D S_\text{fct}^1\, , 		\end{align}
\end{subequations}
where the last part of the last equation is a specific rearrangement possible at the
one-loop level, and where the linearized Slavnov-Taylor operator $b_D$ is defined in analogy to 
$b$ in Eq.\ (\ref{bDefinition}).
The general equations establishing the cancellation of
divergences and symmetry restoration, (\ref{LIMcond}), become
\begin{subequations}
	\label{eq:rencon}
	\begin{align}
	S^1_\text{sct} +\Gamma^1_\text{div} &=0\, ,
	\label{eq:S1sct_1L}
	\\
	\big(\widehat{\Delta}\cdot\Gamma^1_\text{} +
	\Delta_\text{ct}^1\big)_\text{div} &= 0 \, ,
	\label{eq:S1sct_1Lcheck}
	\\
	\mathop{\text{LIM}}_{D \, \to \, 4} \, \big(\widehat{\Delta}\cdot\Gamma^1_\text{} +
	\Delta_\text{ct}^1\big)_\text{fin} &= 0 \, .
	\label{eq:S1fct_1Lv1}
	\end{align}
\end{subequations}
Compared to the general Eq.\ (\ref{LIMcond}), terms that vanish at
one-loop order were dropped.
The quantities that need to be explicitly computed here
are
the one-loop divergences $\Gamma^1_\text{div} $ and the one-loop
diagrams with one insertion of the evanescent operator
$\widehat{\Delta}$, $\widehat{\Delta}\cdot\Gamma^1_\text{}$. 
The first of these equations then determines the divergent one-loop
counterterms $S_{\text{sct}}^1$, and the  second equation provides a
consistency check of the divergences.
In view of Eq.\ (\ref{Delta1rearrange}), the last line contains $b_D
S_{\text{fct}}^1$ and thus determines the symmetry-restoring one-loop
counterterms.

Next we consider the two-loop order.
At the two-loop level, the corresponding  equations for the effective
action and the symmetry breaking of counterterms are
\begin{subequations}\label{eq:rencond2L}
	\begin{align}
	\Gamma_\text{DRen}^{(2)} &= \Gamma_{\text{subren}}^{(2)} + S_\text{sct}^2 + S_\text{fct}^2 \, ,
	\label{eq:GammaDR2_2}
	\\\label{Delta2rearrange}
	\Delta_\text{ct}^2 &= \mathcal{S}_D (S_0+S_\text{ct})^2   = \mathcal{S}_D (S_0+S_\text{ct}^1)^2 +b_D S_{\text{sct}}^2+b_D S_{\text{fct}}^2\,,
	\end{align}
\end{subequations}
where the upper index $^2$ corresponds to extracting the two-loop
terms. The last equation exhibits the appearance of the genuine
two-loop counterterms in a way  specific to the two-loop level.
The equations corresponding to finiteness and symmetry restoration read
\begin{subequations}
	\label{eq:deltadef}
	\begin{align}
	S^2_\text{sct} +\Gamma^2_\text{subren,div} &= 0 \, ,
	\label{eq:S2sct_2L}
	\\
	\label{eq:CheckDeltasct2L} \big(\widehat{\Delta}\cdot\Gamma_\text{\text{subren}}^2 + \Delta_\text{ct}^1\cdot\Gamma_\text{}^{1}
	+ \Delta_\text{ct}^2\big)_\text{div} &= 0 \, ,
	\\
	\label{eq:3rd}
	\mathop{\text{LIM}}_{D \, \to \, 4} \, \big(\widehat{\Delta}\cdot\Gamma_\text{\text{subren}}^2 + \Delta_\text{ct}^1\cdot\Gamma_\text{}^{1}
	+ \Delta_\text{ct}^2\big)_\text{fin} &= 0 \, .
	\end{align}
\end{subequations}
Here we have to calculate first the two-loop divergences to obtain the
two-loop divergent counterterms. Then we have to calculate diagrams
with insertions of $\widehat{\Delta}$ up to the two-loop level (and
including one-loop subrenormalization), as well as one-loop diagrams
with insertions of $b_D$-transformed one-loop counterterms. The second
equation must automatically hold and provides a check. The third
equation then
determines the genuine finite two-loop symmetry-restoring counterterms
$b_D S_{\text{fct}}^2$, which 
appear via Eq.\ (\ref{Delta2rearrange}) in $\Delta_\text{ct}^2$.

In summary the recipe is as follows,
\begin{itemize}
\item
UV-renormalize the theory, previously renormalized up to order $n-1$, at order $n$ to obtain the singular counterterms,
\item
calculate genuine $n$-loop Green functions with one-time insertion of $\widehat{\Delta}$ for their divergent and finite part,
\item
calculate the $k$-loop order insertion into $(n-k)$-loop order graphs and determine their divergent and finite contributions,
\item
check that the divergences thus obtained sum up to zero,
\item
collect the finite contributions and choose monomials $X$ such that $b_D\,X$ cancels them.
This is always possible as discussed in the previous subsections.
\end{itemize}

\subsubsection{The Counterterm Lagrangian in the BMHV~Scheme}
\label{sec:countertermLagrangianBMHV}
The output of the regularization/renormalization programme is the renormalized
effective action and  the required counterterm action consisting of
singular and finite counterterms.
In the context of the BMHV scheme the previous subsections show that
the counterterm action can in general contain five different kinds of terms,
\begin{equation}
S_\text{ct}=S_\text{sct,inv}+S_\text{sct,non-inv}+S_\text{fct,inv}+S_\text{fct,restore}+S_\text{fct,evan}.
\end{equation}
This equation is a more detailed version of the generic decomposition explained
in Sec.\ \ref{sec:DRegGamma} into singular and finite counterterms.
For both the singular and the finite counterterms we may isolate a
symmetry-invariant piece, which has the pattern of symmetric
counterterms discussed in Sec.\ \ref{sec:ctsymmetrypreserving}
and typically corresponds to counterterms generated by a
renormalization transformation
as
\begin{equation}
S_0\stackrel{\text{ren.\ transf.\ (\ref{RenTrafo})}}{\longrightarrow}S_0+S_\mathrm{sct,inv}+S_\text{fct,inv}.
\end{equation}
In general, the conditions of Sec.\ \ref{sec:ctsymmetrypreserving} are
not met and symmetry-violating counterterms are required. Accordingly,
the next type of counterterms
$$S_{\text{sct,non-inv}}$$
corresponds to additional singular
counterterms needed to cancel additional $1/\epsilon$ poles of loop
diagrams that cannot be cancelled by symmetry-invariant counterterms.
They may be evanescent and, starting from $2$-loop order, also
$4$-dimensional (non-evanescent). They cannot be obtained by
renormalization transformations. 
We note that the subtraction of evanescent
$1/\epsilon$ poles is a necessity for the consistency of higher orders
(see also Ref.\ \cite{Gnendiger:2017pys} for a review
discussing this point).

Next,
$$S_{\text{fct,restore}}$$ corresponds to finite counterterms needed
to restore the Slavnov-Taylor identity and thus the underlying gauge
invariance. They are the central 
objects of the present discussion and the outcome of the practical
recipe of Sec.\ \ref{sec:practicalrestorationgeneral}.
Determining these counterterms is one of 
the key tasks in the usage of the BMHV scheme. Once those counterterms
are found, the theory can be considered to be renormalized.

As mentioned before, the symmetry-restoring counterterms are not
unique. Clearly, they may be modified by shifting around any symmetry-invariant
counterterm between $S_{\text{fct,inv}}$ and $S_{\text{fct,non-inv}}$, 
since invariant terms would drop out of
Eqs.\ (\ref{LIMcond},\ref{eq:S1fct_1Lv1},\ref{eq:3rd}). The overall
sum of  $S_{\text{fct,inv}}+S_{\text{fct,non-inv}}$ 
can only be fixed by imposing a renormalization scheme
(such as e.g.\ the on-shell scheme), and the split into $S_{\text{fct,inv}}$ and $S_{\text{fct,non-inv}}$
can only be fixed by picking a convention. To illustrate
this point, let us assume the counterterm Lagrangian must contain a
non-gauge invariant term $z A^\mu\,\square\,A_\mu$ where $z$ is a
coefficient and $A^\mu$ a gauge field.
Two different options for the counterterm Lagrangians would then
be
\begin{subequations}
\begin{align}
  \mathcal{L}_{\text{fct,non-inv}} &=
  z A^\mu\,\square\,A_\mu,
  &
  \mathcal{L}_{\text{fct,inv}} &=
  \delta Z (A^\mu\,\square\,A_\mu+(\partial A)^2),
  \\
  \mathcal{L}_{\text{fct,non-inv}} &=
  -z (\partial A)^2,
  &
  \mathcal{L}_{\text{fct,inv}} &=
  (\delta Z+z) (A^\mu\,\square\,A_\mu+(\partial A)^2).
\end{align}
\end{subequations}
The invariant counterterm here corresponds to an invariant
counterterm generated by a field renormalization from the usual gauge
invariant kinetic term $F^{\mu\nu}F_{\mu\nu}$. According to the
assumption, both options restore the symmetry, and they lead to the
identical renormalized theory. The field renormalization constant
$\delta Z$ can be used to adopt a desired renormalization condition.

Finally,
$$S_{\text{fct,evan}}$$ corresponds to additional counterterms that
are both finite and evanescent. Adding or changing such counterterms
can change e.g.\ a purely 4-dimensional counterterm
$A^\mu\bar{\psi}\bar{\gamma}_\mu\psi$ to a fully $D$-dimensional
counterterm $A^\mu\bar{\psi}{\gamma}_\mu\psi$. These counterterms vanish in the
4-dimensional limit, but they can affect calculations at higher
orders.
They also drop out of
Eqs.\ (\ref{LIMcond},\ref{eq:S1fct_1Lv1},\ref{eq:3rd}). Hence one
viable option is that the symmetry-restoring counterterms
$S_\text{fct,restore}$ are always defined by using strictly
4-dimensional quantities only. However, this is not the only option;
in concrete cases elevating 4-dimensional terms to fully
$D$-dimensional ones may simplify expressions appearing at higher
orders.
At any rate, each such choice generates a different, valid,
renormalized theory. From a practical point of view it is desirable to
make a computationally simple choice. 

\newpage
\section{Practical Treatment of Chiral Gauge Theories in the BMHV
  Scheme of DReg}
\label{sec:ApplicationsChiralQED}

In recent years, the treatment of chiral gauge theories with the
non-anticommuting $\gamma_5$ BMHV scheme has received increasing
interest. Applications in the SM at the multiloop level and in
effective field theories with additional operators involving chiral
fermions have become more important, see e.g.\ the discussions in
Refs.\ 
\cite{Blondel:2018mad,Denner:2019vbn,Fuentes-Martin:2022vvu,Carmona:2021xtq}.
Accordingly
the usefulness of   
regularization/renormalization schemes for which ultimate consistency
is fully established is becoming more appreciated. After the pioneering
one-loop discussion of gauge theories with chiral fermions in
Refs.\ \cite{Martin:1999cc,Sanchez-Ruiz:2002pcf}, Ref.\ \cite{Belusca-Maito:2020ala} extended the
analysis to general chiral gauge theories including scalar fields and
Yukawa couplings to chiral
fermions. Ref.\ \cite{Belusca-Maito:2021lnk} pioneered the application
of the BMHV scheme to chiral gauge theories at the two-loop level with
a first, abelian example. Ref.\ \cite{Cornella:2022hkc} extended the one-loop
analysis to the case of the background field gauge fixing and to the
full gauge--fermion sector of the electroweak SM.

In this section, we give concrete illustrations of how to treat chiral
gauge theories in the BMHV scheme with non-anticommuting
$\gamma_5$.
The discussion is based on our results in Refs.\
\cite{Belusca-Maito:2021lnk,Belusca-Maito:2020ala}.
The following subsection \ref{sec:overview} provides an extended
overview of the procedure and a guide for the present section.

\subsection{Overview and Guide to the Present~Section}
\label{sec:overview}

In Sec.\ \ref{sec:setup}, we discussed the basic defining gauge
invariance of gauge theories and reformulated it in terms of BRST
symmetry and Slavnov-Taylor and Ward identities. In Sec.\ \ref{sec:algren}
we explained how these symmetry identities are elevated to
defining properties of the renormalized theory at higher orders. For
the gauge theories we study here, it is known that these defining
symmetry identities can be fulfilled in any consistent
regularization/renormalization procedure, by appropriately defining
counterterms. In Sec.\ \ref{sec:DReg}
we explained the definition of dimensional regularization and the
BMHV scheme for $\gamma_5$, which in general breaks gauge invariance
in the presence of chiral fermions. In Sec.\ \ref{sec:renormalization} we explained
the proof that dimensional regularization constitutes one of the
consistent regularization/renormalization procedures.

As a result it
is in principle established that the dimensional regularization
including BMHV scheme for $\gamma_5$ may be used for chiral gauge
theories. Further, Sec.\ \ref{sec:AlgRenInDReg} also provided a
blueprint how to determine the required
counterterm structure in concrete calculations. In this section, we carry out such
concrete calculations and illustrate all required steps in detail. 

In the Abelian chiral gauge theory defined below, we expect the
validity of simple QED-like Ward identities; the simplest one
corresponds to the transversality of the photon self-energy. It turns
out that in the BMHV scheme, the actual one-loop self-energy violates this transversality (see Eq.\ (\ref{eq:QGBSE})). The
violation affects both the divergent and the finite part in the BMHV
scheme of dimensional regularization. The breaking, however, is a
polynomial in the momentum; hence it can be cancelled by adding a
local counterterm to the Lagrangian --- in line with the general
existence statement mentioned above. After adding this counterterm the
required transversality is fulfilled. The concrete required form of
the counterterm can be found in Eqs.\ (\ref{eq:SingularCT1LoopPhotonSE},\ref{eq:FiniteCT1LoopPhotonSE}). 

A question is then what is the most efficient way to determine such
symmetry breakings in general. Answers were given in
Sec.\ \ref{sec:SymmetryAndSymmetryBreakingInDReg} and can be
illustrated as follows. One way in principle is to explicitly
evaluate all Green functions and test the validity of all Ward and
Slavnov-Taylor identities between all Green functions. The explicit
computation of the non-transverse terms in
Eq.\ (\ref{eq:QGBSE}) provides an example. Given that there
are in principle infinitely many identities between Green functions
and given that the computation of Green functions involves also
complicated non-local terms that cannot contribute to the symmetry
violation, this strategy is not the most efficient.

Sec.\ \ref{sec:SymmetryAndSymmetryBreakingInDReg} also explained a
shortcut that is based on the regularized quantum action 
principle of dimensional regularization discussed in
Sec.\ \ref{sec:QAPDReg}. Staying with the example of the photon self-energy, the terms violating the transversality in
Eq.\ (\ref{eq:QGBSE}) and then in Eq.\ (\ref{WIphotonSEbreaking}) may be
equivalently obtained by computing one special Feynman diagram, shown
in Eqs.\ (\ref{DiagramForDeltaDotGammaAc},\ref{DeltaDotGammaAc}). This diagram involves an insertion of
the operator $\widehat{\Delta}$ which reflects the breaking of chiral
gauge invariance in $D$ dimensions, and the quantum action principle
guarantees that the evaluation of this diagram reproduces directly the
breaking of the transversality of the photon self-energy. The
simplification is threefold: First and foremost, since
$\widehat{\Delta}$ is evanescent, only the ultraviolet divergent part of the
diagram can contribute --- hence the evaluation is  simpler
(the degree of simplification dramatically increases for more
complicated Green functions and at higher orders). Second, in the
general case there are much fewer diagrams with insertions of
$\widehat{\Delta}$ than ordinary diagrams. Third, since only divergent
parts contribute it is clear that the symmetry breaking/restoration
procedure requires only the computation of power-counting
divergent diagrams with insertions of $\widehat{\Delta}$.

This more efficient but less obvious strategy based on the quantum
action principle was applied to chiral gauge theories at the
one-loop level in
Refs.\ \cite{Belusca-Maito:2020ala,Martin:1999cc,Sanchez-Ruiz:2002pcf,Cornella:2022hkc}
with and without scalar
sector and to an Abelian chiral gauge theory at the two-loop level in
Ref.\ \cite{Belusca-Maito:2021lnk}. It was also applied to the case of supersymmetric
gauge theories in the context of dimensional reduction at the two- and
three-loop level in Refs.\ \cite{Hollik:2005nn,Stockinger:2018oxe}.

In the largest part of the present section we focus on the simpler
case of an Abelian chiral
gauge theory. We begin in Sec.\ \ref{sec:DefOfAbelianChiralGaugeTheoryForApplications} by defining the considered model and collecting
all relevant symmetry identities. Then we discuss the subtleties in
the continuation to $D$ dimensions and determine the insertion
operator $\widehat{\Delta}$. Sec.\
\ref{sec:symmetryrestorationrequirements} provides a more technical overview of the
procedure to determine the symmetry-restoring counterterms than the
previous remarks. In Sec.\
\ref{sec:explicitAbelian} we then discuss the explicit computations in
the abelian model in detail. We begin with the case of the photon self-energy mentioned above and illustrate both strategies to determine the
symmetry-restoring counterterms, then we progress to other Green
functions and to the two-loop level. Thereafter
Sec.\ \ref{sec:explicitNonabelian} discusses the
case of non-Abelian Yang-Mills theories and presents
explicit calculations and results  at the one-loop level.

\subsection{Definition of an Abelian Chiral Gauge~Theory}
\label{sec:DefOfAbelianChiralGaugeTheoryForApplications}
Here we define a concrete Abelian chiral gauge theory which will be
used in explicit calculations. It is first defined in $4$ dimensions
along with its symmetry requirements
in Sec.\ \ref{sec:ChiralQEDIn4Dim}; then the definition is extended to
$D$ dimensions 
within
the framework of the BMHV $\gamma_5$ scheme and resulting BRST symmetry
breaking is exhibited in Sec.\ \ref{sec:DefChiralQEDDReg}.

\subsubsection{Chiral Electrodynamics in 4 dimensions}
\label{sec:ChiralQEDIn4Dim}
Following Sec.\ \ref{sec:AbelianPeculiarities}, the 4-dimensional classical Lagrangian for  quantum electrodynamics (QED) is given by
\begin{equation}
\label{eq:lagrangianOrdinaryQED}
\mathcal{L}_{\text{QED}} =i \overline{\psi}_i \slashed{D}_{ij} {\psi}_j- \frac{1}{4} F^{\mu\nu} F_{\mu\nu} - \frac{1}{2 \xi} (\partial_\mu A^\mu)^2-\bar{c}\partial^2c+ \rho^\mu s{A_\mu} + \bar{R}^i s{{\psi}_i} + R^i s{\overline{\psi}_i},
\end{equation}
with U(1) ghost and external BRST sources included.\footnote{In
  contrast to Sec.\ \ref{sec:AbelianPeculiarities}, we already
  integrated out the Nakanishi-Lautrup field 
  $B(x)$, i.e. we used $B = - (\partial_{\mu} A^{\mu}) /\xi$
  already in the Lagrangian.}
The only generator
in this theory is the real and diagonal charge
$Q_{ij}=Q_i\delta_{ij}$, so that the covariant derivative reads
\begin{equation}
D_{ij}^\mu = \partial^\mu \delta_{ij} + i e A^\mu {Q}_{ij} \, .
\end{equation}

We now define a similar, but chiral Abelian gauge theory.
We separate the fermionic content into left-handed and right-handed chirality parts,
\begin{equation}
\psi_{R/L}=\mathbb{P}_{R/L}\psi, \quad \mathbb{P}_{R/L} = \frac{\mathbbm{1}\pm\gamma_5}{2},
\end{equation}
and allow only purely right-handed fermions to appear as dynamical
fields.\footnote{This is a choice made to simplify the
  discussion. E.g.\ the
  U(1)$_Y$ sector of the SM contains both left-handed and right-handed
  fermions with different gauge quantum numbers. It could be treated similarly.}  The 4-dimensional and purely
right-handed classical Lagrangian of the model then reads
 \begin{equation}
 \label{eq:lagrangianChiralQED}
 \mathcal{L}_{\chi\text{QED}} =i \overline{\psi_R}_i \slashed{D}_{ij} {\psi_R}_j- \frac{1}{4} F^{\mu\nu} F_{\mu\nu} - \frac{1}{2 \xi} (\partial_\mu A^\mu)^2-\bar{c}\partial^2c+ \rho^\mu s{A_\mu} + \bar{R}^i s{{\psi_R}_i} + R^i s{\overline{\psi_R}_i},
 \end{equation}
 where 
  the interaction, coupling only to the right-handed fermions, is defined by the covariant derivative as
 \begin{equation}
 D_{ij}^\mu = \partial^\mu \delta_{ij} + i e A^\mu {\mathcal{Y}_R}_{ij} \, .
 \end{equation}
Emphasizing the similarity with the U(1)$_Y$ sector of the Standard
Model we call the generator $ {\mathcal{Y}_R}_{ij}=
{\mathcal{Y}_R}_{i}\delta_{ij}$ the hypercharge.
It can be seen that the left-handed fermions $\psi_L$ are now decoupled from the theory.
In order to avoid triangle anomalies we need to impose the following additional anomaly cancellation condition to the hypercharge,
\begin{equation}
\label{eq:AnomCond}
\text{Tr}(\mathcal{Y}_R^3) = 0 \, .
\end{equation}
Following Sec.\ \ref{sec:setup}, the non-vanishing BRST transformations for this model are 
\begin{subequations}
	\label{eq:BRST4}
	\begin{align}
	s{A_\mu} &= \partial_\mu c\, , \\
	s{\psi_i} &= s{{\psi_R}_i} = -i \, e \, c \,{\mathcal{Y}_R}_{ij} {\psi_R}_j \, , \\    { }
		s{\overline{\psi}_i} &= s{\overline{\psi_R}_i} = -i \, e \, \overline{\psi_R}_j c {\mathcal{Y}_R}_{ji} \, \\
		s{\overline{c}}&=B\equiv -\frac{1}{\xi}\partial A,\label{Bdefinition}
	\end{align}
\end{subequations}
where $s$ is the nilpotent generator of the BRST transformations, which acts
as a fermionic differential operator.
This 4-dimensional tree-level action
\begin{equation}
S_0^{(4D)} =  \int d^4 x \, \mathcal{L}_{\chi\text{QED}}
\end{equation}
satisfies the following Slavnov-Taylor identity
\begin{equation}
\label{eq:STIS0}
\mathcal{S}(S_0^{(4D)}) = 0 \, ,
\end{equation}
where the Slavnov-Taylor operator, with the field content we consider, was already given in Eq. (\ref{eq:STIAbelianCase}).
At this point, we  emphasize two additional functional identities
that hold in 4 dimensions and that were derived and discussed in
Sec.\ \ref{sec:AbelianPeculiarities}. 
The first is the ghost equation,  
\begin{align}
\label{eq:GhostEq}
\bigg(\frac{\delta}{\delta\bar{c}} + \partial_\mu\frac{\delta}{\delta \rho_\mu}\bigg) S_0^{(4D)} &= 0 \, .
\end{align}
The second is the functional form of the abelian Ward
identity\footnote{Here we  keep the Nakanishi-Lautrup field $B(x)$
  explicitly. However, one could integrate it out here as well using
  $B = - (\partial_{\mu} A^{\mu}) /\xi$.} 
\begin{align}
\label{TreelevelWI}
\bigg(\partial^\mu\frac{\delta}{\delta A^\mu(x)}
+ie {\mathcal{Y}}_{R}^{j}
\sum_{\Psi}(-1)^{n_{\Psi}}\Psi(x)\frac{\delta}{\delta\Psi(x)}
\bigg)
S_0^{(4D)}
&=-\Box B(x)
\,,
\end{align}
suitably adapted to the present theory $\chi$QED and its field content.
The summation extends over the charged fermions and their sources,
$\Psi\in\{{\psi_R}_j,\overline{\psi_R}_j,R^j,\bar{R}^j\}$ and $n_{\Psi}\in\{0,1,0,1\}$. 

Functional relations such as the ghost equation and the local Ward
identity are part of the definition of our theory in 4
dimensions. Once we perform the regularization and renormalization
procedure, the requirement that those identities still hold imposes
important restrictions as we will soon see in the explicit loop
calculations. But first we extend the model to $D$ dimensions and
examine the consequences of this extension. 
\subsubsection{Definition of Chiral Electrodynamics in DReg}
\label{sec:DefChiralQEDDReg}
We can immediately see that the extension of $\chi$QED to $D$ dimensions is not unique 
due to the right-handed
chiral current
$\overline{\psi_R}_i \gamma^\mu {\psi_R}_j$. The extension to $D$ dimensions of this term has 
three \emph{inequivalent choices}, each of them \emph{equally correct}:
\begin{align}
\overline{\psi}_i \gamma^\mu \mathbb{P}_{R} \psi_j \, , &&
\overline{\psi}_i \mathbb{P}_{L} \gamma^\mu \psi_j \, , &&
\overline{\psi}_i \mathbb{P}_{L} \gamma^\mu \mathbb{P}_{R} \psi_j \, .
\label{eq:inequivalentchoicesApplications}
\end{align}
They are different because $\mathbb{P}_{L} \gamma^\mu \neq \gamma^\mu \mathbb{P}_{R}$ in $D$ dimensions.
Each of these choices leads to a valid $D$-dimensional extension of the model that is renormalizable using dimensional regularization and the BMHV scheme and is expected to produce the same final results in physical 4 dimensions after the renormalization procedure is performed. However, the intermediate calculations and the $D$-dimensional results will differ, depending on the choice for this interaction term. The third option, which is equal to
\begin{equation}
\overline{\psi} \mathbb{P}_{L} \gamma^\mu \mathbb{P}_{R} \psi = \overline{\psi} \mathbb{P}_{L} \overline{\gamma}^\mu \mathbb{P}_{R} \psi = \overline{\psi_R} \overline{\gamma}^\mu \psi_R \, ,
\label{eq:thirdoptionforchiralcurrent}
\end{equation}
is the most symmetric one and leads to the simplest intermediate expressions.
Notice that this choice is actually the most straightforward one since
it preserves the information that right-handed fermions were present
on the left and on the right sides of the interaction term before the
extension, see also the review \cite{Jegerlehner:2000dz}.

The second, more serious problem, is that as it stands the pure fermionic kinetic term
$i \overline{\psi_R}_i \slashed{\partial} {\psi_R}_i = i \overline{\psi}_i \mathbb{P}_{L} \slashed{\partial} \mathbb{P}_{R} \psi_i$
projects only the purely $4$-dimensional derivative, leading to
a purely 4-dimensional propagator
\begin{equation}
    \frac{i\,\mathbb{P}_{R} \, \slashed{p} \, \mathbb{P}_{L}}{\bar{p}^2},
\end{equation}
and to unregularized loop diagrams. As discussed in
Sec.\ \ref{sec:relationL}, the \textit{only} valid choice for the
propagator in the $D$-dimensional theory in the context of dimensional regularization is
\begin{equation}
    \frac{i\, \slashed{p} }{p^2},
\end{equation}
so we are thus led to consider the full Dirac fermion $\psi$ with both a left and right-handed component, and use
instead the fully $D$ dimensional covariant kinetic term
$i \overline{\psi}_i \slashed{\partial} {\psi}_i$.
It can be re-expressed in terms of projectors as follows:
\begin{equation}
\begin{split}
i \overline{\psi}_i \slashed{\partial} {\psi}_i&=i \overline{\psi}_i \overline{\slashed{\partial}} {\psi}_i + i \overline{\psi}_i \widehat{\slashed{\partial}} {\psi}_i\\ &= i (\overline{\psi}_i \mathbb{P}_{L} \slashed{\partial} \mathbb{P}_{R} {\psi}_i + \overline{\psi}_i \mathbb{P}_{R} \slashed{\partial} \mathbb{P}_{L} {\psi}_i) + i (\overline{\psi}_i \mathbb{P}_{L} \slashed{\partial} \mathbb{P}_{L} {\psi}_i + \overline{\psi}_i \mathbb{P}_{R} \slashed{\partial} \mathbb{P}_{R} {\psi}_i)
\end{split}
\end{equation}
Notice that the fictitious, \textit{sterile} left-chiral field $\psi_L$ is introduced, which appears only within the kinetic term and nowhere else, it does not interact so it does not couple in particular to the gauge bosons of the theory, and we enforce it to be invariant under gauge transformations.

Unfortunately, the choice of the $D$-dimensional propagator, crucial for loop regularization, that led to the introduction to the left-handed component in the kinetic term, breaks the gauge invariance of the fermionic part of the Lagrangian, which is evident if we separate it in this way:
\begin{subequations}
\begin{align}
\label{eq:Lfermionsplit}
\mathcal{L}_\text{fermions}
&=\mathcal{L}_\text{fermions,inv} + \mathcal{L}_\text{fermions,evan} \, , \\
\mathcal{L}_\text{fermions,inv}
&= i \overline{\psi}_i \overline{\slashed{\partial}} {\psi}_i
- e {\mathcal{Y}_R}_{ij} \overline{\psi_R}_i \slashed{A} {\psi_R}_j \, , \\
\mathcal{L}_\text{fermions,evan}
&= i \overline{\psi}_i \widehat{\slashed{\partial}} {\psi}_i
\, ,
\end{align}
\end{subequations}
where the first term contains purely
4-dimensional derivatives and gauge fields and preserves the gauge and BRST invariance, since
the fictitious left-chiral field $\psi_L$ is a gauge singlet.
The invariant term can also be written as a sum of purely left-chiral and
purely right-chiral terms involving the 4-dimensional covariant derivative as
\begin{subequations}
\begin{align}
\mathcal{L}_\text{fermions,inv}
&= i \overline{\psi_L}_i \overline{\slashed{\partial}} {\psi_L}_i
+ i \overline{\psi_R}_i \overline{\slashed{\partial}} {\psi_R}_i
-e {\mathcal{Y}_R}_{ij} \overline{\psi_R}_i \slashed{A} {\psi_R}_j
\\
&= i \overline{\psi_L}_i \overline{\slashed{\partial}} {\psi_L}_i
+ i \overline{\psi_R}_i \overline{\slashed{D}} {\psi_R}_i
\, ,
\end{align}
\end{subequations}
where the gauge invariance is obvious.
The second term in \cref{eq:Lfermionsplit} is purely evanescent, i.e.\ it vanishes
in 4-dimensional limit. If we rewrite the evanescent term as
\begin{align}
  \label{evanescentLfermion}
\mathcal{L}_\text{fermions,evan} &=
i \overline{\psi_L}_i \widehat{\slashed{\partial}} {\psi_R}_i
+
i \overline{\psi_R}_i \widehat{\slashed{\partial}} {\psi_L}_i
\, ,
\end{align}
it can be easily seen that it mixes left- and right-chiral fields
with different gauge transformation properties. This causes \textit{the
breaking of gauge and BRST invariance} --- the central difficulty of
the BMHV scheme.\footnote{  We remark that the problem is not specific to the case where the
  left-handed fermion is sterile. As Eq.\ (\ref{evanescentLfermion})
  shows the problem generally exists if the left-handed and
  right-handed fermions have different gauge quantum
  numbers. Refs.\ \cite{Martin:1999cc,Cornella:2022hkc} consider this
  case and end up with essentially the same breaking of BRST
  invariance in $D$ dimensions and the same further consequences.
  }

We can summarize this symmetry property and the symmetry breaking as
\begin{subequations}
\begin{align}
  s_D\mathcal{L}_\text{fermions,inv}
  &=0
  \,,\\
  s_D\mathcal{L}_\text{fermions,evan}&\ne0\,,
\label{eq:sDActingOnFermionLagrangian}
\end{align}
\end{subequations}
where $s_D$  is the obvious extension of the BRST operator (\ref{eq:BRST4}) to
$D$ dimensions.
  
Since the extension of BRST transformation of fields in $D$ dimensions is straightforward, our $D$ dimensional action is then
\begin{equation}
\label{eq:S0Def2}
\begin{split}
S_0 =\;& \int d^D x \bigg( i \overline{\psi}_i \slashed{\partial} {\psi}_i + e {\mathcal{Y}_R}_{ij} \overline{\psi_R}_i \slashed{A} {\psi_R}_j - \frac{1}{4} F^{\mu\nu} F_{\mu\nu}
- \frac{1}{2 \xi} (\partial_\mu A^\mu)^2 \\
&-\bar{c} \partial^2 c + \rho^\mu (\partial_\mu c) + i \, e \, \bar{R}^i c \,{\mathcal{Y}_R}_{ij} {\psi_R}_j + i \, e \, \overline{\psi_R}_i c {\mathcal{Y}_R}_{ij} R^j  \bigg)
\, \\
\equiv\;&
\sum_i S^i_{\overline{\psi}\psi} + \sum_i \overline{S^i_{\overline{\psi}_R A \psi_R}}
+ S_{AA}
+ S_\text{g-fix} + S_{\bar{c} c}
+ S_{\rho c}  + S_{\bar{R} c \psi_R} + S_{R c \overline{\psi_R}}
\, ,
\end{split}
\end{equation}
where also useful abbreviations for the individual terms were
introduced.
Similar to the fermion Lagrangian, the full $D$-dimensional
action may be written as the sum of two parts, an ``invariant'' and
an ``evanescent'' part,
\begin{subequations}
\label{eq:SeparatedDdimAction}
	\begin{align}
	S_0 &= S_{0,\text{inv}} + S_{0,\text{evan}} \, , \\
	S_{0,\text{evan}} &= \int d^D x \, i
	\overline{\psi}_i \widehat{\slashed{\partial}} {\psi}_i \, .\label{S0evan}
	\end{align}
\end{subequations}
The second part $S_{0,\text{evan}}$ consists solely of one single,
evanescent fermion kinetic term, the remnant of the $D$-dimensional propagator.

Now we quantify the symmetry breaking caused by the BMHV scheme, the
non-anticommuting $\gamma_5$ and the resulting evanescent term in the action.
Acting with the $D$-dimensional BRST operator on the $D$-dimensional tree-level action \cref{eq:S0Def2} gives:
\begin{equation}
s_D S_0 = s_D S_{0,\text{inv}} + s_D S_{0,\text{evan}} = 0 + s_D \int d^D x \, i \overline{\psi}_i \widehat{\slashed{\partial}} {\psi}_i\equiv \widehat{\Delta},
\end{equation}
where the non-vanishing integrated breaking term $\widehat{\Delta}$ is given by
\begin{equation}
\label{eq:BRSTTreeBreaking}
\widehat{\Delta} =  -\int d^D x
e\,  {\mathcal{Y}_R}_{ij} \, c \, \left\{
\overline{\psi}_i \left(\overset{\leftarrow}{\widehat{\slashed{\partial}}} \mathbb{P}_{R} + \overset{\rightarrow}{\widehat{\slashed{\partial}}} \mathbb{P}_{L}\right) \psi_j
\right\}
\equiv \int  d^D x \, \widehat{\Delta}(x).
\end{equation}
 Acting with the $D$-dimensional Slavnov-Taylor operator $\mathcal{S}_D$ on the tree-level action, we obtain
\begin{equation}
\mathcal{S}_D (S_0) = {s}_D S_{0,\text{inv}} + {s}_D S_{0,\text{evan}} = 0 + \widehat{\Delta}\, ,
\end{equation}
hence the Slavnov-Taylor identity in $D$ dimensions is violated by the
same  BRST breaking term at  tree-level.\footnote{  \label{footnotelinearEQs}
  The simpler linear equations
  (\ref{eq:AntiGhostEquationAbelianCase}) to
  (\ref{eq:GaugeFixingAbelianCase}) specific for Abelian theories are
  manifestly valid also in $D$ dimensions. We will not discuss them
  further, but they have the consequence that higher-order
  corrections, including counterterm actions, cannot depend on the
  ghost/antighost and source fields. For this reason, the linearized
  Slavnov-Taylor operator here reduces to BRST transformations, $b_D=s_D$.
  }

As mentioned in the overview Sec.\ \ref{sec:overview}, this breaking term
will be a crucial tool in practical calculations. 
This breaking will be used as a composite operator insertion in
Feynman diagrams. It generates an interaction vertex whose Feynman
rule (with all momenta incoming and derived from the combination $i\widehat{\Delta}$) is:
\begin{equation}
\begin{tabular}{rl}
\raisebox{-40pt}{\includegraphics[scale=0.6]{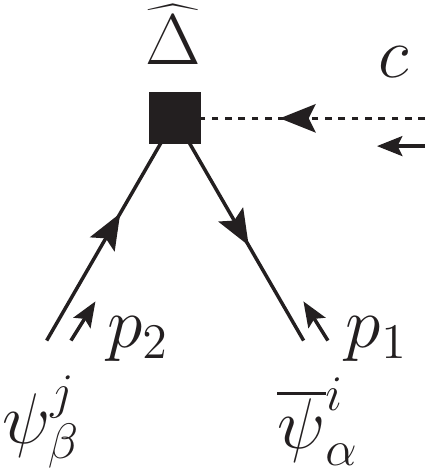}} &
$\begin{aligned}
&= -\frac{e}{2} \, {\mathcal{Y}_R}_{ij} \left((\widehat{\slashed{p_1}} + \widehat{\slashed{p_2}}) + (\widehat{\slashed{p_1}} - \widehat{\slashed{p_2}}) \gamma_5 \right)_{\alpha\beta} \\
&= -e \, {\mathcal{Y}_R}_{ij} \left(\widehat{\slashed{p_1}} \mathbb{P}_{R} + \widehat{\slashed{p_2}} \mathbb{P}_{L} \right)_{\alpha\beta}
\, .
\end{aligned}$
\end{tabular}
\label{FeynmanruleDelta}
\end{equation}
As discussed in the context of Eq.\ (\ref{OdotGammaDef}), in this way the
functional derivatives of $i\widehat{\Delta}\cdot\Gamma$ correspond to
1PI Feynman diagrams with one insertion of the Feynman rule
(\ref{FeynmanruleDelta}). 
An analogous Feynman rule is derived for charge-conjugated fermions.

It is important to notice that this breaking $\widehat{\Delta}$ is evanescent, i.e. it
vanishes in the 4-dimensional limit.  This results from the evanescent
original term (\ref{S0evan}) and has the consequence that insertions
of $\widehat{\Delta}$ can only contribute in power-counting divergent Feynman diagrams.

\subsection{Symmetry Restoration~Requirements} 
\label{sec:symmetryrestorationrequirements}
Before beginning the explicit calculations we recall and collect the required
symmetry identities and the strategy for symmetry restoration in a
more technical way than in the overview Sec.\ \ref{sec:overview}.
We begin by collecting the required symmetry identities.

Symmetry identities expressing gauge/BRST invariance are considered
part of the definition of the theory. Hence they are required to be fulfilled
at all orders, see Secs.\ \ref{sec:AbelianPeculiarities} and
\ref{sec:AlgRenormalization} for  detailed discussions.

The symmetry requirements are defined for the renormalized and finite 4-dimensional effective action of the form
\begin{equation}
\Gamma_\text{ren} = S_0^{(4D)} + \mathcal{O}(\hbar),
\end{equation}
where we again highlight that the effective action coincides with the
classical action up to higher-order corrections, and that loop
corrections are of higher order in $\hbar$, see
Eq.\ (\ref{GammaEffectiveAction}) and Sec.\ \ref{sec:DReg}.
The first symmetry requirement is BRST  (and underlying gauge) invariance, which is
expressed as the Slavnov-Taylor identity
\begin{equation}
\label{GeneralSTI}
\mathcal{S}(\Gamma_\text{ren}) = 0,
\end{equation}
for the renormalized theory. Notice that in $\chi$QED the fields $c$, $\bar{c}$ and $\rho^\mu$ do not have higher order corrections, so relations 
\begin{align}
\label{TrivialIdentities}
\frac{\delta\Gamma_\text{ren}}{\delta c(x)}
&= \frac{\delta S_0^{(4D)}}{\delta c(x)} \, ,
&
\frac{\delta\Gamma_\text{ren}}{\delta\bar c(x)}
&= \frac{\delta S_0^{(4D)}}{\delta \bar c(x)} \, ,
&
\frac{\delta\Gamma_\text{ren}}{\delta \rho^\mu(x)}
&= \frac{\delta S_0^{(4D)}}{\delta \rho^\mu(x)} \, .
\end{align}
hold trivially, since the respective derivatives of the tree-level
action
are linear in the dynamical fields as described in Sec.\ \ref{sec:AbelianPeculiarities}.
The fact that the ghost does not have higher loop corrections will play a part in reducing the number of diagrams appearing in higher orders, compared to an analogous Yang-Mills theory.
The local Ward identity
\begin{equation}\label{eq:GeneralWI}
\bigg(\partial^\mu\frac{\delta}{\delta A^\mu(x)}+ie {\mathcal{Y}}_{R}^{j}\sum_{\Psi}(-1)^{n_{\Psi}}\Psi(x)\frac{\delta}{\delta\Psi(x)}\bigg)\Gamma_\text{ren} = -\Box B(x) \, ,
\end{equation}
is an automatic consequence of the Slavnov-Taylor identity as we have shown in Sec.\ \ref{sec:AbelianPeculiarities}.

We record here the application of the Ward identity to the photon self-energy as an example that will later be illustrated in explicit
computations.
If we rewrite the Ward identity in the momentum-space representation
and take a variation with the respect to photon field, we obtain the requirement
\begin{equation}
\label{eq:Ward1}
i p_\nu \frac{\delta^2{\widetilde{\Gamma}_\text{ren}}}{\delta{A_\mu(p)} \delta{A_\nu(-p)}} = 0 \,,
\end{equation}
which  corresponds to the  transversality of the photon self-energy.

All previous symmetry identities  must hold after
regularization and renormalization at each loop order.
If the
  symmetries are broken in the intermediate regularization
  procedure, as is the case when we use the BMHV scheme, they must be
  restored order by order in perturbation theory, by adding suitable
  counterterms.
  
The symmetry identities are covered by the general analysis of
algebraic renormalization discussed in
Sec.\ \ref{sec:algebraicrenormalizationDetails}, and the theory has no
gauge anomaly, see Eq.\ (\ref{eq:AnomCond}). This guarantees that
the procedure of symmetry restoration works at all orders.

Now we recapitulate the practical strategies for the concrete determination of
symmetry-restoring counterterms, following the detailed outline given
in  Sec.\ \ref{sec:AlgRenInDReg}. The application will be discussed in
the subsequent subsections, where we treat not only the chiral model
$\chi$QED but also compare it with the familiar case of ordinary QED to
highlight the features of the BMHV treatment of $\gamma_5$.

The first obvious difference is that  ordinary QED is a  vector-like
gauge theory, and DReg preserves all relevant symmetry identities
manifestly at each step: the counterpart to the tree-level breaking $\widehat{\Delta}$ in
Eq.\ (\ref{eq:BRSTTreeBreaking}) vanishes as already discussed in
Sec.\ \ref{sec:QAPexamples}.
Hence generating counterterms by a renormalization transformation is
sufficient, see the discussion in Sec.\ \ref{sec:ctsymmetrypreserving}
and Eq.\ (\ref{RenTrafo}).

For the case of  $\chi$QED,  the existence of a tree-level
symmetry breaking, $\widehat{\Delta}\ne0$,  necessitates
symmetry-restoring counterterms. Hence, generating counterterms by a
renormalization transformation is not sufficient, and the general
structure is the one discussed in
Sec.\ \ref{sec:countertermLagrangianBMHV}, i.e.\ the combination
\begin{equation}
\label{eq:CT_structure}
S_{\text{sct,inv}}
+  S_{\text{sct,non-inv}}
+  S_{\text{fct,inv}}
+  S_{\text{fct,restore}}
+  S_{\text{fct,evan}} \, .
\end{equation}

Sec.\ \ref{sec:AlgRenInDReg}
presented two basic strategies to carry out the required computations
of the crucial symmetry-restoring counterterms $S_{\text{fct,restore}}$.
The first
is based on the explicit computation of ordinary Green functions and
explicitly checking symmetry identities. Its essential equation is
Eq.\ (\ref{SdGammasubrenDREG}), which requires computing
$$\mathcal{S}_D(\Gamma_\mathrm{subren}^{(n)}+S_{\text{sct}}^n)$$
at each
new order $n$. If this expression is non-zero, finite counterterms
have to be found and added to the action such that the symmetry
breaking is canceled.

The second strategy is based on using
the regularized quantum action principle and represented by
Eq.\ (\ref{LIMcond}),
$$
\mathop{\text{LIM}}_{D \, \to \, 4} \, \left(\widehat{\Delta}\cdot\Gamma_\mathrm{DRen}^n+\sum_{k=1}^{n-1}\Delta^k_\mathrm{ct}\cdot\Gamma^{n-k}_\mathrm{DRen}+\Delta^n_\mathrm{ct}\right)=0.
$$
The computation of full Green functions and evaluating 
Slavnov-Taylor identities is replaced by the computation of Green
functions with insertions of breaking operators such as
$\widehat{\Delta}$. This equation is
specialized to Eqs.\ (\ref{eq:rencon}) and
(\ref{eq:deltadef}) at the one- and two-loop level.

In the following subsections
we will illustrate Feynman diagrammatic computations for both
strategies. The more efficient second strategy is illustrated also at
the two-loop level. We will then see how the desired
symmetry-restoring counterterms are determined.

\subsection{Explicit Calculations and Results in the Abelian Chiral
 Gauge~Theory}
\label{sec:explicitAbelian}
In this section explicit calculations in the Abelian chiral gauge theory defined above in
Sec.\ \ref{sec:DefOfAbelianChiralGaugeTheoryForApplications} are performed in the
BMHV scheme of DReg and all necessary counterterms are provided up to the two-loop level. 
In particular, the evaluation of the photon self-energy 
at the one-loop (sections \ref{sec:oneloopphotonSE} and \ref{sec:oneloopdirect}) 
and the two-loop level (Sec.\ \ref{sec:TwoLoopPhotonSelfEnergyChiralQED})
is highlighted 
and 
the results are 
compared to ordinary QED.
As announced in Sec.\ \ref{sec:overview}, there are two different ways of determining symmetry-restoring
counterterms.
While the method in Sec.\ \ref{sec:oneloopphotonSE} amounts to the explicit evaluation of the full
photon self-energy, i.e.\ a full Green function, 
including its finite part,
Sec.\ \ref{sec:oneloopdirect} employs the direct
method based on the regularized
quantum action principle where the symmetry breaking
is determined via special Feynman diagrams with an insertion of the $\widehat{\Delta}$-operator, 
which reflects the breaking of chiral gauge invariance. 
Sec.\ \ref{sec:oneloopdirect} then concludes
by providing the full one-loop counterterm action for chiral QED in the BMHV scheme.
Similarly, in Sec.\ \ref{sec:TwoLoopPhotonSelfEnergyChiralQED} the two-loop counterterms for the photon
self-energy are obtained using the latter method 
based on the regularized quantum action principle
but are verified by comparing with the explicit result for the full photon self-energy including its
finite part.
Concluding, Sec.\ \ref{sec:Full2LoopRenormalizationOfChiralQED} provides the full two-loop
renormalization
of chiral QED in the BMHV scheme.

\subsubsection{One-Loop Photon Self-Energy and Symmetry-Restoring~counterterms}\label{sec:oneloopphotonSE}

To better understand the features of the BMHV scheme, we now
focus on explicit loop calculations. We take the photon self-energy
and compare its results in ordinary QED and chiral QED. The photon
self-energy is subject to the simplest Ward identity (\ref{eq:Ward1})
--- it must be
transverse, to guarantee the correct physical interpretation of the
theory describing a massless spin 1 particle with two transverse
polarizations.

The photon self-energy is denoted as
\\
\begin{equation}
\begin{tabular}{rl}
$\begin{aligned}
i \widetilde{\Gamma}_{AA}^{\nu \mu}(p) \, = \, 
\end{aligned}$
\raisebox{-26pt}{\includegraphics[scale=0.6]{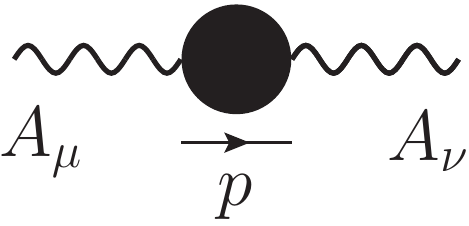}} &
$\begin{aligned}
.
\end{aligned}$
\end{tabular}\nonumber
\end{equation}
We use the notation explained in
Sec.\ \ref{sec:GreenFunctions}, corresponding to the 1-particle irreducible diagrams with
external fields and momentum as indicated.\footnote{  However, in this subsection we use a slightly simpler notation than 
  in Sec.\ \ref{sec:DRegGamma} for unrenormalized/subrenormalized
  expressions.
  We drop the subscript $_{\text{subren}}$ and simply write $\Gamma^1$
  for the unrenormalized one-loop effective action and $\Gamma^2$ for
  the subrenormalized two-loop effective action. Accordingly, the
  following equations correspond to the unrenormalized one-loop photon
  self-energy.\label{footnoteSimplerGammaExplicit}
  }

We begin by recalling the well-known one-loop result of ordinary QED
with massless fermions as defined in Eq.\ (\ref{eq:lagrangianOrdinaryQED}),
\begin{subequations}
\begin{align}
i \widetilde{\Gamma}_{AA}^{\nu \mu}(p)|^{1}_\text{div,QED} = &
\frac{i e^2}{16 \pi^2 \epsilon}  \frac{4 \, \text{Tr}(Q^2)}{3} (p^\mu p^\nu - p^2 g^{\mu\nu})
\, ,\label{eq:QEDphotonSE1Ldiv}
\\i \widetilde{\Gamma}_{AA}^{\nu \mu}(p)|^{1}_\text{fin,QED} =&
\frac{i e^2}{16 \pi^2} \frac{2 \, \text{Tr}(Q^2)}{3}
\left[ \left(\frac{10}{3}-2\ln(-p^2)\right) ({p}^\mu {p}^\nu - {p}^2 {g}^{\mu\nu}) \right]
\, .\label{eq:QEDphotonSE1Lfin}
\end{align}
\end{subequations}
Here and in all following results we set $D=4-2\epsilon$ and suppress
the dimensional regularization scale $\bar{\mu}^2 = \mu^2 \, 4 \pi e^{-\gamma_{E}}$ in
dimensionful logarithms. We see that the result is transverse and
satisfies the Ward identity (\ref{eq:Ward1}) both in its divergent and
finite parts.
Adding the counterterm action
\begin{align}
\label{eq:SingularCT1LoopQEDPhotonSE}
S_\text{sct,QED}^{1} =\;&
\frac{-\hbar \, e^2}{16 \pi^2 \epsilon} 
\frac{4 \text{Tr}(Q^2)}{3} S_{AA}
+\ldots
\,,
\end{align}
where 
the dots denote terms unrelated to the photon self-energy,
cancels the divergences and preserves the validity of the Ward
identity. The factor $\hbar$ was explicitly restored to highlight
that the counterterm action is of one-loop order. As is well known,
this counterterm action can be generated via a photon field
renormalization transformation.

In comparison, the result for the one-loop photon self-energy diagram
in $\chi$QED with massless fermions as defined in Eq.\ (\ref{eq:lagrangianChiralQED}) reads
\begin{subequations}
\label{eq:QGBSE}
\begin{align}
i \widetilde{\Gamma}_{AA}^{\nu \mu}(p)|^{1}_\text{div,$\chi$QED} =&
\frac{i e^2}{16 \pi^2 \epsilon}  \frac{2 \, \text{Tr}(\mathcal{Y}_R^2)}{3} \left[(\overline{p}^\mu \overline{p}^\nu - \overline{p}^2 \overline{g}^{\mu\nu}) - \frac{1}{2}  \widehat{p}^2 \overline{g}^{\mu\nu}\right]
\, ,\label{eq:QGBSE1Ldiv}
 \\
i \widetilde{\Gamma}_{AA}^{\nu \mu}(p)|^{1}_\text{fin,$\chi$QED} =&
\frac{i e^2}{16 \pi^2} \frac{\text{Tr}(\mathcal{Y}_R^2)}{3} \Bigg[
  \left(\frac{10}{3}-2\ln(-p^2)\right) (\overline{p}^\mu
  \overline{p}^\nu - \overline{p}^2 \overline{g}^{\mu\nu})
  \nonumber\\
  &\qquad\qquad\qquad-  \bigg(\overline{p}^2 + \widehat{p}^2\Big(\frac{8}{3} - \ln(-p^2)\Big)\bigg)\overline{g}^{\mu\nu} \Bigg]
\, .\label{eq:QGBSE1Lfin}
\end{align}
\end{subequations}
From this illustrative example, we can extract several  interesting
comments. First, and most obviously, transversality is violated by the
last terms in Eqs.\ (\ref{eq:QGBSE}). This will be our main focus.
But also the transverse part
shows two differences compared to ordinary QED. 
Since the interaction vertex in  $\chi$QED differs from the one given in standard QED
by 
\begin{align}
  V_{\text{QED}}&\to-i e \gamma^{\mu}Q_{ij}, & V_{\text{$\chi$QED}}&\to-i e \bar{\gamma}^{\mu}\mathbb{P}_{R}\mathcal{Y}_{R,ij},
\end{align}
it  projects the fermion loop content, so the transverse part becomes
purely $4$-dimensional, explaining the appearance of the covariants
$\overline{g}^{\mu\nu}$ and $\overline{p}^\mu$ in
Eqs.\ (\ref{eq:QGBSE}). Further, due to this projection only half the
number  of  fermionic degrees of freedom appear in the loop for the
chiral case, resulting in the relative factor of 2 with respect to ordinary
QED.

Let us now focus on the breaking of transversality in the photon self-energy. The divergent breaking term in Eq.\ (\ref{eq:QGBSE1Ldiv}) is
proportional to $\widehat{p}^2$, i.e.\ it is evanescent. In contrast,
the finite breaking term in Eq.\ (\ref{eq:QGBSE1Lfin}) contains
finite expressions that do not vanish in the 4-dimensional limit. The
finite breaking also contains evanescent terms that vanish in
the $\mathop{\text{LIM}}_{D \to 4}$; these will be ignored in the
following.

We can exhibit the breaking explicitly by plugging the photon self-energy into the Ward identity (\ref{eq:Ward1}); we obtain
\begin{align}
i p_\nu  \widetilde{\Gamma}_{AA}^{\nu \mu}(p)|^{1}_\text{div+fin,$\chi$QED}
&=
\frac{i e^2}{16 \pi^2} \frac{\text{Tr}(\mathcal{Y}_R^2)}{3} \Bigg[
 - \frac{1}{\epsilon}\widehat{p}^2 \overline{p}^{\mu} -  \overline{p}^2\overline{p}^{\mu} \Bigg]
\ne 0 \, .
\label{WIphotonSEbreaking}
\end{align}
Here we have ignored the finite, evanescent term, as announced.
In line with the derivation of the Ward identity from the
Slavnov-Taylor identity via derivatives with respect to a ghost field,
see Eq.\ (\ref{STIabelianWIderivation}), the result is equivalent to the violation of the
Slavnov-Taylor identity
\begin{align}
  [\mathcal{S}(\Gamma)]_{A^\mu c}^1 &=
\frac{i e^2}{16 \pi^2} \frac{\text{Tr}(\mathcal{Y}_R^2)}{3} \Bigg[
 - \frac{1}{\epsilon}\widehat{p}^2 \overline{p}^{\mu} -
 \overline{p}^2\overline{p}^{\mu} \Bigg]\,,
\label{STIphotonSEbreaking}
\end{align}
where the left hand side denotes functional derivatives in momentum
space, similarly to the notation of $\Gamma_{AA}$.

A decisive feature of the breaking terms is their locality --- the
breaking terms in all the previous equations are polynomials of the
momentum in momentum space, and this translates into local expressions
on the level of the (effective) action. This locality is in line with
the general statement discussed in
Sec.\ \ref{sec:algebraicrenormalizationDetails} which forms
the basis of algebraic renormalization. This means that
a local counterterm can be defined that cancels the symmetry
breaking.

In view of the explicit results, the required counterterms for the
sector of the photon self-energy can be read off as follows. We first
discuss the divergent counterterms. The
divergent counterterms can be split into an invariant and a
non-invariant part as in Eq.\ (\ref{eq:CT_structure}) as $S_\text{sct} =S_{\text{sct,inv}}
+  S_{\text{sct,non-inv}}
$ such that the one-loop parts relevant for the photon self-energy in
$\chi$QED read
\begin{subequations}
\label{eq:SingularCT1LoopPhotonSE}
  \begin{align}
S_{\text{sct,inv,$\chi$QED}}^{1}&=
\frac{-\hbar \, e^2}{16 \pi^2 \epsilon} 
\frac{2 \text{Tr}(\mathcal{Y}_R^2)}{3} \overline{S_{AA}} +\ldots 
\,,\\
S_{\text{sct,non-inv,$\chi$QED}}^{1} &=
\frac{-\hbar \, e^2}{16 \pi^2 \epsilon} 
 \frac{\text{Tr}(\mathcal{Y}_R^2)}{3} \int d^D x \, \frac{1}{2} \bar{A}_\mu \widehat{\partial}^2 \bar{A}^\mu +\ldots
\, ,
\end{align}
\end{subequations}
where the dots denote terms unrelated to the photon self-energy. As
in the case of ordinary QED, the divergences are canceled, and the
invariant counterterm can be generated via a photon field
renormalization transformation. In
contrast to ordinary QED, however, the non-invariant term is required,
and it cannot be obtained from a renormalization transformation but
must be read off by hand.

Obviously, adding these counterterms does not only cancel the
divergences of the photon self-energy but it also cancels the
divergences in the breaking of the Ward/Slavnov-Taylor identities
(\ref{WIphotonSEbreaking},\ref{STIphotonSEbreaking}). Specifically
adding the counterterms to the action 
modifies the Slavnov-Taylor identity $\mathcal{S}(\Gamma)$ to
$\mathcal{S}(\Gamma+
S_\text{sct,$\chi$QED}^{1})=\mathcal{S}(\Gamma)+s_D
S_\text{sct,$\chi$QED}^{1}+\ldots$, where the dots denote higher-order
terms and where
\begin{equation}
\label{eq:bdSsct1L}
\begin{split}
  s_D  S_\text{sct,$\chi$QED}^{1}=\Delta_\text{ct}^1\big|_{\text{div}}
&= -\frac{\hbar}{16 \pi^2 \epsilon }\frac{e^2 \text{Tr}(\mathcal{Y}_R^2)}{3}  \, \int d^D x \,(\overline{\partial}_\mu c) \, (\widehat{\partial}^2 \bar{A}^\mu)
\, .
\end{split}
\end{equation} 
In momentum-space, with incoming $A^\mu$ momentum $p$, this is
precisely  the negative of the divergent term in
Eq.\ (\ref{STIphotonSEbreaking}). This is an automatic consequence of
the finiteness.

Now we discuss the required finite counterterms to the photon self-energy.
The explicit result (\ref{eq:QGBSE1Lfin}) shows that the transversality is restored
by the following finite counterterm:
\begin{align}
\label{eq:FiniteCT1LoopPhotonSE}
  S^1_\text{fct,$\chi$QED} & =
\frac{\hbar}{16\pi^2} \int d^4 x 
\frac{-e^2 \text{Tr}(\mathcal{Y}_R^2)}{6} \bar{A}_\mu
\overline{\partial}^2 \bar{A}^\mu
+\ldots
\end{align}
In momentum space this counterterm cancels the non-transverse
$\overline{p}^2$-term of (\ref{eq:QGBSE1Lfin}) (we recall that the
remaining non-transverse finite terms are evanescent and vanish in the
$\mathop{\text{LIM}}_{D \to 4}$). On the level of the Slavnov-Taylor
identity, adding the finite counterterm 
modifies the Slavnov-Taylor identity $\mathcal{S}(\Gamma)$ by the term
\begin{align}
  \label{eq:sdSfct}
  s_D
  S^1_\text{fct,$\chi$QED} & =
-\frac{\hbar}{16\pi^2} \int d^D x 
\, \frac{e^2 \text{Tr}(\mathcal{Y}_R^2)}{3} (\overline{\partial}_\mu c)  (\overline{\partial}^2 \bar{A}^\mu)\, .
\end{align}
In momentum space, this is the negative of the finite term in
Eq.\ (\ref{STIphotonSEbreaking}).

In total, after adding all counterterms
(\ref{eq:SingularCT1LoopPhotonSE},\ref{eq:FiniteCT1LoopPhotonSE}) to
the photon self-energy 
and taking the 
$\mathop{\text{LIM}}_{D \to 4}$, the renormalized one-loop photon self-energy is
\begin{align}
i \widetilde{\Gamma}_{AA}^{\nu \mu}(p)|^{1}_\text{ren, $\chi$QED} =&
\frac{i e^2}{16 \pi^2} \frac{\text{Tr}(\mathcal{Y}_R^2)}{3} \Bigg[
  \left(\frac{10}{3}-2\ln(-p^2)\right) (\overline{p}^\mu
  \overline{p}^\nu - \overline{p}^2 \overline{g}^{\mu\nu})
 \Bigg]\,.
\end{align}
It is finite, defined in 4 dimensions, and it is properly
transverse. One may still add further, finite, symmetric
counterterms. These can be derived from usual field and parameter
renormalization but are not our focus here.

\subsubsection{One-Loop Photon Self-Energy --- Direct Computation of
 Symmetry~Breaking}
\label{sec:oneloopdirect}

In the previous subsection we determined the required
counterterms
(\ref{eq:SingularCT1LoopPhotonSE},\ref{eq:FiniteCT1LoopPhotonSE})  
by carrying out an explicit
computation of a Green function, including its finite part, and by
explicitly evaluating the breaking of the relevant symmetry
identity. We now show how the determination of the counterterms can be
performed in a simpler way. We still illustrate it for the one-loop photon self-energy, but the advantage of that simplification will
become more and more prominent for higher orders and more complicated
Green functions.

Instead of evaluating the full photon self-energy including its finite
part (\ref{eq:QGBSE}), the following is sufficient: First we need the divergent part
of the photon self-energy, i.e.\ only (\ref{eq:QGBSE1Ldiv}). This of
course determines the divergent 
counterterms (\ref{eq:SingularCT1LoopPhotonSE}) unambiguously.

Second, we need the violation of the symmetry, expressed in terms of
Eq.\ (\ref{STIphotonSEbreaking}). This violation can be obtained in a
more direct way, by using the 
regularized quantum action principle discussed in Sec.\ \ref{sec:QAPDReg}. This
tells us that the violation $\mathcal{S}(\Gamma)\ne0$ is directly
given by diagrams with insertions  
of the composite operator $\widehat{\Delta}$, corresponding to the
tree-level violation of the Slavnov-Taylor identity in $D$
dimensions. For the photon self-energy, the violation
(\ref{STIphotonSEbreaking}) can be  
obtained directly by computing the Green function $ [\widehat{\Delta}
  \cdot \widetilde{\Gamma}_{Ac}^{\mu}]$, i.e.\ the 1-particle
irreducible Green function with an insertion of $\widehat{\Delta}$ and
external $A^\mu$ and $c$ fields.
  
At one-loop order there is only one diagram.
\\
\begin{minipage}{0.3\textwidth}
	\begin{equation*}
	i [\widehat{\Delta} \cdot \widetilde{\Gamma}_{Ac}^{\mu}]^{(1)}\, =
	\end{equation*}
\end{minipage}
\begin{minipage}{0.2\textwidth}
	\centering
	\includegraphics[scale=0.6]{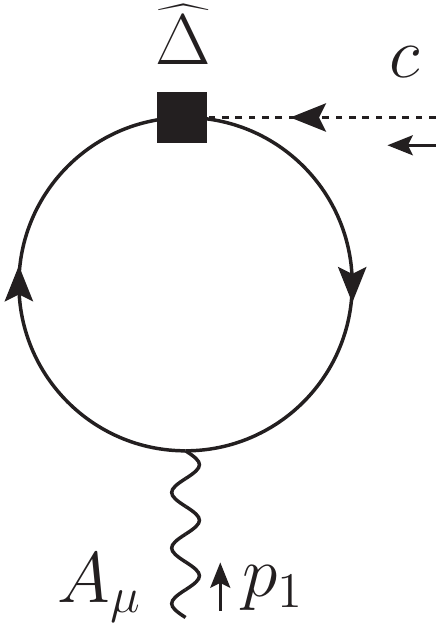}
\end{minipage}
\hfill\begin{minipage}{0.2\textwidth}
\begin{equation}\label{DiagramForDeltaDotGammaAc}
\end{equation}
\end{minipage}
\\
The result of this single diagram is
\begin{subequations}
  \begin{align}
  i [\widehat{\Delta} \cdot
    \widetilde{\Gamma}_{Ac}^{\mu}]^{1}_\text{div}
  &=
  \frac{ e^2 }{16 \pi^2 \epsilon}  \, \frac{\text{Tr}(\mathcal{Y}_R^2)}{3}\widehat{p_1}^2 \overline{p_1}^\mu
	\, ,\\
i \,[\widehat{\Delta}\cdot \widetilde{\Gamma}_{Ac}^{\mu}]^{1}_{\text{fin}} 
&=
\frac{ e^2}{16\pi^2} \frac{\text{Tr}(\mathcal{Y}_R^2)}{3}
\overline{p}_1^2 \overline{p}^\mu_1\;. 
\end{align}\label{DeltaDotGammaAc}
\end{subequations}
We see that the result of this diagram indeed agrees with the
right-hand side of Eq.\ (\ref{STIphotonSEbreaking}), as it is
guaranteed by the regularized quantum action principle.

The important point is the technical simplification: the computation
of this diagram is technically easier than the 
computation of the finite part of the photon self-energy since only
power-counting divergent parts of the loop integrals are relevant. We
reiterate that the technical advantage is much more dramatic at higher
orders and for more complicated Green functions.

It is instructive to rewrite the result in coordinate space,
\begin{subequations}
  \begin{align}
    [\widehat{\Delta} \cdot \Gamma]^{(1)}_\text{div} &= \frac{      e^2}{16 \pi^2 \epsilon}  \, \frac{\text{Tr}(\mathcal{Y}_R^2)}{3}\int d^D
x (\overline{\partial}_{\mu}c) ( \widehat{\partial}^2 \bar{A}^{\mu})+\ldots,
\\
  [\widehat{\Delta} \cdot \Gamma]^{(1)}_\text{fin} &=
  \frac{    e^2 }{16\pi^2} \, \frac{\text{Tr}(\mathcal{Y}_R^2)}{3} \int d^D x 
\,  (\overline{\partial}_\mu
c)  (\overline{\partial}^2 \bar{A}^\mu)+\ldots
\, .
\label{eq:Delfct1L}
\end{align}
  \label{DeltaDotGammaAcx}
\end{subequations}
The dots denote terms unrelated to the photon self-energy.

The divergent part provides no independent information but a check. As
discussed after Eq.\ (\ref{eq:bdSsct1L}), the expression $s_D
S_\text{sct,$\chi$QED}^{1}$ must automatically cancel the divergent
part of the symmetry breaking. Using our new result, this means that $s_D
S_\text{sct,$\chi$QED}^{1}+[\widehat{\Delta} \cdot
  \Gamma]^{(1)}_\text{div}=0$
must automatically hold.
Clearly, this is true, and the check is passed.

The important new information is in the finite part of the
$\widehat{\Delta}$-insertion diagram
Eqs.\ (\ref{DeltaDotGammaAc},\ref{DeltaDotGammaAcx}).
Its result is equal to the finite part of the violation of the
Slavnov-Taylor identity (\ref{STIphotonSEbreaking}), thus eliminating
the need to explicitly evaluate the Slavnov-Taylor identity.

The finite, symmetry-restoring counterterm may now be obtained from
solving the equation
\begin{align}
  s_D   S^1_\text{fct,$\chi$QED} & =
-
[\widehat{\Delta} \cdot \Gamma]^{(1)}_\text{fin} \,.
\label{definingcondition}
\end{align}
For the sector of the photon self-energy, the result is the one given
in Eq.\ (\ref{eq:FiniteCT1LoopPhotonSE}). In summary: There, the result was
  obtained from inspecting the finite part of the photon self-energy;
  here, the result can be obtained from evaluating
  Eq.\ (\ref{DeltaDotGammaAc}) and then solving
the defining condition (\ref{definingcondition}).

To conclude the section we summarize the full one-loop results for the
counterterm structure of $\chi$QED. First, all divergences of all
one-loop diagrams need to be evaluated, generalizing
Eq.\ (\ref{eq:QGBSE1Ldiv}). The negative of these results define
unambiguously the one-loop divergent counterterms, generalizing
Eq.\ (\ref{eq:SingularCT1LoopPhotonSE}). The result reads
\begin{align}
\label{eq:SingularCT1Loop}
\begin{split}
S_\text{sct,$\chi$QED}^{1} =\;&
\frac{-\hbar \, e^2}{16 \pi^2 \epsilon} \left(
\frac{2 \text{Tr}(\mathcal{Y}_R^2)}{3} \overline{S_{AA}}
+\xi \, \sum_j (\mathcal{Y}_R^j)^2 \left( \overline{S^j_{\overline{\psi}\psi_R}} + \overline{S^j_{\overline{\psi_R} A \psi_R}} \right)
\right.\\
&\left. + \frac{\text{Tr}(\mathcal{Y}_R^2)}{3} \int d^D x \, \frac{1}{2} \bar{A}_\mu \widehat{\partial}^2 \bar{A}^\mu \right)
\, .
\end{split}
\end{align}
Most terms are similar to their counterparts in ordinary QED and can
be obtained by a renormalization transformation of fields and
parameters as in Eq.\ (\ref{RenTrafo}), where it is noteworthy that
only the physical, right-handed fermion is renormalized, while the
sterile left-handed fermion is not. However, this renormalization
transformation does not generate the last term involving
the $\widehat{\partial}^2$ operator, and it generates the full
$D$-dimensional photon kinetic term ${S_{AA}}$ instead of its
4-dimensional version $\overline{S_{AA}}$. Hence the
$\widehat{\partial}^2$-term and the difference
$\overline{S_{AA}}-S_{AA}$ correspond to symmetry-breaking singular
counterterms. These counterterms become particularly important in the
context of two-loop calculations where they are necessary for the
proper subrenormalization.

Second, all one-loop symmetry breakings need to be determined,
generalizing either Eq.\  (\ref{STIphotonSEbreaking}) or
Eq.\ (\ref{DeltaDotGammaAcx}). We use the method based on the
regularized quantum action principle. In this case, the full symmetry
breaking is given by the complete set of all one-loop diagrams with a
$\widehat{\Delta}$ insertion. Since only power-counting divergent
diagrams can provide non-vanishing contributions, there are only
precisely four contributing diagrams: with external fields $cA$,
$cAA$, $cAAA$, or $c\bar\psi\psi$. One of them vanishes due to the
anomaly cancellation condition (\ref{eq:AnomCond}). The full result of the
symmetry breaking is
\begin{align}
\label{eq:STIBreakingAt1Loop}
\widehat{\Delta}\cdot\Gamma^1 =& \,
\frac{1}{16\pi^2} \int d^Dx \bigg[
               \, \frac{e^2 \text{Tr}(\mathcal{Y}_R^2)}{3}\left( \frac{1}{\epsilon}(\overline{\partial}_\mu c) \, (\widehat{\partial}^2 \bar{A}^\mu)
	+ (\overline{\partial}_\mu c)  (\overline{\partial}^2 \bar{A}^\mu) \right)\\ \notag
&+\,\frac{e^4 \text{Tr}(\mathcal{Y}_R^4)}{3} \, c \, \overline{\partial}_\mu  (\bar{A}^\mu \bar{A}^2)
		\\ \notag
		&- \,\frac{(\xi+5)e^3}{6}\sum_{j}(\mathcal{Y}_R^j)^3  \, c \, \overline{\partial}^\mu (\overline{\psi}_j \overline{\gamma}_\mu \mathbb{P}_{R} \psi_j)\bigg]\, .
\end{align}
And using the defining condition (\ref{definingcondition}) for the
finite, symmetry-restoring counterterms, we obtain
\begin{equation}
\label{eq:Sfct1L}
\begin{split}
S^1_\text{fct} =
\frac{\hbar}{16\pi^2} \int d^4 x &\Bigg\{
\frac{-e^2 \text{Tr}(\mathcal{Y}_R^2)}{6} \bar{A}_\mu \overline{\partial}^2 \bar{A}^\mu
+ \frac{e^4 \text{Tr}(\mathcal{Y}_R^4)}{12} \bar{A}_\mu \bar{A}^\mu \bar{A}_\nu \bar{A}^\nu
\\
& + \frac{5+\xi}{6} e^2 \sum_j (\mathcal{Y}_R^j)^2 i \overline{\psi}_j \overline{\gamma}^\mu \overline{\partial}_\mu \mathbb{P}_{R} \psi_j \Bigg\}
\, .
\end{split}
\end{equation}
This is the complete result for the symmetry-restoring counterterms of
the $\chi$QED model at the one-loop level. Each of the terms has a
clear and simple interpretation.
The first finite counterterm restores the
transversality of the photon self-energy as discussed before.
The second term restores a similar transversality identity for the
photon 4-point function. And the last term restores the QED-like Ward
identity relating the fermion self-energy with the fermion--photon
three-point function.

These three counterterms must be inserted in higher-order
calculations. They give additional contributions to loop diagrams
compared to the renormalization in vector-like theories or to a naive
$\gamma_5$ treatment where gauge invariance is manifestly preserved.

\subsubsection{Two-Loop Photon Self-Energy and Corresponding Breaking~Diagram}
\label{sec:TwoLoopPhotonSelfEnergyChiralQED}
Now we illustrate the determination of two-loop  counterterms in
$\chi$QED using the BMHV scheme. We immediately follow the more direct
strategy explained in Sec.\ \ref{sec:oneloopdirect} based on diagrams
with $\widehat{\Delta}$-insertions.

  At the 2-loop level, diagrams contributing to the subrenormalized photon self-energy
  are on the one hand genuine 2-loop diagrams and on the other hand
  1-loop diagrams with counterterm insertions. Both the singular
  counterterms (\ref{eq:SingularCT1Loop}) as well as finite symmetry-restoring
  counterterms (\ref{eq:Sfct1L}) must be used.   The result for the divergent part of the subrenormalized two-loop
  photon self-energy is given by\footnote{We still use the simplified notation described in footnote
\ref{footnoteSimplerGammaExplicit} where $\Gamma^2$ denotes the
subrenormalized two-loop effective action.
    }
\begin{subequations}
	\begin{align}
	\label{eq:QGBSE2L}
	i \widetilde{\Gamma}_{AA}^{\nu \mu}(p)|^{2}_\text{div,$\chi$QED} &=
	\frac{i e^4}{256 \pi^4} \frac{\text{Tr}(\mathcal{Y}_R^4)}{3} \left[ \frac{2}{\epsilon} (\overline{p}^\mu \overline{p}^\nu - \overline{p}^2 \overline{g}^{\mu\nu}) + \left( \frac{17}{24 \epsilon} - \frac{1}{2 \epsilon^2} \right) \widehat{p}^2 \overline{g}^{\mu\nu} \right]
	\, ,\intertext{which can be compared to the result in ordinary
          QED}
	i \widetilde{\Gamma}_{AA}^{\nu \mu}(p)|^{2}_\text{div,QED} &=
	\frac{i e^4}{256 \pi^4 \epsilon} 2 \, \text{Tr}(Q^4) (p^\mu p^\nu - p^2 g^{\mu\nu})
	\, .
	\end{align}
\end{subequations}
Notice again that the transverse part for QED is fully $D$-dimensional
but projected to 4 dimensions in the chiral case, and in the chiral
case an evanescent term is present, again spoiling gauge and BRST
invariance. Unlike at the 1-loop level, the global factor in front of
the chiral transversal part is not half  of the QED
case, since the additional diagram with finite 1-loop counterterm
insertion spoils this relationship.

From this singular part of the two-loop diagrams we reconstruct an
equivalent result in coordinate space, 
\begin{equation}
\Gamma^{2,AA}_\text{div} =
\frac{e^4}{256 \pi^4} \frac{\text{Tr}(\mathcal{Y}_R^4)}{3} \Big[\frac{1}{\epsilon}
\overline{A}_\mu(\overline{\partial}^2 \overline{g}^{\mu\nu}-\overline{\partial}^\mu \overline{\partial}^\nu)\overline{A}_\nu
+ \overline{A}_\mu \widehat{\partial}^2\overline{A}^\mu \Big(\frac{1}{4\epsilon^2} - \frac{17}{48\epsilon}\Big)\Big]
\, ,
\end{equation}
which results in the required singular countertem of the form
\begin{equation}
  S^2_\text{sct} =
  -\Bigg(\frac{\hbar \, e^2}{16 \pi^2}\Bigg)^2 \frac{\text{Tr}(\mathcal{Y}_R^4)}{3} \left[ \frac{2}{\epsilon} \overline{S_{AA}} + \left( \frac{1}{4\epsilon^2} - \frac{17}{48\epsilon} \right) \int d^D x \overline{A}_\mu \widehat{\partial}^2\overline{A}^\mu \right]+\ldots\,,
\end{equation}
which cancels the divergences.
Clearly, this counterterm also breaks  BRST symmetry  at the 2-loop level by
\begin{equation}
\label{eq:BRSTtransfoSsct2}
\begin{split}
\Delta_\text{sct}^2 =
s_D {S_\text{sct}^2} =&
\frac{-\hbar^2 e^4}{256 \pi^4} \frac{\text{Tr}(\mathcal{Y}_R^4)}{6}
\left( \frac{1}{\epsilon^2} - \frac{17}{12\epsilon} \right) \int d^D x (\overline{\partial}_\mu c) (\widehat{\partial}^2\overline{A}^\mu)+\ldots
\, .
\end{split}
\end{equation}

Now we use the regularized quantum action principle and determine the
symmetry breaking at the two-loop level in the photon self-energy
sector. Hence we need to evaluate the Green function
$\left( [\widehat{\Delta} + \Delta_\text{ct}^{1}] \cdot
\widetilde{\Gamma} \right)^{2}_{A_\mu c}$ at the two-loop
level.

\begin{figure}[t]
	\centering
	\begin{tabular}{*{3}{>{\centering\arraybackslash}m{0.3\textwidth}}}
		\includegraphics[scale=0.6]{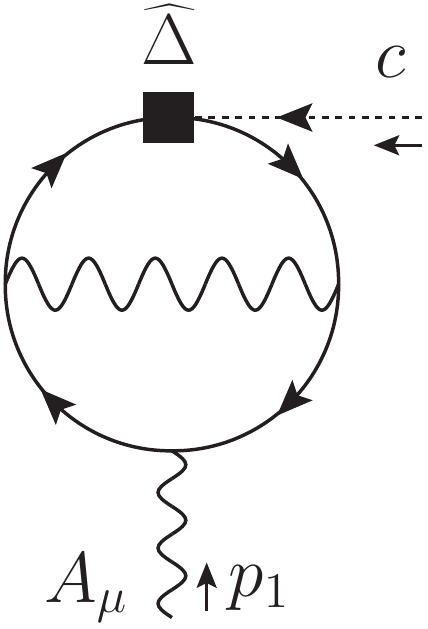}
		&
		\includegraphics[scale=0.6]{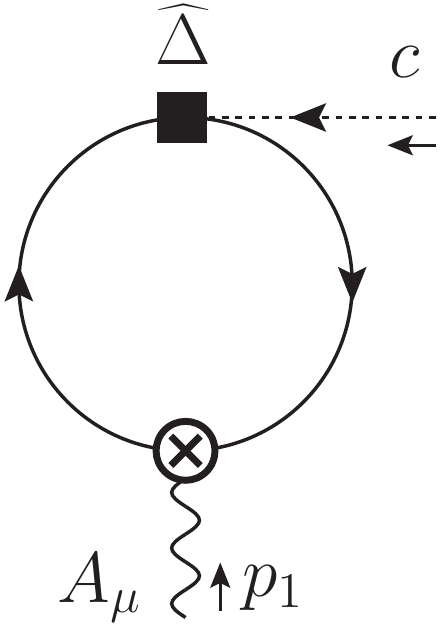}
		&
		\includegraphics[scale=0.6]{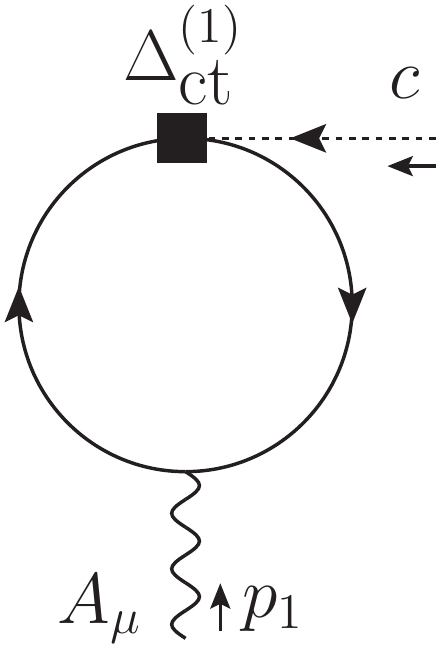}
	\end{tabular}
	\begin{tabular}{*{3}{>{\centering\arraybackslash}m{0.3\textwidth}}}
		\includegraphics[scale=0.6]{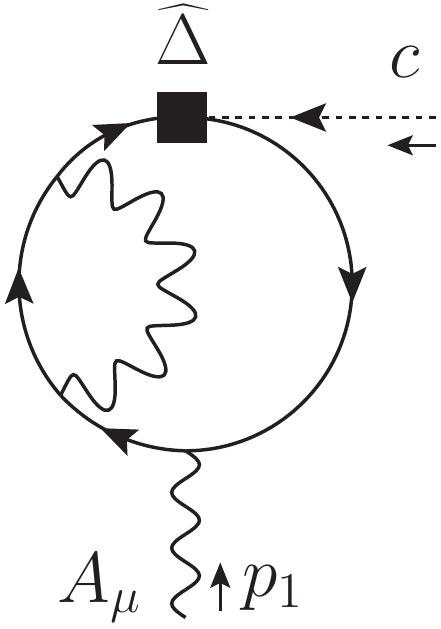} \newline
		+ loop on the other fermion propagator.
		&
		\includegraphics[scale=0.6]{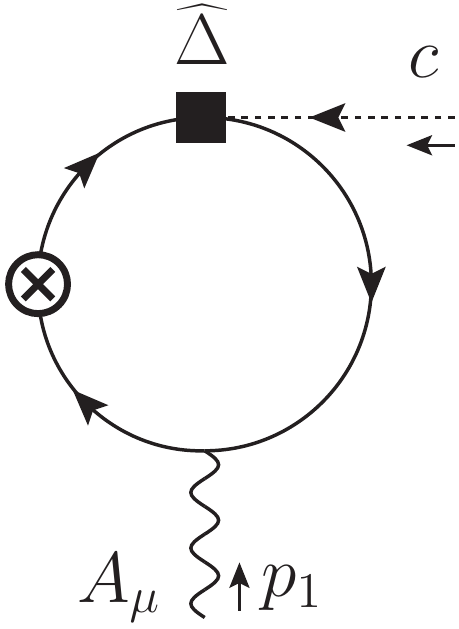} \newline
		+ fermion counterterm on the other fermion propagator.
		&
		\includegraphics[scale=0.6]{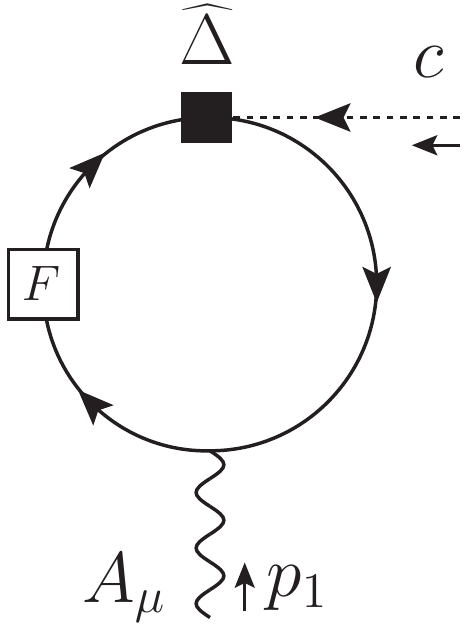} \newline
		+ fermion finite counterterm on the other fermion propagator.
	\end{tabular}
	\caption{\label{fig:2LBRSTDGcA}
		List of Feynman diagrams for the ghost--photon breaking contribution given in \cref{2LBRSTDGcA}.
}\end{figure}

Compared to the one-loop level, there are several new features.
There are four types of two-loop level diagrams, see \cref{fig:2LBRSTDGcA}. 
The diagrams in the first column of the figure are genuine two-loop
diagrams with one insertion of the tree-level breaking
$\widehat{\Delta}$.  The diagrams in the second column are one-loop
diagrams with one insertion of a one-loop singular counterterm,
denoted as a circled cross.  The third column contains a one-loop
diagram with an insertion of a one-loop symmetry-restoring counterterm
obtained from the fermion self-energy operator, denoted by a boxed
$F$, and a one-loop diagram with an insertion of the one-loop breaking
$\Delta_\text{ct}^{1}$.

The total two-loop breaking in this sector, i.e.\ the result of the
diagrams in  \cref{fig:2LBRSTDGcA} is
\begin{equation}
\label{2LBRSTDGcA}
i \left( [\widehat{\Delta} + \Delta_\text{ct}^{1}] \cdot \widetilde{\Gamma} \right)^{2}_{A_\mu c} =
\frac{1}{256 \pi^4} \frac{e^4 \text{Tr}(\mathcal{Y}_R^4)}{6}
\left[
\left( \frac{1}{\epsilon^2} - \frac{17}{12\epsilon} \right) \hat{p}_1^2 \overline{p}_1^\mu
- \frac{11}{4} \overline{p}_1^2 \overline{p}_1^\mu + \mathcal{O}(\hat{.})
\right]
\, .
\end{equation}
The result contains $1/\epsilon^2$ poles and $1/\epsilon$ poles with
local, evanescent coefficients and a finite, non-evanescent
term.

Like at the one-loop level, we first use the result to check the
cancellation of the UV divergences as prescribed by
\cref{eq:deltadef}.\footnote{In the Abelian case considered here, \cref{eq:deltadef} can be
simplified. For reasons mentioned in footnote \ref{footnotelinearEQs},
$b_D=s_D$ and we simply have $\Delta_\text{ct}^2=s_D S_{\text{ct}}^2$. }
As expected, this cancellation  with
$s_D {S^{2}_\text{sct}}$ given in \cref{eq:BRSTtransfoSsct2} indeed occurs as
\begin{equation}
\label{eq:sct2cancellation}
\Delta_\text{sct}^2 = s_D {S^{2}_\text{sct}}
= - \left( [\widehat{\Delta} + \Delta_\text{ct}^{1}] \cdot \Gamma \right)^{2}_\text{div} \, .
\end{equation}
The remaining finite part can then be evaluated in strictly 4
dimensions, 
\begin{equation}\label{eq:BRSTtransfoSfct2}\begin{split}
\Delta_\text{fct}^2
=\;&-\mathop{\text{LIM}}_{D \to 4} \left\{ \left( [\widehat{\Delta} + \Delta_\text{ct}^{(1)}] \cdot {\Gamma} \right)^{(2)} + s_D {S^{(2)}_\text{sct}} \right\}
\\
=&
\frac{  e^4}{256 \pi^4} \text{Tr}(\mathcal{Y}_R^4) \; s\left( \frac{11}{48} \int d^4 x \bar{A}_\mu \overline{\partial}^2 \bar{A}^\mu \right)+\ldots\,.
\end{split}\end{equation}
The defining relation for the finite, symmetry-restoring counterterm
is then
\begin{align}
\mathop{\text{LIM}}_{D \to 4}s_D S^{2}_\text{fct}=-\Delta_\text{fct}^2\,.
\end{align}
From this we reconstruct the corresponding finite counterterm as
\begin{equation}
S^{2}_\text{fct} =
\left(\frac{\hbar}{16\pi^2}\right)^2 \int d^4x \, 
e^4 
\text{Tr}(\mathcal{Y}_R^4) \frac{11}{48} \bar{A}_\mu
\overline{\partial}^2 \bar{A}^\mu
+\ldots\,.
\end{equation}
As before we only display terms related to the photon self-energy. Adding this counterterm restores the photon self-energy
transversality at the 2-loop level.

At this point the determination of the two-loop counterterms of this
sector is complete, and the counterterms of other sectors can be
determined analogously. The required computations were the ones of the
divergent part of the photon self-energy and of the finite part of the
diagrams of \cref{fig:2LBRSTDGcA}.

Nevertheless we now confirm the result by comparing with the explicit
result for the finite part of the photon self-energy.
The finite part of the photon self-energy at the two-loop level
(including one-loop counterterms 
but excluding two-loop counterterms)  is given by 
\begin{equation}
\begin{split}
\left. i \widetilde{\Gamma}_{AA}^{\mu\nu}(p) \right|^{2}_\text{fin} &=
\frac{i e^4}{256 \pi^4} \frac{\text{Tr}(\mathcal{Y}_R^4)}{3} \\&\left[ \left( \frac{673}{23} -6 \log(-\overline{p}^2)-24\zeta(3) \right) (\overline{p}^\mu \overline{p}^\nu - \overline{p}^2 \overline{g}^{\mu\nu}) + \frac{11}{8} \overline{p}^\mu \overline{p}^\nu \right]
\, .
\end{split}\label{twoloopphotonSE}
\end{equation}
Similar to the one-loop result (\ref{eq:QGBSE}), the non-local $\log(-\overline{p}^2)$
and transcendental $\zeta(3)$ parts are by 
themselves transversal and so do not break  gauge invariance. The last
term breaks the transversality, but this breaking term is local.

Plugging the result into the Ward or Slavnov-Taylor identity
we obtain
\begin{subequations}
\begin{align}
\left. i \, p_{\nu} \, \widetilde{\Gamma}_{A(-p)A(p)}^{\mu\nu}
\right|^{2}_\text{fin}
&
=
\frac{i e^4}{256 \pi^4} \frac{\text{Tr}(\mathcal{Y}_R^4)}{6}
\frac{11}{4}  \overline{p}^2 \overline{p}^{\mu}
\\
&= -\left( [\widehat{\Delta} + \Delta_\text{ct}^{1}] \cdot \widetilde{\Gamma} \right)^{2}_{\text{fin},\, A_{\mu}(-p)c(p)}
\, .
\end{align}
\end{subequations}
The first of these equations is obtained by direct computation
using the finite parts in Eq.\ (\ref{twoloopphotonSE}). The second
  equation is then observed by comparison with
  \cref{2LBRSTDGcA}.
Hence we confirm 
that the violation of the symmetry is  restored by our finite
counterterm evaluated from breaking diagrams.

\subsubsection{Full Two-Loop Renormalization of Chiral~QED}
\label{sec:Full2LoopRenormalizationOfChiralQED}

In the previous sections we performed the full one-loop renormalization with singular and finite, symmetry-restoring counterterms (\ref{eq:SingularCT1Loop}) and (\ref{eq:Sfct1L}), respectively, and studied the photon self-energy and the corresponding breaking at the two-loop level, cf.\ Sec.\ \ref{sec:TwoLoopPhotonSelfEnergyChiralQED}. In this section we present the full two-loop renormalization of chiral QED 
based on  our results in Ref.\ \cite{Belusca-Maito:2021lnk}.

A list of all divergent 1PI two-loop Green functions together with the
individual results is to be found in chapter $7$ of Ref.\ \cite{Belusca-Maito:2021lnk}. From the singular part of these Green functions we obtain the singular counterterm action at the two-loop level
\begin{align}
\label{eq:Ssct2loopchiralQED}
    S_{\text{sct}}^{2} = &- \left(\frac{\hbar e^2}{16\pi^2}\right)^2 \frac{\text{Tr}(\mathcal{Y}_R^4)}{3} \bigg[ \frac{2}{\epsilon} \overline{S_{AA}} + \left(\frac{1}{4 \epsilon^2} - \frac{17}{48 \epsilon}\right) \int d^{D}x \, \bar{A}_{\mu} \widehat{\partial}^2 \bar{A}^{\mu} \bigg]\nonumber\\
    &+ \left(\frac{\hbar e^2}{16\pi^2}\right)^2 \sum_{j} (\mathcal{Y}_R^j)^2 \bigg[ \left( \frac{1}{2\epsilon^2} + \frac{17}{12 \epsilon} \right) (\mathcal{Y}_R^j)^2 - \frac{1}{9 \epsilon} \text{Tr}(\mathcal{Y}_{R}^2) \bigg] \left( \overline{S^j_{\overline{\psi}\psi_{R}}} + \overline{S^j_{\overline{\psi_{R}}A\psi_{R}}} \right)\nonumber\\
    &- \left(\frac{\hbar e^2}{16\pi^2}\right)^2 \sum_{j} \frac{(\mathcal{Y}_{R}^j)^2}{3 \epsilon} \bigg( \frac{5}{2} (\mathcal{Y}_{R}^j)^2 - \frac{2}{3} \text{Tr}(\mathcal{Y}_{R}^2) \bigg) \overline{S^j_{\overline{\psi} \psi_{R}}}
\end{align}
which cancel the divergences. Comparing (\ref{eq:Ssct2loopchiralQED}) with its one-loop counterpart in Eq.\ (\ref{eq:SingularCT1Loop}), we see that its structure is the same up to the term in the last line, which breaks BRST invariance by a non-evanescent amount and is thus a new feature emerging at the two-loop level.

This two-loop counterterm action (\ref{eq:Ssct2loopchiralQED}) generates the BRST breaking
\begin{align}
\label{eq:Ssct2loopBRSTBreaking}
    \Delta_{\text{sct}}^2 &= s_{D} S_{\text{sct}}^{2}\nonumber\\
    &= - \frac{\hbar^2e^4}{256 \pi^4} \frac{\text{Tr}(\mathcal{Y}_R^4)}{6} 
    \left( \frac{1}{\epsilon^2} - \frac{17}{12\epsilon} \right) \int d^D x \, (\overline{\partial}_\mu c) (\widehat{\partial}^2\bar{A}^\mu)\\
    &\hspace{0.44cm} - \frac{\hbar^2e^5}{256 \pi^4} \frac{1}{3 \epsilon} \sum_{j} (\mathcal{Y}_{R}^j)^3 \bigg( \frac{5}{2} (\mathcal{Y}_{R}^j)^2 - \frac{2}{3} \text{Tr}(\mathcal{Y}_{R}^2) \bigg) \int d^Dx \, c \, \overline{\partial}_{\mu} \big( \overline{\psi} \bar{\gamma}^{\mu} \mathbb{P}_{R} \psi \big). \nonumber
\end{align}
Compared to the previous section \ref{sec:TwoLoopPhotonSelfEnergyChiralQED} we this time provided the full two-loop result explicitly and see that, in contrast to the one-loop case (\ref{eq:bdSsct1L}), this BRST breaking contains a non-evanescent contribution given by the last line of (\ref{eq:Ssct2loopBRSTBreaking}).

Following the restoration procedure described in
Sections~\ref{sec:AlgRenInDReg} or
\ref{sec:symmetryrestorationrequirements} and analogous to the
ghost--gauge boson contribution (\ref{2LBRSTDGcA}) in the previous
Section~\ref{sec:TwoLoopPhotonSelfEnergyChiralQED}, we additionally
need to calculate $([\widehat{\Delta} + \Delta_\text{ct}^{1}] \cdot
\widetilde{\Gamma})^{2}$ for the ghost--fermion--fermion,
the~ghost--double gauge boson, and the ghost--triple gauge boson
contributions (i.e.\ with external fields $c\psi\bar\psi$, $cAA$,
$cAAA$, respectively). It turns out that the ghost--double gauge boson
contribution 
vanishes and the ghost--triple gauge boson contribution does not
contain UV divergences, but only finite terms. In total the result  is
\begin{align}
\label{eq:STIBreakingAt2Loop}
    \left(\big[\widehat{\Delta} + \Delta_\text{ct}^{1}\big] \cdot \Gamma\right)^{2} &= \frac{e^4}{256 \pi^4} \int d^Dx \, \nonumber\\
    &\bigg\{ - \frac{\text{Tr}(\mathcal{Y}_{R}^4)}{6} \bigg[ \left( \frac{1}{\epsilon^2} - \frac{17}{12 \epsilon} \right) c \, \overline{\partial}_{\mu} \widehat{\partial}^2 \bar{A}^{\mu} - \frac{11}{4} c \, \overline{\partial}_{\mu} \overline{\partial}^2 \bar{A}^{\mu} \bigg]\nonumber\\
    &+ e \sum_{j} \frac{(\mathcal{Y}_{R}^j)^3}{3} \bigg[ \frac{1}{\epsilon} \left( \frac{5}{2} (\mathcal{Y}_{R}^j)^2 - \frac{2}{3} \text{Tr}(\mathcal{Y}_{R}^2) \right)\\
    &\hspace{2.3cm} + \frac{127}{12} (\mathcal{Y}_{R}^j)^2 - \frac{1}{9} \text{Tr}(\mathcal{Y}_{R}^2) \bigg] c \, \overline{\partial}_{\mu} \big( \overline{\psi}_j \bar{\gamma}^{\mu} \mathbb{P}_{R} \psi_{j} \big) \nonumber \\
    &+ \frac{3 e^2 \text{Tr}(\mathcal{Y}_{R}^6)}{2} c \, \overline{\partial}_{\mu} \big( \bar{A}^{\mu} \bar{A}_{\nu} \bar{A}^{\nu} \big) \bigg\} + \mathcal{O}(\hat{.}) \nonumber
\end{align}
for the full two-loop breaking of the Slavnov-Taylor identity of two-loop subrenormalized 1PI Green functions. Comparing this with the corresponding one-loop contribution (\ref{eq:STIBreakingAt1Loop}), we see that the structure of the terms is the same.

For the symmetry restoration at the two-loop level, we first note that $\Delta_{\text{sct}}^2$ in Eq.\ (\ref{eq:Ssct2loopBRSTBreaking}) completely cancels the UV divergent terms in Eq.\ (\ref{eq:STIBreakingAt2Loop}). In addition to that, we need to determine the finite, symmetry-restoring counterterms at the two-loop as indicated in Eq.\ (\ref{eq:BRSTtransfoSfct2}). Thus, our choice for the full finite counterterm action, which restores the Slavnov-Taylor identity at the two-loop level, is
\begin{align}
\label{eq:Sfct2Loop}
    S_{\text{fct}}^{2} &= \left(\frac{\hbar}{16 \pi^2}\right)^2 \int d^D x \, e^4 \bigg\{ \text{Tr}(\mathcal{Y}_{R}^4) \frac{11}{48} \bar{A}_{\mu} \overline{\partial}^{2} \bar{A}^{\mu} + 3 e^{2} \frac{\text{Tr}(\mathcal{Y}_{R}^6)}{8} \bar{A}_{\mu} \bar{A}^{\mu} \bar{A}_{\nu} \bar{A}^{\nu}\nonumber\\
    &\hspace{2.2cm} - \sum_{j} (\mathcal{Y}_{R}^j)^2 \left(\frac{127}{36} (\mathcal{Y}_{R}^j)^{2} - \frac{1}{27} \text{Tr}(\mathcal{Y}_{R}^{2})\right) \Big(\overline{\psi}_{j} i \bar{\slashed{\partial}} \mathbb{P}_{R} \psi_{j}\Big) \bigg\}.
\end{align}
Similar to its one-loop counterpart in Eq.\ (\ref{eq:Sfct1L}),
$S_{\text{fct}}^{2}$ consists of three kinds of terms, or in other
words, the same three field monomials are involved. These three terms
correspond to the restoration of the Ward identity relations for the
photon self-energy, the photon $4$-point function and the fermion
self-energy/photon-fermion-fermion interaction.
Ref.\ \cite{Belusca-Maito:2021lnk} also gave a discussion
of the explicit results for these three Ward identity relations,
similar to the discussion at the end of
Sec.\ \ref{sec:TwoLoopPhotonSelfEnergyChiralQED}.
In all cases, the breaking terms of the Ward identity are explicitly
exhibited and the cancellation with the symmetry-restoring
counterterms (\ref{eq:Sfct2Loop}) is made manifest.
 
\subsection{Non-Abelian Chiral Yang-Mills Theory and Comparison with the Abelian Chiral Theory at the One-Loop Level}
\label{sec:explicitNonabelian}
In this section we review the application of the BMHV scheme to
non-abelian chiral gauge theories and present the differences to the
abelian chiral QED discussed above. In particular, we study a massless
chiral Yang-Mills theory at the one-loop level based on
Refs.\ \cite{Belusca-Maito:2020ala, Martin:1999cc}. {Note that
  in our publication \cite{Belusca-Maito:2020ala} the considered
  theory also contained real scalar fields. Here, similar to
  Ref.\ \cite{Martin:1999cc}, scalar fields are omitted in order to focus on the key-points of the BMHV scheme in the framework of chiral gauge theories and the differences compared to the abelian case discussed above.}

As discussed in section \ref{sec:YMGaugeTheoriesSetup}, the group
generators of Yang-Mills theories satisfy the non-trivial commutation relations (\ref{LieAlgebra}); in particular, they are not simultaneously diagonalizable. These algebraic structures of the non-Abelian gauge group of Yang-Mills theories lead to new effects, such as more interactions terms and non-linear BRST transformations of the gauge fields and the ghosts, compared to the Abelian case, cf.\ Sec.\ \ref{sec:AbelianPeculiarities}. Especially, gauge boson self-interactions, interactions of the Faddeev-Popov ghosts with the rest of the theory and the renormalization of the BRST transformations distinguish non-Abelian Yang-Mills theories from the Abelian case above. 

The outline of this section is analogous to the Abelian case discussed above.
First, we briefly introduce the Lagrangian of the theory and the BRST transformations using the notations from section \ref{sec:setup}. Second, we discuss the analytical continuation of the theory to $D$ dimensions in DReg treating $\gamma_{5}$ with the BMHV scheme and comment on the BRST breaking induced by this scheme. Finally, we present the results for the singular and the symmetry restoring counterterms at the one-loop level, cf.\ \cite{Belusca-Maito:2020ala,Martin:1999cc}, necessary to consistently renormalize the theory, and discuss the differences to the Abelian theory.

\subsubsection{Definition of the Non-Abelian Chiral Yang--Mills~Theory}
\label{sec:DefChiralYMLagrangian}
Following the conventions of Sec.\ \ref{sec:BRSTSTI}, the Lagrangian in $4$ dimensions can be written as
\begin{align}
    \La_{\chi\text{YM}} = \La_{\text{inv}} + \La_{\text{fix,gh}} + \La_{\text{ext}}\,.
\end{align}
The physical part of the Yang-Mills Lagrangian reads
\begin{align}
\label{chiralYMLagrangianInv}
    \La_{\text{inv}} = - \frac{1}{4} G^{a}_{\mu\nu} G^{a, \mu\nu} + i \, \overline{\psi_{R}}_{i} \slashed{D}_{ij} {\psi_{R}}_{j}\,,
\end{align}
with covariant derivative $D^{\mu}_{ij} = \partial^{\mu} \delta_{ij} + i g G^{a, \mu} \, {T^{a}_{R}}_{ij}$ and field strength tensor $G^{a}_{\mu\nu} = \partial_{\mu} G^{a}_{\nu} - \partial_{\nu} G^{a}_{\mu} - g f^{abc} G^{b}_{\mu} G^{c}_{\nu}$, leading to three- and four-point gauge boson self-interactions. The gauge-fixing and ghost Lagrangian, already presented in Eq.\ (\ref{GaugeFixingGeneral}), is
\begin{align}
    \La_{\text{fix,gh}} = s \left[\cbar^{a} \left((\partial^\mu G^{a}_{\mu}) + \frac{\xi}{2}B^{a}\right)\right] = B^{a} (\partial^\mu G^{a}_{\mu}) + \frac{\xi}{2} B^{a} B^{a} - \cbar^{a}\partial^\mu D^{ab}_{\mu}c^{b},
\end{align}
with $D^{ab}_{\mu} = \partial_{\mu} \delta^{ab} + g f^{abc} G_{\mu}^{c}$, implying ghost-antighost-gauge boson interactions which is a consequence of the nonlinear gauge transformations of the gauge fields $G^{a}_{\mu}$ as shown below in (\ref{NonAbelianBRSTTrafosChiralYM}).
The Lagrangian of the external sources, as introduced in Sec.\ \ref{sec:BRSTSTI}, is
\begin{align}
    \La_{\text{ext}} = \rho^{a, \mu} s G_{\mu}^{a} + \zeta^{a} s c^{a} + \bar{R}^{i} s {\psi_{R}}_{i} + R^{i} s \overline{\psi_{R}}_{i}\,.
\label{eq:LaExtChiralYM}
\end{align}
The BRST transformations are given by
\begin{subequations}
\label{NonAbelianBRSTTrafosChiralYM}
\begin{align}
sG^{a}_{\mu}(x) & = D_{\mu}^{ab} c^{b}(x) = \partial_{\mu} c^{a}(x) + g f^{abc} c^{b}(x) G^{c}_{\mu}(x)\,,\\
s\psi_{i}(x) & = s {\psi_{R}}_{i}(x) = - i g {T^{a}_{R}}_{ij} c^{a}(x) {\psi_{R}}_{j}(x)\,,\\
s\overline{\psi}_{i}(x) & = s\overline{\psi_{R}}_{i}(x) = - i g \overline{\psi_{R}}_{j}(x) c^{a}(x) {T^{a}_{R}}_{ji}\,,\\
sc^{a}(x) & = \frac{1}{2} g f^{abc} c^{b}(x) c^{c}(x)\,, \\
s\cbar^{a}(x) & =  B^{a}(x)\,,\\
sB^{a}(x) & = 0\,.
\end{align}
\end{subequations}
In contrast to the Abelian case, the BRST transformations of the gauge boson $G^{a}_{\mu}$ and the Faddeev-Popov ghost $c^{a}$ are non-linear, which means that non-trivial quantum corrections are expected.

Hence, the tree-level action of the considered chiral Yang-Mills theory in $4$ dimensions is given by
\begin{align}
\label{chiralYMin4D}
    S_0^{(4D)} = \intx \, \La_{\chi\text{YM}}
\end{align}
and satisfies the tree-level Slavnov-Taylor identity
\begin{align}
    0 &= \mathcal{S}\big(S_0^{(4D)}\big)\nonumber\\
    &= \intx \Bigg(
        \frac{\delta S_0^{(4D)}}{\delta \rho^a{}^\mu(x)} \frac{\delta S_0^{(4D)}}{\delta G^a_\mu(x)} +
        \frac{\delta S_0^{(4D)}}{\delta \zeta^a(x)} \frac{\delta S_0^{(4D)}}{\delta c^a(x)}\\
        &\hspace{1.45cm} +
        \frac{\delta S_0^{(4D)}}{\delta \bar{R}^i(x)} \frac{\delta S_0^{(4D)}}{\delta {\psi_{R}}_i(x)} +
        \frac{\delta S_0^{(4D)}}{\delta {R}^i(x)} \frac{\delta S_0^{(4D)}}{\delta \overline{\psi_{R}}_i(x)} +
        B^a(x) \frac{\delta S_0^{(4D)}}{\delta \cbar^a(x)} \Bigg), \nonumber
\end{align}
which just manifests the BRST invariance of $S_0^{(4D)}$.

The different group invariants, which will be employed in the
following results below, follow the notations of
  \cite{Belusca-Maito:2020ala}
and are provided by
\begin{align}
    C_{2}(R) \, \mathbbm{1} &= T^{a}_{R} T^{a}_{R}\,,
    &
    S_{2}(R) \, \delta^{ab} &= \text{Tr}\big(T^{a}_{R} T^{b}_{R}\big),
\end{align}
with an irreducible representation $R$ of the gauge group for the right-handed fermions with corresponding Hermitian group generators $T^{a}_{R}$. The adjoint representation of the gauge group is denoted by $G$ and its Casimir index is $C_{2}(G)$.

\subsubsection{Chiral Yang--Mills Theory in~DReg}

To regularize the theory we employ dimensional regularization, treating $\gamma_5$ with the BMHV scheme.
Analogous to the Abelian case above, there are two problems regarding the continuation of the chiral Yang-Mills theory (\ref{chiralYMin4D}) to $D$ dimensions, as already discussed in Sec.\ \ref{sec:DefChiralQEDDReg} for the Abelian case and extensively discussed in \cite{Belusca-Maito:2020ala} for chiral Yang-Mills theories. 

First, there is an ambiguity in extending the fermion-gauge interaction term in Eq.\ (\ref{chiralYMLagrangianInv}), which involves the right-handed chiral current $\overline{\psi_{R}}_{i} \gamma^{\mu} {\psi_{R}}_{j}$, to $D$ dimensions. Again, there are three inequivalent choices for the $D$-dimensional version of this chiral current, cf.\ Eq.\ (\ref{eq:inequivalentchoicesApplications}), which are all equally correct. Analogous to the Abelian case above, we resolve this problem by choosing the most symmetric version cf.\ Eq.\ (\ref{eq:thirdoptionforchiralcurrent}).

Second, the purely fermionic kinetic term $i \overline{\psi_{R}}_{i} \slashed{\partial} {\psi_{R}}_{i}$ projects only the purely $4$-dimensional derivative, leading to a purely $4$-dimensional propagator and thus to unregularized loop diagrams, as explained above in Sec.\ \ref{sec:DefChiralQEDDReg}. Hence, we again introduce a gauge-singlet left-chiral field $\psi_{L}$ with trivial BRST transformations
\begin{align}
    s {\psi_{L}}_{i}(x) &= 0,
    &
    s \overline{\psi_{L}}_{i}(x) &= 0,
\end{align}
which appears solely in the fermionic kinetic term and nowhere else
and which  is thus completely decoupled from the rest of the
theory. Using it we obtain a fully $D$-dimensional covariant kinetic term $i \overline{\psi}_{i} \slashed{\partial} \psi_{i}$.

Finally, we can again separate the $D$-dimensional fermionic Lagrangian into an invariant and an evanescent part, analogous to Eqs.\ (\ref{eq:Lfermionsplit}) to (\ref{eq:sDActingOnFermionLagrangian}). Hence, we may write the $D$-dimensional action as
\begin{align}
    S_{0} &= S_{0,\text{inv}} + S_{0,\text{evan}}\nonumber\\
    &= (S_{GG} + S_{GGG} + S_{GGGG}) + \sum_{i} \left(S^{i}_{\overline{\psi}\psi} + \overline{S^{i}_{\overline{\psi}_{R} G \psi_R}}\right) + S_\text{g-fix}\\
    &\hspace{0.44cm} + (S_{\bar{c} c} + S_{\bar{c} G c}) + (S_{\rho c} + S_{\rho G c} + S_{\zeta c c} + S_{\bar{R} c \psi_R} + S_{R c \overline{\psi_R}})\,, \nonumber
\end{align}
having it separated into an ``invariant'' and an ``evanescent'' part in the first line, cf.\ Eq.\ (\ref{eq:SeparatedDdimAction}) in Sec.\ \ref{sec:DefChiralQEDDReg},
and having used the notation of \cite{Belusca-Maito:2021lnk,Belusca-Maito:2020ala} and of Eq.\ (\ref{eq:S0Def2}) to present the $D$-dimensional action as a sum of its integrated field monomials in the last two lines.

Similar to the Abelian case in Sec.\ \ref{sec:DefChiralQEDDReg}, we quantify the symmetry breaking caused by the BMHV scheme, the non-anticommuting $\gamma_5$ and the evanescent term $S_{0,\text{evan}}$ by acting with the $D$-dimensional BRST operator $s_{D}$ on the $D$-dimensional tree-level action $S_{0}$. Thus, for the BRST breaking we obtain
\begin{align}
    s_{D} S_{0} = s_{D} S_{0,\text{inv}} + s_{D} S_{0,\text{evan}} = 0 + s_{D} \int d^Dx \, i \overline{\psi}_i \widehat{\slashed{\partial}} {\psi}_i \equiv \widehat{\Delta}\,,
\end{align}
which leads to a breaking of the Slavnov-Taylor identity of the form
\begin{align}
    \mathcal{S}_{D}\big(S_{0}\big) &= \widehat{\Delta}\,,
\end{align}
with the non-vanishing integrated breaking
\begin{equation}
\label{eq:BRSTTreeBreakingYM}
\widehat{\Delta} =  -\int d^D x
g\,  {T^a_R}_{ij} \, c^{a} \, \left\{
\overline{\psi}_i \left(\overset{\leftarrow}{\widehat{\slashed{\partial}}} \mathbb{P}_{R} + \overset{\rightarrow}{\widehat{\slashed{\partial}}} \mathbb{P}_{L}\right) \psi_j
\right\}
\equiv \int  d^D x \, \widehat{\Delta}(x).
\end{equation}
As in the Abelian case, this breaking term
will be a crucial tool in practical calculations and will be used as a composite operator insertion in
Feynman diagrams. It generates an interaction vertex whose Feynman rule (with all momenta incoming) is:
\\
\begin{equation}
\begin{tabular}{rl}
\raisebox{-40pt}{\includegraphics[scale=0.6]{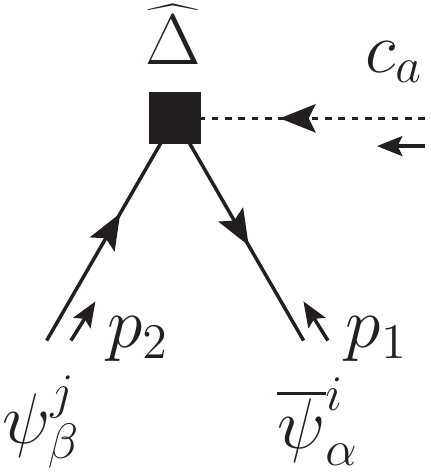}} &
$\begin{aligned}
&= -\frac{g}{2} \, {T^a_R}_{ij} \left((\widehat{\slashed{p_1}} + \widehat{\slashed{p_2}}) + (\widehat{\slashed{p_1}} - \widehat{\slashed{p_2}}) \gamma_5 \right)_{\alpha\beta} \\
&= -g \, {T^a_R}_{ij} \left(\widehat{\slashed{p_1}} \mathbb{P}_{R} + \widehat{\slashed{p_2}} \mathbb{P}_{L} \right)_{\alpha\beta}
\, .
\end{aligned}$
\end{tabular}
\end{equation}
For charge-conjugated fermions an analogous Feynman rule can be derived.

\subsubsection{One-Loop Singular Counterterm and Symmetry-Restoring Counterterm Action in Chiral Yang--Mills~Theory}

In this subsection, we present the results of the one-loop renormalization of the above introduced chiral Yang-Mills theory based on the results of \cite{Belusca-Maito:2020ala}, but also already discussed in \cite{Martin:1999cc}.\footnote{Note the different sign convention w.r.t.\ the covariant derivative $D^{\mu}_{ij}$ in this review compared to \cite{Belusca-Maito:2020ala}. This influences some signs, such as the relative sign in Eq.\ (\ref{eq:bDVariationGGTerm}) and relative sign in the bracket of the last term of Eq.\ (\ref{eq:DeltaDotGammaDivChiralYM}).}

The basic renormalization procedure is the same as in the Abelian theory discussed above. The difference is that there are more interaction terms; in particular, the gauge bosons interact with themselves and with the Faddeev-Popov ghosts. The fact that the ghosts now participate in interactions, and thus may propagate as internal particles in loop diagrams, leads to a non-trivial renormalization of the field monomials including external sources. Besides this, the renormalization procedure is also more demanding than in an Abelian theory, due to the larger number of loop diagrams and the more complicated algebraic structures of the non-Abelian gauge group.

After computing all UV divergent one-loop 1PI Feynman diagrams, which can be found in section $5$ of \cite{Belusca-Maito:2020ala} with detailed individual results, the singular one-loop counterterm action is given by 
\begin{align}
\label{YangMillsOneLoopSingularCountertermAction}
    S_{\text{sct}}^{(1)} 
        &= \frac{\hbar g^2}{16 \pi^2 \epsilon} \bigg\{-\frac{2 S_{2}(R)}{3} \big( \overline{S_{GG}} + \overline{S_{GGG}} + \overline{S_{GGGG}} \big) - \xi C_{2}(R) \big( \overline{S_{\overline{\psi} \psi_{R}}} + \overline{S_{\overline{\psi} G \psi_{R}}} \big)\nonumber \\
    &\hspace{0.5cm} + \frac{13- 3 \xi}{6} C_{2}(G) S_{GG} + \frac{17 - 9\xi}{12} C_{2}(G) S_{GGG} + \frac{2 - 3\xi}{3} C_{2}(G) S_{GGGG}\nonumber \\
    &\hspace{0.5cm} - \frac{3 + \xi}{4} C_{2}(G) \overline{S_{\overline{\psi} G \psi_{R}}} + \frac{3 - \xi}{4} C_{2}(G) \big( S_{\cbar c} + S_{\rho c} \big) \\
    &\hspace{0.5cm} - \frac{\xi C_{2}(G)}{2} \big( S_{\cbar G c} + S_{\rho G c} + S_{\zeta c c} + S_{\bar{R} c \psi_{R}} + S_{R c \overline{\psi_{R}}} \big) \bigg\}\nonumber \\
    &\hspace{0.5cm} - \frac{\hbar g^2}{16 \pi^2 \epsilon} \frac{S_{2}(R)}{3} \intx \, \frac{1}{2} \bar{G}^{a, \mu} \widehat{\partial}^2 \bar{G}^{a}_{\mu}\nonumber,
\end{align}
such that it cancels all UV-divergences. The structure has
similarities with the Abelian counterpart,
Eq.\ (\ref{eq:SingularCT1Loop}). Again, most terms can be obtained by
a renormalization transformation  of the kind (\ref{RenTrafo}), and
only the right-handed fermions renormalize. But again also
non-symmetric singular counterterms appear.

Comparing Eqs.\ (\ref{YangMillsOneLoopSingularCountertermAction}) and
(\ref{eq:SingularCT1Loop}) in detail, we can see many additional contributions. Only the $\overline{S_{GG}}$, $\overline{S_{\overline{\psi} \psi_{R}}}$ and $\overline{S_{\overline{\psi} G \psi_{R}}}$ terms in the first line of the RHS of (\ref{YangMillsOneLoopSingularCountertermAction}) as well as the explicit evanescent operator in last line of (\ref{YangMillsOneLoopSingularCountertermAction}) have Abelian counterparts. All other terms in (\ref{YangMillsOneLoopSingularCountertermAction}) do not appear in the Abelian theory, and are thus new effects of the non-Abelian Yang-Mills theory due to additional interaction terms, as mentioned above. In particular, we can see new contributions to the field monomials including the Faddeev-Popov ghosts and the external sources in the last term of the third line and the penultimate line of (\ref{YangMillsOneLoopSingularCountertermAction}), as announced at the beginning of this subsection.

Similar to the Abelian result (\ref{eq:SingularCT1Loop}), we have just
one explicit evanescent operator in the last line of
(\ref{YangMillsOneLoopSingularCountertermAction}) in the considered
Yang-Mills theory, generating the Feynman rule $-i \widehat{p}^{2}
\overline{g}_{\mu\nu} \delta^{ab}$. This is specific to our choice for
the fermion-gauge interaction term, corresponding to the most
symmetric version of Eq.\ (\ref{eq:thirdoptionforchiralcurrent}). We would have obtained many more evanescent operators, if we used another $D$-dimensional choice instead.

Following the algebraic renormalization procedure described in section \ref{sec:algren}, as well as in section $6$ of \cite{Belusca-Maito:2020ala} specifically for the considered case, we need to check that
\begin{align}
    0 = \mathop{\text{LIM}}_{D \, \to \, 4} \, \bigg( \big[\widehat{\Delta} \cdot \Gamma^{(1)} \big]_{\text{div}}^{(1)} + b_{D} S_{\text{sct}}^{(1)} + \big[\widehat{\Delta} \cdot \Gamma^{(1)} \big]_{\text{fin}}^{(1)} + b_{D} S_{\text{fct,restore}}^{(1)} \bigg).
\label{eq:SymmetryRestorationRequirementYMOneLoop}
\end{align}
In other words, we need to check that the $b_{D}$-variation of the singular counterterms (\ref{YangMillsOneLoopSingularCountertermAction}) cancels the divergent part of the symmetry breaking $[\widehat{\Delta} \cdot \Gamma^{(1)}]_{\text{div}}^{(1)}$ and we need to determine finite symmetry-restoring counterterms $S_{\text{fct,restore}}^{(1)}$ whose $b_{D}$-variation cancel the finite part of the symmetry breaking $[\widehat{\Delta} \cdot \Gamma^{(1)}]_{\text{fin}}^{(1)}$.

The $b_{D}$-variation of the singular counterterms (\ref{YangMillsOneLoopSingularCountertermAction}), calculated in \cite{Belusca-Maito:2020ala}, is provided by 
\begin{align}
    b_{D} S_{\text{sct}}^{(1)} = \frac{-\hbar}{16 \pi^2 \epsilon} \bigg\{ g^2 \frac{\xi C_{2}(G)}{2} \widehat{\Delta} + g^2 \frac{S_{2}(R)}{3} b_{D} \int d^{D}x \, \frac{1}{2} \bar{G}^{a, \mu} \widehat{\partial}^{2} \bar{G}^{a}_{\mu} \bigg\},
\label{eq:bDSsct1}
\end{align}
where, in the last term, $b_{D}$ acts like the BRST transformation, leading to
\begin{align}
    b_{D} \int d^{D}x \, \frac{1}{2} \bar{G}^{a, \mu} \widehat{\partial}^{2} \bar{G}^{a}_{\mu} &= \int d^{D}x \, \big( s_{D} \bar{G}^{a, \mu} \big) \widehat{\partial}^{2} \bar{G}^{a}_{\mu}\nonumber\\
    &= \int d^{D}x \, \Big( \overline{\partial}^{\mu} c^{a} - g f^{abc} \bar{G}^{b, \mu} c^{c} \Big) \widehat{\partial}^{2} \bar{G}^{a}_{\mu}\,.
\label{eq:bDVariationGGTerm}
\end{align}
Indeed, (\ref{eq:bDSsct1}) is a pure $1/\epsilon$ singular term and perfectly cancels the non-vanishing contribution
\begin{align}
    \big[\widehat{\Delta} \cdot \Gamma \big]_{\text{div}}^{(1)} = \frac{1}{16 \pi^2 \epsilon} \bigg\{ g^2 \frac{\xi C_{2}(G)}{2} \widehat{\Delta} + g^2 \frac{S_{2}(R)}{3} \int d^{D}x \, \Big( \overline{\partial}^{\mu} c^{a} - g f^{abc} \bar{G}^{b, \mu} c^{c} \Big) \widehat{\partial}^{2} \bar{G}^{a}_{\mu}  \bigg\},
\label{eq:DeltaDotGammaDivChiralYM}
\end{align}
as explicitly shown in \cite{Belusca-Maito:2020ala}.

Now, the finite symmetry-restoring counterterms $S_{\text{fct,restore}}^{(1)}$ need to be determined following (\ref{eq:SymmetryRestorationRequirementYMOneLoop}) in order to cancel the remaining finite part of the symmetry breaking, which was explicitly performed in section $6$ of \cite{Belusca-Maito:2020ala} with the result
\begin{align}
    S_{\text{fct, restore}}^{(1)} &= \frac{\hbar}{16\pi^2} \bigg\{ g^2 \frac{S_{2}(R)}{6} \bigg( 5 S_{GG} - \int d^{4}x \, G^{a, \mu} \partial^{2} G^{a}_{\mu} \bigg)\nonumber\\
    &\hspace{0.44cm} + g^2 \frac{(T_{R})^{abcd}}{3} \int d^{4}x \, \frac{g^2}{4} G^{a}_{\mu} G^{b, \mu} G^{c}_{\nu} G^{d, \nu} + g^2 \bigg(1 + \frac{\xi - 1}{6}\bigg) C_{2}(R) S_{\overline{\psi}\psi}\nonumber \\
    &\hspace{0.44cm} + g^2 \frac{S_{2}(R)}{6} S_{GGG} - g^2 \frac{\xi C_{2}(G)}{4} \big( S_{\bar{R} c \psi_{R}} + S_{Rc\overline{\psi_{R}}} \big) \bigg\},
\label{eq:YMSfctrestore}
\end{align}
where $(T_{R})^{a_{1} \cdots a_{n}} \equiv \text{Tr}[T^{a_{1}}_{R} \cdots T^{a_{n}}_{R}]$.
Comparing (\ref{eq:YMSfctrestore}) with the Abelian result (\ref{eq:Sfct1L}), we can again see that only the first two lines of (\ref{eq:YMSfctrestore}) have Abelian counterparts, whereas the terms in the last line of (\ref{eq:YMSfctrestore}) do not appear in an Abelian theory. The new terms in the last line of (\ref{eq:YMSfctrestore}) are due to triple gauge boson contributions and contributions including external sources. The latter implies that again Green functions with external sources have to been evaluated, this time with a $\widehat{\Delta}$-vertex insertion, which stands in contrast to the Abelian case.

These finite counterterms (\ref{eq:YMSfctrestore}) are necessary and sufficient to restore the BRST symmetry at the one-loop level in the BMHV scheme, if the (non-spurious) anomalies cancel, which are given by \cite{Belusca-Maito:2020ala} 
\begin{align}
    &- \frac{g^2}{16 \pi^2} \bigg(-\frac{S_{2}(R)}{3} d_{R}^{abc} \int d^{4}x \, g \epsilon^{\mu\nu\rho\sigma}c^{a} \big(\partial_{\rho} G^{b}_{\mu}\big) \big(\partial_{\sigma} G^{c}_{\nu}\big)\nonumber\\
    &\hspace{2.4cm} + \frac{\mathcal{D}_{R}^{abcd}}{3\times3!} \int d^{4}x \, g^2 c^{a} \epsilon^{\mu\nu\rho\sigma} \partial_{\sigma} \Big( G^{b}_{\mu} G^{c}_{\nu} G^{d}_{\rho} \Big) \bigg),
\end{align}
with fully symmetric $d_{R}^{abc} \equiv \text{Tr}[T_{R}^{a}
  \{T_{R}^{b}, T_{R}^{c}\}]$ and fully antisymmetric
$\mathcal{D}_{R}^{abcd} \equiv (-i) 3! \text{Tr}[T_{R}^{a} T_{R}^{[b}
    T_{R}^{c} T_{R}^{d]}]$ for the $R$-representation.
This result of course agrees with the general result
(\ref{ABJexplicit}) obtained by the analysis of algebraic
renormalization, and it provides an explicit result for the
coefficient $L$ appearing there.
To ensure the renormalizability of the theory the fermionic content and
their associated group representations have to be chosen such that
these anomalies cancel, i.e.\ such that the expression
(\ref{LFanomaly}) vanishes, which equivalently means that
$d_{R}^{abc}$ vanishes. This then also implies the vanishing of
$\mathcal{D}_{R}^{abcd}$, see Eq.\ (\ref{DabcdRelation}). It becomes apparent that also the possible anomalies are more complex than in the Abelian model. 

These finite counterterms (\ref{eq:YMSfctrestore}), purely $4$-dimensional and non-evanescent, are not gauge-invariant. They modify all self-energies, as well as some specific interactions: the gauge-boson self-interactions and the interactions between gauge-bosons and fermions.

Concluding, we see that the resulting counterterm action, not only for the Abelian case at the one- and two-loop level, but also for non-Abelian Yang-Mills theories, may be written in a relatively compact way. Thus, treating $\gamma_5$ rigorously in the BMHV scheme does not lead to extraordinarily lengthy or complicated results, but in fact to counterterms which can easily be implemented in computer algebra systems.

\newpage
\section{Acknowledgments} 

A.I., H.B.\ and M.M.  acknowledge  financial support from the Croatian
Science Foundation (HRZZ) under the project ``PRECIOUS'' (``Precise
Computations of Physical Observables in Supersymmetric Models'')
number \verb|HRZZ-IP-2016-06-7460|. P.K., D.S.\ and M.W.\ acknowledge financial support by the German Science 
Foundation DFG, grant STO 876/8-1.

\bibliography{bibliography}{}
\bibliographystyle{ieeetr}
\end{document}